\documentclass[longauth]{aa}   %%% Institutes at end of document

\usepackage{hyperref}

\usepackage{graphicx}
\usepackage{txfonts}
\usepackage{upgreek}
\usepackage{natbib}

\bibpunct{(}{)}{;}{a}{}{,} % to follow the A&A style

\begin{document}

\title{ATOMIUM: Probing the inner wind of evolved O-rich stars with new, highly excited  H$_2$O and OH lines}

\titlerunning{H$_2$O and OH inner gas layers probed with ALMA}
\authorrunning{A. Baudry et al.}

%\subtitle{}

\author{A. Baudry \inst{1},  K.T. Wong \inst{2,9}, S. Etoka \inst{3}, A.M.S. Richards  \inst{3}, H.S.P.  M\"uller \inst{4}, F. Herpin \inst{1}, T. Danilovich \inst{5,15}, M.D. Gray \inst{3,6}, S. Wallstr\"om \inst{5}, D. Gobrecht \inst{5,11}, T. Khouri \inst{7}, L. Decin \inst{5}, C.A. Gottlieb  \inst{8},  K.M. Menten \inst{10}, W. Homan \inst{5},  T.J. Millar \inst{12}, M. Montarg\`es \inst{13}, B. Pimpanuwat \inst{3},  J.M.C. Plane \inst{14}, P. Kervella \inst{13}}

\institute{
Laboratoire d'Astrophysique de Bordeaux, Univ. Bordeaux, CNRS, B18N, all\'ee Geoffroy Saint-Hilaire, 33615 Pessac, France.
        \email{alain.baudry@u-bordeaux.fr}
                        \and 
                Institut de Radioastronomie Millim\'etrique, 300 rue de la Piscine, 38406 Saint-Martin-d'H{\`e}res, France
                         \and
              The University of Manchester, Jodrell Bank Centre for Astrophysics, Manchester  M13 9PL, United Kingdom
              \and
              Universit\"at zu K\"oln, I. Physikalisches Institut, 50937 K\"oln, Germany
               \and 
                Institute of Astronomy, KU Leuven, Celestijnenlaan 200D, 3001 Leuven, Belgium
              \and
              National Astronomy Research Institute of Thailand, 260 Moo 4, Chiangmai 50180, Thailand 
              \and
              Chalmers University of Technology, Onsala Space Observatory, 43992 Onsala, Sweden
              \and
              Harvard-Smithsonian Center for Astrophysics, Cambride MA 02138, USA      
              \and
              Department of Physics and Astronomy, Uppsala University, Box 516, 75120 Uppsala, Sweden
              \and
              Max-Planck-Institut f\"ur Radioastronomie, 53121 Bonn, Germany
              \and
              Department of Chemistry and Molecular Biology, University of Gothenburg, 40530 G\"oteborg, Sweden 
              \and
              Astrophysics Research Centre, School of Mathematics and Physics, Queen's University Belfast, 
              Univ. Road, Belfast BT7 1NN, UK
              \and
              LESIA, Observatoire de Paris, Univ. Paris 5, CNRS, 5 place Jules Janssen, 92195 Meudon, France 
              \and
              University of Leeds, School of Chemistry, Leeds LS2 9JT, UK
              \and
              School of Physics and Astronomy, Monash University, Wellington Road, Clayton 3800, Victoria, Australia
              }

\date{\today}

\abstract 
  % context heading (optional)
  {Water (H$_2$O) and the hydroxyl radical (OH) are major constituents of the envelope of O-rich late-type stars. Transitions involving energy levels that are rotationally or vibrationally highly excited (energies $\gtrsim$4000\,K) have been observed in both H$_2$O and OH. These and more recently discovered transitions  can now be observed at a high sensitivity and angular resolution in the inner wind close to the stellar photosphere with the Atacama Large Millimeter/submillimeter Array (ALMA).}
   %leave it empty if necessary
{Our goals are: (1) to identify and map the emission and absorption of H$_2$O in several vibrational states, and of OH in $\Lambda$-doubling transitions with similar excitation energies; and (2) to determine the physical conditions and kinematics in gas layers close to the extended atmosphere  in a sample of  asymptotic giant branch stars (AGBs) and red supergiants (RSGs). }
  % methods heading (mandatory)
   {Spectra and maps of H$_2$O and OH lines observed in a 27~GHz aggregated bandwidth   and with an angular resolution of $\sim$0\farcs02$-$1\farcs0 were obtained at two epochs with 
the main ALMA array. Additional observations with the Atacama Compact Array (ACA) were used to check for time variability of water transitions. Radiative transfer models of H$_2$O were revisited to characterize  masing conditions. Up-to-date chemical models were used for comparison  with the observed OH/H$_2$O abundance ratio.}
  % results heading (mandatory)
   {Ten rotational transitions of H$_2$O with 
excitation energies $\sim$4000$-$9000~K were observed in vibrational states up to ($\varv_1$,$\varv_2$,$\varv_3$) = (0,1,1).  
All but one are new detections in space, and from these we have derived accurate rest frequencies. Hyperfine split  $\Lambda$-doubling transitions in $\varv = 0, J = 27/2$ and 29/2 levels of the $^2\Pi_{3/2}$ state, as well as  
$J = 33/2$ and 35/2 of the $^2\Pi_{1/2}$ state of OH with excitation energies of $\sim$4780$-$8900~K were also observed. Four of these transitions are new detections in space. Combining our measurements with earlier observations of OH, the $\varv = 0$ and $\varv = 1$ $\Lambda$-doubling frequencies 
 have been improved.  Our H$_2$O maps  show compact emission 
toward the central star and extensions up to  twelve  stellar radii or more. The 268.149~GHz emission line of water in the $\varv_2 = 2$ state is time variable, tends to be  masing with dominant radiative pumping, and is widely excited in AGBs and RSGs. 
The widespread but weaker 262.898~GHz water line in the $\varv_2 = 1$ state also shows signs of maser emission. 
The OH emission  is weak and quasithermally excited.  
Emission and absorption features of H$_2$O and OH reveal an infall of matter  and complex kinematics  influenced by binarity.  From the OH and H$_2$O column densities derived with nonmasing transitions in a few sources, we obtain OH/H$_2$O abundance ratios of $\sim$(0.7$-$2.8)$\times10^{-2}$.
} 
 % conclusions heading (optional)
 %leave it empty if necessary
 {}
 
 \keywords {stars: AGB and post-AGB -- supergiants -- stars: circumstellar matter -- line: identification -- molecular data -- masers--techniques: interferometric}
   \maketitle

%
%________________________________________________________________

\section{Introduction}

Water and the hydroxyl radical are formed from two of the three most abundant elements in the Universe. Many H$_2$O and OH lines have now been observed in the radio, infrared, or visible domains in a broad range of astronomical objects ranging from the  planetary and cometary atmospheres of our Solar System to the envelopes of evolved stars or the star-forming regions of our Galaxy. We note that H$_2$O and OH are also present in the disks or nuclei of nearby and distant galaxies. 
The first radio identification of OH was reported at an 18-cm wavelength by \citet[][]{weinreb1963} in absorption toward the supernova remnant Cassiopeia~A, and the 22.235 GHz (1.35 cm) radio signature of H$_2$O was first reported by  \citet[][]{cheung1969} in star-forming regions  and  by \citet[][]{knowles1969} in the red supergiant (RSG) VY CMa.  These two centimeter-wave transitions often give rise to a remarkably bright cosmic maser emission which has been observed throughout the Universe in many different regions, including those near the massive black holes of active galactic nuclei \citep[e.g., 22.235~GHz image of NGC~3079, ][]{kondratko2005}. 
In addition to the strong H$_2$O and OH 1.35- and 18-cm wave radiation, various  rotational transitions of water  in the ground and vibrationally excited states  have been identified in several Galactic late-type stars from ground-based observatories  \citep[e.g.,][]{menten1989,menten1990a,melnick1993,menten1995,gonzalezalfonso1998} or from space  \citep[e.g.,][]{justtanont2012}. Recently, \citet[][]{khouri2019} used their own and archival Atacama Large Millimeter/submillimeter Array (ALMA) data to identify highly excited OH transitions in a few O-rich evolved stars. 
In star-forming regions of the Galaxy, various rotationally excited  transitions of H$_2$O and a few low-lying energy transitions of OH have  also been reported  \citep[e.g.,][]{menten1990a,menten1990b,baudry1997,harvey2005,hirota2012,hirota2014}. 
Due to the absorption of water vapor from the Earth's atmosphere, astronomical observations of H$_2$O  are difficult. 
Nevertheless, high-lying energy transitions of water are accessible from Earth's dry sites or from airborne telescopes (see e.g., references above, review by \citet{humphreys2007} or Tables~1 and 2 in \citet{gray2016} predicting that several 
H$_2$O lines are observable by (sub)millimeter telescopes).  
Many lines of water vapor remain, however, inaccessible inside the terrestrial atmosphere (e.g., the $1_{1,0}$--$1_{0,1}$ transition at 556.936~GHz) and can only be observed from space observatories, such as  the Infrared Space Observatory \citep[ISO;][]{neufeld1996, barlow1996}, the Submillimeter Wave Astronomy Satellite \citep[SWAS;][]{harwit2002}, the Odin satellite \citep{justtanont2005}, and the Herschel Space Observatory \citep{decin2010}. The role played by these  space missions in our understanding of the interstellar water chemistry is described in \citet{vandishoeck2013}, for example.

In parallel with the observational work, chemical models  have been developed to 
explain the ubiquitous presence of H$_2$O and OH. In the general interstellar medium, the review work of \citet{vandishoeck2021} demonstrates that  these two molecular species are essential to explain the O budget of the molecular products observed in star-forming regions. Furthermore, H$_2$O and OH are also known to play a central role in  the production of many other species observed in the envelopes of evolved O-rich stars where they are the principal oxidizing agents  \citep[e.g.,][]{cherchneff2006}, and they are essential to the dust nucleation processes leading, ultimately, to the formation of circumstellar dust grains \citep[e.g.,][]{gobrecht2022}. 

For the purposes of this study,  we primarily used our discovery of highly excited H$_2$O and OH radio lines (above $\sim$4000~K) to probe the photospheric environment and the dust formation zone of O-rich late-type stars. 
As the nuclear burning reactions diminish in the stellar core, the O-rich stars, depending on their initial masses, evolve along the asymptotic giant branch (AGB) or the red supergiant (RSG) branch. The late evolutionary stages of these stars stem from complex mechanisms involving convection, stellar pulsation-driven wind, and shocks that can levitate stellar material above the photosphere. 
Many millimeter and submillimeter radio observations have shown that shocks and  stellar winds offer the favorable conditions that stimulate an active gas chemistry \citep [e.g.,][]{justtanont2012, alcolea2013, velillaprieto2017, gottlieb2022}, including  OH and H$_2$O that are formed near the photosphere. The Herschel-HIFI observations also provided the abundance of cool water in M-type AGB stars \citep{khouri2014, maercker2016}. Clearly, the nonequilibrium conditions observed beyond the photosphere facilitate the formation of dust-forming clusters    
in O-rich stars \citep[][]{gobrecht2016, boulangier2019, gobrecht2022} which later form  the circumstellar dust grains.

The size of the dust formation zone around evolved stars was first estimated by infrared interferometry \citep [e.g.,][]{danchi1994}. It has later been refined with radio interferometers in the continuum and in the SiO lines which give the size of the molecular shell centered on the photosphere \citep [e.g.,][]{cotton2004}. The dust formation zone encompasses the warm molecular envelope invoked by \citet{tsuji1997} to explain the observations made with the ISO grating spectrometer and the molecular layers observed in near-infrared molecular bands and modeled by \citet{perrin2004}. This zone is within  the radio photosphere first described in \citet{reid1997}.  Beyond the dust formation region, the radiation pressure on the dust particles accelerates the gas flow to outer circumstellar layers \citep [e.g.,][]{hofner2018}, extending to hundreds of stellar radii (R$_{\star}$).

The present work focuses on the identification and  interpretation of new H$_2$O and OH radio lines excited at energy levels in the range $\sim$3900 to 9000 K ($\sim$2700 to 6300 cm$^{-1}$).  These lines, observed in O-rich late-type stars, allowed us to probe the hot and dense gas above the photosphere and in the dust formation zone of the inner circumstellar wind.  
We adopted the inner wind terminology to include regions covering from the stellar surface to a few R$_{\star}$ and up to $\sim$30  R$_{\star}$  within which the dust was formed, the wind was launched, and where active gas and dust interactions were observed. Our data were acquired during the ALMA Cycle 6 Large Program 2018.1.00659.L \citep{decin2020-science, gottlieb2022}. The main objectives of this program, named  {\sc atomium}, include the study of the molecular paths leading to the formation of the dust precursors as well as the study of the inner (\la 30 R$_{\star}$) and intermediate ($\sim$30 to hundreds of R$_{\star}$) stellar wind morphology.

 In Sect. \ref{sec:observations} we present the observed sources (Sect. \ref{sec:sample}), main ALMA array observations (Sect. \ref{r}), and supplementary ALMA Compact Array (ACA) observations of H$_2$O and other molecules (Sect. \ref {aca_observations}). Table~\ref{primarysource_list} gives the {\sc atomium} source sample and the radio detection (or not) of highly excited  H$_2$O and OH transitions in the stellar atmosphere of the {\sc atomium} stars. Identification of  H$_2$O and OH  transitions as well as some spectroscopy background for these two species are given in Sect.~\ref {sec:spectro}. Several stars exhibit very rich H$_2$O and OH spectra (Tables~\ref{fluxdensity_source_list}  and \ref{OH_size_peak_intensity}) and a few of them display all or most of the H$_2$O and OH lines reported in this work (e.g., R~Hya). The  source spectra and maps, and a first analysis of our data are presented in Sects.~\ref{sec:line_properties} and  \ref {sec:H2O_analysis} for H$_2$O, and in Sects.~\ref {sec:detect_OH} and \ref {sec:OH_analysis} for OH. The widespread 268.149 and 262.898~GHz H$_2$O emissions and H$_2$O maser modeling are discussed in Sects. \ref{sec:268_H$_2$O_maser} and  \ref{sec:maser}, respectively. Furthermore, H$_2$O and OH chemical considerations and the OH/H$_2$O abundance ratio in the inner wind of AGBs are discussed in Sect. \ref {sec:chemistry}. Concluding remarks are given in Sect. \ref {sec:conclusion} specifying the (sub)sections where the main results are acquired. The Appendices provide the OH  $\Lambda$-doubling transitions and further H$_{2}$O and OH spectra and maps.

\section{Source sample and observations}
\label{sec:observations}

\subsection{Source sample }
\label{sec:sample} 
The {\sc atomium} source sample includes seventeen O-rich evolved AGB or RSG stars covering a relatively broad range of properties in terms of variability type and mass-loss rate (Table \ref{primarysource_list}). Sources are ordered by increasing mass-loss rate noting, however, that this rate is uncertain especially for the distant RSGs AH~Sco, KW~Sgr and VX~Sgr. The source coordinates are obtained from the emission peak of 2D-Gaussian fits to the stellar continuum observed around 241.8~GHz with the ALMA extended configuration, see Table~E.2  in \citet [][]{gottlieb2022}. The astrometric accuracy is determined by the accuracy of phase referencing and of fitting to the stellar peak; it is in the range 5 to 10~mas for both factors.

Most of the adopted distances to the Mira and SR variables  in Table \ref{primarysource_list} are extracted from the Gaia DR3 catalog \citep{gaia2022} which, however,  must be used with care if the parallax uncertainties exceed $\sim$20\% \citep{andriantsaralaza2022}. This is not the case here with the exception of GY~Aql whose distance has been revised to 410~pc by \citet{andriantsaralaza2022}. 
For U~Her and IRC$-$10529 we have also adopted the best distance estimates from \citet{andriantsaralaza2022} and, for 
the three RSGs and IRC$+$10011, the distances are taken from VLBI radio measurements or other works as mentioned in  \citet{gottlieb2022}. All other source distances in  our Table \ref{primarysource_list} are from Gaia DR3\footnote{DR3 measurements are robust, although photospheric structures could affect the photocenter position. Data processing of the next data release will be improved with more astrometric measurements.} \citep{gaia2022}.  
The mass-loss rates in Table \ref{primarysource_list} are taken from \citet{gottlieb2022}; they are updated for GY~Aql and IRC$-$10529 because of their revised distances. The Local Standard of Rest (LSR) velocity used at the time of the {\sc atomium} observations of each star is given in the eighth column of Table \ref{primarysource_list}; the ninth column gives the LSR velocity based on  a sample of various  lines according to  \citet{gottlieb2022}. Table \ref{primarysource_list}  also indicates whether at least one highly excited H$_2$O and OH transition  is observed in the stellar atmosphere of each source, while the second last and last columns of Tables~\ref{H2O-line-list} and  \ref{OHline_list}  give the number of detected sources for each transition. 

\subsection{Main array observations, data reduction, and products}
\label{r}

We observed in three array configurations,
extended, mid and compact configurations providing angular resolutions
of approximately 0\farcs02$-$0\farcs05, 0\farcs15$-$0\farcs3 and 1\farcs0, respectively, 
henceforth high, mid and low resolutions.  In the present paper, we identify and analyze the H$_2$O and OH lines 
falling in the observed bandwidth within the total frequency range covering from 213.83 to 269.71~GHz.  
We observed 16 spectral windows, or cubes, within this range using four tunings (four spectral windows per tuning), giving an actual bandwidth of 26.8~GHz for the extended and mid configurations and 12.9~GHz for the compact configuration. 
The central frequency and velocity width of the 16 spectral windows are given in Table~2 of \citet{gottlieb2022}. The exact spectral coverage of each window depends on the adjustment to the adopted LSR velocity on the dates of the observations.  The frequency tunings for the three array configurations are also specified in \citet{gottlieb2022}. 
The highest spatial resolution allows us to resolve the inner few, to few tens of stellar radii for sources
in our sample. The extended, mid and (where available) compact configuration data were also combined to provide higher sensitivity to the overlapping angular scales, and could be weighted to provide a range of resolutions.

The maximum recoverable angular sizes are $\sim$0\farcs4$-$0\farcs6, $\sim$1\farcs8$-$3\farcs0 and $\sim$10\farcs0 
 for the high-, mid- and low-resolutions, respectively. Hence, emission which is smooth on
larger scales would not be detected. The H$_2$O and OH emission or
absorption lines studied here are probably not affected, being much 
more compact than $\sim$3 arcsec,
as they are excited at very high energies
and all maps are dominated by compact structures within the inner
layers of the stellar envelope. This is verified by comparing the total flux density detected in different configurations.

Our data were acquired between 2018 and 2021. The extended
configuration observations were all performed in June and/or July 2019 under good atmospheric conditions 
with low precipitable water vapor. The exact dates of
observations for each line and each source in our sample can be
retrieved from Table E.1 in \citet{gottlieb2022}. All observations
were calibrated,  imaged and continuum-subtracted 
in CASA\footnote{https://casa.nrao.edu/} as described in Sect. 3.2 of \citet{gottlieb2022}. All the data for a specific star and configuration were combined, and aligned on the stellar peak of the first epoch present. When combining configurations, the most accurate measurements, at the highest resolution, were used. After subtracting continuum (mostly stellar) emission, spectral image cubes were made, adjusted to constant frequency in the LSR frame.  All imaging is in total intensity (both observed polarizations combined).  Standard cubes were made for each tuning with the angular resolution ranges described above, the exact resolutions depending on observing frequency, target elevation and exact array configuration.  In most cases, these spectral cubes were used for analysis but, where appropriate, we made additional cubes around specific lines (see, e.g., Sect.~\ref{sec:atomium_OH}),
optimizing the trade-off between sensitivity and synthesized beam size. The continuum and line clean beams are given in Tables E.2 and E.3 of \citet{gottlieb2022}.

Our channel maps are measured in flux density per clean beam which is a surface brightness over an area converted to Gaussian beam units.  For simplicity and consistency with many previous publications, we label the map intensity scale  
as "flux density" in mJy per beam.
The typical r.m.s.  noise outside the line-emission channels is in the 
range 0.5$-$1 and $\sim$2~mJy/beam for the high and medium resolutions, respectively. 
The conversion of mJy per beam to brightness temperature in degrees Kelvin is explained in Sect. \ref{sec:H2O_conditions}.  
The spectral channel separation  of 976.6~kHz in the ALMA correlator gives a velocity resolution  $\sim$1.1 to $\sim$1.3~km\,s$^{-1}$ depending on the observing frequency.  The flux density scale errors are typically around 10\% per
array configuration. However, total uncertainties may exceed 10\%
due to various amplitude and phase noise effects generated during the
data reduction, especially when combining observations made at
different times. We adopt here a conservative 15\% flux density scale
uncertainty.

We extracted  H$_2$O  and OH spectra from our image cubes for various aperture sizes
(larger than the synthesised beam and smaller than the maximum recoverable scale), using circular apertures of diameters typically 0\farcs08, 0\farcs4 and 4\farcs0 
 for the high, mid and low resolution cubes. Comparisons between these (and with the ACA data, Sec. \ref{aca_observations}) showed that in most cases no additional flux was detected in larger apertures except where variability is likely (Sec. \ref{sec:variability_268}), confirming that we are not resolving out significant emission from these lines. In  some cases, the flux density may appear higher at the highest resolution, due to a combination of the lower map noise in these data and possibly less effective "cleaning" of weak emission much smaller than the lower resolutions.

The H$_{2}$O and OH spectra and maps referred to in this paper are shown in Sects. \ref{sec:line_properties} and \ref {sec:detect_OH} and in Appendices \ref{sec:h2o_channel_maps}, \ref{sec:254.23_vxsgr_ahsco}, \ref{sec:oh_channel_maps} and \ref{sec:append_mira_ohspec}. For both species the extracted spectral lines are non-Gaussian but the brighter channels within a line profile are well identifiable and used in this work (e.g., Table \ref{fluxdensity_source_list}) with an uncertainty of a few mJy for the high resolution data. The velocity extent of the identified H$_{2}$O lines is determined from the  blue and red line wing velocities at the 2.5$\sigma$ level of each detected line as in the {\sc atomium} molecular line inventory (Wallstr\"om et al., in preparation). The velocity uncertainty is on the order of one channel, $\sim$1.1$-$1.3~km\,s$^{-1}$, for each line wing determination. 

The angular extent of the emitting or absorbing H$_{2}$O and OH regions is determined without beam de-convolution from our channel maps or from the velocity-integrated intensity maps (zeroth moment or mom~0 maps) using the velocity ranges identified in the spectra or  channel maps. For simplicity again, the moment~0 maps intensity scale is labeled as integrated intensity in Jy/beam~km/s.
The angular extent in our clean images is often irregular and cannot be modeled with simple Gaussian or uniform disk  profiles. However, typical or maximum H$_{2}$O and OH extents can be estimated from the maximum and minimum dimensions within the 3$\sigma$ contour of our channel maps or mom 0 maps. For the most compact H$_{2}$O emission sources, we have also used in the AIPS (Astronomical Image Processing System) package\footnote{http//www.aips.nrao.edu}  a specific task  to fit Gaussian models by least squares to our images (see Sect. \ref{sec:small_scaleh2o}). In OH, despite often irregular emission or absorption contours,  we have used a 3$\sigma$ contour mask in CARTA\footnote{https://cartavis.org/}  to fit 2D Gaussians to the observed regions (see Sects. \ref{sec:OHchannmaps} and \ref{sec:OHabsorption}).

\subsection{ACA observations}
\label{aca_observations}

To follow up on the widespread detection of the 268.149~GHz H$_2$O line in the {\sc atomium} sample, we performed standalone observations with the ACA toward a number of {\sc atomium} sources. The main goals of the ACA observations are to cover H$_2$O lines at 268.149, 254.040, and 254.053~GHz (see Table \ref {H2O-line-list}) at a higher spectral resolution and to measure the H$_2$O line flux densities at an additional epoch. Three high-spectral resolution windows are placed at 268.15 (H$_2$O), 254.04 (H$_2$O), and 255.48~GHz ($^{29}$SiO $\varv=1$) at the spectral resolution of 61\,kHz (0.07\,km\,s$^{-1}$)   with a bandwidth of 0.25~GHz (at 268.15~GHz) or 0.125~GHz (at 254.04 and 255.48~GHz).
In addition, two wide band windows of 2~GHz each 
 are centered at 252.0 and 266.5 GHz at the resolution of 976~kHz (1.1--1.2~\,km\,s$^{-1}$). Observations were carried out under the ALMA project 2019.2.00234.S in September 2021 during the Return to Operations phase of Cycle 7 toward three stars: R~Aql, GY~Aql, and VY~CMa. In this paper, we focus only on the results obtained for R~Aql and GY~Aql. All ACA data were reduced with the Cycle 8 ALMA pipeline (version 2021.2.0.128) in CASA 6.2.1-7. The products are essentially the same as those delivered to the ALMA Archive after the QA2 process, except that we have manually identified the continuum spectral ranges in our post-delivery pipeline reduction. Furthermore, we have  carried out one round of phase-only self-calibration on the continuum of R~Aql and GY~Aql using the target scan length as the solution interval. The achieved angular resolutions at the 268.149~GHz H$_2$O line are roughly $7{\farcs}4\times4{\farcs}3$ and $6{\farcs}9\times4{\farcs}4$ for R~Aql and GY~Aql, respectively.

\begin{table*}
\caption{Main properties of the {\sc atomium} stellar sample and H$_{2}$O, OH detection.}       
\label{primarysource_list}      
\centering                          
\begin{tabular}{llllcccccc} 
     
 \hline\hline      
Star & Variability$^{a}$ & RA(radio)$^{b}$ & Dec(radio)$^{b}$ & Distance$^{c}$ & Mass-loss rate$^{d}$ & 2 R$_{\star}$$^{e}$  & $\varv$$^{\rm obs}_{\rm LSR }$$^{f}$ & $\varv$$^{\rm new}_{\rm LSR }$$^{g}$ & H$_{2}$O, OH$^{h}$
\\   
                     & Spect. type & ICRS (h m s) & ICRS ($^{\circ}$ $^{\prime}$ $^{\prime\prime}$)  &  (pc) &   (M$_{\sun}$yr$^{-1}$)   & (mas) & (km\,s$^{-1}$) & (km\,s$^{-1}$) & (yes/no)   \\
                      \hline
                     
 S Pav & SRa & 19 55 14.0055 & $-$59 11 45.194  & 174 & 8.0 x 10$^{-8}$  & (11.6) &  $-$20.0 & $-$18.2 & y, y  
 \\
 & M8 & & & & & \\
 
  T Mic &  SRb & 20 27 55.1797 &  $-$28 15 39.553 &  186 &  8.0 x 10$^{-8}$ & (9.3) & 25.3  & 25.5 &  y, y       
  \\
  & M7$-$8 & & & & & \\
  
U Del  & SRb &  20 45 28.2500 &   +18 05 23.976  & 335 & 1.5 x 10$^{-7}$ & 7.9 &  $-$6.4 & $-$6.8 &  y, n   
   \\ 
   & M4$-$6 & & & & & \\
   
 RW Sco & Mira &   17 14 51.6867 &    $-$33 25 54.544  & 578 & 2.1 x 10$^{-7}$ & (4.9) &   $-$72.0 & $-$69.7 & y, y 
 \\ 
 & M6e & & & & & \\
 
V PsA & SRb &   22 55 19.7228 &    $-$29 36 45.038  & 304 & 3.0 x 10$^{-7}$ & (11.4) &   $-$11.1  & $-$11.1 &   y, n
 \\ 
 & M7$-$8 & & & & & \\
 
 SV Aqr & LPV &  23 22 45.4003 &   $-$10 49 00.187  & 431 & 3.0 x 10$^{-7}$ & (5.7) & 8.5  & 6.7 &  y, n  
  \\  
  & M8 & & & & & \\
  
 R Hya & Mira & 13 29 42.7021 &  $-$23 16 52.515 & 148 & 4.0 x 10$^{-7}$& 23.7 &  $-$11.0 & $-$10.1&  y, y  
 \\
 & M6$-$9e & & & & & \\
 
 U Her & Mira & 16 25 47.4514 & +18 53 32.666 & 271 & 5.9 x 10$^{-7}$ & 11.2 & $-$14.5 & $-$14.9 & y, y
 \\
 & M6.5$-$8e & & & & & \\
 
 $\pi^{1}$ Gru & SRb & 22 22 44.2696 & $-$45 56 53.007 &162 & 7.7 x 10$^{-7}$ & 18.4  & $-$13.0 & $-$11.7 & n, n
  \\
  & S5,7 & & & & & \\
  
 AH Sco  & SRc & 17 11 17.0159 & $-$32 19 30.764 & 2260 & 1.0 x 10$^{-6}$ & 5.8 & $-$4.0 & $-$2.3 & y, y 
    \\
  & M5Ia$-$7Ib & & & & & \\
  
  R Aql  & Mira & 19 06 22.2567 &  +08 13 46.678 & 234 & 1.1 x 10$^{-6}$ &10.9  & 47.0 & 47.2 & y, y  
  \\
  & M6.5$-$9e & & & & & \\
  
  W Aql  & Mira & 19 15 23.3781 & $-$07 02 50.331 & 374 & 3.0 x 10$^{-6}$ &11.6 & $-$25.0 & $-$23.0 &  n, n \\
  & S6, 6e & & & & & \\

KW Sgr  & SRc & 17 52 00.7282 & $-$28 01 20.572 & 2400 &  5.6 x 10$^{-6}$ & 3.9 &  4.0 & $-$4.4  &  y, n 
  \\
  & M4Ia & & & & & \\
  
   IRC$-$10529  & OH/IR & 20 10 27.8713 & $-$06 16 13.740 & 930 &  6.7 x 10$^{-6}$ & (6.5) & $-$18.0 & $-$16.3 & y, n 
  \\
  & M? & & & & & \\
  
 IRC$+$10011 & Mira & 01 06 25.9883 &  +12 35 52.849 & 740 & 1.9 x 10$^{-5}$ & (6.5) & 10.0 & 10.1 & y, y 
  \\
  & M8 & & & & & \\
  
   GY Aql  & Mira & 19 50 06.3148 & $-$07 36 52.189 & 410 & 3.0 x 10$^{-5}$ & (20.5) &  34.0 & 34.0 & y, n  
  \\
  & M8 & & & & & \\
  
 VX Sgr & SRc & 18 08 04.0460 & $-$22 13 26.621 & 1560 & 6.1 x 10$^{-5}$ & 8.8 &  5.3 & 5.7 & y, y 
  \\
  & M8.5Ia & & & & & \\
  
\hline                                  
\end{tabular}
\tablefoot{$^{(a)}$ SR stands for semi-regular pulsations; SRa and SRb are AGB stars with persistent and poorly defined periodicity, respectively; SRc is used for the supergiants. $^{(b)}$ Stellar peak measured in the continuum around 241.8 GHz for the extended configuration \citep{gottlieb2022}; astrometric accuracy $\sim$5$-$10 mas (see Sect. \ref{r}). The position is corrected for proper motion at the time of the observations. $^{(c)}$ See Sect. \ref{sec:sample} for adopted distances. $^{(d)}$ Mass-loss rate from references in Table~1 of \citet{gottlieb2022} with a few revisions (see Sect.~\ref{sec:sample}). $^{(e)}$ Angular diameter from Table~1 of \citet{montarges2023}. Figures in brackets are for calculated and not measured diameters; the derived optical diameter for GY~Aql appears to be  too large as our radio continuum ($\sim$250~GHz) uniform disk size is 14.0~mas.  $^{(f)}$ Local standard of rest velocity used as input for the {\sc atomium} observations. $^{(g)}$ Newer LSR velocity derived from a sample of lines with well-behaved line profiles \citep{gottlieb2022}.  $^{(h)}$ y or n indicates detection or not of at least one transition of H$_{2}$O (first entry) and OH (second entry). 
}
\end{table*}
%%%%%%%%%%

\section{H$_2$O  and OH spectroscopy, line identification}
\label{sec:spectro}

This Section briefly describes the principles leading to the spectroscopic determination of H$_2$O line frequencies (Subsect. \ref{spec_backgr})  and our line frequency detection criteria  (Subsect. \ref{sec:detectwater}). A similar approach is followed for OH in Subsects. \ref{sec:spec_backgr_OH} and \ref{sec:identif_OH}. 
Identification of which H$_2$O and OH lines are observed in which source is given in Tables~\ref{fluxdensity_source_list}  and \ref{Astro-Lines-Fit}, respectively.

\subsection{The water molecule, spectroscopy background}
\label{spec_backgr}

%%%%%%%%%%%%%%%%%%%%%%

\begin{figure}
\centering
\includegraphics[width=5.8cm,angle=270]{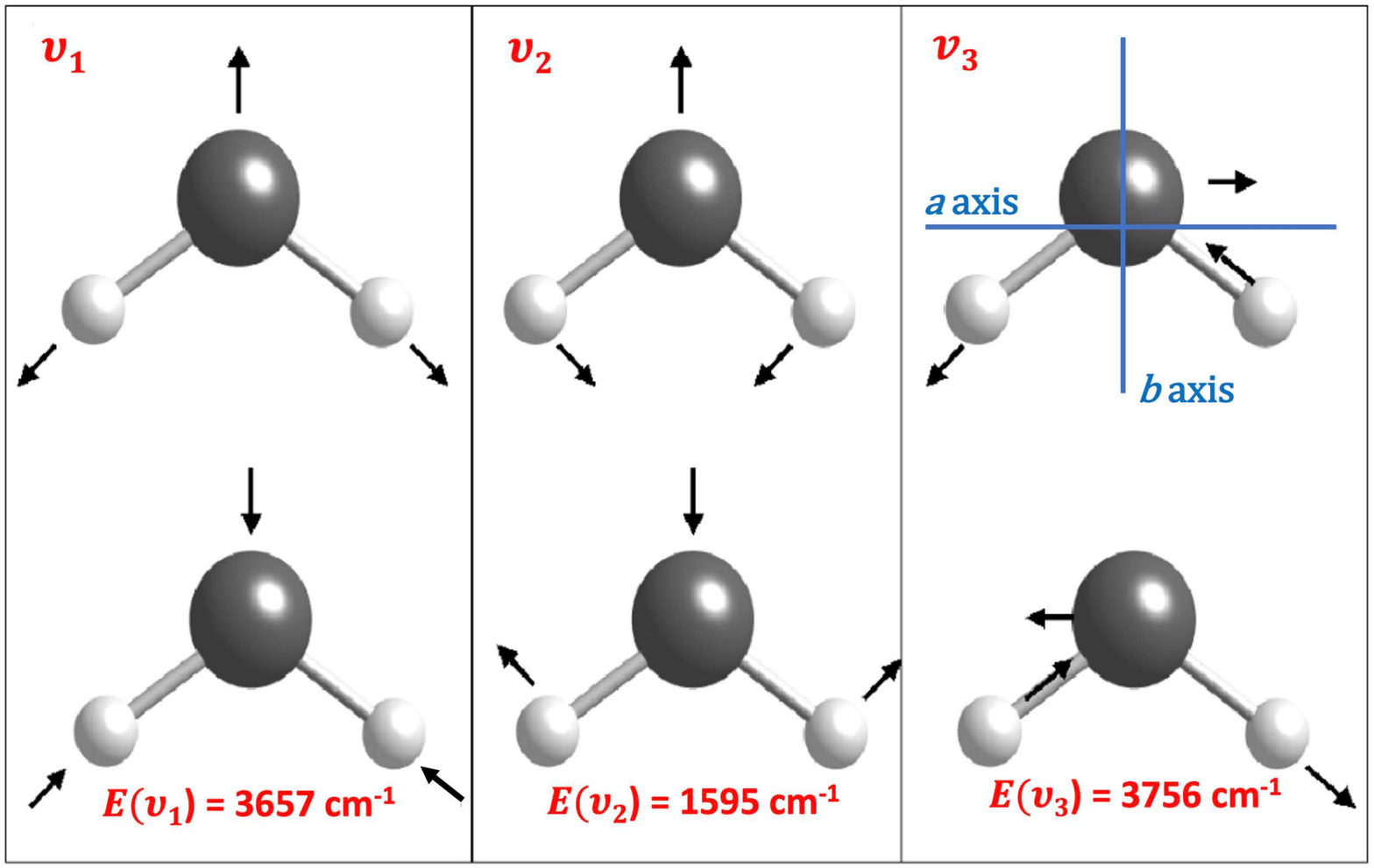}

\caption{Three fundamental vibrational modes of water vapor. They are denoted $\nu_1$ and $\nu_3$  for the symmetric and asymmetric stretchings, and $\nu_2$   for the symmetric bending. The arrows simulate the direct and reciprocal vibrational motions of the O and  the two H atoms \citep[adapted from][]{schroeder2002}. The O to H bond length is nearly 0.1~nm and the H-O-H average angle is 104$^{\circ}$. The $a$ and $b$ axes discussed in Sect. \ref{spec_backgr} intersect at the center of mass of the molecule. The energy of the three fundamental vibrational states $\nu_1$, $\nu_2$, $\nu_3$ are 3657, 1595 and 3756~cm$^{-1}$, respectively; the equivalent state temperature and wavelengths are 5261, 2294 and 5404 K (see also Fig. \ref{H2O_energy-levels}) and 2.73, 6.27 and 2.66~$\mu$m.  
  }
\label{H2O_vibration}
\end{figure}

%%%%%%%%%%%%%%%%%

The rotational and rovibrational energy levels of light hydrides, molecules 
consisting of one or more H atoms and at most one light non-H atom, are 
often difficult to describe by a conventional Watson-type Hamiltonian
because of the large effects of centrifugal distortion  \citep{Euler_H2O_2005}. 
The water molecule, H$_2$O, is a prototype in this regard and an overview of alternative models developed to fit rotational and 
rovibrational spectra of H$_2$O is presented in  \citet{Euler_H2O_2005}. 

%%%%%%%%%%%%%%%%%%%%%%%%%%%%%%%%%%%

\begin{figure}
\centering
\includegraphics[width=7.2cm,angle=270]{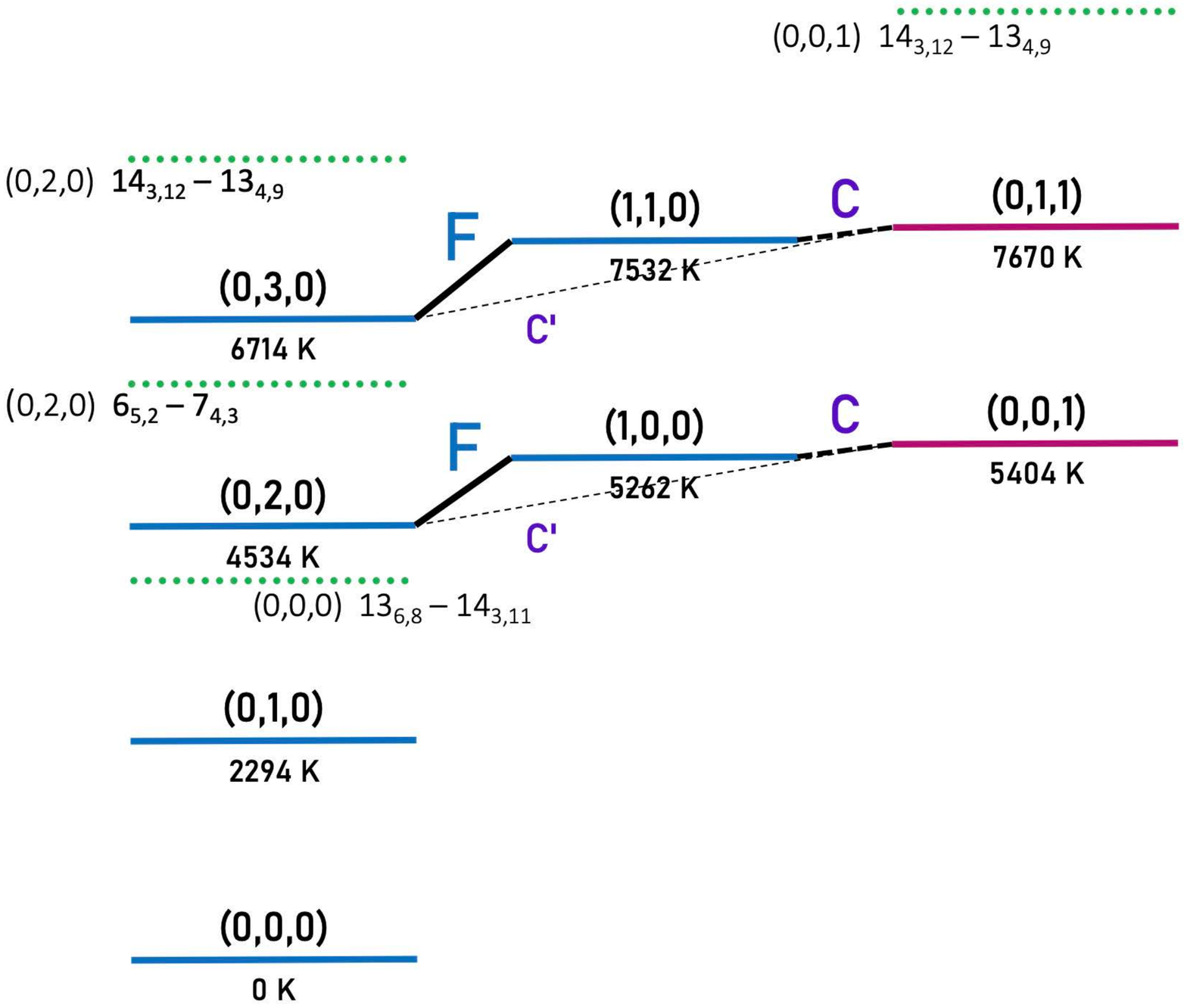}

\caption{Lowest eight vibrational states of water vapor and their quantum numbers ($\varv_1$,$\varv_2$,$\varv_3$). States of $a_1$  symmetry are referred to by horizontal blue lines, states of $b_2$ symmetry by aubergine lines. Vibration-rotation interactions are indicated by lines connecting the interacting states. The labels F, C and C' symbolize Fermi, Coriolis, and rotational (or Coriolis-type) interactions. The energy of each vibrational band origin is given in Kelvin below each horizontal solid line. The rotational and ro-vibrational transitions in the range of energy covered by the {\sc atomium} project are listed in Table~\ref{H2O-line-list}. Four examples, corresponding to lines 10, 14, 8 and 4 in Table \ref{H2O-line-list}, 
are shown with their quantum identification, in the form of dotted horizontal green lines; they are  ordered by increasing energy.
}
\label{H2O_energy-levels}
\end{figure}

%%%%%%%%%%%%%%%%%%%%%%

H$_2$O is an asymmetric rotor with energy levels described by $J_{K_a,K_c}$ where $J$ is the rotational quantum number and $K_a$ and $K_c$ are the projections of the total angular momentum along two of the three axes, $a$, $b$ and $c$ used to derive the three  principal moments of inertia of water. The $a$-axis is parallel to the H to H direction and orthogonal to the $b$-axis that crosses the O atom  and along which the H$_2$O permanent  dipole moment is observed (see right panel in Fig. \ref{H2O_vibration}). A precise value of the dipole moment along the $b$-axis was obtained by \citet{H2O_HDO_dip_1991}, $\mu_0 = 1.855$~D. We point out that vibration and distortion corrections to the dipole moment are required to accurately model intensities and derived quantities  
\citep{H2O_HDO_dip_1991, H2O_dipole-surface_1991, H2O_dipole-surface_2012}. 
H$_2$O has three fundamental vibrational modes, the symmetric stretching mode $\nu_1$ 
and the bending mode $\nu_2$ both of $a_1$ symmetry,  and the antisymmetric stretching mode $\nu_3$ of $b_2$ symmetry. 
A common shorthand notation to describe any vibrational states is a triplet which indicates 
the degree of excitation of each fundamental mode in the form ($\varv_1$,$\varv_2$,$\varv_3$). 
The three fundamental modes of vibration  are schematically represented  in Fig. \ref{H2O_vibration} and the lowest eight vibrational states of H$_2$O are displayed in Fig.~\ref{H2O_energy-levels} together, as an example, with the  ($\varv_1$,$\varv_2$,$\varv_3$) $J_{K_a,K_c}$ upper energy levels of four transitions detected in this work (see Fig. \ref{H2O_energy-levels} caption and Table \ref{H2O-line-list}). 
The vibrational states (0,2,0), (1,0,0) and (0,0,1) as well as (0,3,0), (1,1,0) and (0,1,1) are quite close in energy (Fig.~\ref{H2O_energy-levels}). As a consequence, rotational levels with the same total angular momentum $J$ may interact if they are nearly degenerate and obey certain selection rules (i.e., there may be mutual interaction of nearby, unperturbed levels). Fermi interaction may occur between rotational levels involving the (0,2,0) and (1,0,0), and (0,3,0)  and (1,1,0) vibrational states; the quantum numbers of the $J_{K_a,K_c}$ levels need to differ in $K_a$ and $K_c$ by even numbers. 
The interaction effects are usually largest if $K_a$ and $K_c$ are identical. Coriolis interaction 
of $c$-type may occur between levels involving (1,0,0) and (0,0,1), and (1,1,0) and (0,1,1); the quantum numbers 
need to differ in $K_a$ by an odd number and in $K_c$ by an even number. The interaction effects are usually largest if $K_a$ differs by one. 
The interaction between (0,2,0) and (0,0,1), and (0,3,0) and (0,1,1) is called rotational or Coriolis-type, 
or, frequently, just Coriolis (C$'$ in Fig. \ref{H2O_energy-levels}). This is appropriate as the operators 
describing the interaction are the same as for a proper Coriolis interaction.
However, this type of interaction is of higher order and usually has relatively small effects.

The energy difference between two interacting rotational levels 
is larger than in the noninteracting case and tends to mix levels.
One consequence in the case of the Fermi interaction between (0,2,0) and (1,0,0) is that 
if a transition from one level in (0,2,0) to an interacting level in (0,2,0) is allowed, 
then a transition from the first level in (0,2,0) to the corresponding interacting level 
in (1,0,0) is also allowed; the strength of the latter transition depends on the degree 
of the mixing between the two interacting levels. Such transitions can also occur for 
other types of vibration-rotation interaction. An additional effect in the rotational 
spectrum of H$_2$O is that the presence of two equivalent H nuclei leads to spin-statistical 
weights of three and one for levels of ortho and para H$_2$O, 
respectively. The ortho and para states for vibrations with $a_1$ 
symmetry are described by $K_a + K_c$ being odd and even, respectively, while it is the 
opposite for vibrations with $b_2$ symmetry. The states of $a_1$ and $b_2$ symmetry are labeled in blue and aubergine in Fig. \ref{H2O_energy-levels}, respectively.

The rotational spectrum of H$_2$O in low-lying vibrational states has been investigated  extensively. 
A fairly recent and extensive study of THz  and far IR transitions of water in the lowest 
five vibrational states was published by \citet{H2O_1st-Triad_2012}. Their analysis included 
numerous transitions  in the range 293$-$2723~GHz (determined with $\sim$1 to 100 kHz accuracy) and in the 50$-$600~cm$^{-1}$ far IR region (with accuracy up to a few tens of MHz). These data, taking into account earlier data, are the current basis  for the JPL catalog \citep{JPL-catalog_1998} entries of H$_2$O in its ground vibrational 
state and in its next four excited vibrational states, (0,1,0), (0,2,0), (1,0,0) and (0,0,1). 
Shortly thereafter, \citet{H2O_2nd-Triad_2013} determined transition frequencies 
for the next three vibrational states (the second triad) and redetermined some frequencies in the first five states. \citet{H2O_2nd-Triad_2014} provided additional transition frequencies of the lowest eight vibrational states in the far IR region; they cover, in particular, highly rotationally excited states. They also presented a catalog file for the second triad consisting of the (0,3,0), (1,1,0), and (0,1,1) states. Unfortunately, no frequencies have been calculated below 500~GHz.

%%%%%%%%TABLE #2
\begin{table*}
\begin{center} 
 \caption{Observable transitions of H$_2$$^{16}$O covered by the {\sc atomium} program.}

\label{H2O-line-list}
\renewcommand{\arraystretch}{1.10}
\begin{tabular}[t]{cr@{}lr@{}lcccr@{}lr@{}lccl}
\hline \hline
Line number& \multicolumn{2}{c}{$\nu^{a}$} & \multicolumn{2}{c}{Unc.$^{a}$} & QN$_{\rm vib}$$^{b}$ & QN$_{\rm rot}$$^{c}$ & \multicolumn{1}{c}{$S_{\rm p}$$^{d}$} &
\multicolumn{2}{c}{$E_{\rm up}$$^{e}$}  &
% \multicolumn{1}{c}{$E_{\rm up}$}  &
& ${A}$$^{f}$ &   \multicolumn{1}{c}{$n_{\rm det.}$$^{g}$}    &  \multicolumn{1}{c}{Reference$^{h}$} \\
%&   \multicolumn{1}{c}{\#$_{So}$}    
  &  \multicolumn{2}{c}{(MHz)} &  \multicolumn{2}{c}{(MHz)} & &  &   & (cm$^{-1}$)  &   \multicolumn{3}{c}{($\times$10$^{-6}$s$^{-1}$)}  \\
  &   & & &  &   & &   & ($\it K$)  &   \\
\hline
1  & 222014&.12  &   4&.53 & (0,3,0)$-$(0,3,0) &   $8_{3,6} - 7_{4,3}$   & $o$ & 5790&.1  & &  15.6   & &  W2020\\
   & 222017&.31    &       1&.50        &           &                         &     &  $\it8331$   &   &  &    &        8    & TW\\
   \\
2  & 227780&.09 &  0&.29 & (1,0,0)$-$(0,2,0) &   $7_{0,7} - 8_{3,6}$   & $o$ & 4232&.2 & & 0.003  &  0    & JPL\\
&    &           &           &      &      &             &     &  $\it6089$   &   &  &    &           & \\

\\
3  & 230191&.53  & 0&.53  &  (0,0,1)$-$(0,2,0) & $8_{3,5} - 9_{5,4}$  &  $o$ & 4792&.3 & & 0.03 &  0  & JPL\\
&    &           &           &      &      &             &     &  $\it6895$   &   &  &    &           & \\
\\

4  & 236805&.40 &  1&.82 & (0,0,1)$-$(0,0,1) & $14_{3,12} - 13_{4,9}$  & $p$ & 6263&.9  &   & 2.11     &     & JPL\\
   & 236797&.0$^i$    &       2&.00      &           &            &                 &  $\it9012$   & &   &  &   1$^i$    & TW\\\\

5  & 244330&.37  &  3&.44 & (1,1,0)$-$(0,1,1) &   $4_{2,2} - 3_{2,1}$   & $p$ & 5557&.9 & &   4.84  &         & W2020\\
   &abso.$^j$ &     & &      &     &      &         &      $\it7997$     &        &              &              & 4$^j$ & TW\\
 \\
 
6  & 252172&.25 &  0&.38 & (1,0,0)$-$(0,2,0) &   $7_{4,3} - 8_{5,4}$   & $o$ & 4572&.4  & & 0.55 &       & JPL\\
   & 252170&.56  &   2&.00  &                   &                         &     &  $\it6579$   &   &  &        &  8    & TW\\
\\
7  & 254039&.88 &  0&.15 & (0,0,1)$-$(0,0,1) &   $3_{1,3} - 2_{2,0}$   & $o$ & 3895&.6 & & 9.55  &       & JPL, P1995\\
   & 254039&.75  &    2&.00   &                   &                         &     &  $\it5605$   &   &  &         &  9   & TW\\
\\
8  & 254052&.55 &  3&.42 & (0,2,0)$-$(0,2,0) & $14_{3,12} - 13_{4,9}$  & $o$ & 5790&.4 & & 6.76  &       & JPL\\
   & 254055&.18  &   2&.00   &                   &                         &     &     $\it8331$  &  &    &      &  6    & TW\\
\\
9  & 254234&.65  &  3&.00 & (0,1,1)$-$(1,1,0) &   $7_{3,4} - 6_{5,1}$   & $p$ & 6179&.8  & & 2.06  &       & W2020\\
     & & &      &    &                   &                        &  &     $\it8891$  &   &  &         & 2?$^k$    & \\
   \\
10  & 259952&.18 &  0&.20 & (0,0,0)$-$(0,0,0) &  $13_{6,8} - 14_{3,11}$ & $p$ & 2748&.1  & & 3.24  &       & JPL, A1991\\
   & 259951&.87  &   2&.00   &                   &                         &     &   $\it3954$  &   &  &          &  10   & TW\\
   \\
 11 & 262555&.55 &  1&.42 & (0,2,0)$-$(1,0,0) &  $10_{7,3} - 11_{4,8}$  & $p$ & 5473&.8   & &   0.09  & 0     & JPL\\
 &    &           &           &      &      &             &     &  $\it7876$   &   &  &    &           & \\
 \\
12 & 262897&.75 &  0&.15 & (0,1,0)$-$(0,1,0) &   $7_{7,1} - 8_{6,2}$   & $p$ & 3109&.9 & & 3.97 &       & JPL, P1991\\
   & 262897&.87  &    1&.50   &                   &                         &     &   $\it4475$  &   &  &         &  12   & TW\\
   \\
13 & 266574&.10 &  1&.85 & (0,0,0)$-$(0,0,0) & $21_{4,17} - 20_{7,14}$ & $o$ & 5748&.1 & & 4.32  &       & JPL\\
   & 266567&.52  &    2&.00  &                   &                         &     &    $\it8270$ &   &  &         &  7    & TW\\
   \\
14 & 268149&.12 &  0&.15 & (0,2,0)$-$(0,2,0) &   $6_{5,2} - 7_{4,3}$   & $o$ & 4197&.3  & &15.3 &       & JPL, P1991\\
   & 268148&.51  &    1&.50   &                   &                         &     &  $\it6039$  &   &  &    & 15   & TW\\
\hline
\end{tabular}
\end{center}
\tablefoot{
$^{(a)}$ Transition frequency and uncertainty for the given line number: first entry is from the reference given in the last column; second entry  is the frequency determination in this work (TW in last column) from observed spectra (see Sect. \ref{sec:detectwater}). No second entry means no detection. Uncertainty refers to: JPL laboratory measurements or calculations (first entry) and to our own estimates from this work  (second entry).
$^{(b)}$ Vibrational quantum number: ($\varv_1'$,$\varv_2'$,$\varv_3'$)$-$($\varv_1''$,$\varv_2''$,$\varv_3''$). 
$^{(c)}$ Rotational quantum number: $J'_{K_a',K_c'}$--$J''_{K_a'',K_c''}$. 
$^{(d)}$ Spin modification: $ortho$ or $para$ H$_2$O.  
$^{(e)}$ Upper state energy in cm$^{-1}$ and K (italics below cm$^{-1}$). 
$^{(f)}$ Spontaneous emission rate from the HITRAN data base (https://lweb.cfa.harvard.edu/HITRAN/) except for line 3  where it is not available; uncertainty is larger for transitions between different vibrational states because of limited wavenumber precision in data base. Spontaneous emission rate in line~3 is taken from https://splatalogue.online// from which line strengths ($S\mu^2$ in D$^2$) can also be retrieved for most lines listed here.
$^{(g)}$ Number of  {\sc atomium} sources detected in H$_2$O transition. 
$^{(h)}$ JPL catalog \citep{JPL-catalog_1998}, W2020 \citep{H2O_W2020}, P1991 \citep{hot-H2O_JCP_1991}, P1995 \citep{Pearson_1995}, A1991 \citep{exc-H2O_Amano_1991}, TW (This Work). 
$^{(i)}$ Transition identified in R~Hya only (uncertain identification in S~Pav) and rest frequency derived from extended array data. See R Hya combined array spectra in Appendix \ref{sec:254.23_vxsgr_ahsco}.
$^{(j)}$ 244.330~GHz line absorption observed in R~Hya, S~Pav, R~Aql and IRC$+10211$ (Sect. \ref{sec:absomaps_H2O}.).
$^{(k)}$ Uncertain identification of line 9 near strong $\varv$=0 $^{30}$SiO  line wing in AH Sco and VX Sgr (see spectra and discussion in Appendix \ref{sec:254.23_vxsgr_ahsco}). 
} 
\end{table*}

%%%%%

\subsection{Identification of highly excited water lines}
\label{sec:detectwater}

Water vapor has a rich spectrum of pure rotational transitions as well as many rovibrational transitions spanning a broad range of wavelengths from the IR to the submm/mm domain. 
The first excited states of the symmetric and asymmetric stretching modes observed around  2.7~$\mu$m ($\nu_1$ and $\nu_3$ bands) and the first excited state of the bending mode observed at 6.27~$\mu$m ($\nu_2$ band) are the most important IR transitions of water (Fig. \ref{H2O_vibration}). Other vibrational transition bands have long been identified in the low dispersion astronomical spectra of Mira stars, \citep[e.g.,][]{spinrad1965, hinkle1979}.

The high sensitivity and high spectral resolution achieved with ALMA allow us to search for various ro-vibrational or pure rotational transitions of water in different vibrational states so that a broad range of energies and physical conditions can be probed with an appropriate selection of water transitions. Using the JPL catalog \citep{JPL-catalog_1998}\footnote{https://spec.jpl.nasa.gov/ftp/pub/catalog/}  and the W2020 data base  \citep{H2O_W2020}  and, limiting ourselves to energy levels up to $\sim$6500~cm$^{-1}$ (9400~K), we have searched for all  pure rotational or ro-vibrational dipolar electric transitions of water in our frequency settings, without any a priori spectral line intensity cut-off. We found fourteen transitions of the main isotopic species of water with energy up to $\sim$9000~K; they are listed in Table~\ref{H2O-line-list}. Ten are safe detections  in the present work and were identified in different targets (see last two columns in Table~\ref{H2O-line-list} for number of detected sources and spectroscopic references). Six are ortho H$_2$$^{16}$O  and four para H$_2$$^{16}$O  transitions; one ortho H$_2$$^{16}$O transition is uncertain (line 9 in Table~\ref{H2O-line-list}; see Appendix~\ref{sec:254.23_vxsgr_ahsco}). The first entry in the second and third columns in Table \ref{H2O-line-list} gives the rest frequency and uncertainty from the JPL catalog (c018003 and c018005 files) or from the W2020 data base. The second entry in the second and third columns of Table~\ref{H2O-line-list} for lines 1, 4, 5, 6, 7, 8, 10, 12, 13 and 14 gives our own rest frequency measurements and estimated maximum uncertainties (see discussion below).  Using the same line selection criterion as above, there are one H$_2$$^{17}$O and four H$_2$$^{18}$O transitions in our frequency setup. 
 None of them are in the ground vibrational state and no signal was observed  in the vicinity of the expected  frequencies. 
 Moreover, the predicted line intensities are too weak for reliable identification. 
 H$_2$O, and later OH, without superscript in this article, always refer to H$_2$$^{16}$O and  $^{16}$OH. 

We have assigned to  H$_2$O the high signal-to-noise ratio (S/N) features identified in our spectra when such features, once corrected for the systemic velocity of the star,  coincide within a few MHz with transitions of  H$_2$O in the JPL or W2020 data bases. The spectra used for this identification  have been extracted from both the high and mid resolution 
data cubes for different aperture diameters (0\farcs08 and 0\farcs4 diameters are used in general for  high and mid, respectively). The number of sources for which we have a spectral identification as defined above, varies from fifteen (line14 in Table \ref{H2O-line-list}) to four (line number 5, in absorption) or just one or perhaps two (line 4, see Appendix \ref{sec:254.23_vxsgr_ahsco}). Line identification did not suffer from spectral confusion problems. In addition, we used the CDMS\footnote{https://cdms.astro.uni-koeln.de/classic/entries/} \citep{muller2005, endres2016} data base to check for possible misidentifications due to the spectral proximity with the molecular species in the {\sc atomium} chemical inventory (Wallstr\"om et al., in preparation). Lines 4, 6 and 8 in our Table \ref{H2O-line-list} lie close to SO$^{18}$O at 236.805, 252.185 and 254.067 GHz but only SO$_2$ and $^{34}$SO$_2$ are identified in the {\sc atomium} inventory. We note the frequency proximity of the H$_2$O line 1 at 222.014 GHz with the $\varv = 0$,~$J_{K_a,K_c} = 8_{3,6}$--$7_{4,3}$ transition of SiC$_2$ at 222.009 GHz, but this species is identified only in the S-type star W Aql which has no water emission.

Despite uncertainties discussed below,  we have used the observed emission line peak for a given transition (or the average of a few line channels for flat-emission features) to estimate our own line rest frequency and confirm line identification. The average of our frequency measurements in different stars for a given transition,  corrected for the stellar velocity used during the observations, is our observed rest frequency. It is shown in the second column of Table~\ref{H2O-line-list} below the JPL or W2020 rest frequency; we add TW (for This Work) as appropriate in the last column of Table~\ref{H2O-line-list}. (We have not made a frequency estimate for line~5 seen in absorption  near 244.330~GHz.) There are a few caveats to our line frequency estimates. Line opacity effects or line profiles skewed by line wing absorption due to gas infall, for example, may eventually bias our measurements. However, the high energy  transitions of water studied here are not optically thick in general (Sect. \ref{sec:pop_diagram}); for line 14 which tends to be masing in some sources  (Sect. \ref{sec:268_H$_2$O_maser}) specific velocity components could be enhanced, however. We have used as much as possible "well-behaved" line profiles with the hope that averaging independent frequency measurements in different stars minimizes the errors.  
Our main sources of error most probably come from limited spectral resolution, $\sim$1 MHz, and the stellar velocity uncertainty. The latter uncertainty is small as suggested by comparing the velocities given in the  eighth and ninth columns of Table~\ref{primarysource_list}. There are eight sources with differences below 0.4~km\,s$^{-1}$, or less than 0.36~MHz at 268.149~GHz, and eight other sources with differences $\la$0.9$-$2.0~km\,s$^{-1}$ or below 1.8~MHz.  
Despite all potential errors we find that, for those transitions for which we have a significant number of independent measurements, the variance of our frequency calculations is $\la$1~MHz.  We adopt 1.5 or 2~MHz as our total frequency uncertainty in Table \ref{H2O-line-list} and note that in spite of various uncertainties, the frequency discrepancy between our calculations  and the JPL catalog  remains within $\sim$0.2$-$3 MHz, except for the $\sim$9000 and 8300 K high energy lines 4  and 13 where it is larger. 
We further note that our rest frequency determinations are in good agreement with those  
measured in the laboratory; this is well verified for lines 7, 10, 12 and 14 in Table~\ref{H2O-line-list}. In the case of relatively high uncertainties for the calculated rest frequencies in catalogs our rest frequency determinations could be better, especially for lines 1, 8 and 13.

We stress that, as far as we know, nine out of the ten water transitions detected in this work are new radio detections in space. Line 14  in the $\varv_2 = 2$ state at 268.149~GHz is the only transition that was first observed as a strong emission in  VY~CMa \citep{tenenbaum2010} and a weak line in IK Tau \citep{velillaprieto2017}. We find here that this line is excited  in  twelve AGBs and three RSGs of the {\sc atomium} sample. The only two targets without  any  H$_2$O  transition, are the two S-type stars  in Table \ref{primarysource_list} with a water abundance expected to be lower than in the O-rich M-type stars (compare e.g., the $1.5 \times 10^{-5}$ water abundance derived by  \citet{danilovich2014} in W Aql with the higher water abundance obtained by \citet{khouri2014} and \citet{maercker2016} in M-type stars).

The  268.149 and  262.898~GHz transitions in the (0,2,0) and (0,1,0) states are widespread in evolved stars and will be analyzed in Sect.~\ref{sec:268_H$_2$O_maser}. Line 1 in the (0,3,0) state at 222.014 GHz is the first radio detection of water in such a high vibrational state; it also seems to be widespread in O-rich stars. Finally, we note that the three ro-vibrational transitions without detection in stars, lines~2, 3 and 11,  have low spontaneous emission rates (${A}$ value) implying weak spectroscopic intensities. However,  the  ${A}$ value for line 6, with eight sources detected, is not much larger than that of line 11 with no detection. We also point out that line 4 with the highest energy levels observed in this work is identified in at most two stars (see footnote in Table \ref{H2O-line-list}).

\subsection{The OH radical, spectroscopy background}
\label{sec:spec_backgr_OH}

\begin{table*}
%\begin{small}
\caption{Observable $\varv=0$ and 1,  ${\Delta} J = 0$, ${\Delta} F = 0$ transitions of OH in the {\sc atomium} frequency line setting (excluding  ${\Delta} F = \pm 1$ and very high $N$, $J$ transitions).}       
\label{OHline_list}      
\centering                           
\begin{tabular}{lcccccccccc}       
 \hline\hline

Line  & JPL Frequency$^{a}$ & Unc.$^{b}$ & $\varv$, $N, J^{c}$ & $F_{\rm up}-F_{\rm low}$$^{c}$ & $E_{\rm up}$$^{d}$ & $A$$^{e}$ &  $n_{\rm det.}$$^{f}$  \\
  & (MHz) & (MHz) &  &  & (cm$^{-1}$,  K) & (x10$^{-6}$s$^{-1}$) &    \\
  
\hline

 1 & 221335.49  & 0.29 & 0, 13, 27/2 & $13^{-} -13^{+}$ & 3319.3, \it4776 & 2.39 &  6   

  \\

 2 & 221353.48 & 0.29 &    & $14^{-} -14^{+}$ &  &  & 7        

  \\
 
 3 & 223683.92 & 1.53 & 1, 17, 33/2& $17^{-} -17^{+}$ &  8889.8, \it12791 & 0.90 & 0
 
 \\
 4 & 223713.79 &  1.53 &  & $16^{-} -16^{+}$&  &  & 0 
 
 \\
  5 & 236328.12 & 0.56  &  0, 17, 33/2 &  $17^{-} -17^{+}$  & 5533.5, \it7962  &   1.07 &  4$^{g}$
  
  \\
 6 & 236359.64 & 0.56 & & $16^{-} -16^{+}$ & &    & 2$^{h}$
 
  \\
 
7 & 252127.10 & 0.43 &  0, 14, 29/2 &   $14^{+} -14^{-}$    & 3819.1, \it5495 & 3.03 & 6

  \\
 
 8 & 252145.35 & 0.43 & & $15^{+} -15^{-}$ &   & & 9
 
  \\
  
 9 & 265734.66 &  0.80 &   0, 18, 35/2 & $18^{+} -18^{-}$  & 6157.7, \it8860 & 1.38 &2$^{i}$
 
  \\

 10 & 265765.32 & 0.80 & & $17^{+} -17^{-}$ &  &  & ?$^{i}$

 \\
 
11 & 269530.55 & 1.09 & 1, 15, 31/2 & $15^{-} -15^{+}$ & 7752.6, \it11155  &  3.23 & 0
 
 \\
 12 & 269547.76 & 1.09 &  & $16^{-} -16^{+}$&  & & 0

 \\ 
\hline                                  
\end{tabular}
\tablefoot{
$^{(a)}$~Frequency taken from version 5 of the JPL catalog; frequency and uncertainty from observations of {\sc atomium} stars are given in Table \ref{Astro-Lines-Fit}. 
$^{(b)}$Calculated~JPL catalog uncertainty derived from laboratory measurements. 
$^{(c)}$~Quantum numbers defined in Sect. \ref{sec:spec_backgr_OH}.
$^{(d)}$~Upper energy level in cm$^{-1}$ and K (in italics). For a same value of $J$, the $F_{\rm up}$ and $F_{\rm low}$ level energies  are identical within one tenth of a cm$^{-1}$. 
$^{(e)}$~Einstein $A$ spontaneous emission probability; see also last row in Appendix \ref{sec:Lambda_doubling_freqs} Tables. 
$^{(f)}$~Number of detected OH sources in ATOMIUM program. 
$^{(g)}$~Unambiguous assignment of hyperfine transition for R Hya, S Pav and R Aql; in U Her assignment to $F'-F'' = 17 - 17$ from OH line stacking (see Sect. \ref{sec:OHstackmaps}).
$^{(h)}$~Unambiguous hyperfine identification for R Hya and identification from OH line stacking for T Mic.
$^{(i)}$~Line~9  is observed at the $\sim$5$\sigma$ level in R~Hya (Sect.~\ref{sec:OHstackmaps}) and in Mira (Fig.~\ref{OH_Mira_Band6}). The $J = 35/2$ line~10 is detected at low ($\sim$2.5$\sigma$) significance in R~Hya and blended with  TiO$_{2}$ in Mira (Fig.~\ref{OH_Mira_Band6}).
} 
\end{table*}

The hydroxyl radical, OH, exhibits a complex spectrum because of its unpaired electron and coupling with the nuclear spin of the hydrogen atom. The electronic ground state of OH is a $^{ 2}{   \Pi }$ state according to the value of one for its electronic orbital angular momentum projection along the OH internuclear axis. Accounting for the spin-orbit coupling, the total electronic angular momentum along the OH axis is described by the quantum number  ${ \Omega}$ which takes the value 3/2 or 1/2. The OH states are designated  $^{ 2}{   \Pi_{ 3/2} }$ and  $^{ 2}{   \Pi_{ 1/2} }$,  $^{ 2}{   \Pi_{ 3/2} }$ being lower in energy. The spin-orbit splitting of OH is  
 appropriately described by Hund's case~$(a)$ for lower quantum numbers while Hund's case~$(b)$, where the spin is not coupled to the internuclear axis, is more appropriate for higher quantum numbers. In addition, the energy of the weak coupling of the OH rotational angular momentum with the total electronic angular momentum depends on their respective sense of rotation. Hence, each degenerate rotational energy level is split into ${\Lambda}$-doublets with nearby energy levels and different parity.
 For the higher rotational levels, where case~$(b)$ applies, the orbital electronic momentum and the rotational molecular momentum form a resultant vector described by the scalar number $N$ which, combined with the scalar value of the spin vector $S = \pm1/2$, gives $J = N + 1/2$ and  $N - 1/2$ in the $^{ 2}{   \Pi_{ 3/2} }$ and  $^{ 2}{   \Pi_{ 1/2} }$ states. Hund's case~$(b)$ is appropriate for the {\sc atomium} high$-$$J$ OH observations.

Another weak magnetic coupling between the unpaired electron spin of OH and the hydrogen nuclear spin, described by the total quantum number $F = J \pm 1/2$, is observed at high spectral resolution both in the laboratory and in space. The ${\Delta} F = 0$ hyperfine transitions within a given ${\Lambda}$-doublet (${\Delta} J = 0$) are called principal (or main) lines because their  local thermodynamic equilibrium intensities relative to the so-called satellite lines (for which ${\Delta} F$ = $\pm1$) are 5:1 and 9:1 in the $^{ 2}{   \Pi_{ 3/2} }, J = 3/2$ ground state; these relative intensities can reach much higher values at higher rotational levels.

There are twelve ${\Delta} F = 0$ lines of OH in the $\varv$ = 0 and 1 vibrational states falling in the {\sc atomium} frequency range. They are given in Table~\ref{OHline_list} and ordered by increasing frequency. We have not included in Table~\ref{OHline_list} the very weak  ${\Delta} F = \pm 1$ satellite lines nor the five 
very weak $ J = 41/2-43/2$, ${\Delta} F = 0, \pm 1$  transitions in the $\varv$ = 0 and 1 vibrational states, although they fall in our  frequency setup. (Other weak $^{ 17}$OH and $^{ 18}$OH transitions, which are not observed in our data, are not discussed either.) 
The rest frequencies, based on rotational spectra of OH and isopotologs \citep{drouin2013}, and associated errors rounded to ten kHz are taken from the JPL catalog (c017001 file, version~5) and given in the second and third columns of Table \ref{OHline_list} together with some relevant quantum numbers, the upper energy level  and  Einstein-$A$ coefficient. The number of OH sources detected per transition in this work is shown in the last column of Table \ref{OHline_list}.  We  note that lines in Tables \ref{OHline_list} and  \ref{Astro-Lines-Fit} (see below) can also be identified from their $^{ 2}{\Pi_{ 3/2} }$ or $^{ 2}{\Pi_{1/2} }$ electronic ground state. Lines 5 and 8 for example, observed in four and nine different sources, correspond to   $^{ 2}{\Pi_{ 1/2} }$, $\varv=0$, $J = 33/2, F'-F'' = 17^{-}-17^{+}$ and   $^{ 2}{\Pi_{ 3/2} }$, $\varv=0$, $J = 29/2, F'-F'' = 15^{+}-15^{-}$ near 236.328 and 252.145~GHz, respectively.

The $\Lambda$-doubling frequencies of OH derived from the ALMA observations in this work and in \citet{khouri2019}  are gathered in Table~\ref{Astro-Lines-Fit}. The observed OH frequencies (third column in Table~\ref{Astro-Lines-Fit}) may sometimes differ from the JPL frequencies (second column in Table~\ref{OHline_list}). This discrepancy is commented on in Sect.~\ref{sec:identif_OH} where we also derive new OH $\Lambda$-doubling frequencies. 
The O$-$C column in Table \ref{Astro-Lines-Fit} corresponds to the difference between the frequency derived from the astronomical observations and our new frequency calculation described in Sect. \ref{sec:identif_OH} and Appendix \ref{sec:Lambda_doubling_freqs}.

%%%%%%%%%%%%%%%%

\begin{table*}
\begin{center}
\caption{Frequencies of OH $\Lambda$-doubling transitions from astronomical observations and comparison with calculated frequencies.}
\label{Astro-Lines-Fit}
\begin{tabular}[t]{ccccccccccccr@{}lr@{}lr@{}ll}
\hline \hline
$\varv$, $N$, $J^{a}$ &  $F_{\rm up}-F_{\rm low}^{a}$ & Frequency$^{b}$ & Unc.$^{b}$ &  O$-$C$^{c}$ & Observed source$^{d}$  & Line$^{e}$                \\
 & &   (MHz) & (MHz) & (MHz) &    &     \\
\hline
 0, 12, 23/2 &   $12^{+} -12^{-}$  & 107037.11 & 0.72 &    0.2923 & R Dor  &   \\
 0, 12, 23/2 &    $11^{+} -11^{-}$ &  107073.45 & 0.81 &    0.3731 & R Dor  &   \\
 \\
 0, 13, 25/2 &  $13^{-} -13^{+}$ & 130078.59 & 0.68 &    0.3580 & W Hya  &   \\
 0, 13, 25/2  & $12^{-} -12^{+}$ &  130113.79 & 0.67 &    0.3499 & W Hya   &  \\
 \\
 0, 13, 27/2 &  $13^{-} -13^{+}$  & 221333.34 & 1.50  & $-$0.6893 & R Hya, R Aql, S Pav  &  1 \\
 0, 13, 27/2 &  $14^{-} -14^{+}$ &  221351.17  & 1.50  & $-$0.8575 & R Hya, R Aql, S Pav & 2  \\
 \\
  0, 17, 33/2 &  $17^{-} -17^{+}$ & 236327.29  & 1.50  &    0.5849 & R Hya, R Aql, S Pav   & 5  \\
  0, 17, 33/2  & $16^{-} -16^{+}$ &  236356.12  & 2.00  & $-$2.0998 &  R Hya  & 6   \\
\\
 0, 14, 29/2 &   $14^{+} -14^{-}$ &  252123.22  & 1.50  & $-$1.6962 & R Hya, R Aql, S Pav & 7  \\
 0, 14, 29/2 &   $15^{+} -15^{-}$ &  252141.78  & 1.50  & $-$1.3937 & R Hya, R Aql, S Pav & 8 \\
 0, 14, 29/2 &  $14^{+} -14^{-}$ &  252124.11 & 1.63 & $-$0.8102 & IK Tau  &  \\
 0, 14, 29/2 &   $15^{+} -15^{-}$ &  252143.25 & 1.66 &    0.0764 & IK Tau    & \\
 0, 14, 29/2 &   $14^{+} -14^{-}$ &  252124.49 & 1.65 & $-$0.4272 & R Dor  &   \\
 0, 14, 29/2 &   $15^{+} -15^{-}$ &  252143.01 & 1.65 & $-$0.1597 & R Dor &  \\
 \\
 0, 18, 35/2 &   $18^{+} -18^{-}$ &  265731.67  & 2.00  & $-$0.9819 & R Hya &  9 \\
 0, 18, 35/2 &    $18^{+} -18^{-}$ &  265733.29 & 1.38 &    0.6351 & W Hya  &  \\
 0, 18, 35/2 &   $17^{+} -17^{-}$ &  265763.71 & 1.37 &    0.3956 & W Hya &  \\
\\
 0, 16, 33/2 &   $16^{+} -16^{-}$ & 317390.81 & 1.85 & $-$1.5320 & R Dor &   \\
 0, 16, 33/2 &    $17^{+} -17^{-}$ &  317408.68 & 1.60 & $-$2.2051 & R Dor  &  \\
 \\
 0, 17, 35/2 &   $17^{-} -17^{+}$ &  351583.57 & 1.78 & $-$3.4769 & R Dor  &  \\
 0, 17, 35/2 &  $18^{-} -18^{+}$  &  351602.50 & 1.77 & $-$3.1278 & R Dor  &  \\
 \\
 1, 10, 21/2 &    $10^{+} -10^{-}$ &  130639.84 & 0.82 &    0.9539 & W Hya  &  \\
 1, 10, 21/2  &  $11^{+} -11^{-}$  & 130653.94 & 0.74 & $-$0.4141 & W Hya  &  \\
 1, 17, 35/2 &  $17^{-} -17^{=}$ &  333386.59 & 1.71 & $-$1.7442 & W Hya   & \\
 1, 17, 35/2 &  $18^{-} -18^{+}$ &  333405.24 & 1.87 & $-$0.4388 & W Hya   & \\
 \\
\hline
\end{tabular}
\end{center}
\tablefoot{
$^{(a)}$~Quantum numbers  defined in Sect. \ref{sec:spec_backgr_OH}.
$^{(b)}$~Observed frequency and uncertainty, rounded to 2 dp, from this work (line number in last column) and from \citet{khouri2019}. 
$^{(c)}$~O$-$C derived from observed frequency (third column in this Table) minus calculated frequency (see frequency column in $\varv = 0,1$ OH frequency Tables in Appendix \ref{sec:Lambda_doubling_freqs}) and rounded to 4 dp.   
$^{(d)}$~Observed sources are: R Hya, R Aql and S Pav (this work) and,  IK Tau, R Dor, W Hya \citep{khouri2019}. 
$^{(e)}$~Line number as in Table \ref{OHline_list}.
}  
\end{table*}

%%%%%%%%%%%%%%%%

\subsection{Identification of OH lines, improving $\Lambda$-doubling frequencies}
\label{sec:identif_OH}

Identification of the OH lines in the {\sc atomium} sample primarily rests on the JPL catalog line frequencies. 
 We have used the observation of a spectral feature at the expected frequency and the observation in our channel maps and spectra of two nearby features with a frequency separation as predicted from an OH  ${\Lambda}$-doublet as a  secure identification of a ${\Lambda}$-doublet. Four OH ${\Lambda}$-doublets in $J = 27/2, 29/2, 33/2$ and $35/2$ corresponding to seven, and possibly eight, different ${\Delta} F$~=~0 hyperfine transitions have been identified. 
 As for water in  Sect. \ref{sec:detectwater}, we have verified that there is no misidentification with the lines in the {\sc atomium} chemical inventory. OH line~10 in $J = 35/2$ is still uncertain in R~Hya (see footnote in Table \ref{OHline_list}). 
In Mira  (Fig.~\ref{OH_Mira_Band6}), like in R~Hya, the  $J = 35/2$ line~9 is weakly detected  while 
 line~10 is  blended with the relatively strong  $\varv = 0$, $J=24_{2,22} - 24_{1,23}$ transition of TiO$_{2}$ at 265770.5~MHz. (This TiO$_{2}$ transition is not present in our R~Hya data.)
We also note the proximity, but without spectral confusion, of the  OH $J = 29/2$, $F'-F'' =15-15$ transition (line 8) with the (1,0,0)$-$(0,2,0)  $J_{K_a,K_c}=7_{4,3} - 8_{5,4}$ transition of water at 252.172 GHz (see right-hand panels in Fig. \ref{rhya-spav-raql-OH}). The last column in Table \ref{OHline_list} shows that the most frequently detected hyperfine line pairs are observed in the $J$ = 27/2 and 29/2 levels.

For a closer comparison of the observed OH transitions  with the JPL catalog frequencies  we have derived the hyperfine transition frequencies in the $J$ = 27/2 and 29/2 states after we have extracted the OH spectra for an aperture diameter of  0\farcs08 from the high resolution data cubes. The OH line profiles are often slightly asymmetric. However, 
our stronger OH sources, R~Hya, S~Pav or R~Aql, show well-identified line peaks which we have used with the LSR stellar velocity adopted during the observations to derive the observed rest frequencies. We have also verified that in R~Hya, the observed peak frequencies are identical within 0.5 MHz for spectra extracted for a total size region larger than about 
0\farcs1. Optical depth effects of these weak OH lines are not expected to play any significant role in our frequency determinations and, to minimize any possible frequency skew  due to gas kinematics, we adopt the average of our frequency measurements obtained from independent observations in different stars as our OH rest frequencies; they are given in Table~\ref {Astro-Lines-Fit} together with an adopted uncertainty of 1.5 MHz. The lines~6 and 9 in $J = 33/2$ and 35/2, were observed in one star only,  R~Hya. In that case the uncertainty is determined from the difference between, the intensity-weighted average of all channels with detected emission, and the direct frequency average of these channels; it is maximized to 2~MHz  (Table~\ref {Astro-Lines-Fit}).

Our OH frequency measurements differ by $\sim$2$-$4 MHz from the JPL frequencies in Table~\ref{OHline_list} thus exceeding the uncertainties estimated by combining our estimated uncertainties with those quoted in the JPL catalog. (The O$-$C in Table \ref{Astro-Lines-Fit} are nearly always smaller than $\sim$1$-$2~MHz.)
Similarly, \citet{khouri2019} have noted deviations of up to a few MHz between the calculated $\Lambda$-doubling 
transitions and their radio observations. These discrepancies are, on average, systematic, increasing with $J$ and $\varv$, and
cannot be explained  by an uncertainty in the observed star's systemic velocity  or by some intrinsic high velocity motions of the gas where the OH lines are excited.  In addition, we do not see in our data that the OH emission comes from a single side of the star's limb. 
We give at the beginning of  Appendix~\ref{sec:Lambda_doubling_freqs} the various limitations in the laboratory spectroscopic measurements that can explain why our observations, as well as those of \citet{khouri2019}, suffer from deviations up to a few MHz between the calculated  and observed $\Lambda$-doubling transitions in high-$J$ levels. 
Details on  our improved $\Lambda$-doubling calculations for energy levels in the $\varv = 0$ and 1 states from $\sim$1300 to 
$\sim$10500~cm$^{-1}$ are presented in four Tables of Appendix~A:  A 1 and  A 2 for the $\varv = 0$, $^2\Pi _{3/2}$ and $^2\Pi _{1/2}$ states; A3 and A4 for the $\varv = 1$, $^2\Pi _{3/2}$ and $^2\Pi _{1/2}$ states\footnote{Further calculations of the OH rotational spectra in $\varv = 0$ to 2 are given in the CDMS catalog \citep{muller2005, endres2016}. OH entries and our fit results and line list with frequencies from the radio astronomy observations are available from https://cdms.astro.uni-koeln.de/classic/entries/archive/OH/.}.

\section{H$_{2}$O source properties} 
\label{sec:line_properties}

\subsection{Observation of highly excited transitions of water}
\label{sec:detect_H2O}

%%%%%%%%%%%%%%%%%%%%%

\begin{table*}
\begin{center}
\caption{H$_{2}$O peak flux density (first entry in mJy) and velocity extent  (second entry in km\,s$^{-1}$, highlighted with italics) of observed lines for an aperture diameter of 0\farcs08 extracted from the extended configuration$^{a}$. The approximate frequency is given in GHz below each line number (quantum numbers and exact frequency given in Table \ref{H2O-line-list}).} 

\label{fluxdensity_source_list} 
\begin{tabular}[t]{lccccccccccccccr@{}lr@{}lr@{}ll}
\hline \hline
Star & \multicolumn{1}{c}{Line 1} &  \multicolumn{1}{c}{Line 4} & \multicolumn{1}{c}{Line 5} & \multicolumn{1}{c}{Line 6} & \multicolumn{1}{c}{Line 7} & \multicolumn{1}{c}{Line 8} & \multicolumn{1}{c}{Line 10} & \multicolumn{1}{c}{Line 12} & \multicolumn{1}{c}{Line 13} & \multicolumn{1}{c}{Line 14} \\
 & 222.014 & 236.805 & 244.330  & 252.172  & 254.040  & 254.053 & 259.952 & 262.898 & 266.574 & 268.149\\
 
\hline
 S Pav &   15  & --- & $-10$$^{b}$ & 18   & 9  &  7  & 28  & 31  & 18 &   173   \\
   & $\it{10.5} $  & &  $\it{11.3}$ & $\it{11.6}$ &  ?$^{d}$ & ?$^{d}$   & $\it{12.4}$ & $\it{13.4}$ & $\it{6.6}$  & $\it{19.7}$ \\
 
 T Mic &    --- &  --- & ---  & 15 & 10  & abso.?$^{e}$  & 16   & 21  & 11 & 99    \\
 &    & & & $\it{10.4}$ & $\it{8.6}$ &    & $\it{7.9}$ & $\it{8.9}$ & $\it{5.5}$  & $\it{22.9}$ \\

 U Del &  ---   &  ---   & ---  &  --- &  ---  &  --- &  --- &  --- &   --- & 5     \\
  & & & & & &  &    &   &  & $\it{14.5}$ \\

 RW Sco &  ---  & ---   &  --- &  --- &   ---  &  --- &  ---  & 25  &   --- & 93 \\
  & & & &    &  &  &  &  $\it{5.6}$ &   & $\it{10.9}$  \\
 
 V PsA  &  ---  & ---   &  --- &  --- &   ---  &  --- &  ---  & 14  &   --- & 23 \\
 &   & &  &  &  &  &   &  $\it{8.9}$ &   & $\it{16.4}$  \\
 
SV Aqr  &  ---  & ---   &  --- &  --- &   ---  &  --- &  ---  & --- & --- &  10  \\
 & & &  &  &  &    &  &   &   & $\it{5.5}$ \\

 R Hya &16 & 5? & $-$8$^{b}$ & 20  & 14  & 10   & 29  &  38  & 23  &  119  \\
 & \it10.5 & \it13.0? & \it12.0 & \it16.3 &  \it12.7 & \it8.1$^{d}$ &   \it15.8 & \it15.6 & \it12.1  & \it22.9 \\

U Her & 10 & ---  &  --- & 10  & 12  & ---  & 16  & 36  & --- & 209    \\
& \it6.6? & &  & \it7.0 & \it6.3 &  &   \it11.3 & \it15.6 &   & \it18.6 \\

AH Sco &  8   & ---   & ---  & 6  & 10 & ---  &33  & 20  & 9  & 68974    \\
&  \it7.9 &  & &  \it4.8 &  \it5.8 &  &  \it5.6 &  \it5.6 &  \it5.5  &  \it29$^{f}$ \\

 R Aql  &  10   & ---   & $-7.5$$^{b}$ & 22  & 10  & 7  &  18  &  43  & 21  & 220  \\
  &  \it14.3 & &  \it6.5 &  \it11.6 &   ?$^{d}$ &  ?$^{d}$ &    \it17.7 &  \it12.3 &  \it8.8  &  \it8.6 \\
 
 GY Aql &  ---  & ---   &  --- &  --- &   ---  &  --- &    25  & 35  & --- &  159  \\
  & & &  & & &    &  \it7.1 &   \it7.8 &   &  \it13.1  \\
 
 KW Sgr &  ---   &  ---   & ---  &  --- &  ---  &  --- &  --- &  --- &   --- &  10  \\
 & &  &  &  & &   &  &   &   &  \it8.5 \\

 IRC$-10529$ &  14  & --- & --- & ---  & 31  &  17  & 39  &  45  & --- &  92  \\
 &  \it6.2 &  &  & &  \it5.8 & ? &    \it9.0 &  \it6.7 &   &  \it6.5 \\
 
 IRC$+10011$ & 37  & ---  & abso.$^{c}$ &  19   & 49  & 13  & 86  & 100  & 43  &  1823   \\
  &  \it6.6 & & &  \it6.9 &   \it10.4 &  \it7.6 &   \it10.1 &  \it11.1 &  \it9.9  &  \it15.3 \\

 VX Sgr   & 8   & --- &  --- & 14 & 11  & ---  &  27  & 49   &  20  &   440   \\
 &  \it6.6 &  & &  \it7.0 &  \it5.8 & &   \it13.7 &  \it17.8 &  \it6.6  &  \it25$^{f}$ \\
\hline
\end{tabular}
\end{center}
\tablefoot{
$^{(a)}$~Typical peak flux density uncertainties are 1$-$4 mJy. The  velocity range is defined as the difference between the red and blue velocities for line emission above 2.5$\sigma$. A question mark indicates an uncertain value. No detection is indicated with ---. 
$^{(b)}~$Absorption line profiles and maps in Figs. \ref{rhya_spav_raql_abso_h2o} and \ref{abso_h2o_244}.
$^{(c)}$~Uncertain line parameters; zeroth moment map shows unresolved 3$\sigma$ absorption with the combined array (Fig.~\ref{abso_h2o_244}). 
$^{(d)}$~Uncertain velocity extent due to absorption on redshifted side of emission line. 
$^{(e)}$~Apparent redshifted absorption in extended and combined configuration data without emission. 
$^{(f)}$~Estimate of the 268.149~GHz velocity extent hindered by a nearby broad SO$_2$ feature. 
}
\end{table*}

\begin{figure*}
\centering \includegraphics[width=19.0cm,angle=0]{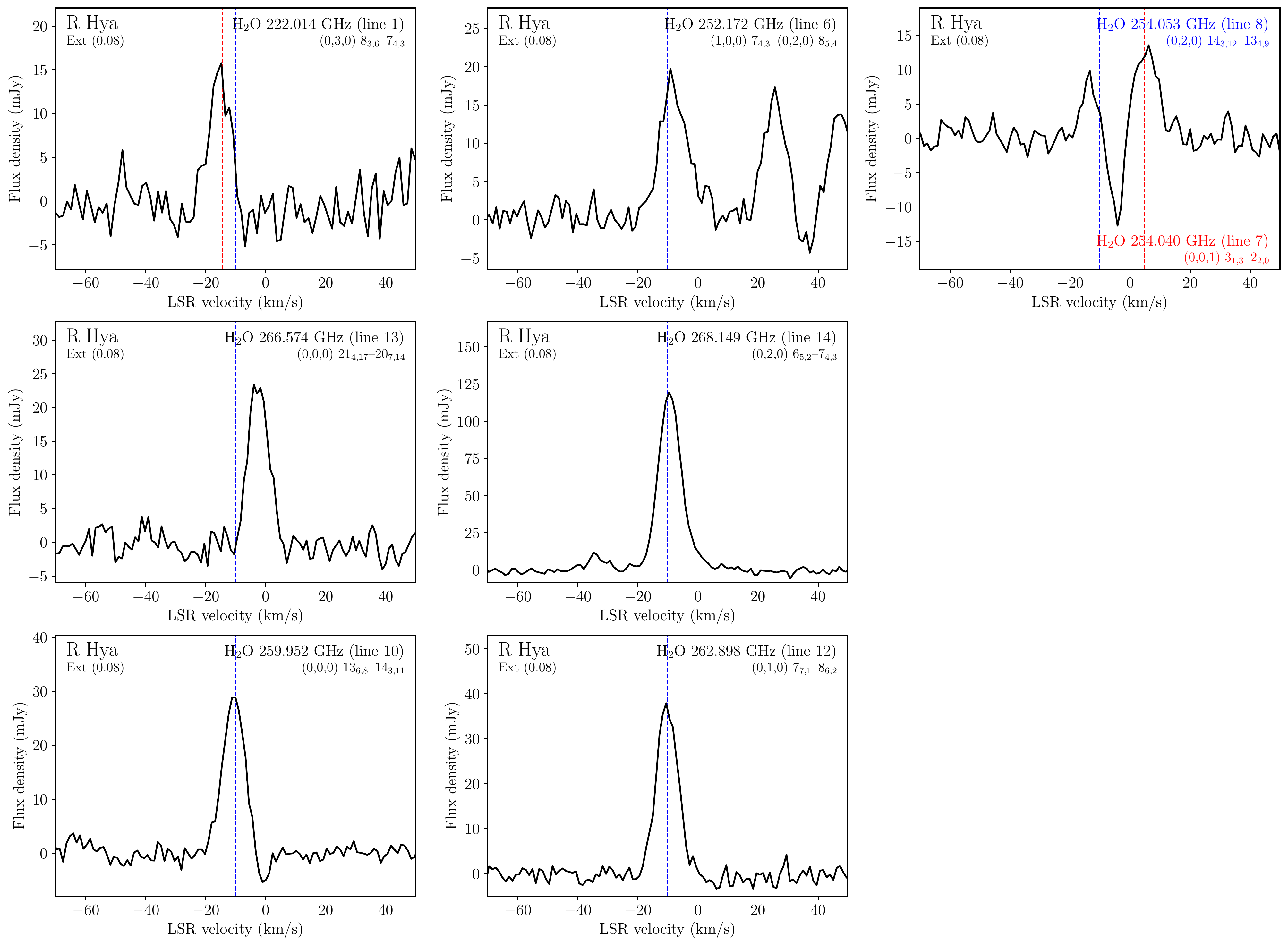}

\caption{Water line profiles extracted from the extended configuration of the main array for an aperture diameter of 0\farcs08 in R~Hya. $\it First$ $two$ $rows$: 
Six transitions of ortho H${_2}$O as defined in Table \ref{H2O-line-list}: lines 1, 6, 8 (including line 7 separated by 12.7~MHz, or 15.0~km\,s$^{-1}$, from line 8) and lines 13 and 14. $\it Last$ $row$:   Two transitions of para H${_2}$O as defined in Table \ref{H2O-line-list}: lines 10 and 12. 
The spectra are converted from the observed frequency to the LSR frame using the H${_2}$O catalog line rest frequencies given in Table \ref{H2O-line-list}. In all spectra, the blue vertical line indicates the adopted new LSR systemic velocity as shown in Table\ref{primarysource_list}. For line 1 (upper left panel), the red vertical line shows the LSR velocity for the slightly different frequency determined in this work. 
Spectral resolution varies from $\sim$1.1~km\,s$^{-1}$ (268 GHz) to $\sim$1.3 (222~GHz)~km\,s$^{-1}$.
}
\label{rhya_h2o_alllines}
\end{figure*}

\begin{figure*}
\centering \includegraphics[width=16cm,angle=0]{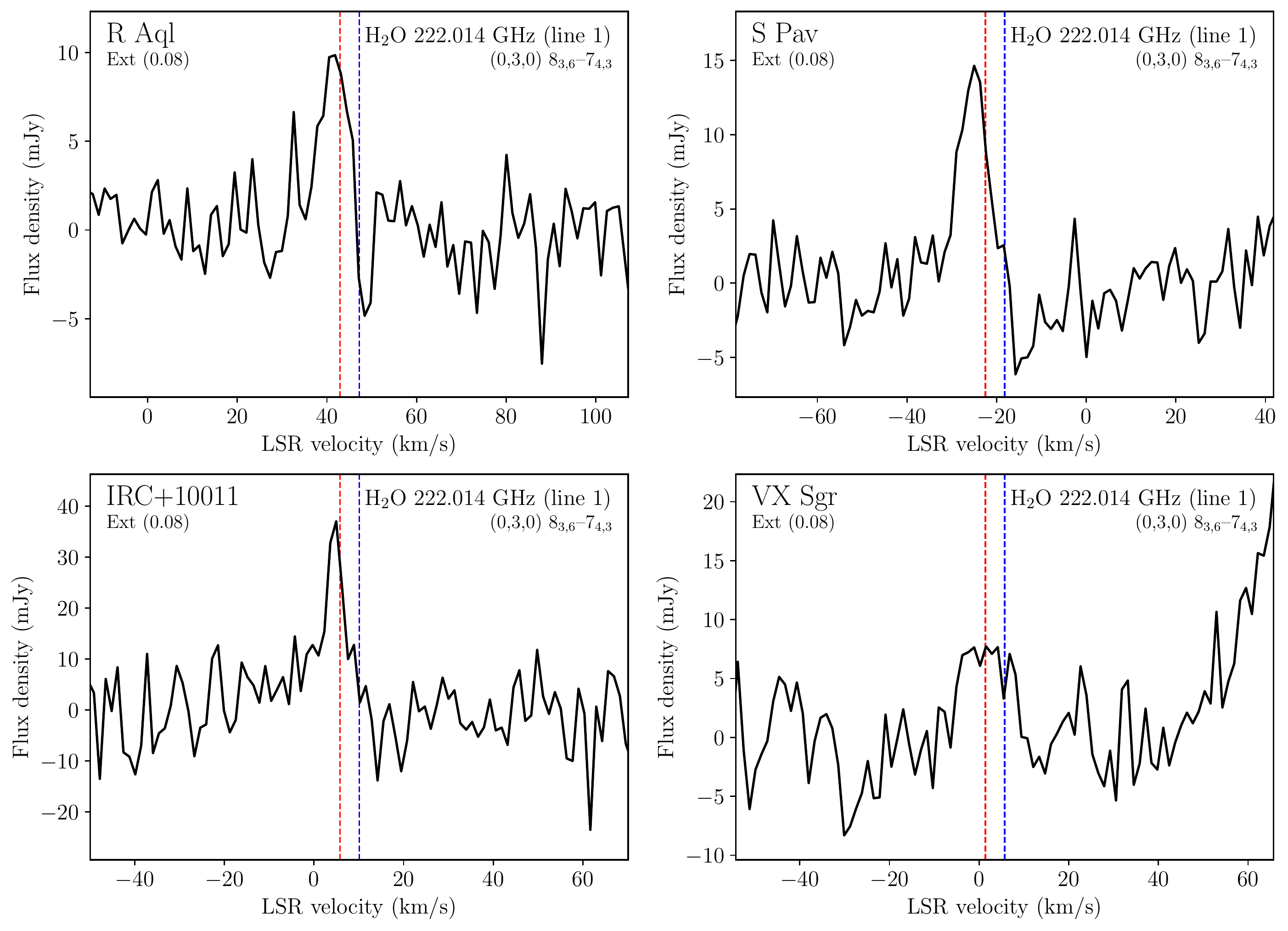}

\caption{Typical line profiles of the  (0,3,0) $8_3,_6-7_4,_3$ transition of H$_2$O 
at 222.014~GHz in R Aql, S Pav, IRC$+$10011 and VX~Sgr. The upper left panel in Fig. \ref {rhya_h2o_alllines} shows the same transition in R~Hya. Spectra are  extracted for an aperture diameter of  0\farcs08 from the extended configuration and  converted from the observed frequency to the LSR frame using the H${_2}$O catalog line rest frequency given in Table \ref{H2O-line-list}. 
The red and blue vertical lines indicate the new LSR systemic velocity (see Table \ref{primarysource_list}) corresponding to our frequency determination and to the catalog frequency, respectively. The spectral resolution is $\sim$1.3~km\,s$^{-1}$.
}
\label{222GHz_h2o}
\end{figure*}

%%%%%%%%%%%%%%%%%%%%%%%

As described in Sect. \ref{sec:detectwater}, ten different transitions of water have been identified on the basis of a close coincidence  of an observed spectral feature with a transition in the JPL or the W2020 catalogs. 
An overview of the H$_2$O spectral line properties extracted from the high resolution image cubes for a circular aperture of 0\farcs08 can be found in Table \ref{fluxdensity_source_list} which brings together the peak line flux density and the velocity extent as defined in Sect.~\ref{r}. Comparing the high resolution spectra with those extracted from the mid resolution data in a 0\farcs4 diameter aperture shows that there is no systematic difference in intensity and we consider that the 0\farcs08 aperture reveals the spectral behavior of the most compact structures  close to the star and within the maximum recoverable angular size. 
We also point out that our mid resolution data  do not reveal any new H$_2$O transition in any source.

Examples of H$_2$O spectra extracted from the high resolution data cubes in R~Hya are presented  in Fig.~\ref{rhya_h2o_alllines}; eight out of ten lines of both the  ortho and para species are strongly detected in this source. 
All spectra in all sources have been gathered in Appendix \ref{plot_H2O_lines} 
except those for the faint line~4 emission discussed in Appendix \ref{sec:254.23_vxsgr_ahsco}. H$_2$O line~5 absorption is shown in Fig.~\ref{rhya_spav_raql_abso_h2o}  and in Appendix \ref{plot_H2O_lines}. As far as we know, the (0,3,0)~$8_{3,6} - 7_{4,3}$ transition at 222.014~GHz (line~1) is the first ever mm-wave transition observed in the (0,3,0) vibrational state toward several AGBs and RSGs  (see Fig.~\ref {222GHz_h2o} and all detections in Appendix  \ref{plot_H2O_lines}). \label{plot_H2O_lines}
(Throughout this paper we only quote once the vibrational triplet for pure rotational transitions whose quantum numbers are specified without adding $J_{K_a,K_c}$; for ro-vibrational transitions we give the two vibrational triplets.)

Because the transitions in Table~\ref{fluxdensity_source_list} arise from high-lying levels between $\sim$4000 to 9000~K (note that line~4 at 236.805~GHz has the highest excitation energy), the observed peak line intensities are generally low. Our new detections significantly increase the number of H$_2$O rotational transitions that are currently observed with ground-based radio telescopes in  evolved stars. The range of excitation energies now extends to much higher values than those observed previously ($\sim$$470~{\rm{to}}~2400$~K). From space, however, water line detections have been reported at high energy levels  (e.g., up to $\sim$7700 K in VY CMa; Alcolea et al. 2013). Apart from line~4 which was observed in only one source (or perhaps two) and line~5 in four sources, all other 
lines in Table~\ref{H2O-line-list} are identified in 4 to 15 {\sc atomium} sources (see Table~\ref{fluxdensity_source_list} and column $n_{det}$ in Table~\ref{H2O-line-list}).

S~Pav, R~Hya, R~Aql and IRC$+10011$ are the sources in our sample with the richest water vapor spectra  
 (Appendix~\ref{plot_H2O_lines}). All H$_2$O lines observed in the present  work are excited 
 in these four sources, except line 4 which is only observed in R~Hya (and perhaps in S~Pav). 
On the other hand, four other sources exhibit only two or three water transitions (RW~Sco, V~PsA, SV~Aqr and GY~Aql) and two stars which are generally line-poor, U~Del and the distant SRc variable KW~Sgr, are weakly detected at 268.149~GHz only. We also note that in IRC$+10011$, despite its large distance, the peak flux density of each detected transition, except at 268.149~GHz, tends to be stronger than in other sources.

Although our source sample is small, it does  not seem  to show any dependence of the line detection rate with physical parameters such as the mass-loss rate; in fact, the star with the lowest mass-loss rate, SPav, exhibits as many detected transitions as our highest mass-loss rate star, VX~Sgr.  This is confirmed by the correlation analysis of Wallstr\"om et al. (in preparation) between several physical parameters and the number of H$_2$O lines. 
We note that the line source detection rate observed here tends to increase as the energy of the transition decreases, that is to say lower states ($\sim$4000$-$5600~K) are detected in nearly two times more sources than in higher states ($\sim$8000$-$9000~K). 
Such  a trend is not surprising ''a~priori'' since we expect the highest energy levels to be less easily populated.

\subsection{Water line absorption}
\label{sec:abso_H2O}

%%%%%%%%%%%%%%%%%%%%%%%%
\begin{figure}
\centering \includegraphics[width=8.8cm,angle=0]{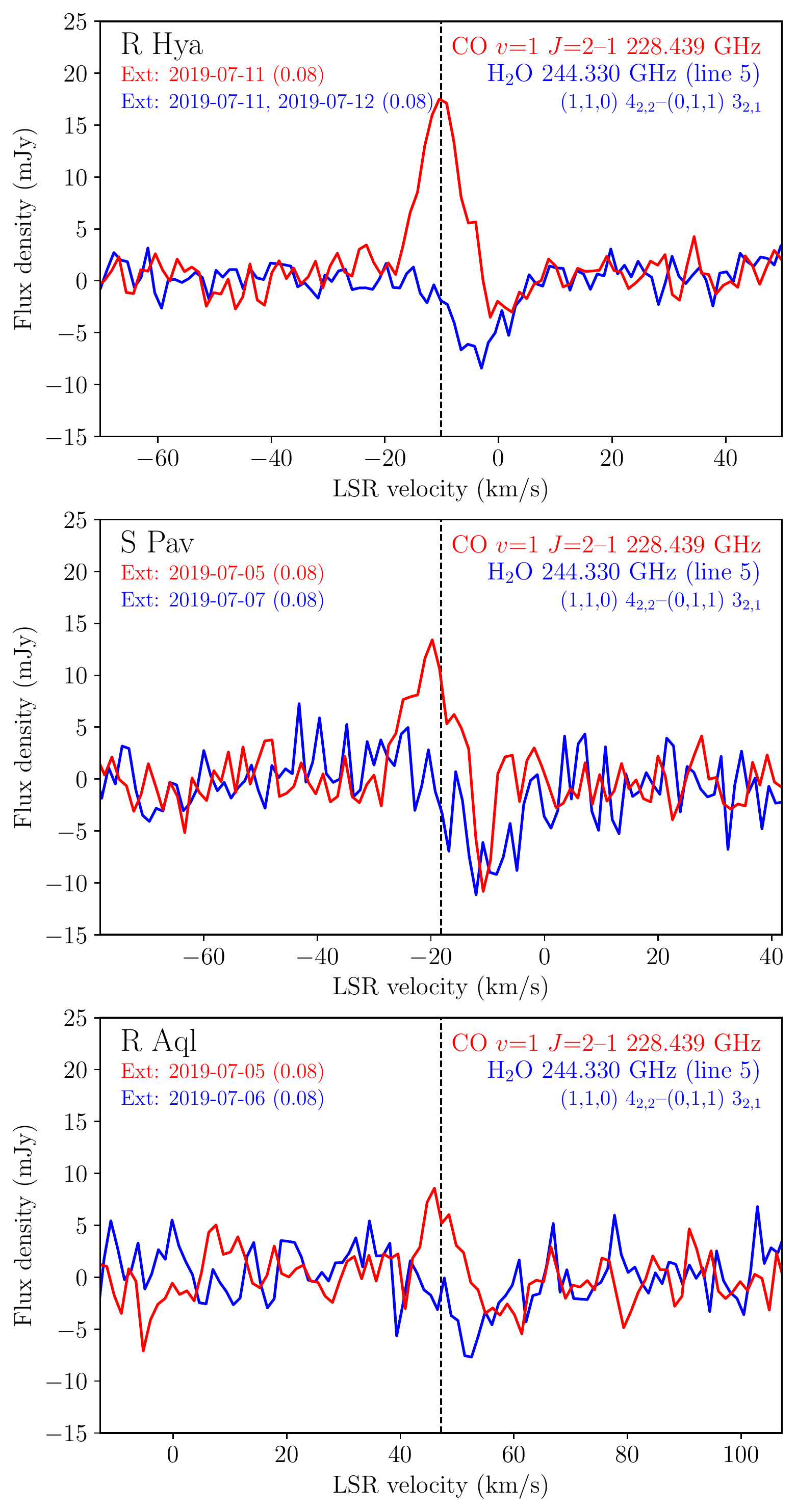}
\caption{
Absorption spectra of para H${_2}$O 
at 244.330~GHz  in R~Hya, S~Pav and RAql (blue profiles) and, emission/absorption spectra of the $\varv=1$ transition of CO(2$-$1) at 228.439~GHz in the same sources (red profiles). The spectra are converted from the observed frequency to the LSR systemic velocity using the H${_2}$O line~5 (Table~\ref{H2O-line-list}) and $\varv=1$, CO(2$-$1) rest frequencies.   All spectra are extracted from the high resolution data cubes for an aperture diameter of 0\farcs08. The vertical black dotted lines indicate the adopted new LSR systemic velocities (see Table \ref{primarysource_list}).
}
\label{rhya_spav_raql_abso_h2o}
\end{figure}

%%%%%%%%%%%%%%%%%%%%%%%%

The absorption of a water line  was  first observed  with the longest baselines of ALMA toward Mira by \citet{wong2016}. Their spatially resolved images of the (0,2,0)~$5_{5,0}$--$6_{4,3}$ 
transition at 232.687 GHz reveal H${_2}$O line absorption against the background continuum and a line emission region that closely corresponds to that of the highly excited $\varv=2$,~SiO line. In our {\sc atomium} high-angular resolution data, we also observe both line emission, which is dominant, and weak absorption. 
 At 259.952~GHz (para H${_2}$O line 10),  absorption is detected at the level of $\sim$5$-$15~mJy for the extended configuration and  0\farcs08 aperture diameter, together with relatively strong emission toward R~Hya (Fig. \ref{rhya_h2o_alllines}), R~Aql, S~Pav and U~Her.  Absorption is also seen in one or two of the nearby frequencies at 254.040 and 254.053 GHz (lines 7 and 8 in two different vibrational states) in the four objects cited above and in IRC$-10529$, IRC$+10011$ and T~Mic. In R~Hya, an absorption feature is observed in Fig. \ref{rhya_h2o_alllines} on the redshifted side of the (0,2,0)~$14_{3,12}$--$13_{4,9}$ main emission line profile at 254.053~GHz  and is confirmed by the absorption map of this feature (see end of Sect.~\ref{sec:absomaps_H2O}). 
The para H${_2}$O line profile at 244.330 GHz (line 5) shows only absorption, which is in contrast with, for example, the 
  $\varv=1$,~CO(2$-$1) line profile, expected to be excited in a similar region and that shows both absorption and emission.  
The spectra of the stronger 244.330~GHz absorption sources are presented  in Fig. \ref{rhya_spav_raql_abso_h2o} and compared with those of  the $\varv=1$, $J= 2 -1$ high energy transition of CO ($\sim$4400~K). 
The water line absorption  is spectrally narrow and lies on the red side of the bulk of the CO emission profiles suggesting an infall of the water-bearing matter (see Sect.~\ref{sec:absomaps_H2O}).

\subsection{Channel maps, angular sizes, ring-like structures} 

\label{sec:channmaps_H2O}

%%%%%%%%%%%%%%%%%
\begin{figure*}
\centering
\includegraphics[width=13.7 cm, angle=270]{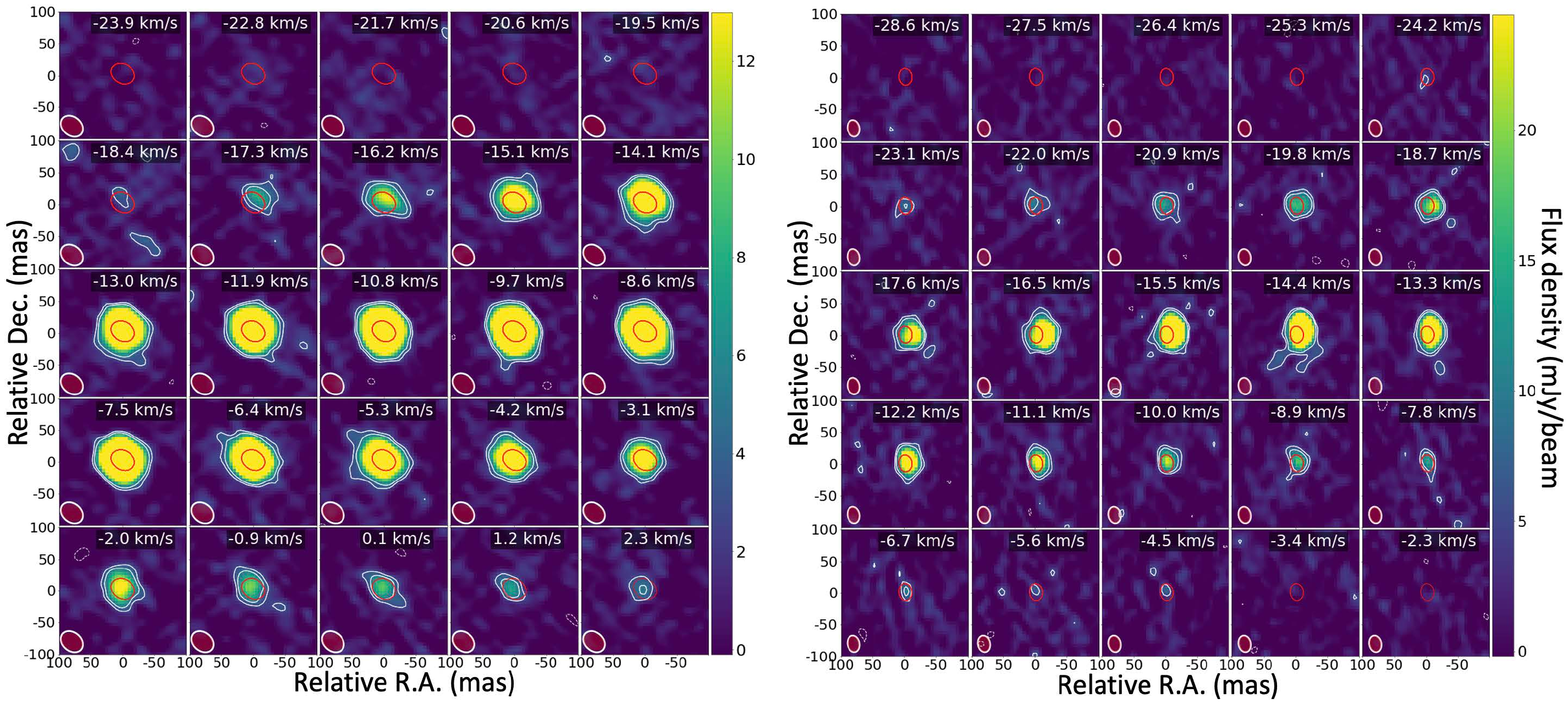}

\caption{High resolution channel maps of H$_2$O (0,2,0)~$6_{5,2}$--$7_{4,3}$ transition at 268.149~GHz in R~Hya and U~Her. Each map (R~Hya and U~Her, left and right panels)  shows offsets in the R.A. and Dec. directions which we call throughout this work "Relative R.A."  and "Relative Dec." The corresponding angular offsets cover 100$\times$100 mas from the continuum emission peak at (0,0) position (coordinates given in Table \ref{primarysource_list}). Each channel velocity is in the LSR frame from  $-$23.9 to 2.3~km\,s$^{-1}$ (R~Hya) and $-$28.6 to $-$2.3~km\,s$^{-1}$ (U~Her). White light contours are at $-3$, 3 and 5$\sigma$. A few negative contours,when present, are dashed. The line peak and typical r.m.s.  noise are 65 and 1 mJy/beam (R~Hya), and  122 and 1.5 mJy/beam (U~Her). The red contour at (0,0) delineates the extent at half peak intensity of the continuum emission. We characterize the elliptical Gaussian clean beams by their major and minor axes and position angle (PA) at half power, hereafter HPBW clean beam parameters. For the line observations of R~Hya and U~Her, these are (38$\times$29)~mas at PA~48$^{\circ}$ and (26$\times$19)~mas at PA~11$^{\circ}$, respectively. The corresponding continuum parameters are (34$\times$25)~mas at PA~67$^{\circ}$ for R~Hya and (24$\times$18)~mas at PA~8$^{\circ}$ for U~Her. The line and continuum beams are shown at the bottom left of each map in white and solid dark-red, respectively. The scale of the line flux density per beam (in mJy/beam) is linear and shown in the vertical bar on the  right side of each channel map.
}
\label{chan_rhya_uher_h2o_268}

\end{figure*}

%%%%%%%%%%%%%%%%

\begin{figure*}
\centering 
\includegraphics[width=13.5cm, angle=270]{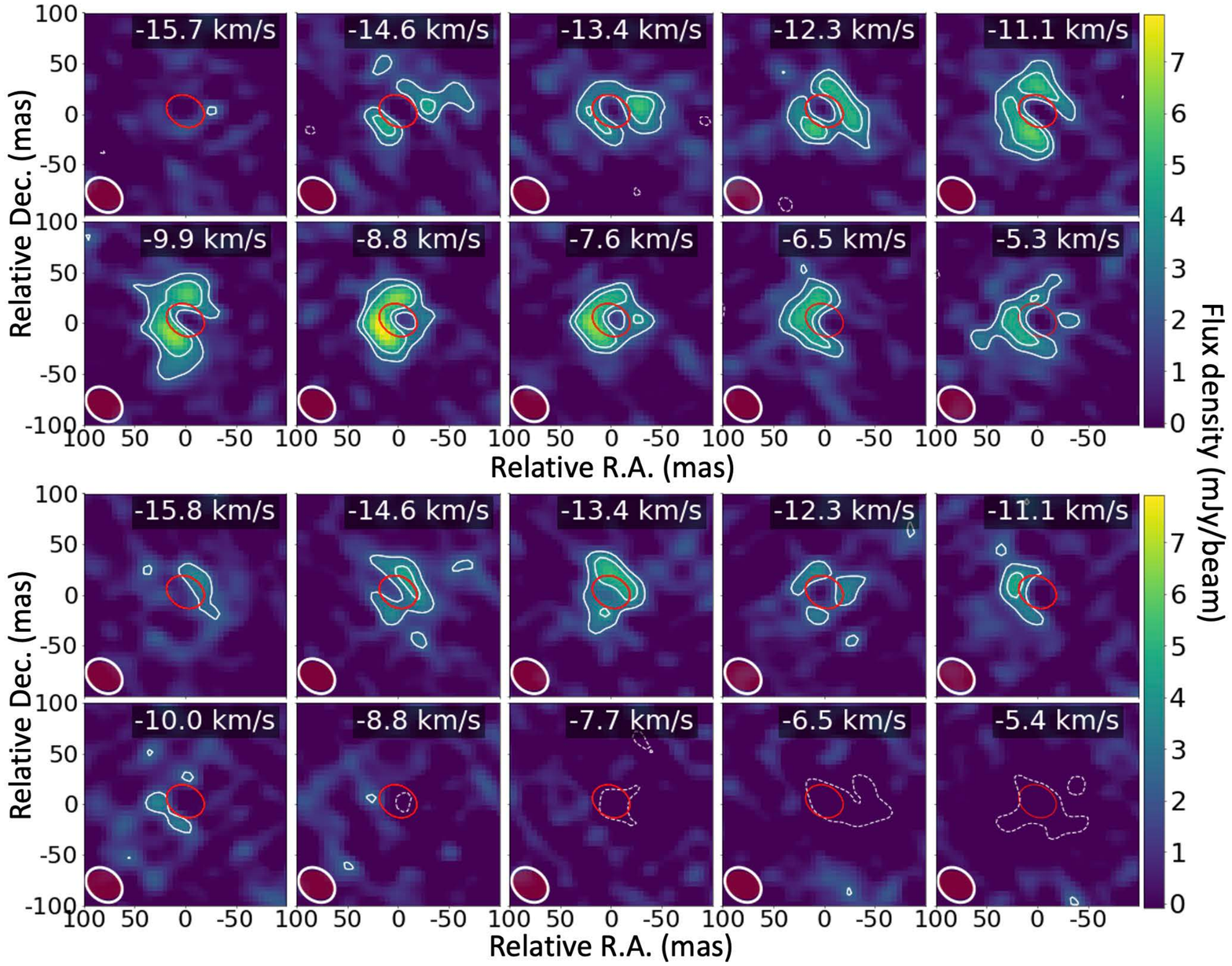}

\caption{High resolution channel maps of the (0,0,1)~$3_{1,3}$--$2_{2,0}$  and (0,2,0)~$14_{3,12}$--$13_{4,9}$ transitions of water near  254~GHz in R~Hya. The upper and lower panels correspond to the 254.040 GHz and 254.053~GHz transitions, respectively (lines~7 and 8 in Table~\ref{H2O-line-list}). Caption as in Fig. \ref{chan_rhya_uher_h2o_268} except for the velocity range and the line peak, 8~mJy/beam; the typical r.m.s. noise is 1~mJy/beam. The HPBW  is (39$\times$30)~mas at PA~49$^{\circ}$ and (34$\times$25)~mas at PA~67$^{\circ}$ for the line and  continuum, respectively. 
}
\label{chan_rhya_h2o_254}
\end{figure*}

%%%%%%%%%%%%%%%

Channel maps  have been produced for each source in our sample for both the extended and mid array configurations and for all detected transitions listed in Table \ref {H2O-line-list}. The angular sizes and structure of the H$_2$O emitting regions observed with the extended configuration are discussed here while the  line~5 absorption is presented in Sects. \ref{sec:absomaps_H2O} and \ref{sec:infall}. 
To illustrate the discussion on the H$_2$O emission,  
we have selected channel maps for transitions  that reflect the variety of excitation conditions, that is maps of the:  
$\it(i)$~strong (0,2,0)~$6_{5,2}$--$7_{4,3}$ emission at 268.149 GHz (line 14), but with relatively high excitation energy levels, in R~Hya and U~Her (Fig. \ref{chan_rhya_uher_h2o_268}), in S~Pav  and  IRC$+10011$ (Figs.~\ref{chan_spav_h2o_268} and \ref{chan_irc+10011_h2o_268}),  and two RSGs, VX Sgr (Fig.~\ref{chan_vxsgr_h2o_268}) and AH Sco (Fig.~\ref{chan_ahsco_h2o_268});
$\it(ii)$~two relatively weak transitions, with close frequencies but with well separated energy levels, (0,0,1)~$3_{1,3}$--$2_{2,0}$ and (0,2,0)~$14_{3,12}$--$13_{4,9}$ (lines 7 and 8)  at 254.040 and 254.053~GHz in R~Hya (Fig. \ref{chan_rhya_h2o_254}); 
$\it(iii$)~relatively low energy transitions 
(0,0,0)~$13_{6,8}$--$14_{3,11}$ at 259.952~GHz (line10 in R~Hya, S~Pav, IRC$+10011$ and VX~Sgr,  Figs.~\ref{chan_rhya_spav_259} and \ref{chan_i10011_vxsgr_259}) and  (0,1,0)~$7_{7,1}$--$8_{6,2}$  at 262.898~GHz (line~12 in R~Hya, U~Her, S~Pav, IRC$+10011$,  VX~Sgr and AH~Sco, Figs.~\ref{chan_rhya_uher_262}, \ref{chan_spav_i+10011_262} and \ref{chan_vxsgr_ahsco_262}).

The majority of the water emission is closely associated with the underlying continuum source although the emission peak may not exactly coincide with the central star; for more details, readers can refer to the channel maps in R~Hya and U~Her at 268.149~GHz (Fig. ~\ref{chan_rhya_uher_h2o_268}) or at 222.014~GHz in R~Hya (Fig. \ref{chan_rhya_spav_222}), for example. Weak, diffuse emission is also observed  away from the central object and, in a few cases, apparent ring-like structures are present in our channel maps (see discussion at the end of this Section).

 The typical angular sizes of the nonmaser line emission regions are obtained in two different ways from the high resolution channel maps without de-convolving the data from the clean beam since the surface brightness distribution is irregular and not Gaussian at the high resolution achieved here. In a first approach, sizes are derived from the geometric mean  of the maximum and minimum angular radii of the emission centered on the star and enclosed within the 3$\sigma$ contour, noting that the fitting accuracy  to this contour is on the order of one third of the beam.
 In the second approach, the radial distances are  obtained from azimuthal averaging of the  emission  within the 3$\sigma$ contour of the moment~0 maps in a manner  similar to that used  by \citet{danilovich2021} in W~Aql.   
 Compiling our H$_2$O measurements  from the  two approaches above we find that the angular radii cover the range $\sim$10$-$50~mas for all stars and all detected lines (reaching $\sim$60~mas for line~14 in AH~Sco and  IRC$+$10011). These radii may  differ by a factor of $\sim$2$-$3, for different lines detected in a given star, and a factor of $\sim$3$-$4 for the three more widespread transitions observed in different stars (lines 10, 12 and 14).

 The largest angular separation from the central star to the 3$\sigma$ contour gives an estimate of the maximum H$_2$O excitation size. It  varies in the range $\sim$15$-$65~mas in general for all detected lines in all stars (except in AH~Sco where it reaches  83~mas at 268.149~GHz). These  maximum sizes correspond to $\sim$2.5$-$12~R$_{\star}$ when they are compared to the optical radius of the central star  and even reach $\sim$20 and 29~R$_{\star}$ at 268.149~GHz in IRC$+$10011 and AH~Sco, respectively.

 The precise size of the actual H$_2$O molecular extent is difficult to assess. Firstly, it is important to note that the radio continuum disk size is larger at millimeter wavelengths than the photosphere and may have nonspherical symmetry or exhibit  outward motions  \citep{vlemmings2019}, suggesting that the inner gas layers within $\sim$15$-$50\,\% of the optical diameter could be  obscured to line emission. This is confirmed by our Band~6 continuum observations (central frequency $\sim$250~GHz)  of the {\sc atomium} stars for which we measure a uniform disk size $\sim$10$-$100\,\% larger than the optical angular diameter. In R~Hya \citep[see also][]{homan2021}, U~Her, and R~Aql, for example, the respective uniform disk sizes are 27.1, 18.5, and 15.0~mas, that is  $\sim$14\,\%, 65 \%, and 38\,\% larger than the optical diameter given in Table~\ref{primarysource_list}.  
 The size discrepancy reaches $\sim$100\,\%  for the two RSGs AH~Sco and VX~Sgr\footnote{Our Band~6 uniform disk sizes are 13.5 and 17.2~mas for AH~Sco and VX~Sgr, respectively. In Betelgeuse the radio continuum diameter \citep{ogorman2017} and the IR/optical diameter \citep{montarges2014} are 58 and 42~mas, respectively.}. 
 Secondly,  we point out that  the maximum radial extents  obtained  from the mid resolution moment~0 maps of water  can be well above 100~mas and may even reach several hundred~mas at 268.149~GHz in the RSG AH~Sco (Wallstr\"om et al., in preparation). A preliminary population diagram analysis in AGBs suggests that, at large distances from the star, H$_2$O tends to be thermally excited and its abundance is well  below that measured  within  the  inner gas layers ($\la$10$-$20~R$_{\star}$). Full analysis of  the sensitivity to low surface brightness emission in the mid resolution data   is beyond the scope of the present work and will be presented elsewhere.

There is not much apparent complexity in the continuum-subtracted high resolution channel maps of water. However,  the (0,0,1)~$3_{1,3}$--$2_{2,0}$ and (0,2,0)~$14_{3,12}$--$13_{4,9}$ transitions at 254.040 and 254.053 GHz exhibit a ring-like structure around R Hya
(Fig.~\ref{chan_rhya_h2o_254}). The outer ring diameter, depending on the channel velocity, 
varies from $\sim$70 mas to $\sim$90 mas and encompasses the continuum emission contour at half the peak intensity 
(red contour centered on the star in Fig. \ref{chan_rhya_h2o_254}) and extends over roughly four times the photospheric diameter. A similar structure is also well delineated in the 254.040~GHz velocity-integrated absorption map of S~Pav (Fig.~\ref {abso_emission_254-spav_rhya}), It is perhaps also present in R~Aql and T~Mic,  but is not apparent in the rather strong 254.040~GHz emission in IRC$+$10011 or VX~Sgr.

Overall, water emission from these high-lying transitions is predominantly detected in a patchy ring around the star. R~Hya is surrounded by $\varv=0$,~SiO  emission but the most compact peaks form a flattened ring with blue- (red-)shifted emission with respect to the stellar velocity in the south and west (north and east). \citet{homan2021} interpret this as a rotating disk inclined at an angle of about 30$^{\circ}$ to the line of sight. The R Hya water emanates from a similar region with some tendencies to a similar offset (Figs. \ref{chan_rhya_uher_h2o_268}, \ref{chan_rhya_h2o_254}, \ref{chan_rhya_spav_259}, 
 \ref{chan_rhya_spav_222}.) Other interpretations are possible for the  R~Hya observations. An emitting hollow shell with appropriate gas motions and geometry could explain the apparent water ring-like structure.  Supposing the hollow shell wind 
is rapidly accelerating (infall or outflow) the emission appears ring-like with the brightest region being close to the stellar velocity, because of greater velocity-coherent paths in the tangential direction, while the front and back caps are weak or not detected.  
The water lines discussed here sometimes show absorption against the central star, for example in Fig. \ref{chan_rhya_h2o_254}, so the front cap is not seen in emission whilst the back cap is obscured by the star. We suggest, although our source sample is small, that the absorption tends to be present in stars with a lower mass-loss rate and thinner circumstellar envelope. This is the case in  R~Hya and S~Pav (see spatial structure and discussion in Sect. \ref{sec:absomaps_H2O}) while in some other stars (e.g., IRC$+$10011 with a higher mass-loss rate, Table \ref{primarysource_list}) the observed  emitting gas is warmer than the underlying continuum source and there is no absorption.

 \subsection{Absorption maps, comparison of absorbing/emitting regions}
\label{sec:absomaps_H2O}

%%%%%%%%%%%%%%%%
\begin{figure*}
\centering \includegraphics[width=12.5cm,angle=270]{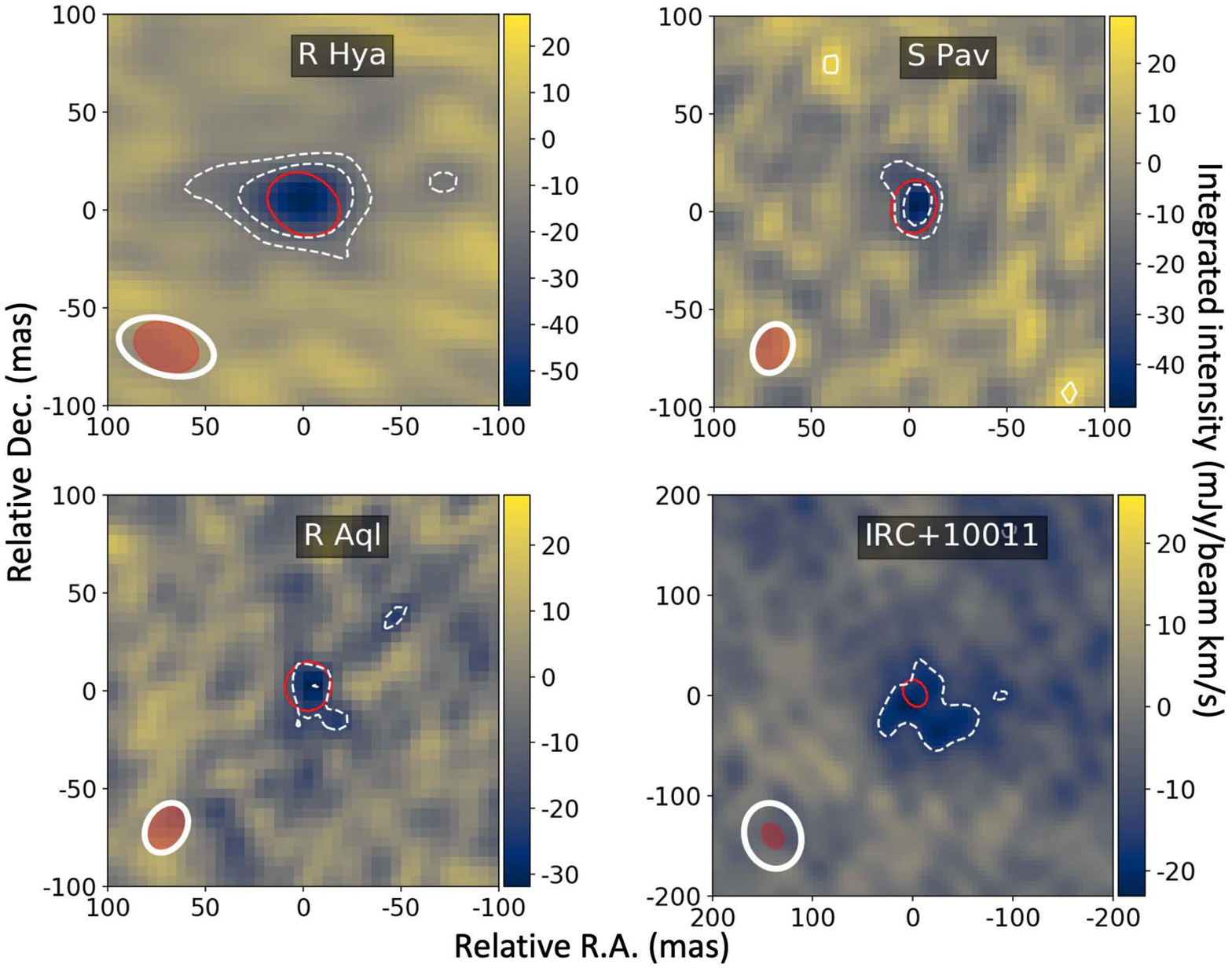}

\caption{Zeroth moment  absorption maps of  the (1,1,0)$-$(0,1,1) $J_{K_a,K_c} = 4_{2,2}$--$3_{2,1}$ transition of para H${_2}$O at 244.330~GHz in R Hya, S~Pav, R~Aql and IRC$+10011$. The extended configuration was used in R Hya, S~Pav and R~Aql while in IRC$+10011$ the  combined extended and mid arrays were used. The map field of view is 100$\times$100 mas except for IRC$+10011$ where it is 200$\times$200 mas. The dotted white lines delineate the $-3$ and 
$-5\sigma$ absorption contours. The red contour at $(0,0)$ position delineates the extent at half intensity of the continuum emission. The noise level is 5.4, 6.9, 5.2 and 6.3 mJy/beam~km/s for R~Hya, S~Pav, IRC$+10011$ and R~Aql, respectively. 
The HPBW line beam (white ellipse) is (50$\times$28)~mas at PA~75$^{\circ}$ in R~Hya, (26$\times$20)~mas at PA~$-$20$^{\circ}$ in S~Pav, (27$\times$20)~mas at PA~$-$30$^{\circ}$ in R~Aql and (59$\times$50)~mas at PA~12$^{\circ}$ in IRC$+$10011. The continuum beam (red ellipse) is (34$\times$25)~mas at PA =~67$^{\circ}$ in R~Hya, (25$\times$20)~mas at PA~$-$13$^{\circ}$ in S~Pav, 
(24$\times$22)~mas at PA~$-$13$^{\circ}$ in R~Aql 
 and (55$\times$44)~mas at PA~34$^{\circ}$ in IRC$+$10011. 
 (The velocity intervals are $-19.0.$ to 4.9, $-19.7$ to 4.3, $15.3$ to 21.2  and 35.3 to 64.1km.s$^{-1}$ for R Hya, S Pav,  IRC$+10011$ and R Aql, respectively.) 
}
\label{abso_h2o_244}
\end{figure*}

%%%%%%%%%%%%%%%%%%%%%%%%%%%%%%%

\begin{figure*}
\vspace{-2.5cm}
\centering \includegraphics[width=12.0cm,angle=270]{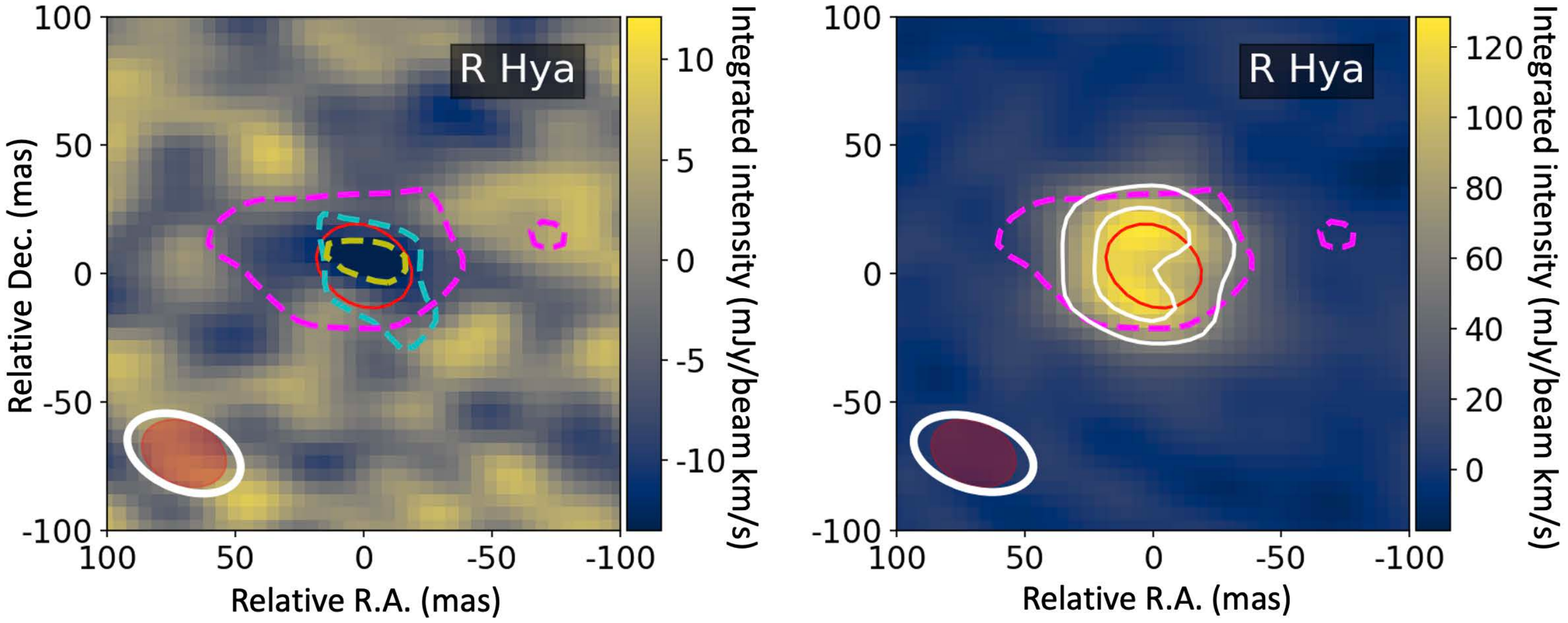}
\vspace{-2.5cm}
\vspace{-0.1cm}
\caption{Comparison of 
absorption and emission lines of water with $\varv=1$, CO(2--1) absorption in R~Hya. $\it Left$ $panel$: Magenta and cyan dashed contours delineate the $-5\sigma$ levels of the 244.330 GHz (line 5) and 259.952 GHz (line 10) mom~0 \emph{absorption} maps. The underlying map is the mom 0 absorption map of $\varv=1$, CO(2--1) with the yellow dashed contour at the $-3\sigma$ level. The line beam width is 50$\times$28~mas with 70$^{\circ}$ orientation (white ellipse in bottom left corner). The red solid contour delineates the 50\% level of the peak continuum emission (the continuum beam width is the dark-red ellipse in the bottom left corner). $\it Right$ $panel$: The magenta dashed contour and red solid contour indicate the 244.330~GHz absorption and mm-wave continuum emission as in the left panel. The underlying map is the 259.952~GHz mom~0 \emph{emission}, with the white solid contours at the 20 and 35$\sigma$ levels. (Line and continuum beam widths  as in the left panel.)
The noise level is 3 mJy/beam~km\,s$^{-1}$ for water in both panels and  4.3 mJy/beam~km\,s$^{-1}$ for CO. The velocity intervals of the mom 0 maps are:  $-13.2$ to 4.3  and $-2.3$ to 1.5~km\,s$^{-1}$ for the 244.330 and 259.952~GHz absorptions; $-18.8$ to $-3.0$~km\,s$^{-1}$ for  the 259.952~GHz emission of water; $-2.2$ to 6.0km\,s$^{-1}$ for the CO(2--1) absorption.
}
\label{abso_244_259_CO}
\end{figure*}

%%%%%%%%%%%%%%%%%%%%

Fig~\ref{abso_h2o_244} shows at 244.330~GHz the mom~0 absorption maps of R~Hya, S~Pav and R~Aql obtained with spatial resolutions in the range $\sim$35~mas (R~Hya) to $\sim$25~mas (S~Pav, R~Aql), together with the IRC$+10011$  mom 0 map obtained with $\sim$55~mas resolution with the combined high and mid resolution arrays. Detection is at the 3$-$5$\sigma$~level in front of the central star and  is essentially unresolved even at  our highest angular resolution. 
       
We have also compared the 244.330~GHz absorption with the mom~0 maps of the emission/absorption of water at 259.952~GHz,  and the mom~0 maps of the $\varv=1$, CO(2$-$1) emission/absorption at 228.439~GHz whose energy levels are close to those of the 259.952~GHz line. This is presented for R~Hya in Fig. \ref{abso_244_259_CO}, left panel,  where  the $\varv=1$, CO(2$-$1) absorption coincides with the stellar continuum and the  259.952~GHz  absorption of water. These absorptions are poorly resolved and coincide with the 244.330~GHz water absorption. 
The right panel in Fig.\ref{abso_244_259_CO} shows that the  259.952~GHz water $emission$, which is slightly resolved, and the $\varv=1$, CO(2$-$1) emission (not shown here for clarity) have a comparable size, $\sim$50 mas,  as the 244.330~GHz water absorption. 

The absorption feature observed in S~Pav and R~Hya between the two transitions of ortho H$_2$O at 254.040 and 254.053~GHz (see spectra in Fig.~\ref{line07_allsrc_ext} and in upper right panel of Fig.~\ref{rhya_h2o_alllines} for R Hya) 
has also been mapped (Fig.\ref{abso_emission_254-spav_rhya}). Our mom~0 maps are sensitive enough to reveal  spatially compact absorption structures with angular sizes as small as $\sim$20$-$40 mas, or slightly more in R~Hya, that are closely associated with the continuum emission from the central star. 
Such absorption features (see other examples in T~Mic or R~Aql (Fig.~\ref{line07_allsrc_ext}) are redshifted by a few km\,s$^{-1}$ with respect 
to the stellar systemic velocity and can be interpreted as being due to the presence of infalling matter (see  Sect.~\ref{sec:infall}).
 
\subsection{Tracing the inner-wind regions at small scales}
\label{sec:small_scaleh2o}

%%%%%%%%%%%%%
\begin{figure*}
\centering \includegraphics[width=13.5cm,angle=270]{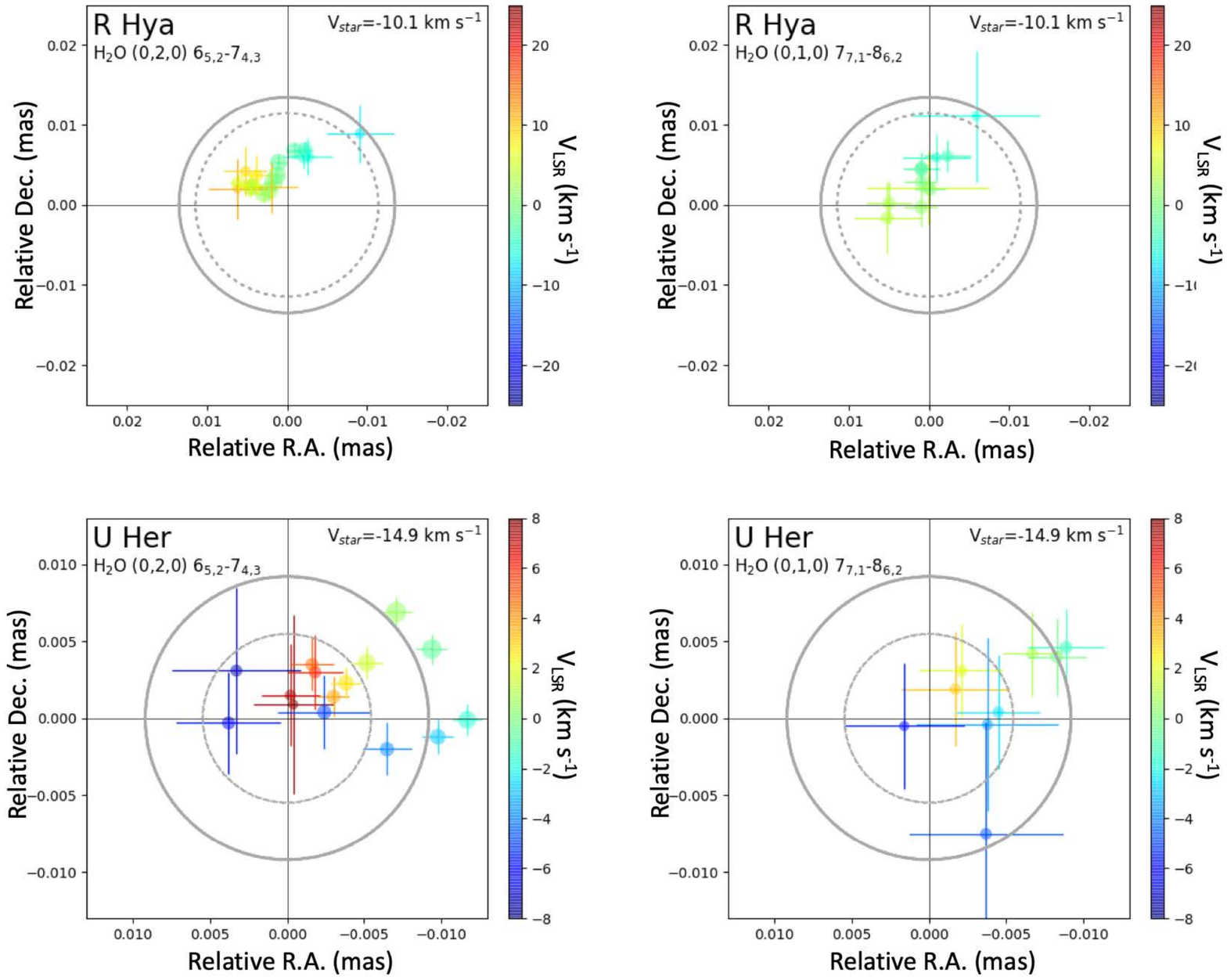}
\caption{Maps of the different velocity components of water identified in the Gaussian-fit procedure at 268.149 and 262.898~GHz toward R~Hya and UHer for the extended configuration of the main array. The size of the colored symbols varies as the square root of the integrated flux density of the Gaussian component; the crosses show the position uncertainty for each component. The velocity scale colors are given on the right side of each map with respect to the stellar system velocity taken to be $-10.1$ and $-14.9$~km\,s$^{-1}$ in the LSR frame for R Hya and U Her, respectively. The dashed gray circle represents the size of the optical photosphere (23.7 and 11.2~mas for R~Hya and U~Her, respectively) and the larger gray circle represents the 250~GHz continuum emission size (27.1 and 18.5~mas for R~Hya and U~Her).
 }
\label{rhya_uher_h2o_component}
\end{figure*}
%%%%%%%%%%%%%

The majority of the H$_2$O emission at 268.149 and 262.898~GHz, the two most widespread transitions, tends to show an organized "position versus velocity" structure in some stars. This is suggested in the channel maps where a slight displacement of the emission peak  is present in different channels  against the continuum emission source (e.g., 268.149~GHz line in R~Hya and U~Her, Fig. \ref{chan_rhya_uher_h2o_268}, or in IRC$+$10011, Fig. \ref{chan_irc+10011_h2o_268}). To better trace the deepest motions of the inner stellar wind, we have conducted 
 a Gaussian analysis of the most compact, nondiffuse regions of water emission mapped in different transitions using 
 the task SAD in AIPS\footnote{SAD (Search And Destroy) enables to search for potential sources within an image and fits Gaussian components. It  is useful for the analysis of complex images provided that they contain compact enough emission \citep[e.g.,][]{etoka2004,baudry1998}.}. 
The component fitting and sorting was performed here  as in  \citet[][]{etoka2004}. We have used 5 times the r.m.s. noise level in the spectral channels as our typical threshold to extract components. We then built tables of velocity, line width and intensity components, applying additional criteria in terms of velocity and spatial "coherence" following the "maser feature" approach described in Sect.~2.2 of \citet{richards2011}. Maser emission usually has a genuinely Gaussian angular profile \citep[e.g.,][]{richards1999} and 2D Gaussian components can be fitted with accuracy determined by the S/N, as $\sim$0.5$-$1$\times$(beam/S/N). In theory the multiplier is 0.5 \citep[see][]{condon1997} but it can be greater   \citep[e.g.,][]{richards1999}. The accuracy is limited by deviations from a theoretical Gaussian and the dynamic range which can exceed 100 here; therefore, the lower limit  to the errors is expected to be well below 1~mas. In practice, a component is deemed to be real if it is part of a spectral feature comprising at least 3 neighboring components with  $\sim$2~km\,s$^{-1}$ velocity coherence.

Fig. \ref{rhya_uher_h2o_component}  presents examples of "position$-$velocity" maps of the various Gaussian-fitted velocity components identified in two prominent H${_2}$O sources, R~Hya and U~Her at 268.149 and 262.898~GHz. The fitted position uncertainty is reflected in the  error bars of Fig.~\ref{rhya_uher_h2o_component}.
In R~Hya, both transitions show a quasilinear position-velocity structure extending along a SE-NW axis $\sim$16~mas long across the 23.7~mas photosphere and well within the 27.1~mas radio continuum size measured at 250~GHz by \citet[][]{homan2021}. On the other hand, the same two transitions in U~Her show a more complex velocity field and a wider extent but, again, with most  of the blueshifted and redshifted components 
seen against the optical photosphere and the 250~GHz radio continuum disk (11.2~mas and 18.5~mas, respectively). 
For both stars, we know that any maser components seen in the direction of the star must be on the near side but we do not know directly the distance along the line of sight.  We do not exclude that an infalling gas clump on the far side of the star's limb overlaps an outflowing
gas clump on the near side (or vice versa) but note that such cases would concern a small fraction of  components, outside the line of sight to the radio continuum disk. 
The complex gas motions observed here, pointing to infall and outflow of gas, are consistent with other observations discussed later in Sect.\ref{sec:infall}. 
 
\section{An initial H$_{2}$O line analysis: Physical conditions}
\label{sec:H2O_analysis}

%%%%%%%%%%%%%%%%%
\begin{figure*}
\centering \includegraphics[width=12.0cm,angle=270]{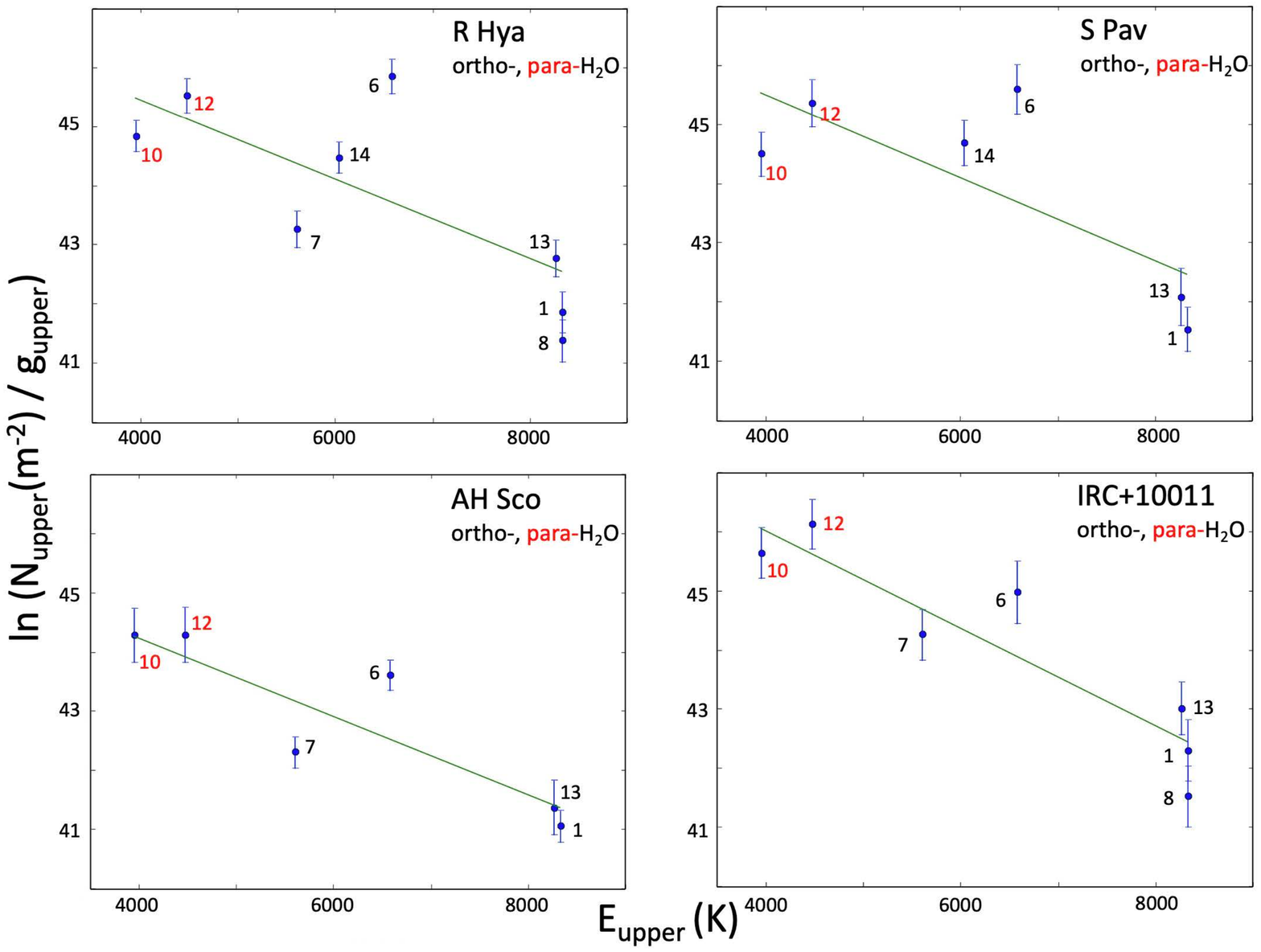}
\caption{Population diagrams for ortho and para H${_2}$O  transitions (black and red numbers, respectively) in R Hya, S Pav, AH Sco and IRC$+$10011. The number near each data point corresponds to the numbering used to identify each line in Table \ref{H2O-line-list}. The vertical bar in each data point includes the $\pm$1$\sigma$  formal error of the integrated flux density and a rough estimate of the filling factor uncertainty. The green line is the regression line across the data points from which the rotational temperature and the column density are derived (see Sect. \ref{sec:pop_diagram}). The strong 268.149 GHz maser line (line 14) is excluded from the IRC$+$10011 and AH Sco plots while it is kept in R Hya and S Pav.
}
\label{h2o_popdiagram}
\end{figure*}

%%%%%%%%%%%%%%%%

We derive here the H$_2$O line brightness temperature, the H$_2$O  column density and address 
questions related to the gas motion near the photosphere. Our calculations and discussion below are supported primarily by  the H$_2$O spectral data in Table~\ref{fluxdensity_source_list} 
which reflect the diversity of physical conditions in different sources for a given water transition and 
the diversity of the line excitation processes  within a source. 

\subsection{Brightness temperature}
\label{sec:H2O_conditions}

%%%%%%%%%%%%%%%%%%%%%%%

\begin{table}
\begin{center}
\caption{Maximum brightness temperatures  of ortho and para water at 268.149 and 262.898~GHz.}       
\label{h2o_brightness}      
\begin{tabular}{lccc}       
 \hline\hline

 Source &  $T_\mathrm{Gauss}(268)$ $^{a}$ & $T_\mathrm{b}(268)$ $^{b}$  & $T_\mathrm{b}(262)$ $^{b}$  \\
 & (K) & (K)  & (K)\\
 
\hline
S Pav & & 3.1$\times$10$^{3}$ $ \pm$ 40  & 535 $ \pm$ 40   \\
  \\
 
 T Mic & & 1.9$\times$10$^{3}$ $ \pm$ 30 & 385 $\pm$ 70   \\ 
  \\
  
RW Sco & & 3.2$\times$10$^{3}$ $ \pm$ 70 & 745 $\pm$ 150  \\ 
  \\
  
 R Hya & & 1.0$\times$10$^{3}$ $\pm$ 15 & 295  $\pm$ 15  \\
  \\

U Her & (1.2$\pm0.3$)$\times$10$^{4}$ & 4.2$\times$10$^{3}$ $\pm$ 70 & 535  $\pm$ 30  \\
 \\

 AH Sco & (9.0$\pm3.0$)$\times$10$^{6}$ & 1.1$\times$10$^{6}$ $\pm$ 20 & 370    $\pm$ 75 \\  
 \\
 
  R Aql & & 3.3$\times$10$^{3}$ $\pm$ 30& 745   $\pm$ 35  \\ 
   \\
   
    GY Aql & & 1.6$\times$10$^{3}$ $\pm$ 100 & 460   $\pm$ 90  \\ 
   \\
   
    IRC$-10529$ & & 0.98$\times$10$^{3}$ $\pm$ 95 & 845 $\pm$ 125  \\ 
   \\
   
   IRC$+10011$  & (1.1$\pm0.3$)$\times$10$^{5}$ & 4.4$\times$10$^{4}$ $\pm$ 75 &  1090 $\pm$ 70   \\ 
   \\

  VX Sgr & & 6.5$\times$10$^{3}$ $\pm$ 50 & 695   $\pm$ 55  \\ 
\end{tabular}
\end{center}
\tablefoot{
$^{(a)}$ Brightness temperature  and uncertainty derived from a Gaussian fit as described in Sect. \ref{sec:H2O_conditions}
 for the maser sources. 
$^{(b)}$ Brightness temperature measured from the high resolution channel map peak flux density and synthesized beam size. 
} 
\end{table}
 
%%%%%%%%%%%%%%%%%%%%%%%%%%%

Brightness temperature is an important quantity for qualifying  the line excitation conditions. It is measured in two different ways.  Firstly, a good approximation to the brightness temperature, $T_\mathrm{b}$,  
of compact sources is derived from the peak flux density per beam  in our channel maps  and the synthesized beam. The uncertainty in $T_\mathrm{b}$ is estimated from the typical noise in maps and the beam area. Our estimates  are  given in the last two columns of Table \ref{h2o_brightness}  for the two strongest water transitions, lines 12 and 14 at 262.898 and 268.149~GHz in $\varv_{2}$ = 1 and 2, respectively.  
The highest 268.149~GHz brightness temperature is observed
in U~Her, VX~Sgr, IRC$+10011$ and in AH~Sco where $T_{\mathrm{b}}$ reaches
1.1$\times$10$^6$~K. The 262.898~GHz transition  is not as strong as
the 268.149~GHz transition reaching, however, 1.1$\times$10$^3$~K in IRC$+10011$. The above temperatures are lower limits to the actual brightness temperature for spatial components smaller than the beam size. Secondly, another approach to estimating the brightness temperature is provided by the de-convolved sizes of the Gaussian components  fitted, as described in Sect.~\ref{sec:small_scaleh2o}, to the 268.149~GHz features with a peak S/N \ga 10. The calculated temperature,   
$T_{\mathrm{Gauss}}$, is given in the second column of Table \ref{h2o_brightness} and its uncertainty is estimated from the flux density error and the component size. 
$T_{\mathrm{Gauss}}$ is roughly 2.5 times larger than $T_{\mathrm{b}}$ derived from the
channel maps for U~Her and IRC$+10011$, and 8 times larger for AH~Sco where it reaches 9$\times$10$^6$~K. These results indicate that the brightest 268.149~GHz emission features  in these three stars are consistent with  strong maser action.

\subsection{Population diagrams, opacity estimates}
\label{sec:pop_diagram}

 Our data provide only one pair of detected lines in the same vibrational state of the   ortho H$_{2}$O species (lines  8  and 14), and there is no similar line pair for the para species. 
Therefore,  line intensity ratios derived from several transitions cannot be used  to establish  possible deviations from the LTE conditions (see also discussion in Sect. \ref{sec:nature_268}).
However, an estimate of the excitation gas temperature and  column density can be made from a population diagram \citep{goldsmith1999} provided that the range of line energy used in such a diagram is large enough and that we are not too far from the optically thin case. (Deviations from the optically thin conditions could eventually be corrected at a later stage although this is uncertain as different transitions can have different opacities; such corrections are not attempted here.) We have constructed population diagrams for a few sources in our sample with at least six lines and with known velocity-integrated flux densities  using the spectra  extracted for an aperture diameter of  0\farcs08 from our spectral cubes. A linear least-squares fit to the data in various vibrational states gives 
the excitation temperature, $T_{\rm ex}$, and the quantity  ln($N$/Q) where $N$ is the column density and Q, the partition function. The fit accounts for the uncertainty in the observed integrated flux density and we assume that each line uniformly fills a circular aperture of  0\farcs08 in diameter with a rough estimate of the filling factor uncertainty. 
The H$_{2}$O partition function includes the spin degeneracy (3/4 and 1/4 for ortho and para H$_{2}$O) and does not separate the rotational from the vibrational partition functions for a better accuracy \citep{polyansky2018}. 

Examples of population diagrams for the observed   ortho and  para H$_{2}$O emission lines are given in Fig. \ref{h2o_popdiagram} using the high resolution  data all observed within a few weeks and thus not affected by variability. 
In AH~Sco  and IRC$+10011$ we have excluded line number 14 which is masing (the corresponding data points lie well above the least squares line fit). Line 5 shows absorption only and is not used in our population diagrams.  
We derive  $T_{\rm ex}$ in the range $\sim$1000$-$1500~K for all sources and column densities $N$(H$_2$O) $\sim$0.6, 2.2, 2.5 and 4.5$\times$10$^{20}$~cm$^{-2}$ in AH~Sco, R~Hya, S~Pav and IRC$+$10011, respectively. In R~Aql where the dispersion of the data points is larger than in the other four sources we derive  $N$(H$_2$O) $\sim$5.8$\times$10$^{20}$cm$^{-2}$. The uncertainty in the least squares fit slope may result in a temperature uncertainty as high as $\sim$700~K at least (and higher in R~Aql) but we note that, because the partition function varies monotonically with the temperature, this does not impact the estimate of the  column density. The actual uncertainty in the projected  density  is difficult to evaluate, however. It is not very sensitive to uncertainties in the integrated line intensity used to derive $N$ whereas the number of transitions and the range of energy covered in the population diagrams have an important impact. Discarding any line in the R~Hya diagram for example, does not change significantly the $N$(H$_2$O) estimates and, from random trials we suggest a statistical uncertainty of the column density at the level of $\sim$$25\%$. 

The column density can  be used to estimate the average water number density, provided the extent of the emission is known. In R~Hya, the water emission has an extent of at least 3 times the continuum  and from maps in different H$_2$O transitions 
the likely depth of the water region is $\sim$60$-$90~mas. At a distance of $\sim$150~pc, the R~Hya extent  corresponds to 9$-$13~au which gives an average density of roughly (1.1$-$1.6)$\times10^{6}$~cm$^{-3}$. For a fractional water abundance of  3$\times10^{-5}$ (used in Sect.\ref{sec:maser}) the H$_2$ density is then (3.7$-$5.3)$\times10^{10}$~cm$^{-3}$ which  is commensurate with the critical density derived in the next Section. Our estimate of the H$_2$ density in the inner regions of R~Hya and in other sources remains  uncertain, however, because of the difficulty in deriving a reliable emission water extent and since the the H$_2$O/H$_2$ fractional abundance used here is debatable.

The opacity at the line center, $\tau${$_c$}, can be estimated from the quantity $N$/Q  and the excitation temperature determined above, the energy of the transition and the observed line width. 
We derive $\tau{_c} \sim (5.145\times10^{-4} \times A_{21}  \int n_{2}dl) / (\nu^{2}\times\Delta\varv_{0} \times T_{12}) $ where $\nu$ is in GHz for the transition from level 2 (upper) to 1 (lower), $A_{21}$ the spontaneous emission rate in s$^{-1}$, $T_{12}$  the line excitation temperature, $\Delta\varv_{0}$ the line width at half intensity and $ \int n_{2}dl $ the integrated density in cm$^{-2}$. The latter quantity is obtained from the integrated Boltzmann population of the upper energy level  using $N$/Q and $T_{12}$ determined from the population diagrams. In RHya, for example, we derive $\tau${$_c$} $\sim$0.25, 0.3, 0.15 and 0.1 for lines 1, 10, 12 and 13,  respectively. We point out that at 268.149~GHz (line 14) where maser excitation can be observed in some stars, our models indicate that negative opacities may become large (see Sect. \ref{sec:maser}). In R~Hya, however, line~14 does not seem to be masing, or is not strongly masing, and we derive from our population diagram $\tau${$_c$} $\sim$0.45  which is larger than the above opacities. In general, $\tau${$_c$} remains relatively small in other sources. However, for the low energy levels of  line 10, we get $\tau${$_c$} $\sim$0.4  in S~Pav and even 0.75 in the thicker shell of IRC$+$10011.  

 Line 5 at 244.330~GHz shows only absorption which can  be used to make another estimate of the gas opacity  in the  region where this line is excited. 
Toward  R~Hya, S~Pav and R~Aql,  the absorption peak flux density reaches $-8$ to $-10$ mJy  against the continuum source (Fig. \ref{rhya_spav_raql_abso_h2o}) and the 
line excitation temperature must be below the  continuum brightness temperature which, for Band~6 in R~Hya, is $\sim$2500~K  \citep[][]{homan2021}. We derive $\tau_{244}$~$\sim$0.015$-$0.08 for an excitation temperature close to gas temperatures in the range 1000$-$2000~K  as predicted in models \citep[e.g.,][]{gobrecht2016}. We also  expect small opacities at 244.330~GHz in S~Pav and R~Aql for continuum and excitation temperatures similar to those in R~Hya. 

\subsection{Critical density}
\label{sec:critical_density}

 A full interpretation of all results in Table \ref{fluxdensity_source_list} requires a detailed modeling of water transitions up to the highest energy levels and is not attempted here (but see Sects. \ref{sec:268_H$_2$O_maser} and \ref{sec:maser} for discussion on maser amplification). However, we can estimate for various H${_2}$O lines 
 the critical density, n${_c}$, above which collisions with H${_2}$ dominate over spontaneous radiative de-excitation. This is done by comparing the spontaneous emission rates and collision rate coefficients for all downward transitions connected to the upper level of the transition under consideration \citep[see discussion in][]{faure2008}.  Adopting the collisional rate coefficients in \citet{faure2008} for a range of different rotational transitions in, for example, the (0,1,0) vibrational state,  we get  n${_c}$ $\sim$5$-$2$\times$$10^{10}$~cm$^{-3}$ for temperatures $\sim$1000$-$4000~K, including the 658.007~GHz transition discussed in Sect. \ref{sec:nature_268}. 
 Another rough estimate  for the widespread transition at 268.149~GHz in the (0,2,0) vibrational state gives n${_c}$ $\sim$$10^{11}$~cm$^{-3}$. These estimates and  the H${_2}$ densities considered in the chemistry models above the photosphere in regions  $\sim$1$-$10~R$_{\star}$ of typical AGBs \citep[e.g.,][]{cherchneff2006, gobrecht2016} suggest that collisions become the prevalent excitation process in the post-shock gas layers where the rate of formation of water from OH is optimal. 
 The above considerations cannot replace of course multilevel radiative transfer models including exact collisional rates of water with H${_2}$. Such models are needed to explain  supra-thermal excitation of water and why some water transitions are masing in the circumstellar envelopes of AGBs (see Sect. \ref{sec:maser}). 
 
\subsection{Gas motion in the inner-wind region}
\label{sec:infall}

The motion of circumstellar gas toward and beyond the photosphere of an evolved late-type star is the result of  complex interactions between shock-driven pulsations, gravity and radiation pressure in the regions where new molecules and dust grains are formed. Radial motions, even if they are  dominant during the pulsation cycles of the central object, can be perturbed by random motions within the inner-wind region or by gravitational interactions with a companion. Some of these questions are briefly discussed below in the light of our water line data. 

In spite of a modest velocity resolution, the Gaussian component analysis of the most compact regions of water emission presented in Sect. \ref{sec:small_scaleh2o}  shows that an outflow or infall of gas  can be traced in the 262.898 and 268.149~GHz lines. We find that the emission can be asymmetrically distributed around the central star  and we observe velocity gradients or gas  motions with an amplitude of several km\,s$^{-1}$. An infall of matter toward the central star is directly suggested from our 244.330~GHz spectra which show a $\sim$6$-$9 km\,s$^{-1}$ redshift with respect to the systemic stellar velocities in R~Hya, S~Pav and R~Aql (Fig. \ref{rhya_spav_raql_abso_h2o}), and from the 244.330~GHz absorption maps in front of the same stars  (Fig.~\ref{abso_h2o_244}). 
The gas infall interpretation is also strengthened by the  inverse P~Cygni profiles of  the $\varv=1$, CO(2$-$1) transition in R~Hya, S~Pav and R~Aql  (Fig.~\ref{rhya_spav_raql_abso_h2o}) and by line maps showing a close spatial association of the CO and water  molecular  species within a few stellar radii (Fig.~\ref{abso_244_259_CO}). We further note that  inverse P~Cygni profiles similar to our  CO(2$-$1) profiles have been observed with ALMA in $\varv=1$, CO(3$-$2) toward R~Aqr  and R~Scl by \citet{khouri2016} who concluded that the gas falls on these objects within $\la$ 3 R$_{\star}$ with a velocity of 7$-$8 km\,s$^{-1}$ (and a  dispersion of 4 km\,s$^{-1}$). 
They also derived, in the molecular layers before the wind starts, densities on the order of a few $10^{10}$~cm$^{-3}$ to $\sim$4$\times$$10^{11}$~cm$^{-3}$ which are comparable to or above the critical density estimated in Sect.~\ref{sec:critical_density}. 
Infall velocities of $\sim$5$-$10~km\,s$^{-1}$ magnitude estimated from our water and CO radio line observations, from the CO lines in  \citet{khouri2016} or, earlier, from  the near-infrared rotation-vibration bands of CO  \citep[e.g.,][]{hinkle1984}, are close to model predictions. In particular, the models developed by \citet{nowotny2010} to synthesize the observed  CO or CN infrared line profiles  and to match the velocity variations observed in the dust-free $\sim$1$-$2~R$_{\star}$ region of AGBs require infall/outflow velocities of $\sim$5$-$10~km\,s$^{-1}$. We also point out that in complex stellar systems, such as the symbiotic star R~Aqr, the gas motions are affected by the presence of the stellar companion \citep[e.g.,][]{bujarrabal2018}. 

The water line velocity of the peak emission extracted from the observations  provides an estimate of the systemic stellar velocity but little information on the atmospheric dynamics. To tentatively compare the observed radial velocity information with models we use the total velocity range between the red and blue extremes of each line, $\Delta\varv$, as defined in Sect. \ref{r} and given in Table~\ref{fluxdensity_source_list}.   We first point out  that in most cases the emission at extreme velocities is within  0\farcs04 of the star and thus, using  for spectra an extraction aperture diameter larger than  0\farcs08 would not change $\Delta\varv$ significantly; in fact, similar $\Delta\varv$ is usually measured in an aperture half that size. Occasionally, a smaller $\Delta\varv$ can be measured from a larger aperture or from lower resolution data, which must be due to beam dilution or imaging artifacts, including absorption which sometimes occurs, or, if considering mid resolution data, due to the lower sensitivity of these data. Examination of the channel maps resolves these apparent issues. 
Using Table~\ref{fluxdensity_source_list}, we find that for each detected transition and for each star, $\Delta\varv$ is always $\ga$5~km\,s$^{-1}$ and may reach $\sim$15$-$30~km\,s$^{-1}$ at 268.149~GHz with the highest values in the supergiants AH~Sco and VX~Sgr. Assuming that the line profiles are grossly symmetric about the stellar velocity and that the velocity fields are not strongly asymmetric, the maximum outflow/infall velocity can be estimated from 0.5$\times$$\Delta\varv$ which gives $\la$12~km\,s$^{-1}$ in AGBs and $\la$13$-$15~km\,s$^{-1}$ in AH~Sco and VX~Sgr. Another way to estimate the outflow and infall velocities is to take the most blueshifted and redshifted velocities of the water lines as defined in Sect.~\ref{r} to determine the velocity extent. These velocities vary with the star and the  line being examined (the 268.149~GHz transition  often gives the largest velocities). Overall, the range of outflow and infall velocities  covers around $-$(3$-$15) and (2$-$15)~km\,s$^{-1}$, respectively; there is no clear difference between the AGB stars and AH~Sco and VX~Sgr. 
These outflow/infall velocities are in accord with the radial pulsation modes described in the nonlinear models of \citet{ireland2011} (predicting expansion velocities at such velocities and shocks propagating even faster),  and  with the maximum velocity of $\sim$10~km\,s$^{-1}$ used in the AGB models of \citet{bladh2019}. 
 Comparing the observations with models remains uncertain, however. We note that, in addition to sensitivity limitations,  
 an observed narrow velocity range, for example, could be due to  the clumpy and irregular nature of the mass loss phenomenon that we are sampling. The larger values of   $\Delta\varv$ could also be influenced by  interactions  with a companion (see below). 

In a few stars the blue and red wing line velocities of the 268.149~GHz spectra  exceed the $\varv=0$, CO(2$-$1) line wings although, as shown in \citet{gottlieb2022}, the {\sc atomium} observations provide high sensitivity to low level CO emission. 
Such a comparison may, however, still be hampered by the dependence of the CO velocity range on sensitivity and by the fact that the interpretation of the 268.149~GHz line profile may be complicated by maser action (e.g., variable pumping and beaming conditions over a given region may give rise to asymmetric line wings).  
Nevertheless, we observe that, for the high resolution data,  the 268.149~GHz water line wing emission  
 exceeds that of CO(2$-$1)   in  RW~Sco, T~Mic, U~Her, AH~Sco and RAql. (In the latter source both the red and blue water line wing velocities exceed the CO(2$-$1) emission.) This finding is reminiscent of weak $\varv=1$, SiO(2$-$1) maser features observed by \citet[][]{cernicharo1997}  and \citet[][]{herpin1998} at velocities exceeding the $\varv=0$, CO(2$-$1) velocity range. The high velocity emission of water reported here could be explained in different ways,  including rotation, 
 asymmetrical gas expansion (potentially explaining blueshifted components in front of the star, if observed), turbulence or kinematic perturbations by  a companion. The surface rotational velocities of evolved stars are expected to be small although models \citep[e.g.,][]{ireland2011} consider that the stellar surface velocities can reach $\sim$10$-$20~km\,s$^{-1}$. Slow surface rotation is observed indeed in R~Dor for which \citet{vlemmings2018}  suggest  a probable acceleration  by a nearby companion and, in Betelgeuse and its close envelope which exhibit a slow equatorial velocity of a few km\,s$^{-1}$ \citep[][]{kervella2018}. The high velocity emission present in our data is most probably due to the impact of a stellar or planetary companion  on the local gravity and/or to the complex and irregular nature of the inner wind where the 268.149~GHz line of water is excited.  Binarity, depending on the mass ratio and other model parameters, was shown to be dominant in shaping the wind of the AGB stars and RSGs and can lead to a high speed emission of the gas close to the star out to a hundred or more stellar radii \citep[][]{gottlieb2022,decin2020-science}.\footnote{It is worth adding that interpretation of the ALMA observations in terms of binarity  is  well documented in L$_{2}$~Pup and $\pi^{1}$~Gru   \citep[][]{kervella2016,homan2020}.
 } 
 
 \section{Widespread   emission at 262.898 and 268.149 GHz}
\label{sec:268_H$_2$O_maser}

Lines 12 and 14  at 262.898~GHz and 268.149~GHz in the vibrational states (0,1,0) and (0,2,0) are the two strongest and most widespread  transitions in this work (see Table~\ref{fluxdensity_source_list} and Appendix~\ref{plot_H2O_lines}). Their line peak and line profile variability are discussed below.

\subsection{Time variability}
\label{sec:variability_268}
Time variability in any of the circumstellar water lines observed here, once demonstrated, suggests changing physical conditions. In the pulsating and shocked environment of the  {\sc atomium} O-rich stars we expect indeed changing line excitation conditions and, hence, some variations in the  water line peak intensity and/or profile. In weak lines, however, small deviations from the quasi-LTE excitation conditions may be difficult to demonstrate. Nevertheless, based on their integrated line intensity measurements in $\varv_2=1$ and~2 transitions of water, \citet{velillaprieto2017}  reported  tentative line variability in IK~Tau. We point out that while variations with quasi-LTE excitation conditions are difficult to prove, variations with maser-type  excitation conditions may be easier to identify. In masers, time variability of the line emission peak and profile become more probable since changing physical conditions in the circumstellar gas  may result in amplified variations of the stimulated emission, provided that the line is not highly saturated. 
(Analysis of the molecular light polarization would also help to characterize any maser action; this is not possible here and would deserve dedicated observations with ALMA or other facilities.)

\subsubsection{268.149 GHz line variability}
\label{sec:line_variability_268}

%%%%%%%%%%%%%
\begin{table*}

\caption{Dates of observations with the main and ACA arrays and associated optical phases (in parentheses) of the variable 268.149~GHz H${_2}$O sources.}     
\label{268_phase_variability}    
\centering  
\begin{tabular}{lcccccc} 
  
 \hline\hline 

 Star &  Pulsation period$^{a}$       & \multicolumn{2}{c}{Date and optical phase} & & Date and optical phase \\
 &(days) & Mid config. array$^{b}$ & Extended config. array$^{b}$ & Mid/Ext peak flux ratio$^{c}$ & ACA$^{b}$\\
 
\hline
  
U Her &  402 & 2018 10 14 (0.13) & 2019 07 06 (0.80) & 34 & $-$  \\ 
          &         & 2019 08 24 (0.92) & & \\
          
 GY Aql & 468 & 2018 11 13  (0.65) & 2019 07 08 (0.15) & 1.7 & 2021 09 14 (0.80) \\

AH Sco & 738 & 2018 11 22 (0.32) &  2019 07 08   (0.73) & 0.1 & $-$  \\ 

VX Sgr &  790 & 2018 11 22  (0.20) &  2019 07 09 (0.45) & 20 & $-$ \\  
 
RW Sco & 389 & 2018 11 22 (0.65) & 2019 07 06 (0.20) & 1 &  $-$ \\

R Aql & 268 & 2018 11 18 (0.38) &  2019 07 06 (0.22) & 0.6 & 2021 09 13  (0.20) & \\
          &         &  2018 11 19 (0.38) & & \\
         
   \hline                                  
\end{tabular}
\tablefoot{
$^{(a)}$ Pulsation period from \citet{gottlieb2022} except for VX Sgr for which the period estimated from the  AAVSO\footnote{https://www.aavso.org.} light curve data is preferred.
$^{(b)}$ Date in year-month-day format and optical phase given in parentheses. Optical phase uncertainty can be $\sim$0.1 (as suggested from uncertain maxima or minima in AAVSO light curves).  
$^{(c)}$ Ratio of the mid array peak flux density to the extended array peak flux density at the dates given in the third and fourth columns for spectra extracted with 0\farcs4 aperture diameter.  
} 
\end{table*}

%%%%%%%%%%%%%%%%%

\begin{figure}
\centering \includegraphics[width=8.5cm,angle=0]{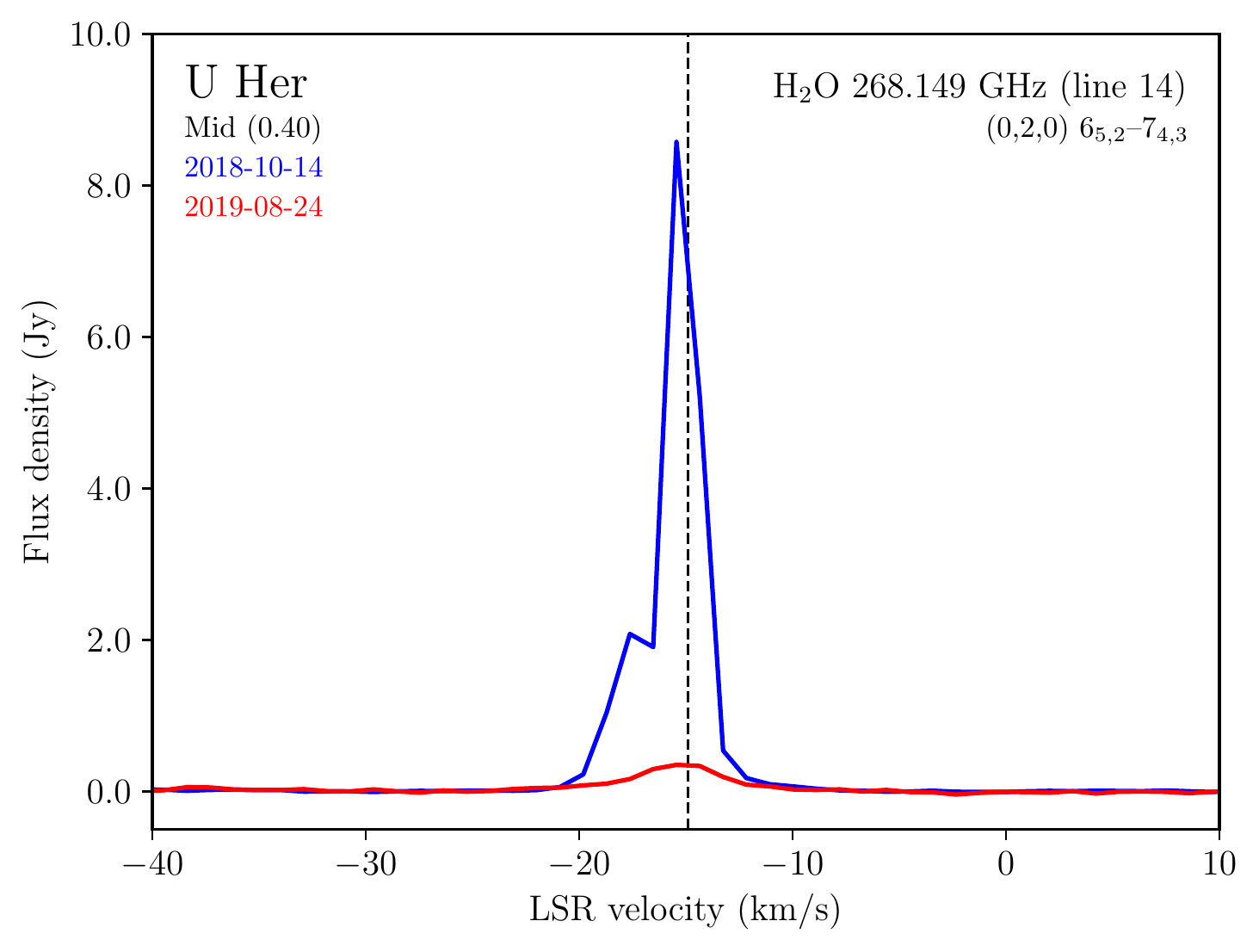}
\caption{Comparison of the 268.149 GHz line profiles in U Her observed with the mid configuration of the main array in Oct~2018 and Aug~2019 (blue and red curves,  respectively).The spectra are extracted for an aperture diameter of 0\farcs4;  the spectral resolution is $\sim$1~km\,s$^{-1}$. The vertical line shows the  systemic stellar velocity. The mom~0 emission maps at these two epochs are presented in Fig. \ref{uher_map_variability_268}.
 }
\label{uher_variab_mid}
\end{figure}

There is no systematic monitoring of the 268.149~GHz line emission in this work. However, time variability can be addressed  by comparing  
 the mid- and extended-configuration spectra of the main array obtained at two different epochs and for the same 0\farcs4 aperture diameter.
We observe markedly different peak flux density ratios at the two epochs of the observations in U~Her, GY~Aql and the two RSGs AH~Sco and VX~Sgr (Table~\ref{268_phase_variability}) and there is a secondary line feature near $-17.1$~km\,s$^{-1}$ in U~Her (Fig. \ref{uher_variab_mid}). Table~\ref{268_phase_variability} gives the dates and optical phases of the observations together with the ratio of the mid to extended peak flux density ratio. RW~Sco is added to Table~\ref{268_phase_variability} despite the peak flux density ratio is equal to one because the line shape is very different at the two epochs of the main array observations. The ACA observations of GY~Aql and R~Aql, also summarized in Table~\ref{268_phase_variability}, are presented later in this Section.  

Variability toward U~Her is the most accurately documented case because a similar mid-configuration of the main array antennas was used in both October 2018 and August 2019. Comparing the spectra at these two epochs (corresponding to  the optical phases $\sim$0.13 and $\sim$0.92), the peak flux changed by a factor of $\sim$24 (Fig.~\ref{uher_variab_mid}). 
In addition, despite our modest $\sim$1.1~km\,s$^{-1}$resolution, the stronger emission spectrum of 2018  exhibits a secondary feature around $-18$~km\,s$^{-1}$ next to the main line emission suggesting multicomponent excitation. Spectral narrowing of the main line feature is discussed in Sect. \ref{sec:narrowing_268}. There is also a clear difference in the angular extent of the 268.149~GHz emission observed in October 2018 and August 2019 as shown in our zeroth moment maps (Fig. \ref{uher_map_variability_268}). The  area within the 5$\sigma$ contour of the  emission in October 2018 is typically $\sim$3.5  times greater than the area within the 5$\sigma$ contour in August 2019 when the emission was weaker. (There is still the possibility, however, that these two different mid-configurations of the array are not equally sensitive to extended weak components.)  In addition, Fig. \ref{uher_map_variability_268} shows an emission blob at the SW of the main source in the October 2018 data when the emission activity is greater\footnote{The {\sc atomium} observations of U~Her show extended thermal SiO emission in this SW blob. Extended emission in similar directions  have also been observed in 22.235~GHz water masers over more than 24~years \citep{richards2012}.}. 

%%%%%%%%%%%%%%%%%%%%%%

%%%%%%%

\begin{figure}
\centering \includegraphics[width=8.5cm,angle=0]{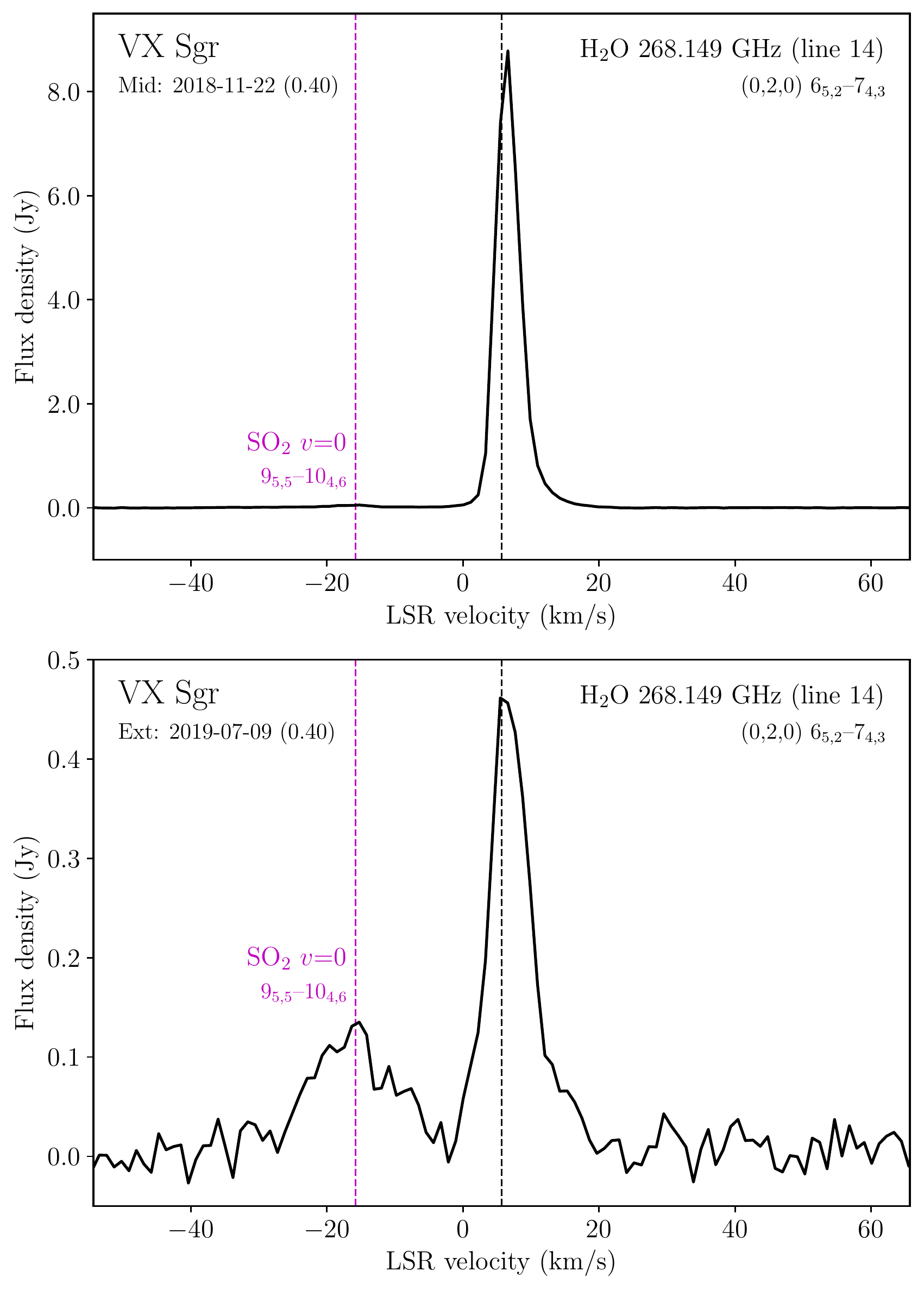}
\caption{Comparison of the 268.149 GHz line profiles extracted for an aperture diameter of 0\farcs4 in VX Sgr observed at $\sim$1~km\,s$^{-1}$ resolution in November 2018 and July 2019 with the mid and extended configurations of the main array (upper and lower panels, respectively). The weak feature near $-20$~km\,s$^{-1}$ in the 2019 spectrum corresponds to the $\varv=0$, $J_{K_a,K_c} = 9_{5,5}-10_{4,6}$ transition of SO$_2$ at 268168.335~MHz; it is also barely seen in the 2018 mid spectrum. The vertical lines  are at the adopted new LSR systemic velocity. 
 }
\label{vxsgr_variab_mid_exten}
\end{figure}

%%%%%%%%
\begin{figure}
\centering \includegraphics[width=8.9cm,angle=0]{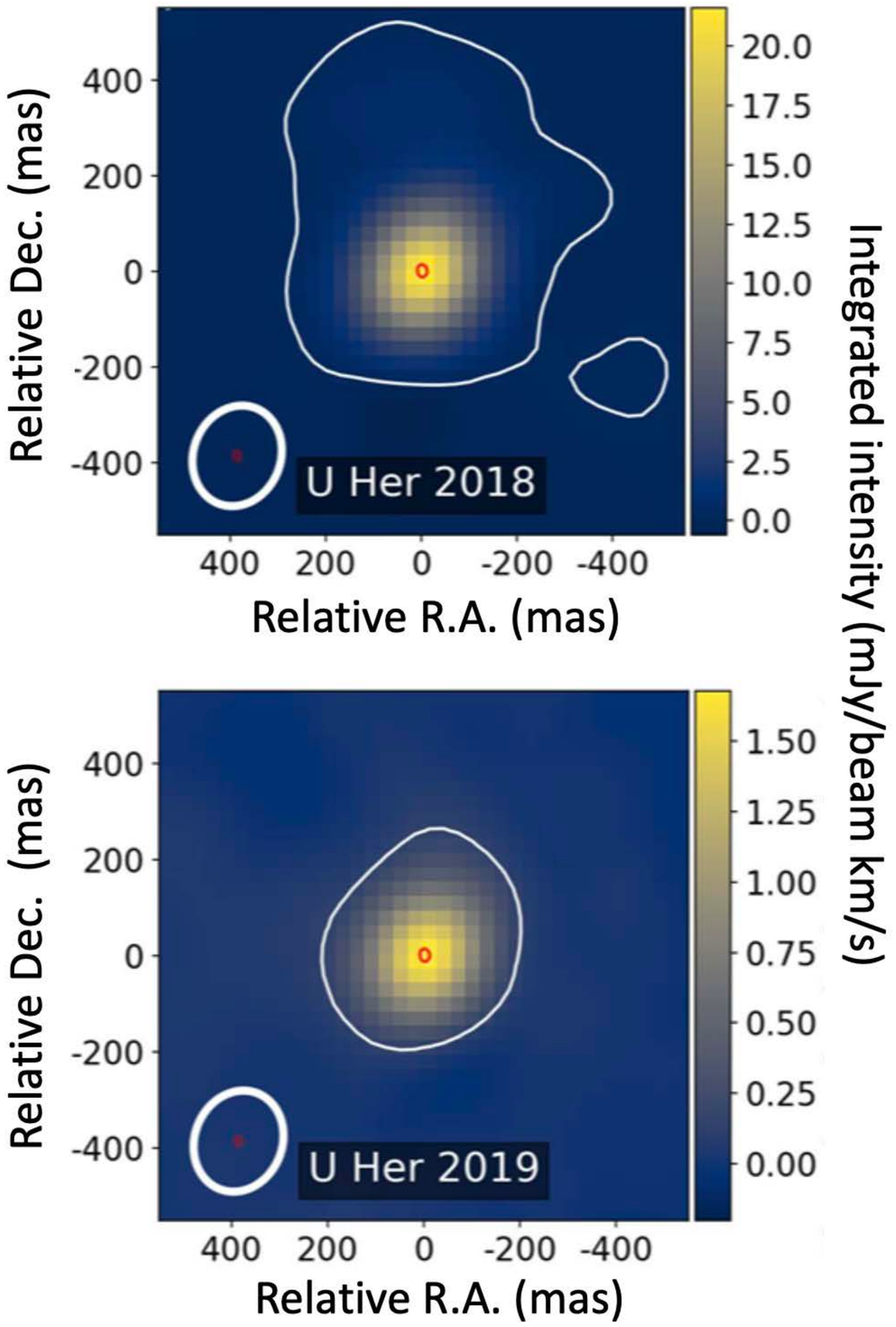}

\caption{Zeroth moment maps of the 268.149~GHz emission of water in U Her obtained with the mid configuration of the main array at two different epochs, 2018, Oct.~14 and 2019, Aug.~24. The emission is integrated over the $-21.6$ to $-9$~km\,s$^{-1}$ velocity range.  The 2019 image has been convolved with the larger 2018 image beam shown in the lower left of each panel (0\farcs21$\times$0\farcs18).    In both maps the noise  is 0.02~Jy/beam~km\,s$^{-1}$ and the 5$\sigma$ emission contour is shown as a white light contour. The map maxima are 21.6 and 1.67~Jy/beam~km\,s$^{-1}$ in 2018 and 2019, respectively. The small red contour at the  image center is the 50\% continuum emission from the high resolution data. Note the SW blob emission in the 2018 map (see Sect.~\ref{sec:line_variability_268}).
}
\label{uher_map_variability_268}
\end{figure}

%%%%fig ACA, GY and R Aql
\begin{figure*}
\centering \includegraphics[width=16.5cm,angle=0]{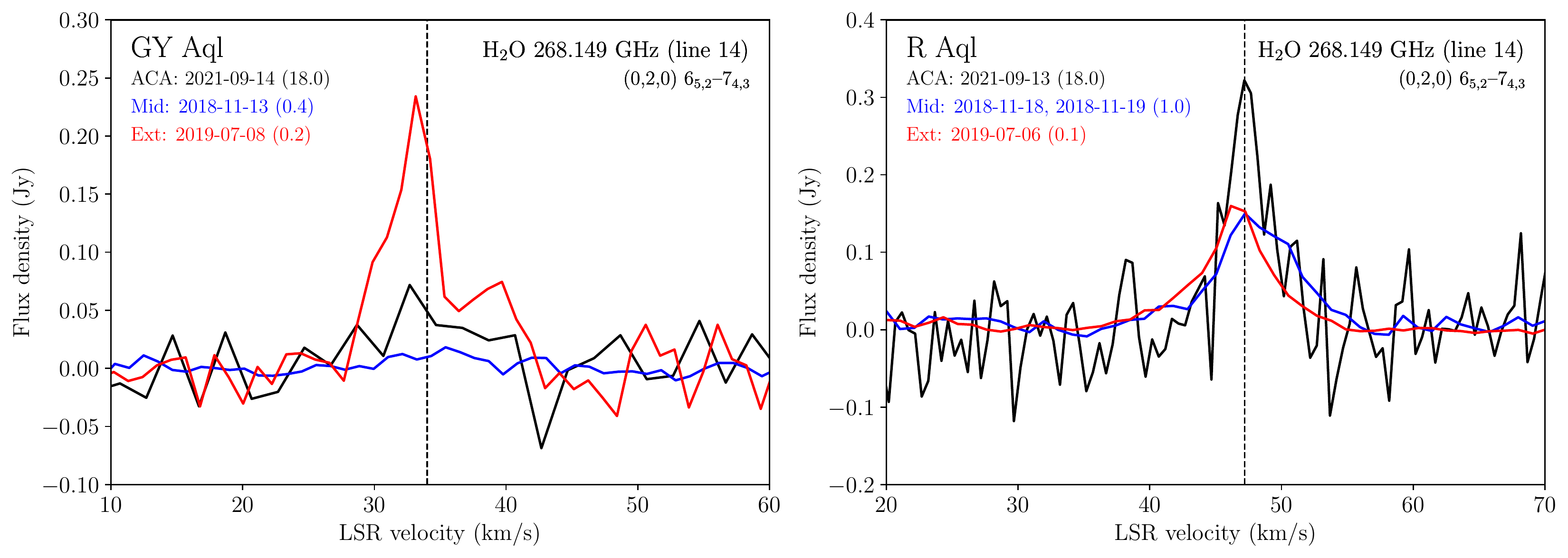}
\caption{Comparison of the 268.149 GHz line profiles in GY Aql and R Aql observed with the ACA and the mid and extended main array configurations. The ACA spectral resolution has been degraded here to 2 and 0.5~km\,s$^{-1}$ (R~Aql and GY~Aql,  respectively) for comparison with the $\sim$1~km\,s$^{-1}$ resolution of the mid and extended configurations of the main array. The epoch of the observations are indicated in the figure labels as well as  the diameter of the extraction aperture given in parentheses (18\farcs0, 0\farcs4 and 0\farcs2 for GY~Aql and 18\farcs0, 1\farcs0 and 0\farcs1 for R~Aql). The vertical lines show the  systemic stellar velocity. 
 }
\label{aca_gyaql_raql}
\end{figure*}

%%%ACA observations
Our 2021 ACA observations of GY~Aql and R~Aql   provide an additional epoch to assess the variability of the 268.149~GHz line emission in these two sources. To support the idea that the ACA and {\sc atomium} main array data can be compared, we have extracted spectra for these two sets of data using different circular apertures encompassing all detected emission. 
The ACA spectral windows   include (see Sect.~\ref{aca_observations}), in addition to H$_2$O lines and  to the probably masing $\varv=1$, $^{29}$SiO(6$-$5) transition, two  relatively strong transitions of  $\varv=0$, HCN(3$-$2) and $\varv=0$, SO($5_{6} - 4_{5}$) at 265.886 and 251.826~GHz, respectively.
We find that the ACA and mid configuration line profiles and peak flux densities in HCN and SO of both GY~Aql and R~Aql are nearly identical, thus demonstrating that the ACA can recover the same total emission as the mid configuration of the main array. A similar conclusion is obtained for GY~Aql in the $\varv$ = 0, SiS(14$-$13) line at 254.103~GHz.  We have thus used these results to compare the 268.149~GHz line emission spectra from the ACA and the mid and extended configuration (Fig. \ref{aca_gyaql_raql}). In GY~Aql, the 268.149~GHz line was only barely detected by the ACA at optical phase 0.8 (Table~\ref{268_phase_variability}). The 3$\sigma$ level flux density, $72 \pm 24$~mJy at ${\sim}32$\,km\,s$^{-1}$, is significantly weaker than our detection with the extended configuration in July 2019 at the optical phase of $\sim$0.2. This line was buried in the noise of the mid-configuration observations in November 2018 at the optical phase of $\sim$0.7 very close to that of our ACA observations. In R~Aql (Fig. \ref{aca_gyaql_raql}), the ACA peak emission at the optical phase of 0.2 (Table \ref{268_phase_variability}) is stronger than the mid and extended configuration peak emission at the optical phases of $\sim$0.4 and 0.2, respectively; the spectrum also peaks at a slightly different LSR velocity. 

\subsubsection{262.898 GHz line variability}
\label{262_variability}

%%%%%%%
\begin{figure}
\centering \includegraphics[width=8.6cm,angle=0]{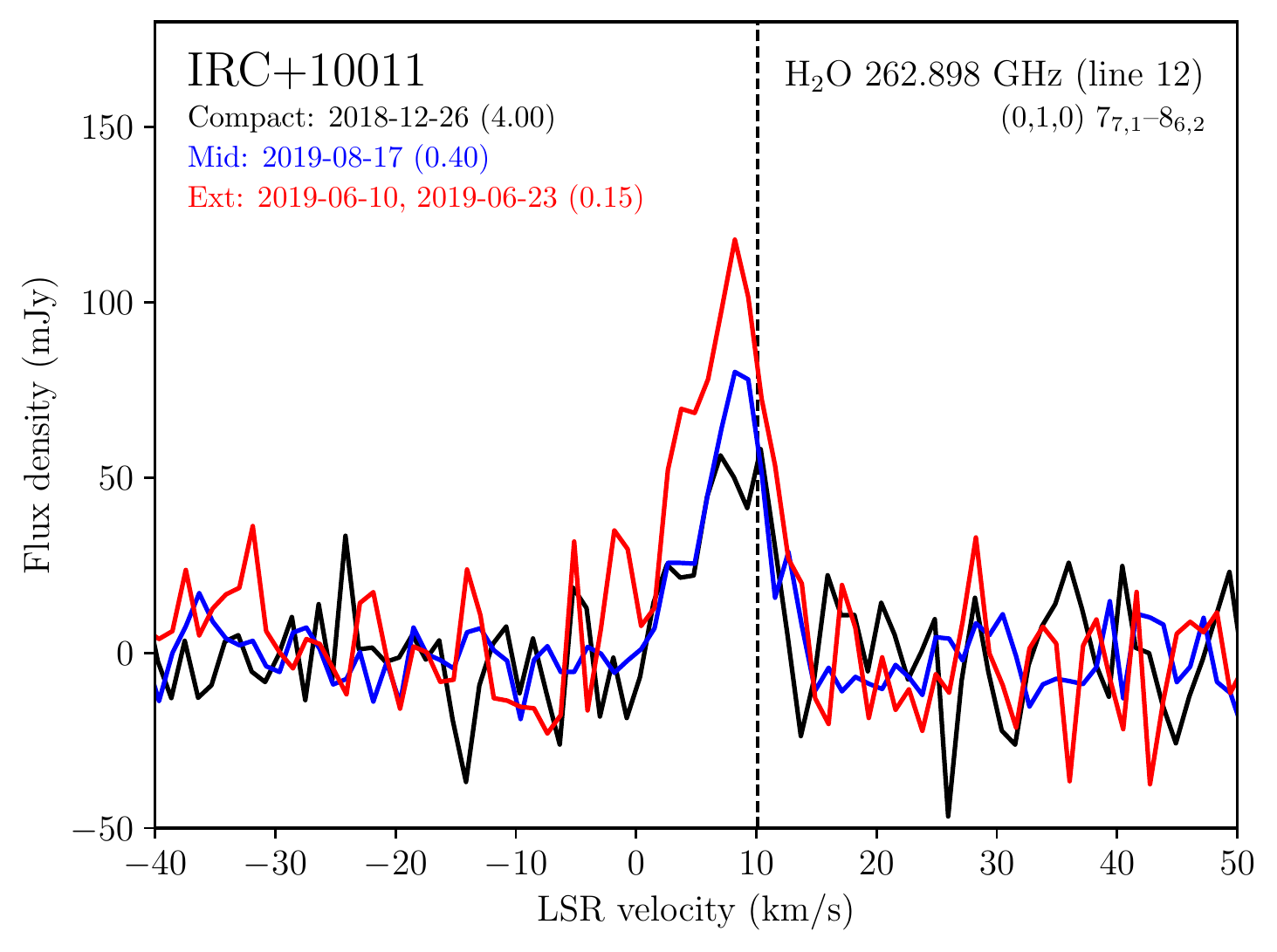}
\caption{Comparison of the 262.898 GHz line profiles in IRC$+10011$ observed at different epochs with the mid, extended and compact configurations of the main array. The epoch of the observations and the  diameter of the extraction apertures (4\farcs0, 0\farcs4 and 0\farcs15) are shown in the figure labels. The spectral resolution is $\sim$1.1 km\,s$^{-1}$. The vertical line shows the  systemic stellar velocity.   }
\label{irc10011_variab_h2o_262}
\end{figure}
%%%%%%

There is no clear evidence of time variability in the {\sc atomium} data  at  262.898~GHz (line 12) when we compare the peak line flux density of detected sources at different epochs. In addition, the line 12 profiles show a dominant quasithermal line excitation in agreement with our population diagram results in Sect. \ref {sec:pop_diagram}. However, toward IRC$+10011$, a close examination of the spectra acquired with the extended, mid and compact configurations of the main array at different dates (observations of June, Aug 2019 and Dec. 2018) shows 
a time-varying emission where the line flux density has changed by a factor of $\sim$2 over the time span of half a year 
(Fig. \ref{irc10011_variab_h2o_262}). 
This result suggests that even though thermal emission  dominates in the line 12 profile, some  maser-like emission may also coexist in our data. Maser models show indeed that there are favorable conditions in the stellar environment to invert the line 12 population levels  in the $\varv_2 = 1$ state (see Sects. \ref{sec:h2o_line12} and \ref{sec:maser}). 

%%%%%%%%%%%
  
\subsection{Line narrowing at 268.149 GHz}
\label{sec:narrowing_268}
In addition to 268.149~GHz  intensity variations in U~Her and VX~Sgr (Figs. \ref{uher_variab_mid} and \ref{vxsgr_variab_mid_exten}), we have observed a clear narrowing of the full width at half maximum, by a factor  of $\sim$2, when the line is stronger. There is no direct evidence of significant line narrowing in other targets, including AH~Sco the strongest 268.149~GHz source which shows similar narrow line widths at the time of the 2018 and 2019 observations. In U~Her (Fig. \ref{uher_variab_mid}), accounting for the 1.1~km\,s$^{-1}$ instrumental line broadening, the intrinsic  line width of the main  268.149~GHz spectral component is around 4.3 and 2.0~km\,s$^{-1}$  in August 2019 and October 2018, respectively. When the emission is stronger (Oct. 2018), the narrowed line width is comparable but slightly below the 2.0 to 2.5~km\,s$^{-1}$ Doppler line width derived for a kinetic temperature in the range  $\sim$1500 to 2500~K. Such a line narrowing suggests maser action in an unsaturated maser. However, since a discussion of maser saturation is impossible with our {\sc atomium} data alone, an estimate of the 268.149~GHz opacity at the line center from these line widths is uncertain. 
In VX~Sgr (Fig.~\ref{vxsgr_variab_mid_exten}), the observed line widths at half intensity, corrected for instrumental broadening, are $\sim$3.7  and 6.4~km\,s$^{-1}$ at the time of the stronger (2018) and weaker (2019) emission, respectively. Such line widths are broader than the expected thermal line profiles and  likely to be blends of emission from individual H$_2$O clumps along the line of sight due to the velocity gradients in the circumstellar material. 

To indirectly identify a possible sign of 268.149~GHz line narrowing in unsaturated masers despite our moderate spectral resolution, we have compared the  line width at half intensity of this transition with other water lines  in the same source. This approach is limited, however, since line widths alone cannot be used to prove or disprove maser action and the conditions leading to line rebroadening or inhibiting line broadening of saturated masers are difficult to assess. 
Nevertheless, in addition to U~Her and VX~Sgr discussed above, we observe that the 
268.149~GHz line widths lie in the range $\sim$2.0$-$4.5  km\,s$^{-1}$ toward AH~Sco, IRC$+$10011, IRC$-$10529, RW~Sco, GY~Aql, and R~Aql while other detected water transitions are generally broader. In AH~Sco, for example, the line width is $\sim$2.5 km\,s$^{-1}$ at 268.149~GHz and $\sim$5$-$7~km\,s$^{-1}$ in all other water transitions. On the other hand, V~PsA  and R~Hya do not exhibit narrow 268.149~GHz line profiles ($\sim$14.5 and  8.5~km\,s$^{-1}$  at half intensity, respectively) and in R~Hya our population diagram (Fig.~\ref{h2o_popdiagram})
shows that there is no strong deviation from LTE (and hence no dominant maser amplification in this source).  
An additional sign of maser action is provided if a secondary narrow spectral feature is identified. This is observed in the 268.149~GHz spectra of GY~Aql, U~Her and S~Pav near 38.1, $-17.1$ and $-6.7$~km\,s$^{-1}$, respectively, without any apparent confusion with other molecular transitions. Maser lines can be narrower than 1~km\,s$^{-1}$, so, together with the extended array, a higher spectral resolution would be desirable to reveal the presence of multiple unresolved spectral components as suggested by the asymmetrical 268.149~GHz line profile in S~Pav.

\subsection{268.149 GHz line excitation}
\label{sec:nature_268}

The (0,2,0)~$6_{5,2}$--$7_{4,3}$ transition of water at 268.149~GHz is widespread in evolved stars since it is observed in 15 out of the 17 objects of the {\sc atomium} sample to which one must add three other sources not included in our sample, VY CMa with strong emission first reported by  \citet{tenenbaum2010}, IK~Tau with weak emission first mentioned  by \citet{velillaprieto2017} and W~Hya with strong maser emission recently reported by Ohnaka et al. (in prep.). In VY~CMa, the line peak flux density reached $\sim$650 Jy which suggests strong maser amplification.  
In all sources of the {\sc atomium} sample the 268.149 GHz transition is always the strongest water line (see Table \ref{fluxdensity_source_list}) and, in AH~Sco and IRC$+$10011, the observed peak flux density reaches $\sim$70 and 2~Jy at the epoch of the observations. The brightness temperature, $T_\mathrm{b}$, derived from the peak flux density in our channel maps is $\ga$1$-$6$\times$10$^{3}\ \mathrm{K}$ and reaches $\sim$4.4$\times$10$^{4}\ \mathrm{K}$ and 1.1$\times$10$^{6}\ \mathrm{K}$ in IRC$+$10011 and AH~Sco, respectively (Table \ref{h2o_brightness}). Even higher values of $T_\mathrm{b}$ are obtained in these two stars and in U~Her from the Gaussian component analysis described in Sect. \ref{sec:small_scaleh2o}. 
Such high brightness temperatures indicate  maser-type amplification of the 268.149~GHz emission since they are greater than  the expected kinetic temperature in the preshock or postshock inner gas layers where this line is excited. 
A trend toward non-LTE conditions is also suggested from the ratio of  the peak flux densities at 268.149~GHz and 254.053~GHz, S$_{268}$/S$_{254}$. Both transitions lie in the same (0,2,0) vibrational state and their LTE line opacity ratio, $(\tau_{268}$/$\tau_{254})_{LTE}$, is $\sim$0.9 for an excitation temperature of $\sim$300~K.
The exact line excitation temperature and opacity of the observed lines are not modeled here but we point out that   
the observed flux density ratio, S$_{268}$/S$_{254}$, is well above the LTE opacity ratio. Based on the values in Table ~\ref{fluxdensity_source_list}, we derive S$_{268}$/S$_{254}$ in the range $\sim$5 (IRC$-$10529) to $\sim$140 (IRC$+$10011), and even much higher  in AH~Sco where the 254.053~GHz transition is not detected. This suggests that, in general, thermal excitation is not the dominant line excitation process at 268.149~GHz.

From these remarks  and our discussion in Sects. \ref{sec:variability_268} and \ref{sec:narrowing_268} we conclude that the widespread 268.149~GHz emission observed in evolved O-rich stars tends to be time-variable (e.g., U~Her) and masing (e.g., AH~Sco) even though quasi LTE line excitation conditions are also observed in some sources (see R~Hya and S~Pav population diagrams where including line 14 is relevant, Fig. \ref{h2o_popdiagram}).

H$_2$O line modeling (see Sect.~\ref{sec:maser}) and any correlation of the 268.149 GHz line parameters with other vibrationally excited H$_2$O lines add new insights into the understanding of the nature of the 268.149~GHz emission. In particular, we note that eight of the fifteen 268.149~GHz sources of ortho H$_2$O in Table~\ref{fluxdensity_source_list}, T~Mic, RW~Sco, R~Hya, U~Her, AH~Sco, R~Aql, IRC$+$10011, VX~Sgr, also exhibit  emission of the same water species in the most widespread (0,1,0)~$1_{1,0}$--$1_{0,1}$ transition of H$_2$O at 658.007~GHz \citep[][]{menten1995, hunter2007, justtanont2012, baudrya2018, baudryb2018}\footnote{VY~CMa, IK~Tau, and W~Hya, not included in our survey, also exhibit emission at both 658.007~GHz \citep[e.g.,][]{hunter2007,baudrya2018}, and 268.149~GHz (Tenenbaum et al. 2010; Velilla-Prieto et al. 2017; K. Ohnaka \& K.T. Wong priv. comms.)}.  
Although both transitions are not in the same vibrational state, and despite time variability may hide any potential relationship between these two lines, the brightest sources at 268.149~GHz tend to show strong emission at 658.007~GHz. 
The  658.007GHz transition has a clear tendency to be masing with minimum brightness temperatures ${T_\mathrm{b}(658)}$ $\sim$10{$^{4-10}$~K for those sources also exhibiting 22.235~GHz masers \citep [][]{baudrya2018}. The latter work and \citet [][]{menten1995} also showed from a comparison of  the velocity ranges of the 658.007~GHz and SiO(2$-$1) maser emissions that the 658.007 GHz transition is excited within or close to the SiO maser-emitting region, that is $\sim$5~R$_{\star}$.

 Interferometric observations at 658.007~GHz were performed by \citet{hunter2007}, but 
direct demonstration that this line can be inverted comes from the ALMA observations of the RSG VY~CMa \citep[][]{richards2014} indicating  compact, $T_\mathrm{b}$ $\sim$ 0.3$-4$$\times$10{$^{7}$}~K components, that is roughly consistent with a spherical shell\footnote{The 658.01 GHz high angular resolution mapping of \citet{asaki2020} delineates a probable shock front around  "clump" C which could have been ejected from the central star, thus highlighting the unusual nature of VY CMa.}.  Mapping of AGBs is  not available yet at 658.007~GHz, but this line is currently known to be excited in about 60 evolved objects according to our source count and latest 658.007 GHz observations  \citep[][]{baudryb2018}. Since eight of the {\sc atomium} sources are also known 658.007~GHz emitters and, because of the relatively large number of identified 658.007~GHz sources, the  actual number of 268.149~GHz stellar sources is likely to be larger than reported in the present work. The general physical conditions leading to 268.1549 and/or 658.007~GHz maser emission are discussed further in Sect.~\ref{sec:maser}.

\subsection{262.898 GHz line excitation }
\label{sec:h2o_line12}

The (0,1,0)~$7_{7,1}$--$8_{6,2}$ transition of para H$_2$O (line 12 in Table \ref{H2O-line-list}), is the only rotational transition covered in our observations that arises from the lowest vibrationally excited state of H$_2$O in the $\varv_2 = 1$ bending mode. Although this line is weaker than the 268.149~GHz transition, the detection rate of both  transitions is comparable (12 and 15 objects, respectively, see Table~\ref{fluxdensity_source_list}). In contrast with the 268.149~GHz line which may show strong maser radiation in some sources, examination of the 262.898~GHz line profiles and the population diagrams in Fig.~\ref{h2o_popdiagram} suggest that line~12  does not exhibit strong non-LTE characteristics. This does not imply, however, that population inversion is impossible as indicated by time variability of some 262.898~GHz spectral features in IRC$+10011$ (see Sect. \ref{262_variability}), and by the general maser line amplification discussion in Sect. \ref{excited_watermasers}.

Rotational transitions within  the (0,1,0) state  are nearly always observed toward evolved stars\footnote{Besides evolved stars, the 232.687~GHz line (not in our set-up) in the (0,1,0) state was observed  with ALMA in Orion \citep[e.g.,][]{hirota2012} and in two other star forming regions  \citep{maud2019,tanaka2020}. \citet{liljestrom1996} also observed  the  96.261~GHz transition in the (0,1,0) state  toward two  young stellar objects.} and, if the (0,1,0)~$7_{7,1}$--$8_{6,2}$ line of para H$_2$O is excited at 262.898~GHz, then one will almost always detect one or more H$_2$O lines from the (0,1,0)~state. This is supported by ground-based observations of late-type stars at:
658.007~GHz (see discussion in the previous Section); 96.261 and 232.687~GHz in VY~CMa and W~Hya \citep{menten1989} and, at 232.687~GHz,  in VY~CMa with ALMA \citep{quintana2023}; 
293.664, 297.439, and 336.228~GHz in VY~CMa \citep{menten2006, kaminski2013} and 336.228~GHz in R~Dor and IK~Tau \citep{decin2018}; 232.687 and  263.451~GHz in RS~Cnc \citep{winters2022}. In addition, several $\varv _2 = 1$ lines have been observed with \emph{Herschel}/HIFI  in various AGB stars and supergiants/hypergiants \citep{justtanont2012,teyssier2012,alcolea2013}.

Except for the  (0,1,0)~$5_{2,3}$--$6_{1,6}$ line at 336.228~GHz, all transitions observed in the $\varv_2 = 1$ state have an upper level with $K_a=J$ and $K_c=0$ or 1, including the strong 658.007~GHz line. 
In parallel with the terminology of  "backbone" levels  used by \citet{dejong1973} in his overpopulation model of the $J=K_c$ levels and in his  22.235~GHz ground state maser model, \citet{alcolea1993} referred to the $J = K_a$ upper levels as on the "transposed backbone" and proposed a pumping mechanism that results in a systematic overpopulation of these levels within the $\varv_2 = 1$ state. They found that, following collisional pumping to the $\varv_2 =1$ state, some infrared radiative decay routes of these transposed backbone levels to the ground state become more optically thick than the lower, nonbackbone levels for gas density, kinetic temperature and H$_2$O column densities typical of the inner circumstellar envelopes. This tends to reduce the radiative decay rates of the levels on the transposed backbone and, hence, results in a systematic overpopulation compared to that in the lower levels. Predicting the efficiency of this mechanism was not possible, however, on the basis of the uncertain collision rates and LVG (large velocity gradient) modeling approximation used in \citet{alcolea1993}.

It is interesting to note that the ortho H$_2$O counterpart of the para H$_2$O line at 262.898~GHz (line 12), the (0,1,0)~$7_{7,0}$--$ 8_{6,3}$ line at 263.451~GHz (not covered in our frequency setup) has been detected in a few evolved stars. The 263.451~GHz ortho line was first mentioned by  \citet{alcolea1993} without presenting the spectrum. An emission line at 263.452~GHz was also detected in the line survey of IK~Tau by \citet{velillaprieto2017}, but the carrier was not identified. Recently, \citet{winters2022}  observed this line in the MS-type AGB star RS~Cnc. Modeling of the 262.898 and 263.451~GHz line pair and other line pairs in the (0,1,0) vibrational state are discussed in Sect. \ref{excited_watermasers}.

\section{H$_2$O maser models}
\label{sec:maser} 

\subsection{General comments }
\label{general_comments_h2omodels}
Competition between radiative excitation and de-excitation of many H$_2$O levels together with
collisional excitation and de-excitation  of H$_2$O with H${_2}$, H and/or electrons\footnote{Excitation by electrons is expected to be small in AGBs. The  mole fraction of electrons estimated from the equilibrium thermochemical data of \citet{agundez2020} is  \la5 10$^{-6}$. The CODEX models \citep{ireland2011}, better suited to a dynamic atmosphere, provide an electron content $\l$10$^{-5}$ except very close to the star where water is not excited.}  
may result in a population inversion of various energy levels of water and, eventually, lead to a predominant stimulated emission of radiation  in  specific transitions; a stimulated emission rate greater than the absorption rate is the prime condition to trigger an astrophysical maser. Population inversion is facilitated in molecules such as water because it has many levels of similar energy (e.g., the  $6_{1,6}$  and $5_{2,3} $ rotational levels in the ground vibrational state giving rise to the emblematic 22.235~GHz maser emission). Therefore, any radiative transfer model of water must take into account potential inversions of the water line levels. Multilevel calculations up to energies $\sim$7000~K in the ground vibrational state were first presented by \citet{neufeld1991} to model the 22.235~GHz maser and predict several other collisionally pumped maser lines throughout the energy ladder. Later, \citet{yates1997} identified radiatively pumped lines of water and showed that dust radiation tends to weaken the collisionally pumped maser lines in the (0,0,0) vibrational ground state. Sensitivity to the dust temperature of maser lines in the (0,0,0) state   is also observed in \citet{bergman2020}. Other models have explored the impact of different collision rates on the maser line opacities  \citep[e.g.,][]{daniel2013} and, recently, \citet[][]{neufeld2021} used their collisional pumping models to explain the observation  of  new THz  water lines in the (0,0,0) state toward evolved stars. 
Infrared radiative pumping in the (0,1,0) vibrational state by hot dust was first considered by \citet{goldreich1974} and, most recently, \citet{gray2022} have shown that radiative pumping to  the (0,1,0) state can also explain the 22.235~GHz maser in the low kinetic temperature regime and at high dust temperatures.

\subsection{Highly excited water masers }
\label{excited_watermasers}

Extensive modeling of water masers was presented in \citet [][]{gray2016} who predict maser emission in evolved stars for many observable transitions. 
Their models  process 411 and 413 levels of ortho and para H${_2}$O 
distributed across the first two excited states of the bending mode, (0,1,0) and (0,2,0), the first excited state of both stretching modes, (1,0,0) and (0,0,1), as well as  the ground vibrational state. Line overlap effects within the ortho and para water transitions are also considered separately in the two species\footnote{\citet{bergman2020} suggest that line overlap between the two H${_2}$O species could explain the strong maser observed in evolved stars around 437.347~GHz in the (0,0,0) state}.
The large number of slab-geometry model outputs allows us to explore a large parameter space in density, n$_{H2}$, kinetic temperature, $T_{\rm K}$,  and dust temperature, $T_{\rm dust}$. It was found that the maser transitions can be divided into  three main groups: one exhibiting both radiative and collisional pumping, one with dominant collisions,  
and another group  with a dominant radiative pumping component. The 268.149~GHz maser emission in the (0,2,0) state  pertains to the latter group when $T_{\rm dust}$ is high (Sect.~\ref{maser_268}).

We reexamine below the maser slab model outputs to better understand the excitation of the three strongest  and most widespread  water lines observed in this work, lines 10, 12 and 14 at 259.952, 262.898 and 268.149~GHz (see Table \ref{fluxdensity_source_list}). We have three main goals in mind: $\it(i)$ explore if there are favorable physical conditions for the cospatial  excitation of the strong 268.149 and 658.007~GHz lines as suggested in Sect. \ref{sec:nature_268};  $\it(ii)$ investigate the conditions leading to the relatively strong line emission observed at 262.898~GHz (this work) and in other $\varv_2=1$ lines (see  references to other published works in Sect. \ref{sec:h2o_line12}), and $\it(iii)$ assess the conditions leading to significant excitation of the 259.952~GHz line in the ground vibrational state. Ortho and para H${_2}$O are treated separately and the H${_2}$ abundance is obtained from the H${_2}$O abundance divided by the fractional abundance of  H${_2}$O/H${_2}$ = 3$\times$$10^{-5}$ that is considered in our models.

\subsubsection{268.149 GHz}
\label{maser_268}

%%%%%%%%
\begin{figure*}
\centering \includegraphics[width=13.5cm,angle=270]{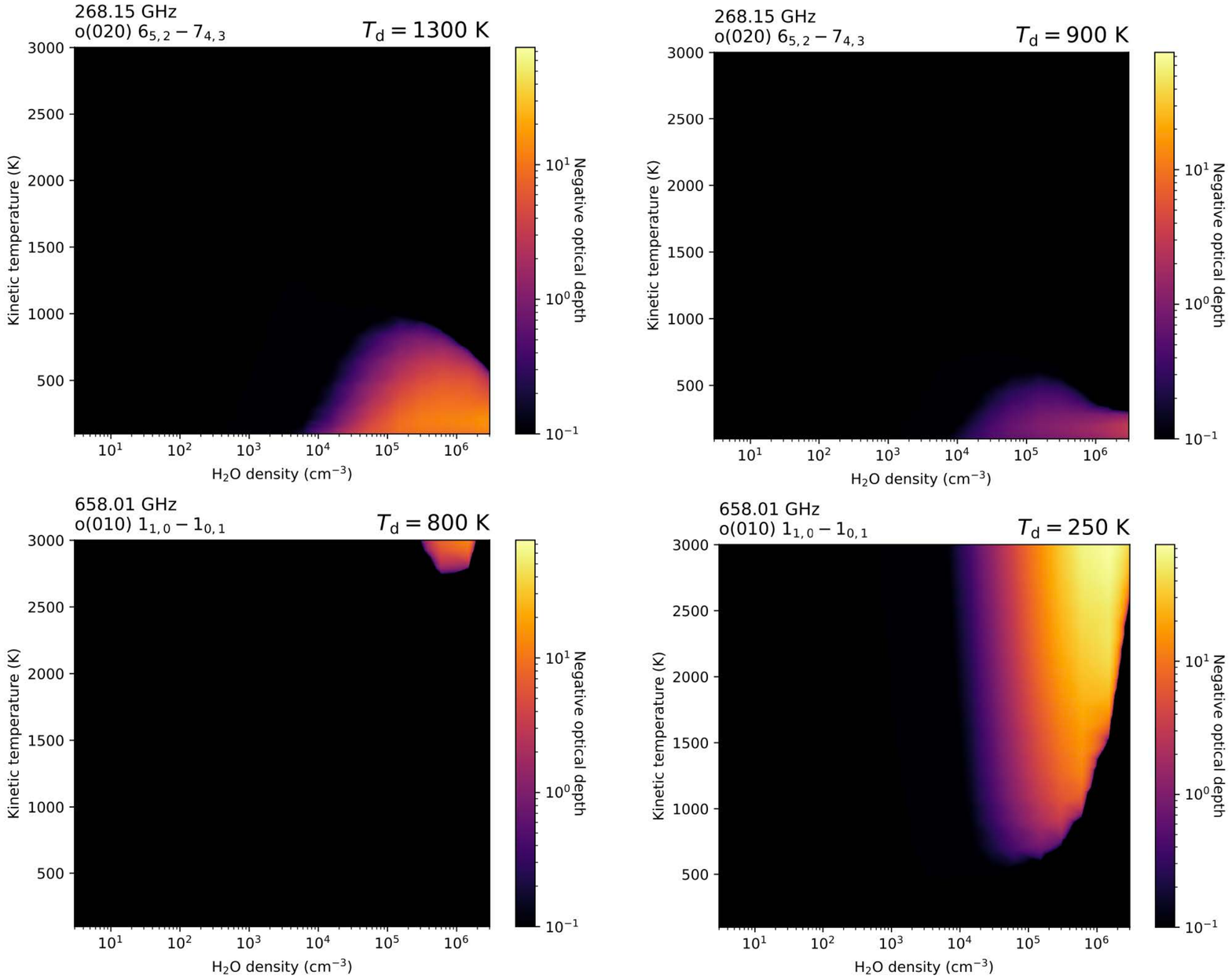}
\caption{Comparison of maser depths of ortho H$_2$O at 268.149 and 658.007 GHz  (upper and lower two panels)  using models in \citet{gray2016}. Strong maser emission at 268.149 and 658.007 GHz is obtained for markedly different kinetic temperatures (vertical axis on left hand-side of each panel)  and dust temperatures $T_{\rm d}$ (given in the upper right-hand side of each panel). The transition frequency is given in the upper left corner of each panel with, below, the letter o  for ortho H$_2$O followed by the vibrational state and the rotational transition. The negative optical depth is specified in the vertical bar on the right-hand side of each panel using a log normal scale (min=$-$0.1 and max=$-$75.0); black means  no inversion. (The molecular H${_2}$ density is obtained by dividing the water abundance by 3$\times$$10^{-5}$.) 
 }
\label{h2o_graymodels}
\end{figure*}

%%%%%%%%

We find that for the high density  conditions favorable to masers, the 268.149 GHz transition is weakly, or not inverted at kinetic temperatures $\ga$500~K for $T_{\rm dust}$  in the range  $\sim$50~K to a few hundreds K.
When $T_{\rm dust}$ reaches several hundreds or  $\ga$1000~K, 268.149~GHz inversion tends to move to higher values of $T_{\rm K}$ in the density vs.  kinetic temperature plane. Increasing $T_{\rm dust}$  from 900 to 1300~K, for example, increases the inversion as shown in the upper two panels of Fig. \ref{h2o_graymodels} for $T_{\rm K}$  $\la$500~K and  $\la$900~K, respectively, around n$_{H{_2}O}$ $\sim$10$^{5}$$-$10$^{6}$~cm$^{-3}$. Inversion is thus expected for n$_{H2}$ $\sim$3$\times$$10^{9}$$-$3$\times$10$^{10}$~cm$^{-3}$ and for  $T_{\rm dust}$ = 900 and 1300~K.  Within the inner 3$-$10~R$_{\star}$ region where we observe the 268.149~GHz line and  where the dust has just formed and is optically thin, the radiation field is dominated by the central star and we expect indeed high dust temperatures and densities.

On the other hand, the radiative pumping component, needed for the 268.149 GHz maser, is not needed at 658.007~GHz for which lower values of $T_{\rm dust}$ (\la 800~K) tend to provide strong inversion for a broad range of $T_{\rm K}$ (see lower two panels in Fig.\ref{h2o_graymodels}). Therefore, the physical conditions prevailing within the inner gas layers where we expect high $T_{\rm dust}$ do not seem a priori favorable to the  658.007~GHz maser excitation. High $T_{\rm dust}$ conditions may nevertheless result in 658.007~GHz line inversion if $T_{\rm K}$~$\ga$3000~K (see lower left panel of Fig.~\ref{h2o_graymodels}) which is plausible in the shocked environment of the expanding stars but, in turn, is not favorable to 268.149~GHz line inversion. We can not exclude, however, that in some peculiar objects, cooler and less dense conditions than required for 268.149~GHz are favorable to the 658.007~GHz maser. We note that in VY~CMa, the 658.007~GHz masers \citep[][]{richards2014, asaki2020} are mostly located at roughly twice the distance from the star where the SiO masers are observed, supporting slightly cooler conditions favorable to the 658.007 GHz line excitation, possibly once opaque dust has formed. (Alternatively, these 658.007~GHz masers could also be excited by a shock around the VY~CMa clump C.)

We conclude that in general, excluding the peculiar VY~CMa object, the potential link between the 268.149  and the 658.007~GHz line emissions mentioned earlier in Sect. \ref{sec:nature_268} (on the basis of source detection rates) is not straightforwardly explained in terms of our current maser models. Of course both lines are not necessarily masing at the same time and position in the same object, and a possible  relation between observed lines may be more difficult to decipher than with a simple search for common masing conditions.

\subsubsection{262.898 GHz }

%%%%%%%%%%%%

\begin{figure}
\centering \includegraphics[width=9.8cm,angle=0]{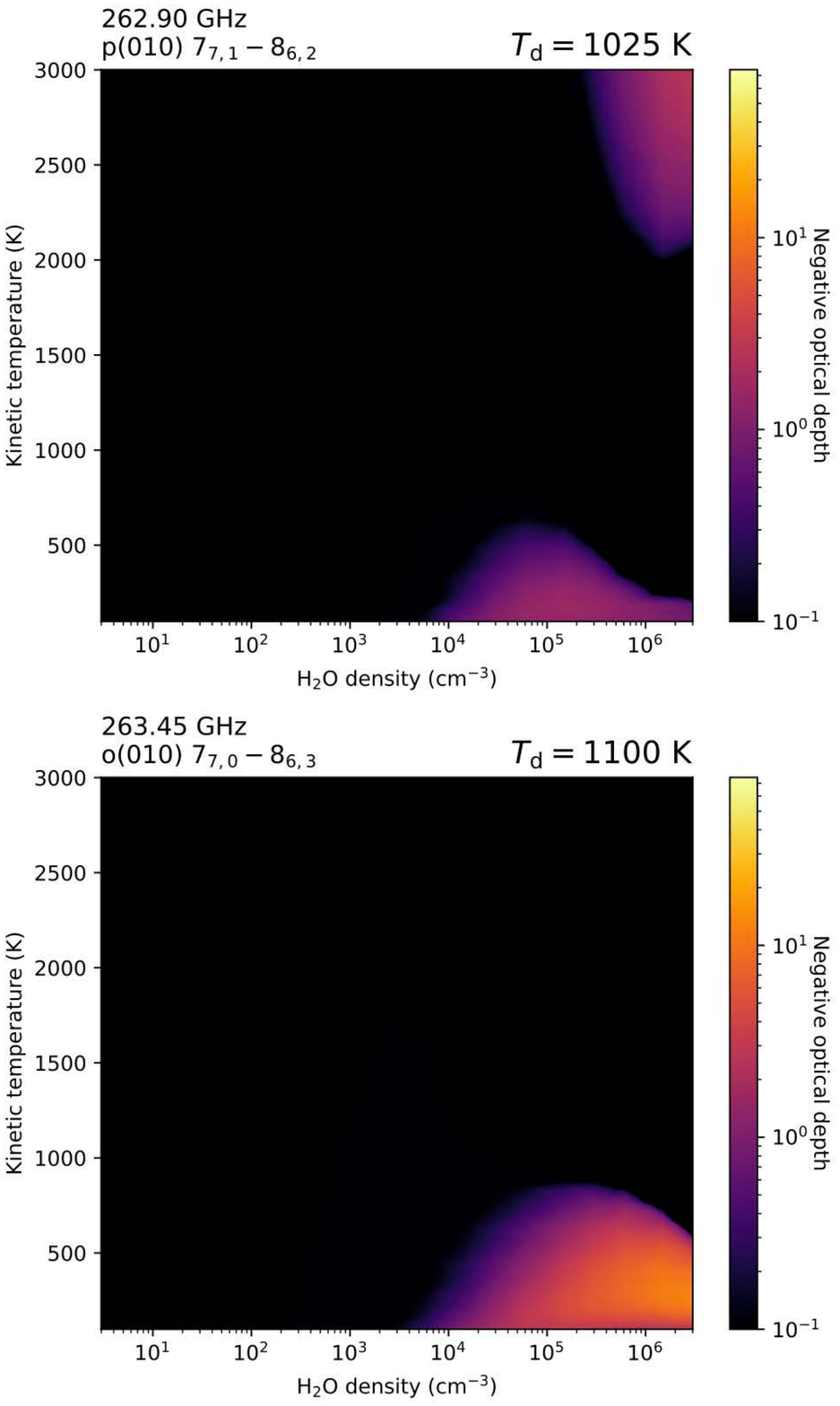}
\caption{Negative optical depths as in Fig. \ref{h2o_graymodels} at 262.898 and 263.451 in the (010) vibrational state. The quantum numbers of these transitions are preceded by the letters p and o for para and ortho H$_2$O. (The molecular H${_2}$ density is obtained by dividing the water abundance by 3$\times$$10^{-5}$.) 
 }
\label{h2o_pairs_graymodels}
\end{figure}

%%%%%%%%%%

We mention at the end of Sect.  \ref{sec:h2o_line12} the importance of the differential radiative trapping scheme for the $J=K_a$ "transposed backbone" levels in the $(0,1,0)$ state. Especially for the para and ortho H${_2}$O line pair at 262.898 (our widespread line 12) and 263.451 GHz observed by \citet{winters2022} (see Sect. \ref{sec:h2o_line12} and below). Our models demonstrate that both lines can be masing but for different physical conditions. Line 12 remains a weak maser in the $T_{\rm K}$ versus density regions considered here for both low and high dust temperatures ($\sim$50 and $\ga$1000 K). 
We find, for example, that the dominant route for 
population inversion at 262.898~GHz for $T_{\rm dust}$ = 1025~K and $n_{\rm H{_2}O}$ $\ga$10$^{6}$ cm$^{-3}$, 
requires $T_{\rm K}$ $\ga$2300~K 
(see the weak negative opacity contours at 262.90~GHz in the  top right-hand corner of the upper panel in Fig. \ref{h2o_pairs_graymodels}) in contrast with the lower $T_{\rm K}$~$\la$700~K required at 263.451~GHz  (see bottom right of the 263.45~GHz maser plot in the lower panel of Fig.  \ref{h2o_pairs_graymodels}).
Our model outputs can be compared with the  observations of the 262.898 and 263.45~GHz line pair. At 262.898 GHz, the brightness temperature derived from the {\sc atomium} data do not give unambiguous evidence of masing (Table \ref{h2o_brightness}) but this could be due to the emission being very compact so that the measured temperature is a lower limit. In addition, the 262.898 GHz line profiles are narrow and there is some evidence for variability (Sect. \ref{262_variability} and Fig. \ref{irc10011_variab_h2o_262}). We conclude that this transition  could be weakly masing  in the inner gas layers with high density and high kinetic temperature. The models also predict 262.898~GHz inversion over a wide range of dust temperature from $\sim$100~K to $\ga$1000~K and for $n_{\rm H{_2}O}$ $\ga$10$^{6}$ cm$^{-3}$ and $T_{\rm K}$ $\ga$2000~K. For the ortho H$_2$O line at 263.451~GHz, the line profile in RS~Cnc appears to be rather broad and stable \citep{winters2022}. Similarly, the unidentified emission feature in IK~Tau at the same frequency \citep{velillaprieto2017}, which we identify as the  ortho H$_2$O line, does not show obvious signs of maser action in its line profile. Observations of other stars should be performed to characterize this emission, but we momentarily conclude  that the maser conditions quoted above for the 262.898~GHz transition do not apply to the 263.451~GHz line (see Fig.\ref{h2o_pairs_graymodels} for $T_{\rm dust}$$\sim$1000~K).  We also note that the low $T_{\rm K}$, high H${_2}$O density regime of these two lines do not exactly overlap (Fig. \ref{h2o_pairs_graymodels})\footnote{This suggests that simultaneous observation of these lines would help to identify nonthermal emission in at least one line if the line intensity ratio deviates from the LTE conditions.}.

Finally, we draw attention to a rather rare occurrence in our models. The  upper panel  in Fig.\ref{h2o_pairs_graymodels} at 262.898~GHz shows that both radiative and collisional pumping regimes can coexist at the same dust temperature whereas most transitions that have both schemes show a clear bifurcation  at  $T_{\rm dust}$ $\sim$750$-$1100~K where the dominant pumping route switches from collisional to radiative. This two-inversion occurrence although seen faintly here at 262.898~GHz can be stronger in a few other maser lines (e.g., at 209.118 and 96.261~GHz).

\subsubsection{259.952 GHz}

We  have not observed any clear sign of time variability in our data at 259.952 GHz (line 10) in the ground vibrational state. However, variability may be hidden by the limited number of epochs  available to us in the {\sc atomium} program.  Line~10 is weaker than the 262.898 and 268.149~GHz lines (lines 12 and 14) but it is  usually stronger than all the other lines (Table \ref{sec:H2O_analysis}).  We find that line 10 has a relatively strong radiative pumping component  for conditions reminiscent of those modeled for the 268.149~GHz maser. This is seen by comparing the 260 GHz plot in Fig.~7 of \citet[][]{gray2016} at a dust temperature of 1025~K, with the upper
panels of our Fig.~\ref{h2o_graymodels} at 268.149~GHz. Line 10 can thus be weakly inverted close to the photosphere but, in contrast with the strong 268.149~GHz emission, it does not show strong signs of maser emission.

We conclude Section~\ref{excited_watermasers}  by stressing that observations toward more evolved stars in the $\varv_2 = 1$ and $\varv_2 = 2$  states as well as coordinated high angular observations of, for example, the 268.149 and 658.007~GHz lines, or  the 262.898 and 263.451~GHz line pair, are desirable for a deeper understanding of  the properties of these high-lying levels.

\section{OH source properties}
\label{sec:detect_OH}

\subsection{Ground state and high-$J$ OH stellar sources}
\label{sec:general_OH}

Before presenting the properties of the high-$J$ OH transitions observed here with ALMA, we briefly recall some characteristics of the well-known  18-cm lines in the $J=3/2$ ground state of OH  observed in many AGB stars and RSGs. 
Strongly non-LTE 18-cm emission of OH from evolved stars was first reported by \citet [][]{wilson1968} who observed that the 1612 MHz "satellite" line is predominantly excited among the four hyperfine transitions of the   $^{ 2}{   \Pi_{ 3/2} }$, $J = 3/2$ ground state. (The ${\Delta} F = \pm 1$ and 0 transitions in a given $J$ state of OH are called satellite and principal or main lines on the basis of their spontaneous emission rate values, see Sect. \ref{sec:spec_backgr_OH}.) The four 18-cm hyperfine transitions of OH excited around 1612, 1665, 1667 and 1720~MHz, have relative LTE  intensity  ratios of 1:5:9:1 which are never observed in stars. Interferometric  18-cm observations have shown that OH emission comes from the expanding circumstellar envelope of evolved M-type stars \citep [e.g.,][]{reid1977} where OH is produced from the H$_2$O photodissociation \citep{goldreich1976}. Excitation models based on OH pumping by FIR photons at $\sim$35 and 53~$\mu$m and on near IR overlaps of OH lines  were proposed to explain the strongly non-LTE 18-cm OH line emission  \citep [e.g.,][]{elitzur1976, collison1994}. This general OH excitation scheme is supported by many ground-based observations of AGB stars and RSGs and is not contradicted by the recent space observations
\citep[see e.g., \emph{Herschel}/HIFI observations of][]{justtanont2012,teyssier2012,alcolea2013}.

Restricting ourselves to ground-based observations of 
evolved stars, we note that until recently, no cm, mm or submm wave detection of OH has ever been reported, as far as we are aware, at high-$J$ levels  well above the  ground state\footnote{In the near IR, however, the OH vibration-rotation overtone bands have long been known in Mira variables; for more details, readers can refer to the CO and OH bands in R~Leo \citep{hinkle1978}, for example.}.
Today, the  ALMA high sensitivity enables us to search for OH lines up to $J = 35/2$ or above in $\varv = 0$ and 1 (i.e., lines with  energy levels up to $\sim$9000~K). High-$J$ OH emission was first reported toward W~Hya and R~Dor by \citet{khouri2019}. In the present work, we  extend the search for high-$J$ OH  to the {\sc atomium} sample.

\subsection{OH main properties}
\label{sec:atomium_OH}

%%%%%%%
 \begin{figure}
  \centering
   \includegraphics[width=8.5cm, angle=0]{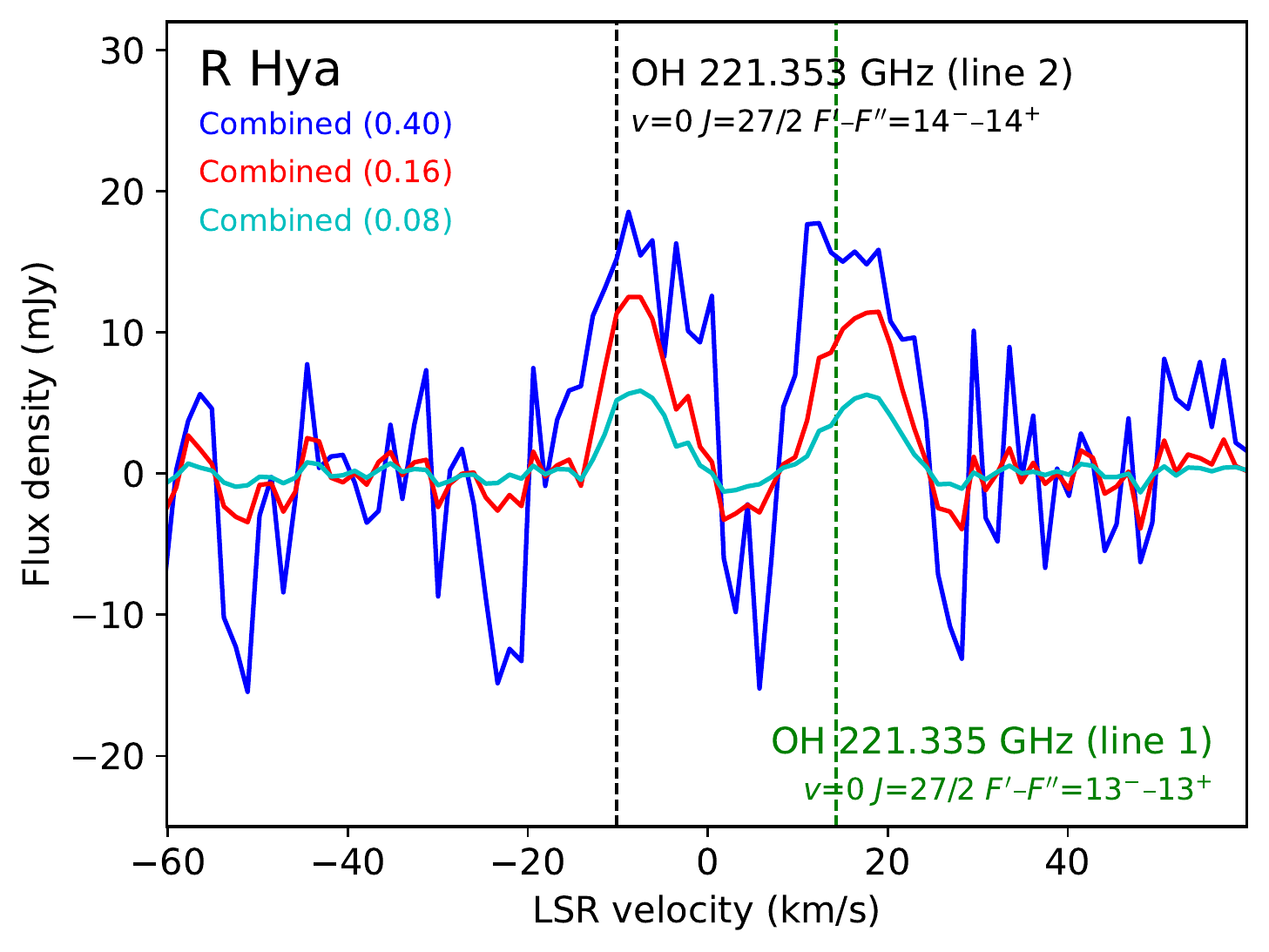}
   
   \caption{R Hya spectra of the $F'-F''$ = 13$-$13 and 14$-$14 hyperfine transitions of OH in the $\varv=0, J=27/2$ rotational transition (lines~1 and 2 in Table~\ref{OHline_list}) extracted from combined high and mid spatial resolution data for an aperture diameter of  0\farcs08, 0\farcs16 and  0\farcs40. 
   The spectra are converted from the observed frequency to the LSR frame using the line~2 rest frequency as the "reference" frequency. The black dotted vertical line (line~2) is at the adopted new systemic velocity (see Table~\ref{primarysource_list}). The green dotted vertical line  is displaced by  $\sim$24.3~km\,s$^{-1}$ compared to line~2 as expected  from  the $\varv =0, J=27/2~\Lambda$-doublet frequency separation; it corresponds to line~1. 
  }
     \label{rhya-OH}
    \end{figure}
    
  %%%%%%%%
    
    \begin{figure*}
 \centering
  \includegraphics[width=16.5 cm, angle=0]{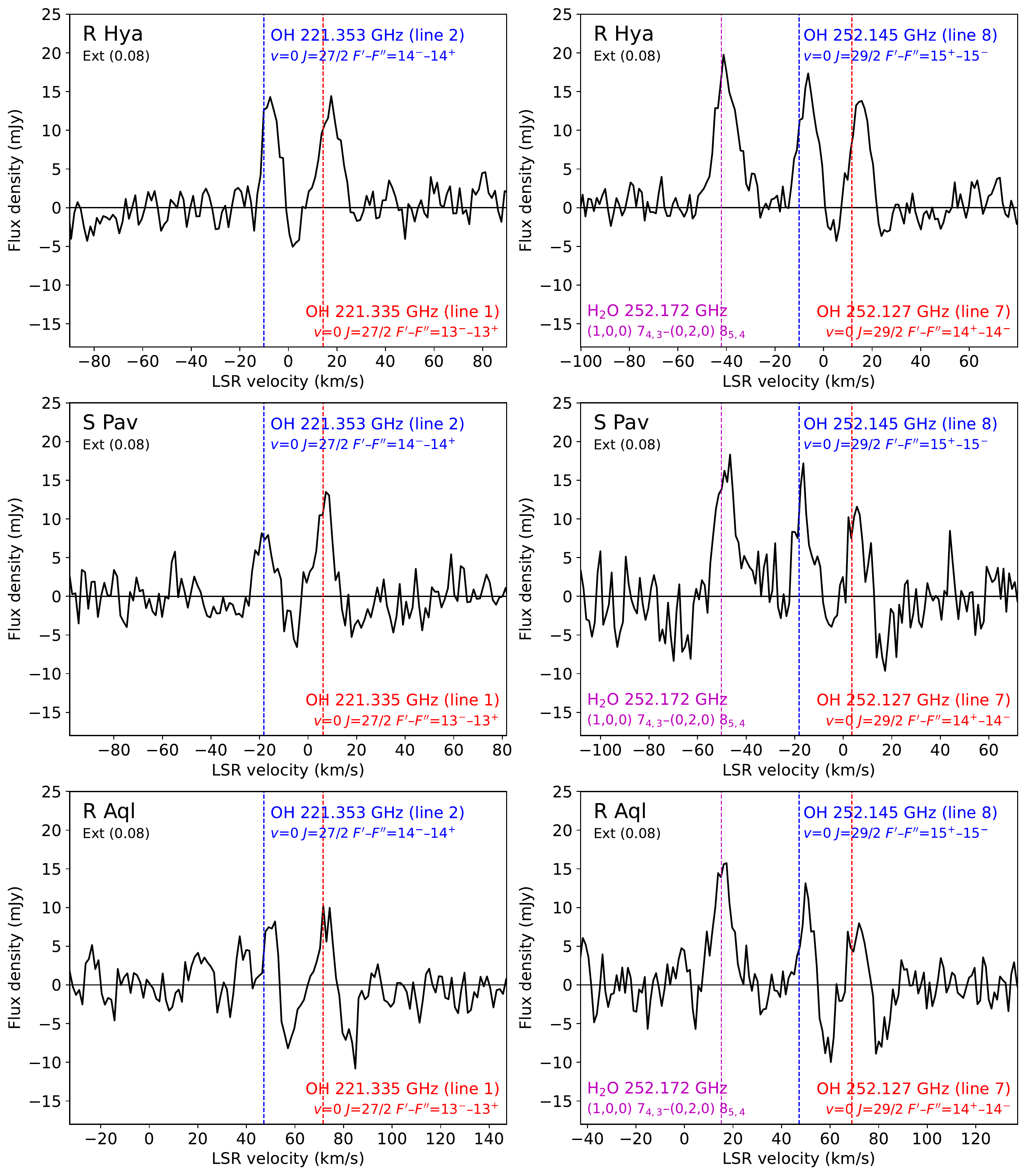}
 \caption{Spectra of R~Hya, S~Pav and R~Aql in the $J$ = 27/2 and 29/2 rotational levels of OH extracted from the high resolution data cubes for an aperture diameter of  0\farcs08. The three left and three right panels correspond to $J$ = 27/2 and 29/2, respectively. In each spectrum, the two nearby emission features correspond to the two hyperfine transitions of each  ${\Lambda}$-doublet (dotted blue and red vertical lines). Note weak absorption features observed in all six spectra. In the $J = 29/2$ spectra, the most negative velocity emission feature (closest to the $F'-F''$ = 15$-$15 transition) is the  252.172~GHz line of water. The spectra are converted from the observed frequency to the LSR frame using the OH catalog line rest frequencies given in Table \ref{OHline_list}. Line 2 at 221353.48~MHz and line~8 at 252145.35~MHz are taken as the reference frequencies for each spectrum in the $J$ = 27/2 and 29/2 states; they are placed at the adopted new systemic velocities (see Table \ref{primarysource_list}). 
 }
     \label{rhya-spav-raql-OH}
            \end{figure*}
            
%%%%%

Secure detection of high-$J$ OH radiation from circumstellar environments relies on the identification of OH ${\Lambda}$-doublets in our spectra and channel maps. For all $\varv =0$, $J= 27/2$, 29/2 or 33/2  spectra and maps we have used the rest frequencies from the JPL catalog (Table \ref{OHline_list}) which are slightly different from the newer rest frequencies derived in this work and given in  Appendix \ref{sec:Lambda_doubling_freqs}.
Spectra have been extracted for different aperture radii from the high and mid  resolution data cubes as well as from the combined data which maximize sensitivity to angular structures $\sim$ 0\farcs1 or larger. The combined spectra  show that the peak flux density increases with the extracted radius suggesting that there is diffuse emission in addition to the bulk of the compact OH emission. This is illustrated in Fig. \ref{rhya-OH} for the $J = 27/2$ rotational line in R~Hya. However, our OH maps show that in addition to moderately extended OH gas material, angularly compact structures are dominant (see e.g., Figs.~\ref{rhya_channmap_OH_29}, \ref{Fig_OH_emission_29_2}). 
Our discussion  is thus primarily based on the high resolution data which 
still can recover scales up to $\sim$0\farcs4$-$0\farcs6. Fig.~\ref{rhya-spav-raql-OH} shows an example of high resolution spectra extracted for an aperture diameter of 0\farcs08 in three prominent OH sources, R~Hya, S~Pav and R~Aql. We stress that with the mid resolution, the absorption features seen  in Fig.~\ref{rhya-spav-raql-OH} at high resolution, are averaged with the surrounding emission (and so are not seen in the mid resolution spectra). These absorption features  are close to the noise level but mom~0 maps made with various velocity intervals show compact absorptions that are discussed in Sect.~\ref{sec:OHabsorption}.

\subsubsection{OH channel maps, angular sizes}
\label{sec:OHchannmaps}

%%%%%    
\begin{figure*}
 \centering
%\vspace{-1cm}
 \includegraphics[width= 16.0 cm, angle=0]{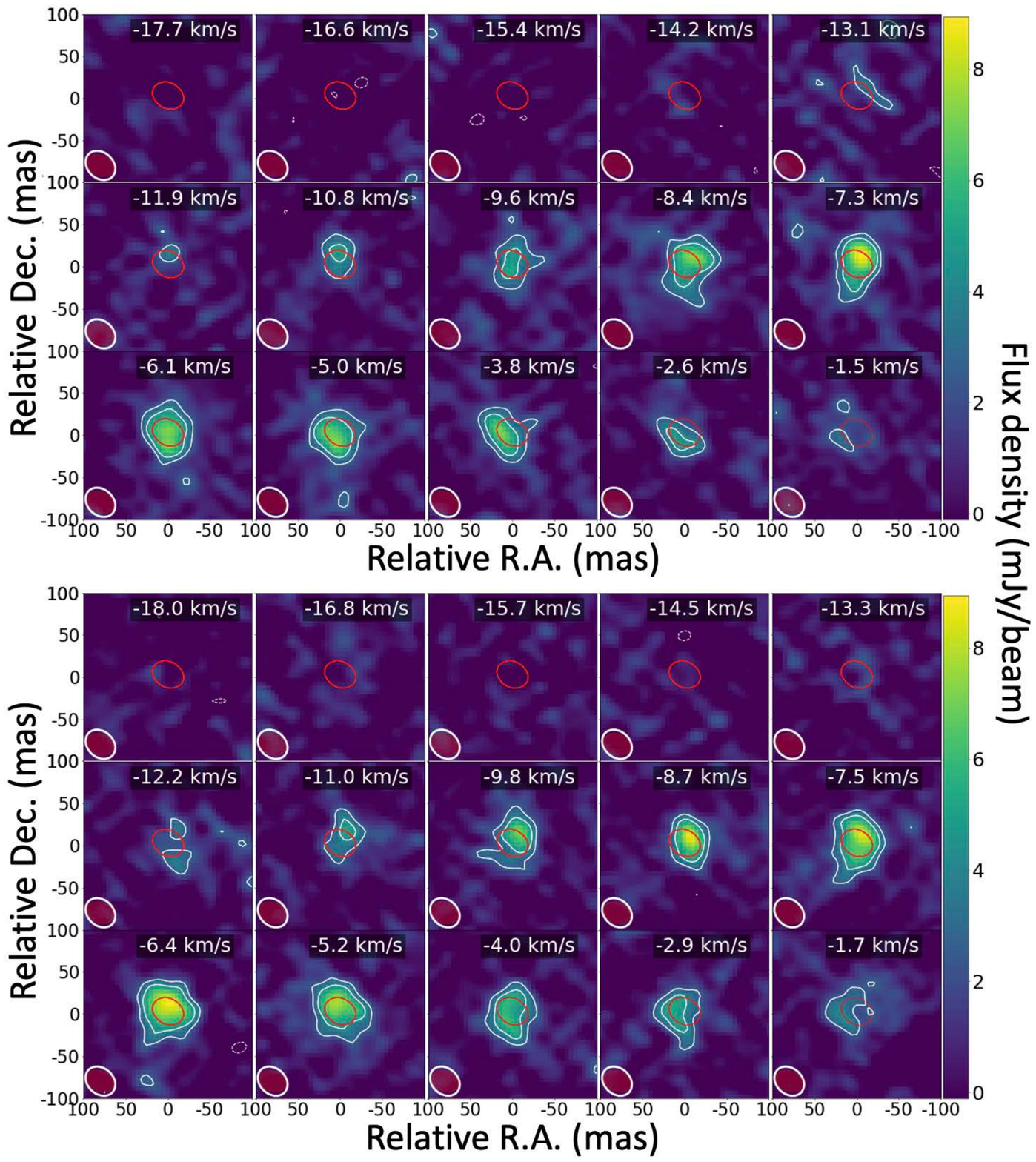}
 
 \caption{High resolution channel maps of $J$ = 29/2, $F'-F'' = 14-14$ and 15$-$15 transitions of OH in R Hya (upper and lower panels, respectively). Each map covers 200$\times$200 mas and is centered on the continuum emission peak at (0,0) position (coordinates given in Table~\ref{primarysource_list}).  Each channel velocity is in the LSR frame using the OH catalog rest frequency in Table~\ref{OHline_list}. 
The white contours are at $-$3,  3 and 5$\sigma$. The line peak flux density  and the typical r.m.s.  noise are 9~mJy/beam and 1~mJy/beam, respectively. The red contour delineates the extent at half peak intensity of the continuum emission. The line and continuum beams are shown at the bottom left of each map in white and dark-red, respectively. The HPBW is (39$\times$31)~mas at PA~46$^{\circ}$ and (34$\times$25)~mas at PA ~67$^{\circ}$ for line and continuum, respectively. 
 }
 
  \label{rhya_channmap_OH_29} 
  \end{figure*}
%%%%%%%

OH channel maps in the $J=27/2, 29/2, 33/2$ and 35/2 rotational  levels have been produced for all sources. They allow us to search for the  signature, at the expected velocities, of the two hyperfine transitions of each mapped $J $  transition. A first example is given for R Hya in the $J = 29/2$ level (Fig. \ref{rhya_channmap_OH_29}). Each hyperfine transition of the $J = 29/2$ $\Lambda$-doublet, with rest frequency taken from the JPL catalog (Table \ref{OHline_list}), exhibits emission within approximately the same velocity range.
More OH maps are shown in Appendix \ref{sec:oh_channel_maps}: $J= 27/2$ in R~Hya and R~Aql (Figs. \ref{rhya_channmap_OH_27} and \ref{raql_channmap_OH_27}) or $J=29/2$ in S~Pav (Fig. \ref{spav_channmap_OH_29}).

All high-$J$ OH sources observed  in this work are weak with peak intensity ranging from $\sim$3$-$7 mJy/beam (S~Pav, T~Mic, U~Her or AH~Sco) to  $\sim$6$-$10 mJy/beam (R~Hya, R~Aql or VX~Sgr). In two other sources, RW~Sco and IRC$+10011$, our channel maps do not show unequivocal OH detection (but see OH stacking in Sect. \ref{sec:OHstackmaps}).

The majority of the OH emission is observed close to the optical photosphere in general; However,  the OH emission peak may not exactly coincide with the continuum peak. Toward R~Hya, in particular, the  emission seems to peak at slightly displaced positions across a few velocity channels (Fig. \ref{rhya_channmap_OH_29}). Nevertheless, we do not see any apparent rotation or any clear velocity signature in the $J$ = 29/2 or 27/2 first moment maps of R Hya where the intensity is weighted by the velocity of each channel for channels centered around the systemic velocity of the star. 

%%%%%%%%%%%%%
\begin{figure*}
 \centering
  \includegraphics[width= 12.8 cm, angle=270]{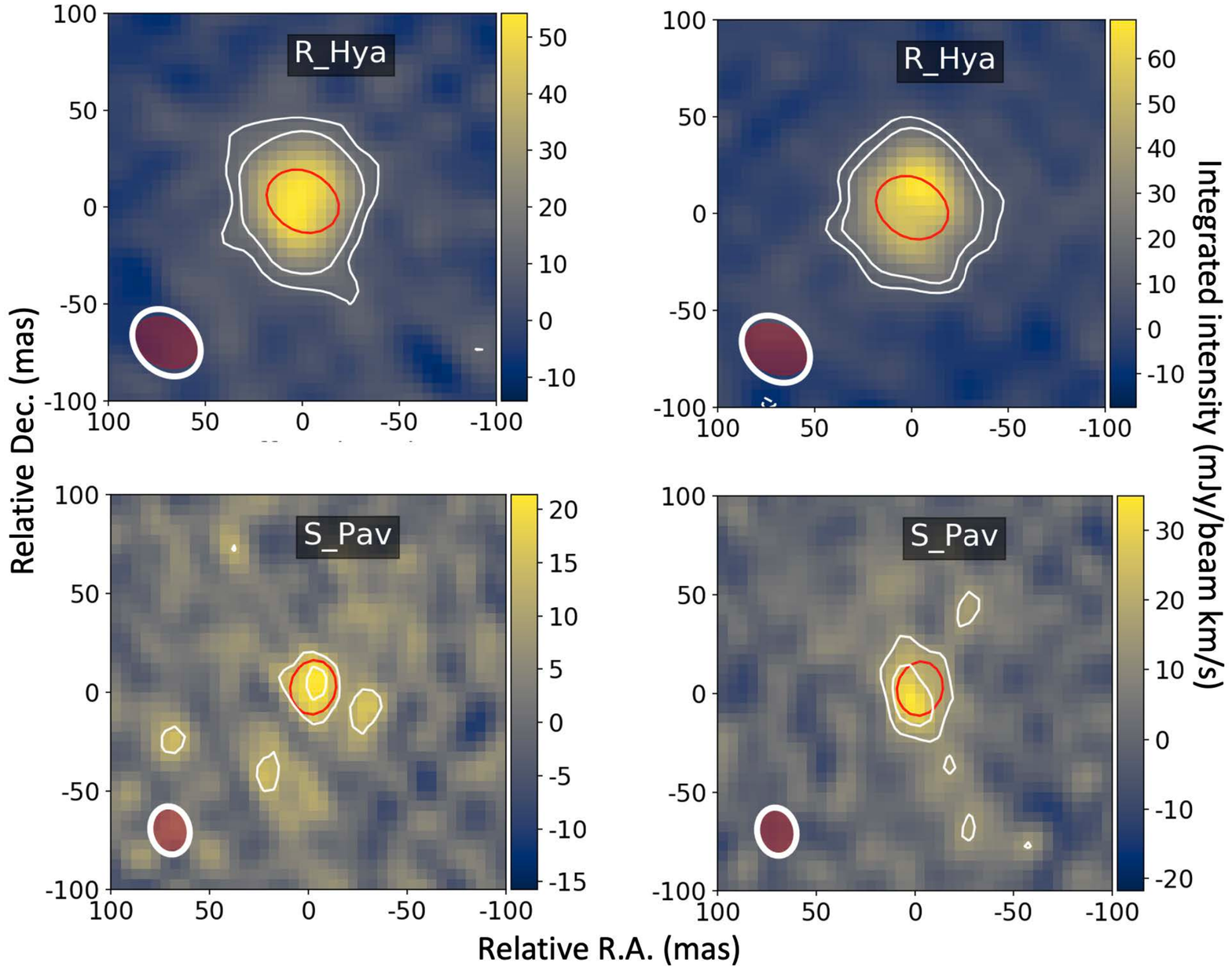}
 \caption{Zeroth moment map of OH emission in the $J$ = 29/2, $F'-F'' = 14-14$  and 15$-$15  transitions. $\it Left~panels$:  $F'-F'' = 14-14$ transition in R~Hya and SPav. $\it Right~panels$:  $F'-F'' = 15-15$ transition in the same sources. Offsets in RA and Dec directions are relative to the coordinates of the peak stellar continuum as given in  Table~\ref{primarysource_list}. 
  The velocity intervals in the $F'-F''=14-14$ and 15$-$15 transitions are: $-$11.8 to $-$2.6~km\,s$^{-1}$ and $-$12.2 to $-$1.7~km\,s$^{-1}$ in R~Hya;  $-21.1$ to $-13.8$~km\,s$^{-1}$ and $-$21.2 to $-$11.9 km\,s$^{-1}$ in S~Pav. 
 In each map the white contours are  at 3 and  5$\times$$\sigma$. The red contour delineates the extent at half peak intensity  of the  continuum emission.  
The line and continuum beams are shown at the bottom left of each map in white and dark-red, respectively. The line HPBW in R~Hya is as in Fig.~\ref{rhya_channmap_OH_29}; in S~Pav the line HPBW is (25$\times$20)~mas at PA~11$^{\circ}$. The continuum beam is (34$\times$25)~mas at PA~67$^{\circ}$ (R~Hya) and (25$\times$20)~mas at PA~$-$13$^{\circ}$ (S~Pav).
 }
     \label{Fig_OH_emission_29_2} 
   \end{figure*}
%%%%%

Weak, diffuse OH emission is visible on larger scales around the stellar continuum sources in all {\sc atomium} sources where OH is observed. Diffuse gas is identified in our mom~0 clean images integrated over the velocity range  corresponding to each OH hyperfine transition thus improving the sensitivity.  
A sample of our mom~0 images  is shown for R Hya and  S Pav  in the two hyperfine transitions $F'-F''=14-14$ and $15-15$ of the $J = 29/2$ level (Fig. \ref{Fig_OH_emission_29_2}). The same transitions in R~Aql and VX~Sgr are presented in Fig.~\ref{Fig_OH_emission_29_2_raql_vxsgr}. The OH mom~0 maps  for these four  stars, but now  in the $J=27/2$ level, are also shown in Fig. \ref{Fig_OH_emission_27_2}. 
The $J = 29/2$ and 27/2 maps exhibit similar structures although the peak intensity ratios of the hyperfine transitions for each one of these two $J$ levels may slightly differ (Table \ref{OH_opacity_ratio}). 
The OH extent of these clean, emission (or absorption, see Sect. \ref{sec:OHabsorption}) images is obtained by fitting 2D Gaussians with CARTA (see end of Sect. \ref{r}) within the 3$\sigma$ level contour of our mom~0 maps; this contour is shown as the first light white contour in Fig.~\ref{Fig_OH_emission_29_2} and in Figs.~\ref{Fig_OH_emission_29_2_raql_vxsgr}  and \ref{Fig_OH_emission_27_2}. The elliptical Gaussian fit dimensions, not de-convolved from the beam, are given for each hyperfine transition of the $J = 27/2$ and 29/2 rotational levels in columns 4 and 8 of Table~\ref{OH_size_peak_intensity} together with the typical noise, $\sigma$,  and the clean beam size. The uncertainties are estimated to be around 10$\%$ from noise variations in the mom~0 maps.

Typical angular sizes of the emitting and absorbing OH regions can be crudely estimated from the geometric mean of the Gaussian dimensions given for a few stars in Table~\ref{OH_size_peak_intensity}.  The OH emission sizes,  are $\sim$80$-$90~mas for R~Hya and $\sim$30$-$50~mas for S~Pav, R~Aql or T~Mic  in the  $J = 27/2$ and 29/2 rotational lines.  These estimates are greater than the radio continuum uniform disk sizes measured at 250~GHz\footnote{Our uniform disk sizes at 250~GHz are 27.1, 20.4, 15.0 and 20.0~mas for R~Hya, S~Pav, R~Aql and T~Mic, respectively. See also Sect.~\ref{sec:channmaps_H2O} for discussion on H$_2$O angular sizes.} and correspond to $\sim$(6.5$-$7.5)$\times$R$_{\star}$ for R~Hya, $\sim$(5$-$8)$\times$R$_{\star}$ for S~Pav and R~Aql, and to $\ge$7$\times$R$_{\star}$ for T~Mic where diffuse emission is observed. The total size encompassing all the OH flux density is uncertain and larger than the above estimates. It would be around three times the full width at half maximum if the gas distribution was Gaussian and centered on the star, but it seems realistic to adopt the 3$\sigma$ contour to roughly define  the OH cloud sizes of  the AGBs in our sample. OH emission is also observed beyond the typical sizes mentioned above; see examples up to $\sim$(7$-$10)$\times$R$_{\star}$ in S~Pav (Figs.~\ref{Fig_OH_emission_29_2}, \ref{Fig_OH_emission_27_2}) and beyond 10$\times$R$_{\star}$ in R~Aql (Fig.~\ref{Fig_OH_emission_29_2_raql_vxsgr}).

The two supergiants AH~Sco and VX~Sgr exhibit  complex  spatial structures. In AH~Sco, the mom~0 maps reveal angularly compact or unresolved emission  in both  $J = 27/2$ and 29/2 (see  Table \ref{OH_size_peak_intensity}), but in the $ J= 29/2, F'-F'' = 15-15 $ transition there is an asymmetric NE emitting region well beyond 10$\times$R$_{\star}$. 
 This  structure is not clearly visible in the OH individual channel maps because of its weakness; however, it is present in the channel maps of water at 252.172~GHz, a frequency very close to the $J = 29/2$ OH emission.
In VX~Sgr, the OH mom~0 image in the $J =29/2$, $F'-F'' = 15-15$ transition also reveals an irregular southern emission structure beyond 10$\times$R$_{\star}$ (Fig.~\ref{Fig_OH_emission_29_2_raql_vxsgr}). AH~Sco and VX~Sgr thus exhibit weak OH gas emission  beyond the central star up to sizes larger than the typical sizes observed in the AGB stars.  

%%%%%%%%%%%%%

 \begin{table*}
\begin{center}
\caption{Peak flux density, r.m.s. noise and angular size in zeroth moment maps of OH emission and absorption.}       
\label{OH_size_peak_intensity}      
\begin{tabular}{lccccccccccc}       
 \hline\hline 

 Source & Mom 0 peak  & $\sigma$$^{a}$  & Size$^{b}$ & Beam$^{c}$ & Mom 0 peak & $\sigma$$^{a}$  & Size$^{b}$  &  Beam $^{c}$  \\
 & (mJy/beam km\,s$^{-1}$ & & (mas) & (mas) & (mJy/beam km\,s$^{-1}$)  & & (mas) & (mas) \\
 \\
 &  $J = 27/2$, 13$-$13  & & 13$-$13 &  & $J = 29/2$, 14$-$14  & & 14$-$14 & \\
 & $J = 27/2$, 14$-$14 & & 14$-$14 & &   $J = 29/2$, 15$-$15 & & 15$-$15 \\
\\
\hline

S Pav & 26.3 & 4.3 & 71 $\times$ 37  & 27 $\times$ 21 & 21.4  & 3.9  & 38 $\times$ 28 & 25 $\times$ 20 \\
 & 33.1 & 4.0 & 48 $\times$ 32  &  & 34.9 & 4.7 & 49 $\times$ 33 \\

$\it absorption$$^{d}$ & $-24.5$  & 4.8  & ? $^{e}$ & & $-34.3 $ & 6.1 &  ? $^{e}$ \\
& $-20.4$ & 3.5 & 32 $\times$ 15  & & $-15?$ & 4.7 & ? \\
\\
 T Mic &  40.7 & 9.5 & ? $^{f}$   & 25 $\times$ 23 & 55.2 & 10.0  & ? $^{f}$ & 28 $\times$ 21  \\ 
  & 27.0 & 6.4. & ? $^{f}$   &  & 35.4 & 6.8 & 43 $\times$ 23 \\
  \\

 R Hya &  61.3 & 5.9 & 100 $\times$ 76  & 41 $\times$ 30 & 54.2 & 3.8  & 98 $\times$ 82 &  39 $\times$ 31\\
& 53.6 & 5.4. & 88 $\times$ 70  &  & 68.6 & 4.0 & 88 $\times$ 90 \\
 %\\
 $\it absorption$$^{d}$ & $-$  & & & & $-15.1$ & 2.9 &  50 $\times$ 20 \\
& $-17.4$ & 3.9 & ? $^{e}$  & & $-$ &  \\
 \\     
U Her & 33.5 & 7.0 & 33 $\times$ 32  & 32 $\times$ 22 & $-$ & \\
& $-$ &   & & & $-$ & \\
 \\
 AH Sco &  18.4 & 3.3 & ? $^{e}$  & 27 $\times$ 23 & $-$ &  & ? &  44 $\times$ 26 \\  
 & 10.0 & 2.5  & ? $^{e}$  &  & 28.0 & 6.7 & 48 $\times$ 19 \\
  \\
  R Aql &  61.3 & 3.5 & 40 $\times$ 26  & 27 $\times$ 22 & 22.8 & 3.3  & 40 $\times$ 38 & 30 $\times$ 20  \\ 
  & 15.6 & 3.1 & ? $^{e}$  &  & 29.2 & 3.5 & 71 $\times$ 43  \\
 $\it absorption$$^{d}$ & $-27.9$ & 3.8  & 37 $\times$ 27 &  & $-27.9$ &  3.1 &  39 $\times$ 36 \\
 & $-25.3$ & 3.0  & 38 $\times$ 26 & & $-36.9$ & 3.1 &  50 $\times$ 34 \\
  \\
  VX Sgr &  36.9 & 8.7 & 38 $\times$ 23  & 31 $\times$ 22 & 35.7 & 7.0  & 44 $\times$ 29 &  35 $\times$ 23\\ 
  & 25.5 & 6.3 & ? $^{e}$   &  & 40.1 & 6.0 & 75 $\times$ 56  \\
  \hline                                  
\end{tabular}
\end{center}
\tablefoot{
$^{(a)} $R.m.s. noise in mom 0 map in mJy.beam$^{-1}$km\,s$^{-1}$
$^{(b)} $Elliptical Gaussian fit to cleaned image above the 3$\times$$\sigma$ contour, not de-convolved from the beam (see Sect. \ref {sec:OHchannmaps}). 
$^{(c)} $Gaussian beam size in mas. 
$^{(d)} $Discussion of absorption features in Sect. \ref {sec:OHabsorption}. 
$^{(e)} $Unresolved absorption or emission. 
$^{(f)} $Diffuse OH gas with uncertain extent (see Sect. \ref{sec:OHchannmaps}).
} 
\end{table*}

%%%%%%%%%%%%%%

 \subsubsection{OH absorption maps and OH gas infall}
\label{sec:OHabsorption}

%%%%%%
\begin{figure*}
 \centering
  \includegraphics[width= 12.8 cm, angle=270]{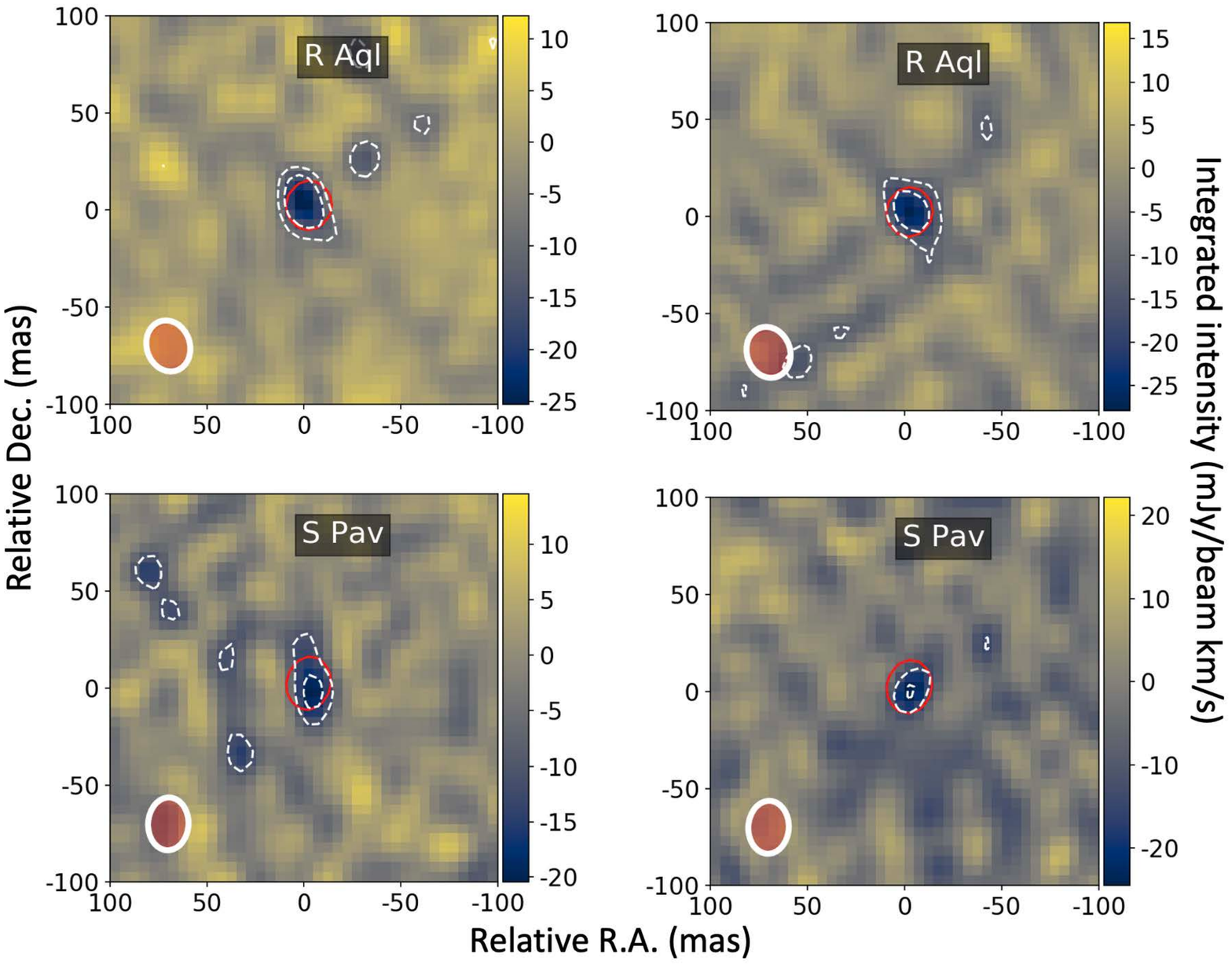}
 \caption{Zeroth moment map of OH $\it{absorption}$ in the $J = 27/2, F'-F'' = 14-14$  and 13$-$13
 transitions. $\it Left~panels$:  $F'-F'' = 14-14$ transition in R~Aql and S~Pav. $\it Right~panels$:  $F'-F'' = 13-13$ transition in the same sources. Offsets in R.A. and Dec. directions are relative to the coordinates of the peak stellar continuum as given in Table~\ref{primarysource_list}. 
 The velocity intervals in the $F'-F''=14-14$ and 13$-$13 transitions are: 
 55.8 to 61.2~km\,s$^{-1}$ and 56.2 to 66.2~km\,s$^{-1}$ in R~Aql;  $-$10.2  to $-3.6$~km\,s$^{-1}$ and $-9.4$ to 3.5~km\,s$^{-1}$
  in S~Pav. 
 The dotted white contours are at  $-$3 and $-$5 $\times$$\sigma$. The red contour delineates the extent at half peak intensity  of the   continuum emission. 
 The line and continuum beams are shown at the bottom left of each map in white and dark-red, respectively. The line HPBW is (27$\times$22~mas) at PA~18$^{\circ}$ 
 and (27$\times$21)~mas at PA~$-$3$^{\circ}$ in R~Aql and S~Pav, respectively. The continuum beam is (24$\times$22)~mas at PA~$-$13$^{\circ}$ (R~Hya) and (25$\times$20)~mas at PA~$-$13$^{\circ}$ (S~Pav). 
 } 
     \label{Fig_OH_absorption_27_2_raql_spav} 
   \end{figure*}
   
%%%%%%%%

The R Hya, S Pav and R Aql spectra extracted from the high resolution data cubes  show that there are absorption features in addition to the two main hyperfine transitions of OH seen in emission in the $J = 27/2$ and 29/2 levels (Fig.~\ref{rhya-spav-raql-OH}). Our images suggest that the absorption observed  at velocities greater than the systemic velocity of each  hyperfine transition in $J = 27/2$ and 29/2  is due to OH gas that is redshifted  with respect to each one of these two transitions. In this scenario, the absorption seen in-between the two hyperfine transitions (Fig.~\ref{rhya-spav-raql-OH}) is interpreted as redshifted gas for the higher frequency transition of the two hyperfine transitions in each $J$ level. (There  is no apparent blueshifted OH component observed in our data for the higher frequency line of each OH $\Lambda$-doublet. This could perhaps be understood as dominant absorption by the cooled gas  when it falls back to the star surface.) In the redshifted gas interpretation the OH gas falls with velocities $\sim$5$-$10~km\,s$^{-1}$ toward the warm gas which excites  each one of the two hyperfine components near the star. This is the same range as the infall velocities of H$_2$O and CO mentioned in Sect. \ref{sec:infall}. OH gas infall is supported by the mom~0 maps of the absorption features 
showing compact OH gas centered on  the star (Fig. \ref{Fig_OH_absorption_27_2_raql_spav}), and  by the $\sim$5$-$12 signal-to-noise ratio observed at the absorption peaks toward R~Aql, S~Pav and R~Hya (Table \ref{OH_size_peak_intensity}). The typical angular extent of the OH absorption above 3$\sigma$ is determined as for the emission region in Sect. \ref{sec:OHchannmaps}. It is smaller than the OH emission extent and on the  order of 30$-$40mas in R~Aql and less or comparable to the beam  toward R~Hya and S~Pav (Table \ref{OH_size_peak_intensity}). 

  \subsubsection{Current and further identification of OH sources}
\label{sec:OHstackmaps}

We have estimated typical OH scale lengths in Sects. \ref{sec:OHchannmaps} and \ref{sec:OHabsorption}, but the $J =27/2$ and 29/2 OH emission or absorption  is  often weak. In addition, and for all sources, the $J = 33/2$ and 35/2 rotational levels were not detected in emission or absorption at the time of our observations. In order to push further the OH detection limit and to better discern the diffuse gas, we have stacked together our data for all four hyperfine transitions with similar energy levels (i.e., $J$ =~27/2 transitions together with 29/2, and $J$ =~33/2 transitions with 35/2). For each group of four transitions, the visibility data were extracted for the same velocity range centered on each hyperfine transition and combined by aligning the velocities.  This allows us to image four stacked hyperfine lines,  reducing the noise by a factor of two, but at the expense of the identification by another approach to know which exact hyperfine transition(s) is(are) excited. If a group of four stacked transitions was detected we also stacked both component pairs separately to identify which rotational transition was detectable. Spectra have also been extracted from the stacked maps to help us identify the hyperfine transitions. 
 The $F'-F''$ assignment of a rotational transition can be safely done only  if the individual hyperfine transitions are   visible in 
channel maps and/or in mom~0 maps. We have  validated this approach by comparing the stacked maps with the single frequency channel maps of our  strongest sources.

 We conclude that there is OH emission and/or absorption in nine sources of the {\sc atomium} sample
 in the $\varv=0$, $J$ = 27/2 and/or 29/2 rotational levels including  weak emission from RW Sco and IRC$+10011$. (In these two sources inspection of channel maps and extracted spectra suggest emission from the $J=29/2$ state only.) Overall, there are slightly more detections in $J=29/2$ than in 27/2 (see last column in Table \ref{OHline_list}). Four sources, R Hya, R~Aql, U~Her and T~Mic are detected in the higher energy rotational levels $\varv=0$, $J = 33/2 $ and only R~Hya is observed  in the 35/2 state. In U~Her and T~Mic, OH is detected when the $\varv=0$, $J = 33/2$ and 35/2 levels are stacked together, so we cannot distinguish whether one or both rotational levels are emitting. However, since the $J = 35/2$ emission is very weak and observed in R~Hya only, we  have arbitrarily assigned the emission observed in U~Her and T~Mic (last column in Table~\ref{OHline_list}) to the lower energy level, that is $J = 33/2$. We further note that the $\varv=0$, $J = 33/2$ level energy ($\sim$8000~K) is close to the energy of the two highest H$_{2}$O levels observed in this work. 
Four $\varv=1$ hyperfine transitions in the $J = 31/2$ and $33/2$ rotational levels (with energy $\sim$11080 and 12800~K)  also fall in our frequency setup but were not detected here. This nondetection  is in contrast with the $\varv=1, J = 21/2$ and 35/2 observations of \citet{khouri2019} in W Hya and is discussed at the end of Sect. \ref{lack_of_v=1}.

At the time of this work,  high-$J$ OH sources have been discovered with ALMA in twelve and probably thirteen AGBs or RSGs. Nine objects  are listed  in Tables \ref{primarysource_list}  and \ref{OHline_list}), three other sources, W Hya, R Dor and IK Tau have been detected by  \citet{khouri2019}, and we further report  in Appendix \ref{sec:append_mira_ohspec} on the potential detection of $J$ = 35/2 and 31/2 OH emission in omi~Cet  (Mira). More high-$J$ OH transitions obtained in future ALMA observations or extracted from ALMA archival data of evolved late-type stars will contribute to our understanding of the role of OH in the formation of water and dust-forming metal oxides/hydroxides.
 
\section{A high-$J$ OH analysis: Physical conditions}
\label{sec:OH_analysis}

As mentioned in Sect. \ref{sec:general_OH}, no hyperfine transitions of the OH radical were ever identified in high-$J$ levels toward evolved stars before ALMA\footnote{Note that  mid-IR spectral resolution observations of the OH $\Lambda$-doubling up to $J=69/2$  have been reported in the young protostellar outflow HH 211 \citep{tappe2008} and modeled \citep{tabone2021}.}. 
 Our discussion below is limited to simple estimates of  the physical conditions in the high-$J$ OH gas observed in the photospheric environment.

\subsection{Line intensity ratios, brightness temperature}
\label{sec:OH_intensity_ratio_brightness}

\begin{table*}
%\begin{small}
\begin{center}
\caption{Observed integrated flux density and peak flux density ratios (Integr. ratio and Peak ratio) and LTE opacity ratio of OH hyperfine transitions in $J= 27/2$, 29/2, 33/2 and 35/2$^{a}$.}       
\label{OH_opacity_ratio}      
\begin{tabular}{lccccc}       
 \hline\hline 

$\varv = 0$, OH level  & $J = 27/2$, 13$-$13/14$-$14 & $J = 29/2$, 14$-$14/15$-$15 & $J = 33/2$, 17$-$17/16$-$16 & $J = 35/2$, 18$-$18/17$-$17  \\
 
   &  Integr. ratio $-$ Peak ratio & Integr. ratio $-$  Peak ratio  & Peak ratio$^{b}$  & Peak ratio$^{b}$  & \\
   
\hline

 R Hya & $0.75\pm0.15$ $-$ $0.96\pm0.10$   & $0.86\pm0.15$ $-$ $0.93\pm0.10$ & $0.95\pm0.40$ & $\ga$$\it1.90$ 

  \\
 R Aql & $\it 1.13 \pm0.15$ $-$ $\it 1.40 \pm0.10$ & $0.70\pm0.15$ $-$ $0.90\pm0.10$   & $ \ga$$\it 1.30$          

  \\
 S Pav & $\it 1.41 \pm0.15$ $-$ $\it 1.60\pm0.15$ $$& $0.81\pm0.15$ $-$ $0.90\pm0.15$ & $\ga$0.90 
 
 \\
 VX Sgr &  & $\it 0.71\pm0.15$ $-$ $\it 0.71\pm0.15$  &  &  
 
 \\
 T Mic & & $\it 0.69\pm0.15$ $-$ $\it 0.83\pm0.15$ & &  
 
  \\
   LTE opacity ratio &  0.93 & 0.94  &  1.06 &  1.06  \\
   
\hline                                  
\end{tabular}
\end{center}
\tablefoot{
$^{(a)}$ Ratios derived from the flux density of the lower frequency of the hyperfine transition in a given rotational level divided by the flux density in the same level of the higher frequency hyperfine transition. The non LTE ratios are shown in italics (see Sect~\ref{sec:OH_intensity_ratio_brightness}). Formal uncertainties lie in the range $\pm 0.10$  to $\pm 0.15$ except in the $J =33/2$ rotational level. The LTE ratios are shown in the last row of this Table.
$^{(b)}$ The lower limits in $J = 33/2$ and 35/2 are derived from the 3$\sigma$ signal level outside the expected line feature for the undetected 16$-$16 ($J = 33/2$) and 17$-$17  ($J = 35/2$) hyperfine transitions. 
} 
\end{table*}

As a first  approach to the understanding of  the OH excitation conditions  we compare the observed ratio of two hyperfine transitions in a given rotational level with the same ratio under LTE conditions. There are several difficulties in estimating such ratios from the  ${\Delta} F = 0$ line parameters. Fitting Gaussians to weak, asymmetrical line profiles is uncertain. Nevertheless, the peak line flux density is often well identified and if not, we can derive the integrated flux density across the line profile above the  noise level.  Our results in five  stars, for both line peak and integrated flux density ratios above  $3\sigma$,  are gathered in Table~\ref{OH_opacity_ratio}  together with the LTE line opacity ratios. All ratios are derived from spectra extracted from the high resolution data cubes for an aperture diameter of 0\farcs08. We have also verified that for R~Hya in  the $J = 27/2$ and 29/2 states, the intensity ratios remain  similar, within the uncertainties, to  those obtained with the mid resolution data. For the higher $J = 33/2$ and 35/2 states, when only one hyperfine transition is detected we give a lower limit based on the $3\sigma$ level outside the expected but undetected line feature. The formal uncertainty of our intensity ratios is estimated to be in the range 0.10$-$0.15 except in the $J=33/2$ state where it is significantly higher due to the lower S/N. 
The line integrated flux density ratio and the peak flux density ratio (see Integr. ratio and Peak ratio columns in Table~\ref{OH_opacity_ratio}, respectively) give comparable results in general. In  R~Hya, for example, the peak ratios are equal to 0.96 and 0.93 for $J = 27/2$ and 29/2, respectively, while the integrated flux density ratios are somewhat smaller but still consistent, within the uncertainties, with the LTE ratios. However, deviations from LTE are significantly above the uncertainties for R~Aql and S~Pav ($\sim$1.1$-$1.6) in the $J =27/2$ state  and for VX~Sgr and T~Mic ($\sim$0.7$-$0.8) in $J= 29/2$. The observed peak intensity ratios are larger than the LTE ratios for $J = 33/2$  and $J = 35/2$ observed in R~Aql and R~Hya. These results show that the OH gas is not excited under full LTE conditions, and there are no strong masers observed in the high-$J$ levels.  Deviations from LTE remain moderate  when compared with the strong  $J=3/2$ ground state masers observed around 1.6GHz when they were first discovered  \citep[][]{wilson1968} and in all other subsequent observations of late-type M stars.

Table~\ref{oh_brightness} gives the OH  peak surface brightness in the $J = 27/2$ and 29/2 rotational levels,  $S_{\rm p}$,  derived from our  channel maps for the extended configuration and for beam widths ranging from $\sim$25$\times$23~mas (T~Mic) to $\sim$41$\times$30 mas (R Hya). For simplicity,  $S_{\rm p}$     is only given for the brightest of the two hyperfine transitions $J = 27/2$, 14$-$14 and $J = 29/2$, 15$-$15. The peak brightness temperature, $T_{\rm b}$, is derived from  $S_{\rm p}$ and  the restoring beam. It gives an estimate of the maximum brightness temperature for OH emitting region sizes on the order of one  or a few beams. $T_{\rm b}$  is  a lower limit to the brightness temperature if the region is smaller than the synthesized beam or made up of multiple clouds (or if maser beaming narrows the measured angular size in individual channels, although masing is unlikely in OH high-$J$ levels). Table \ref{oh_brightness} gives values of  $S_{\rm p}$ and $T_{\rm b}$
 in the range $\sim$3$-$9~mJy/beam and  $\sim$100$-$280~K, respectively. These values of  $T_{\rm b}$ are lower than the local $T_{\rm K}$ and suggest that the OH excitation conditions are close to LTE (in agreement with the intensity ratios in Table \ref{OH_opacity_ratio}), and that the OH lines are optically thin.

 %%%%%%%%%%%%%%%%%%%%

\begin{table}
\begin{center}
\caption{Peak surface brightness ($S_{\rm p}$), peak brightness temperature ($T_{\rm b}$) and uncertainty in the $J$ = 27/2 and 29/2 rotational levels of OH.}       
\label{oh_brightness}      
\begin{tabular}{lccc}       
 \hline\hline 

 Source &  $S_{\rm p}$ $^{a}$   & $T_{\rm b}$ $^{b}$   \\
 & (mJy/beam) & (K)  \\
 & $J $=27/2, $F'-F''=14-14$    \\
 & $J$=29/2, $F'-F''=15-15$   \\

\hline
S Pav & 5.5 & 240 $ \pm$ 45 \\
 & 6.5 & 255 $ \pm$ 40 \\
 
 T Mic &  6 & 260 $\pm$ 45   \\ 
  & 7 & 230 $\pm$ 35 \\
  
RW Sco &  < 3$^{c}$ & < 200$^{c}$ &  \\ 
  & 8 & 280  $\pm$ 70 \\
  
 R Hya & 9 & 180  $\pm$ 20  \\
& 9 & 145    $\pm$ 15  \\

U Her & 7.5 & 270  $\pm$ 70  \\
& 5 &  180  $\pm$ 35 \\

 AH Sco &  2.5 & 100    $\pm$ 40 \\  
 & 5& 85   $\pm$ 15 \\
 
  R Aql &  6 & 255   $\pm$ 40  \\ 
  & 7 & 235    $\pm$ 30  \\
   
   IRC$+10011$  & < 3$^{c}$ & < 100$^{c}$ &   \\ 
  & 5.5 & 200  $\pm$ 75 \\

  VX Sgr & 6 & 220   $\pm$ 35  \\ 
  & 8 & 190   $\pm$ 30 \\
  \hline                                  
\end{tabular}
\end{center}
\tablefoot{
$^{(a)}$ Peak surface brightness for the strongest of the two hyperfine transitions  in the 27/2 and 29/2 rotational levels (see Sect. \ref{sec:OH_intensity_ratio_brightness}).
$^{(b)}$ Brightness temperature derived from the peak surface brightness and restoring beam. Uncertainty derived from typical noise in channel maps $\sim$1$-$2 mJy/beam.
$^{(c)}$ Upper value determined from 3$\times$$\sigma$ in channel maps. 
} 
\end{table}
 
%%%%%%%%

\subsection{Population diagrams and opacity estimates}
\label{OH-opacity-popdiagram}

%%%%%%%%
\begin{figure}
 \centering   

 \includegraphics[width= 10.0 cm, angle=0]{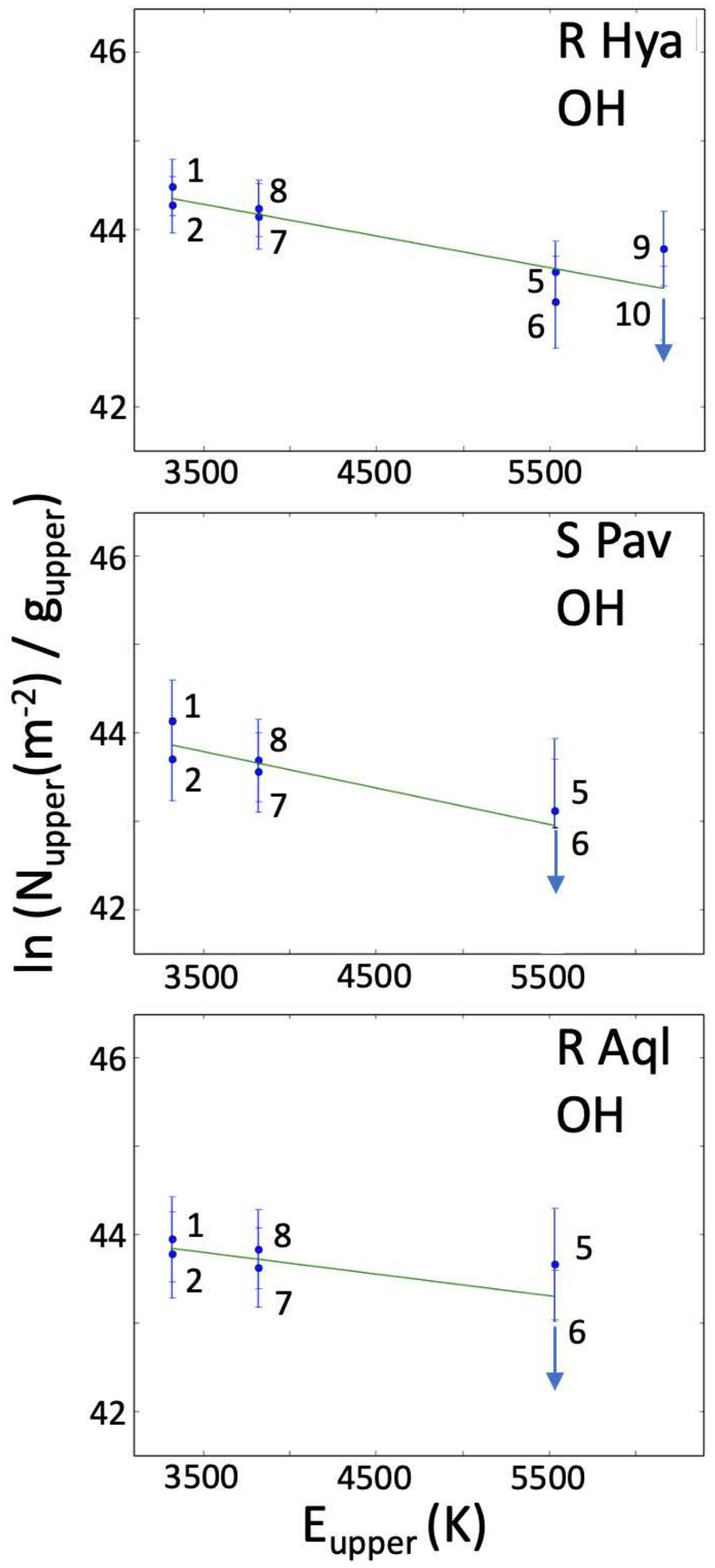}
 
 \caption{OH population diagram for R Hya, S~Pav and R~Aql. The number near each data point is the line number used in Table \ref{OHline_list} of observable OH transitions. 
 The vertical bar in each data point includes the $\pm$1$\sigma$  formal error of the integrated flux density and a rough estimate of the filling factor uncertainty. (An arrow indicates an upper limit.) The green line is the regression line across the data points  (see Sect. \ref{OH-opacity-popdiagram}). 
 } 
 
 \label{oh_popdiagram_rhya_spav} 
  \end{figure}
  %%%%%%%%

An estimate of the OH rotational temperature,  $T_{\rm rot}$, and  column density, $N$(OH), can be obtained from a 
population diagram although this is  uncertain because the detected OH transitions span a moderate range of upper energy levels. As for H$_2$O, we have used the high resolution data, all observed within a few weeks. The S/N of the integrated OH line intensity is on the order of 6$-$10 in R~Hya, S~Pav and R~Aql  in the $J=27/2$ and 29/2 levels and less (or undetected) in $J = 33/2$ and 35/2. We have plotted population diagrams for these three sources (Fig.~\ref{oh_popdiagram_rhya_spav}) but, unfortunately, there are not enough detected OH transitions  in other stars to estimate $N$(OH), including AH~Sco and IRC$+$10011 for which $N$(H$_2$O) is determined (Fig.~\ref{h2o_popdiagram}). The partition function used to derive $N$(OH) is obtained from a direct summation  of the energy levels in different vibrational states and from parameters based on the JPL catalog data; all components due to $\Lambda$ and spin doublings and the hyperfine structure have been included. As for H$_2$O, the partition function varies monotonically with the temperature and $N$(OH) is not very sensitive to uncertainties  in the values of $T_{\rm rot}$. We derive  $T_{\rm rot}$ $\sim$2800 and 3000~K and  
$N$~$\sim$6.1$\times$10$^{18}$ and 4.8$\times$10$^{18}$~cm$^{-2}$ in R~Hya and S~Pav, respectively. In R~Aql, $T_{\rm rot}$ is higher, $\sim$4000~K, and $N$(OH) $\sim$4.3$\times$10$^{18}$~cm$^{-2}$. The main limitations in the determination of $N$(OH) come from the rather restricted energy range and the fact that the observed regions are assumed to have uniform properties. However, tests made with various assumed errors in the observables suggest that  $N$(OH) can vary by at least 30\%.

Using $N$ and $T_{\rm rot}$ obtained above, we get an opacity at the OH emission line center in the range $\tau${$_c$ $\sim$0.02$-$0.03  in both $J =27/2$ and 29/2. (The opacity in a given $J$ level does not vary significantly from one hyperfine transition to the other.) These opacity estimates are in rough agreement with the ratio ($T_{\rm b}$/$T_{\rm rot}$) which suggests that the OH emitting region cannot be much smaller than the beam.

\subsection{Lack of $\varv=1$ OH emission and implications}
\label{lack_of_v=1}     

The lack of  $\varv$ = 1 OH emission from R Hya, our strongest OH source, even after the OH lines have been stacked, is surprising when compared to the relatively strong hyperfine transitions of the $\varv$ = 1, $J = 21/2$ and $J = 35/2$ levels observed by \citet{khouri2019} in the Mira variable W Hya around the optical phase 1.0$-$1.1. 
In R Hya, our  3$\sigma$ upper  limit in the $\varv$ = 1, $J $ = 31/2 and 33/2 spectra  is $\sim$3 mJy (for a total aperture diameter of   0\farcs08) around the optical phase 0.8, whereas the $\varv$ = 0, $J $ = 27/2 and 29/2 emissions peak around 10$-$15 mJy (see Fig.~\ref{rhya-spav-raql-OH}). There is no OH  $\varv=1$ detection toward R~Aql and S~Pav (upper limits around 3 to10 mJy) observed around 0.2 and 0.9 optical phase. Changing the extracted aperture area for our spectra or considering the mid resolution data do not show any $\varv=1$ detection either.

We cannot exclude  that the high-$J$ OH excitation in both $\varv$~=~0 and 1 are time variable and may depend on the stellar optical phase. Our OH observations of R Hya  and other targets  have been  acquired at essentially a  single epoch and with often too low S/N to usefully address time variability issues in individual stars. However, despite the pulsation period of any one star not being well sampled, the whole set of objects observed at several epochs may be considered as covering a range of optical phases and we might have hoped to observe $\varv=1$ OH emission in at least one star\footnote{The approximate optical phases of our OH high resolution observations acquired in 2019 are: 0.8, 0.9, 0.2, 0.5 and 0.7 for R Hya, S Pav, R Aql, VX Sgr and AH Sco, respectively. The phase is not well defined in T Mic, but was $\sim$0.5.}.

It is  interesting to note that time-variable, near IR excitation of OH in the $\varv$ = 0 to $\varv$ = 1 transition at 2.8 $\mu$m 
was invoked by \citet{harvey1974} to explain the 18-cm OH variations observed in late-type stars. But later,  \citet{etoka2000} showed  that the phase delay observed between the 18-cm OH maser peak emission and the near IR maxima suggests that dust in the outer layers contributes in the OH excitation. The pumping mechanism proposed for the ground state of OH, however, cannot explain the potential excitation of high-$J$ OH states in the inner gas layers of O-rich stars.  

The possible role played by the 2.8 $\mu$m radiation  in the excitation of the high-$J$ OH transitions
observed in the inner gas layers, can be crudely appreciated from an estimate of the critical gas density, n${_c}$, for the $\varv$ = 0 to $\varv$ =1 transition. Taking  16.9~s$^{-1}$ for
 the spontaneous emission rate  \citep{brooke2016}, and assuming that the collisional rate coefficient of OH  with H${_2}$ is dominated by rotational transitions and can be represented by $\sim$2$\times$10$^{-10}(T_{\rm K}/100)^{0.5}$cm$^{3}$s$^{-1}$ \citep{klos2017}, we obtain 
 n${_c}$ $\sim$8.5$\times$$10^{10}/(T_{\rm K}/100)^{0.5}$, that is $\sim$1.5$-$5$\times$10$^{10}$~cm$^{-3}$ for $T_{\rm K}$ in the range 300 to 3000~K. A similar density range also applies to the $\varv$ = 2 to $\varv$ = 1  transition with 23.4 s$^{-1}$  spontaneous emission rate. Therefore, we expect that collisions play a significant role in the vibrational line excitation of OH within the  high density regions close to the photosphere where OH is formed after the passage of shocks. 
We also expect that changing physical conditions resulting from stellar pulsations and shocks in regions  where the high-$J$ optically thin OH microwave lines are observed may lead to  both $\varv=0$ and 1 OH line variability. This may perhaps explain why $\varv$ = 1 OH lines observed by \citet{khouri2019}  are not detected in this work. However, questions related to high-$J$ OH time variability deserve dedicated  sensitive mm-wave observations in the future.

\section {Observed OH/H$_2$O abundance ratio and chemical models in the  inner wind}
\label {sec:chemistry}

%%%%%%%%%%%%%
\begin{figure*}
\centering
\vspace{-1.8cm}

\includegraphics[width= 12.5cm, angle=270]{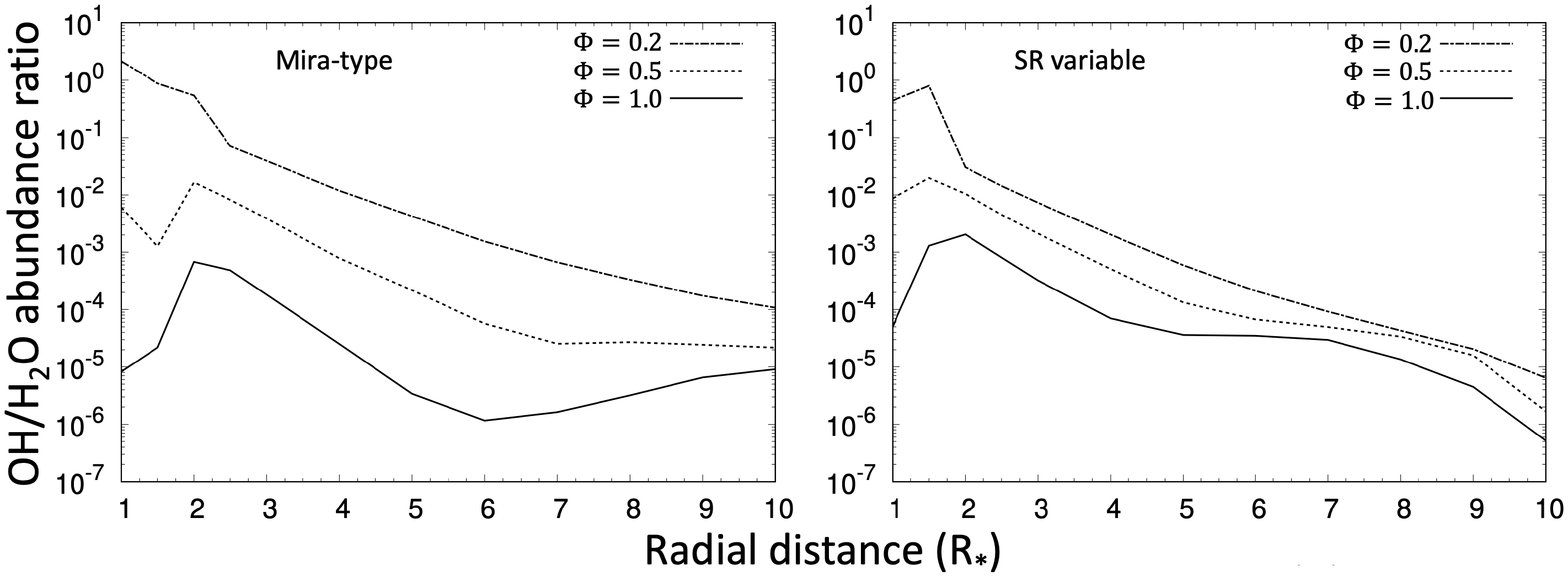}
\vspace{-1.8cm}

 \caption{Model predictions of the OH/H$_2$O abundance ratio versus the radial distance at different pulsation phases of Mira-type and semi-regular variables. The spread in the abundance ratio versus the radial distance and the optical phase is discussed in Sect.~\ref{sec:chemistry}.
  } 
 \label{ohh2o_pulsating_mira_srv} 
  \end{figure*}
%%%%%%%%%%%%%

We primarily  wish here to compare  the OH/H$_2$O abundance ratio derived from our estimates of the 
H$_2$O and OH column densities with the same ratio 
predicted from chemical models. From the high resolution  observation of high-lying energy levels of three AGBs (R~Hya, S~Pav and R~Aql), all performed within a few weeks, and from the population diagram analysis described in Sects. \ref{sec:pop_diagram} and \ref{OH-opacity-popdiagram}, we obtain $N$(OH)/$N$(H$_2$O) $\sim$(0.7$-$2.8)$\times10^{-2}$. Assuming, as suggested from the ALMA maps, that the  observed H$_2$O and OH transitions come from  similar regions in the circumstellar  inner wind, we thus expect OH/H$_2$O $\sim$0.7$-$2.8~$\times10^{-2}$, which will be compared  with our  recent chemical models. 
We first make some general and brief considerations on the chemical context in evolved stars  and  on the model outputs predicting H$_2$O  and OH fractional abundances  for regions up to  \la10~R$_{\star}$.
The chemical equilibrium conditions of an ideal LTE atmosphere \citep[e.g.,][]{tsuji1973} can explain the formation of various molecules observed in cool stars as well as  their dependence on the C/O ratio (i.e., the molecular fractional abundances critically depend on the evolutionary status of the star). However, in the outer stellar atmosphere, convection and stellar pulsations imply that equilibrium conditions are rapidly changing with time or position. Predicting the  chemical composition of the gas  becomes complicated because of the generation and  propagation of shocks, the formation of dust particles and the launch of an inner stellar wind. In such a complex environment, we know that a  rich, nonequilibrium chemistry is possible and is supported  by the observation of  a large variety of molecular species in the IR \citep[e.g.,][]{tsuji1997} and in the mm/submm domains, including the unexpected presence of H$_2$O  in C-rich stars \citep{neufeld2011}  or of HCN in O-rich stars \citep{lindqvist1988, justtanont2012}. Decoupling the hydrodynamics of the inner wind from the chemistry, \citet{willacy1998} and \citet{cherchneff2006} have predicted the molecular abundances of several species including H$_2$O and OH in the $\sim$1$-$5 R$_{\star}$ region. Later, \citet[][]{gobrecht2016} confirmed that periodic shocks in the IK Tau environment provide efficient conditions for the synthesis of various chemical species and dust nucleation in the range $\sim$1$-$10 R$_{\star}$. \citet[][]{boulangier2019} further showed that a nonequilibrium chemistry leading to dust nucleation impacts the dynamics of the wind and, recently, \citet{gobrecht2022} gave new details on dust nucleation scenarios.

The observed abundance ratio, OH/H$_2$O~$\sim$~(0.7$-$2.8)$\times10^{-2}$, despite uncertainties, can be directly compared with various model predictions. In typical O-rich stars  and for C/O ratios $\sim$0.75, \citet{cherchneff2006} obtained a rather flat distribution of the OH and  H$_2$O abundances between $\sim$1.5$-$5 R$_{\star}$ from which we get OH/H$_2$O  in the range 10$^{-3}$ to 2.5 10$^{-5}$; these ratios are far from our observational estimates.  To model the dust formation and the chemical atmosphere of IK Tau,  \citet[][]{gobrecht2016} derived  the molecular content between 1 and 10~R$_{\star}$ as well as the abundance variation with the stellar pulsation phase at different radii. The H$_2$O abundance relative to the total gas density remains flat and around 10$^{-4}$ up to 10 stellar radii. This is slightly smaller than the 4 to 2$\times$10$^{-4}$ H$_2$O fractional abundance obtained by \citet{cherchneff2006} between 1 and 5 R$_{\star}$. 
It is thus useful to revisit the chemical model of  \citet[][]{gobrecht2016} using the more recent kinetic rates given in \citet{gobrecht2022}. 
Our newer calculations for radii between 1 and 10~R$_{\star}$, before the wind is fully accelerated, predict  that for typical Mira-type and semi-regular variables the OH/H$_2$O ratio varies by one to two orders of magnitude  at different pulsation phases (0.2, 0.5 and 1.0 in Fig.~\ref{ohh2o_pulsating_mira_srv}). (We point out that at  the present time we have no model outputs available for RSGs.) 
The decrease observed with the radial distance in the OH/H$_2$O ratio (Fig. \ref{ohh2o_pulsating_mira_srv})  is driven by the decrease in the OH abundance while the H$_2$O  abundance remains roughly at the same  level within the inner gas layers as seen in Fig.~22 of \citet{gobrecht2022}. The larger dispersion in the OH/H$_2$O ratios of the  Mira-type stars with respect to the semi-regular variables is most probably explained by the stronger shock and lower (preshock) temperatures in the Mira-like models. 
Our calculations suggest that in  the range from 2 to 3~R$_{\star}$ the predicted OH/H$_2$O ratios approach those derived from the {\sc atomium} observations at optical phases $\sim$0.5$-$0.2.  However, our current observations, lacking data at different epochs  are unable to  estimate how the OH/H$_2$O ratio varies with the optical phase and if variations match the model predictions. In addition, we note that  there might be some minor phase shift between  pulsation and optical phase \citep{liljegren2016}.

The OH/H$_2$O ratio is primarily controlled by the reaction pair H$_2$ + OH <$-$> H$_2$O + H  ($\Delta$$H_{\rm r}$(0 K) = $-56$ kJ/mol). The forward reaction rate has an activation barrier of  1660~K
whereas the endothermic reverse reaction has a much higher activation energy of 9720~K \citep{baulch1992}. Therefore, 
the larger OH/H$_2$O ratios at pulsation phase 0.2, ranging from $\sim$4$\times$10$^{-2}$ to 2 in Mira-type variables, can be explained by comparatively higher temperatures. At these high temperatures not only the reverse reaction is activated, but also the atomic hydrogen is abundant. In contrast, at later pulsation phases (around 1.0) low temperatures prevail and hydrogen is predominantly molecular. As a consequence, the H$_2$O-forming forward reaction dominates at late pulsation phases and the reverse reaction is negligible, leading to low OH/H$_2$O ratios   $\sim$8$\times$10$^{-6}$  to  $7\times$10$^{-4}$. In general, we note in our models a greater dependence of the OH/H$_2$O ratio  on time (or pulsation phase) than on distance from the star at a given pulsation phase.

 It is interesting to mention that the excitation of high-$J$ levels of OH in the inner gas layers could eventually be triggered by the photodissociation of water from the UV radiation generated in the outflow shocks of pulsating stars. This could lead to the selective formation of high-$J $ OH states as suggested in \citet{tappe2008} 
for the young stellar object HH 211 outflow. However, the shock speeds invoked in this work, at least 40~km\,s$^{-1}$,  
are higher than those observed in AGBs. Therefore, we favor the direct formation of OH from oxygen and H$_2$  after the passage of the circumstellar shock, although special cases are also possible; for example, readers can refer to the UV photon models from a stellar companion impacting the chemical products, for example \citep{vandesande2022}.

\section{Concluding remarks}
\label{sec:conclusion}
  
Ten rotational transitions in the ground  and  excited vibrational states of H$_2$$^{16}$O were observed
in the 213.83$-$269.71~GHz frequency range of the {\sc atomium} Band~6 survey undertaken with the ALMA main array. 
Nine lines are new discoveries and the tenth, reported earlier at 268.149~GHz in two evolved stars, was observed in the 15 O-rich sources of the 17 late-type stars in our sample. High spectral resolution observations at an additional epoch were also undertaken with the ACA around  268~GHz and 254~GHz. 
\noindent The ten  transitions, six in the ortho state and four in the  para  state of H$_2$O, 
include six pure rotational transitions within a single excited vibrational state,  
two ro-vibrational transitions between two nearby vibrational states, and two pure rotational transitions in the ground vibrational state. Together they span a range in excitation energy of between 3950~K in the ground vibrational state
to 9013~K in the ($\varv_1$,$\varv_2$,$\varv_3$)~=~(0,0,1) state. Our observations significantly extend  the number of H$_2$O transitions that have been observed in evolved stars in the radio domain.

In parallel, the hyperfine split $\Lambda$-doubling transitions in the $\varv =0$, $J = 27/2$ and 29/2  high-lying 
rotational levels in the  $X^2\Pi_{3/2}$ state of OH, as well as the $\varv =0$, $J = 33/2$ rotational level in the upper $X^2\Pi_{1/2}$ state have been observed in the same Band~6 spectral line survey. The next higher    $\Lambda$-doubling transition, $J = 35/2$, was observed in one star only (R~Hya). The range  of energy levels covered by the OH lines observed here is $\sim$4780$-$5500 K and  $\sim$8000$-$8900~K. 

\subsection{ H$_2$O}

Some general trends have emerged from our extensive high resolution observations. The sections where our main results are presented and discussed are specified at the end of each paragraph below.

$-$  There is good agreement between the rest frequencies determined from the present observations and the frequencies measured in the laboratory. In those cases where the catalog frequencies were derived with uncertainties exceeding $\sim$0.5~MHz, our rest frequencies  could be better than the catalog frequencies (see Sect.~\ref{sec:detectwater}).

$-$ Overall the H$_2$O emission observed with the extended configuration is compact and observed in regions extending from a few to about 12~R$_{\star}$ from the central star. Possible extensions up to $\sim$20$-$30~R$_{\star}$ were also observed at 268.149~GHz (see Sect.~\ref{sec:channmaps_H2O}).

$-$ The number of stars in which an H$_2$O line was detected is roughly two times higher for transitions that arise 
in lower-lying levels ($\sim$4000$-$5600~K) than in the higher-lying levels ($\sim$8000$-$9000~K). There does not appear to be any correlation between the number of detected transitions and the
physical parameters of the source such as the mass-loss rate (see Sect.~\ref{sec:detect_H2O}).

$-$ The three most intense and widespread transitions in H$_2$O are the ortho line at 268.149~GHz in the (0,2,0) 
vibrational state which was observed in 15 stars, the para lines at 262.898~GHz in the (0,1,0) state which was 
observed in 12 stars, and 259.952~GHz in the ground vibrational state which was observed in ten stars.
A line was also observed in the (0,3,0) vibrational state in eight stars, in the (0,0,1) state in nine stars, and
the ro-vibrational transition (1,0,0)$-$(0,2,0)\,$J_{K_a,K_c} = 7_{4,3} - 8_{5,4}$ was observed in eight stars (see Sect.~3.2).

$-$ Both emission and absorption are observed in our maps and spectra of H$_2$O (and the same applies in OH). In a few
stars, zeroth moment  maps reveal absorbing regions as small as  $\la~$20$-$40~mas that are closely associated with the central star and are redshifted by a few km\,s$^{-1}$ with respect to the gas layers seen in emission.
Infall velocities of 5$-$10~km\,s$^{-1}$ that were estimated  from our H$_2$O observations and from $\varv$ = 1, CO observations are close to model predictions from \citet[][]{nowotny2010} (see Sects.~\ref{sec:absomaps_H2O},~\ref{sec:infall}).

$-$ A Gaussian analysis of the most compact emission regions in two prominent H$_2$O sources in the survey 
 has revealed an organized (R~Hya) and a complex (U~Her) position-velocity emission structure (see Sect.~\ref{sec:small_scaleh2o}).

$-$ High velocity wings of the 268.149~GHz line emission with respect to $\varv$ = 0, CO emission were observed  in several stars. The most probable explanations include local turbulence or kinematic perturbations by a companion (see Sect.~\ref{sec:infall}).

$-$ We have derived  beam-averaged column densities from  our H$_2$O population diagrams in AH~Sco,  R~Aql, R~Hya, S~Pav, and IRC$+10011$. They lie  in the range $\sim$(0.6$-$5)$\times10^{20}$~cm$^{-2}$ (see Sect.~\ref{sec:pop_diagram}).

$-$ The 268.149~GHz transition which is observed in all but the two S-type stars may be excited under quasi-LTE conditions or  is  masing in at least three stars (U Her, AH Sco, and IRC$+$10011) with brightness temperatures in the range 10$^{4}$ to 10$^{7}$ K. Time variability and line profile narrowing have also been observed in our data.  
Among the two other strongest H$_2$O lines in the {\sc atomium} survey, signs of maser emission have also been observed at 262.898~GHz with time variable emission in some velocity channels of IRC$+$10011 (see Sects.~\ref{sec:H2O_conditions},~\ref{sec:variability_268} to \ref{sec:h2o_line12}).

$-$ Based on H$_2$O radiative transfer models in \citet{gray2016}, density, kinetic temperature, and dust temperature conditions leading to maser emission have been revisited. The 268.149~GHz line in the (0,2,0) vibrational state has a strong radiative pumping component similar to several other ortho and para H$_2$O line pairs in  the (0,1,0) state  (see Sect.~\ref{sec:maser}).

\subsection{OH }

Refined OH $\Lambda$-doubling frequencies have been made possible from ALMA high-$J$ OH observations. 
Nine high-$J$ OH sources of the {\sc atomium} sample  have been discovered and their characteristics are summarized below. 

$-$ Based on our OH observations and those published in \citet{khouri2019}, new, accurate  $\Lambda$-doubling frequencies for high-$J$ level transitions have been derived (see Sect.~\ref{sec:identif_OH} and Appendix~\ref{sec:Lambda_doubling_freqs}).

$-$ The main sources of OH observed in this work  in the $\Lambda$-doubling transitions of the $X^2\Pi_{3/2}$, $J = 27/2$ and 29/2 states are: the SR and Mira variables R~Hya, R~Aql, S~Pav, U~Her, and T~Mic; and the RSGs AH~Sco and VX~Sgr. The lines are weak, with peak intensities that range between 3 to 10~mJy/beam. We did not observe  strong deviations from the quasi-LTE line excitation conditions in our data (see Sects.~\ref{sec:OHchannmaps},~\ref{sec:OH_intensity_ratio_brightness}). 

$-$ Very weak OH emission was also identified by stacking the spectra at the expected 
frequencies of the hyperfine split components in RW~Sco and IRC$+10011$. We also extracted weak OH emission in Mira~A from the ALMA archive (see Sect.~\ref{sec:OHstackmaps},  Appendix~\ref{sec:append_mira_ohspec}). 

$-$ Most of the OH emission is centered on the stellar photosphere and there is weak diffuse OH emission that extends well beyond the
stellar continuum sources. In SR and Mira variables, the estimated angular size
 of the OH emission is $\sim$5$-$8~R$_{\star}$, in general, with possible extensions up to $\sim$10~R$_{\star}$ from the photosphere (see Sect.~\ref{sec:OHchannmaps}).
 
$-$ In addition to the two main OH emission profiles in each $\Lambda$ doublet, our spectra and maps strongly suggest that OH absorption is also present with features that are redshifted by $-5$ to $-10$~km~s$^{-1}$. This is comparable to the velocity of the infall layers observed in H$_2$O and in  the $\varv=1$, CO  line. The absorption is barely resolved and typical angular scales are $\la$30$-$40~mas in R~Aql, R~Hya, and S~Pav (see Sect.~\ref{sec:OHabsorption}).  

$-$ We have observed the $\Lambda$-doubling transition in the $J$ =~33/2 rotational level of the  $X^2\Pi_{1/2}$ state
in four sources (R~Hya, R~Aql, U~Her, and T~Mic),
but the  next higher rotational level in the $X^2\Pi_{1/2}$ state ($J = 35/2$) was only detected in R~Hya. 
These lines are very weak and only observed by stacking the spectra 
at the frequencies in the  $J = 33/2$ and $J = 35/2$ states\footnote{We note that the $J = 33/2$ rotational level in the $X^2\Pi_{1/2}$ state of  OH ($E = 7960$~K) is close in energy (within $\lesssim$400~K) to the rotational levels of three high-lying transitions of H$_2$O which were observed in at least six stars in the {\sc atomium} survey:  line \#1 in eight stars,  line \#8 in six stars, and line \#13 in seven stars.} 
(see Sect.~\ref{sec:OHstackmaps}).

$-$ The OH emission from the two RSGs AH~Sco and VX~Sgr exhibits complex emitting regions. In both stars  
weak OH emission extends beyond the central object  up to $\sim$15~R$_{\star}$, which is  larger than the typical OH sizes observed in the AGBs. In both RSGs the mom 0 maps reveal asymmetric extensions in the $J=29/2$ state (see Sect.~\ref{sec:OHchannmaps}).

$-$ Estimates of  the column density of OH in the ground vibrational state were derived from 
rotational temperature diagrams that include the two successive $\Lambda$-doubling transitions 
in the $X^2\Pi_{3/2}$ state, and the lower of the two $\Lambda$ transitions in the $X^2\Pi_{1/2}$ state of S~Pav and R~Aql, and both $\Lambda$-doublet transitions in the $X^2\Pi_{1/2}$ state in R~Hya. We obtained 4.3, 4.8, and 6.1$\times$10$^{18}$~cm$^{-2}$ in R~Aql, S~Pav, and R~Hya, respectively (see Sect.~\ref{OH-opacity-popdiagram}).

\subsection{Chemical abundances and prospective remarks}

From our determinations of the H$_2$O and OH column densities, and on the assumption that both species are excited in similar regions,  we derived -- from R~Hya, S~Pav, and R~Aql -- an OH/H$_2$O ratio  of $\sim$0.7$-$2.8$\times$10$^{-2}$,
thereby establishing a benchmark for predictions of the chemical abundances in the inner wind of oxygen-rich AGBs 
by the chemical kinetic codes.
One of the  caveats in our determination of the OH/H$_2$O ratio is that a more stringent test of the accuracy of the 
chemical kinetic predictions  awaits measurements of the  OH/H$_2$O ratio as a function of the optical phase of the 
pulsating SR and Mira variables. Prior theoretical estimates of the OH/H$_2$O ratio ranged between about 10$^{-3}$ to 10$^{-5}$ \citep[e.g.,][]{cherchneff2006}. In our work  we find, with updated chemical kinetic rates \citep{gobrecht2022}, that the chemistry in the inner wind of a typical SR or Mira variable predicts a strong dependence of the OH/H$_2$O ratio on the pulsation phase of the star. Furthermore, OH/H$_2$O ratio variations by one to two orders of magnitude or even more around a given pulsation phase are possible in the 1$-$10~R$_{\star}$ range.

Despite the whole set of observed stars covering a range of optical phases, we have not, with the  {\sc atomium} data, systematically monitored the H$_2$O and OH  line emission at different epochs in the pulsation cycle,
and hence any evidence for time variability of the OH/H$_2$O ratio cannot be accurately examined. 
However,  time variability is present  in the H$_2$O data. It is well documented in U~Her,  observed in two epochs with the mid-configuration, and in a few other stars observed with the extended configuration and the ACA.

We conclude with two remarks: (1) An in-depth test of the chemical kinetic predictions awaits dedicated monitoring of the H$_2$O and OH emission in R~Hya, S~Pav, and R~Aql, and possibly other SR and Mira variables. (2) An in-depth development of  H$_2$O line excitation models awaits newer collision rates and needs to incorporate higher vibrational states, up to  the (0,3,0), (1,1,0) and (0,1,1) states at least, together with line overlap effects between para and ortho water.

\begin{acknowledgements}
The authors  gratefully thank the referee for his careful reading of the manuscript and his most constructive comments. This paper makes use of the following ALMA Main Array and ACA data: ADS/JAO.ALMA\#2018.1.00659.L (ATOMIUM: ALMA tracing the origins of molecules in dust forming oxygen-rich M-type stars); \#2019.2.00234.S (The remarkable 268~GHz line of water: a new tracer of the inner wind of evolved stars?); \#2017.1.00393.S and \#2018.1.00749.S (Bands~6 and 7 data used for the OH spectra in Mira,  Appendix~\ref{sec:append_mira_ohspec}). 
The standard ALMA pipeline products are available in the ALMA Science Archive (ASA).  The enhanced products prepared by the {\sc atomium} consortium, once ingested, will become available during or soon after publication of this paper.
ALMA is a partnership of ESO (representing its member states), NSF (USA) and NINS (Japan), together with NRC (Canada), NSC and ASIAA (Taiwan), and KASI (Republic of Korea), in cooperation with the Republic of Chile. The Joint ALMA Observatory is operated by ESO, AUI/NRAO and NAOJ.
A.B. and F.H. acknowledge financial support from 'Programme National de
Physique Stellaire' (PNPS) of CNRS/INSU, France and ANR PEPPER.
T.D. acknowledges support from the Research Foundation Flanders (FWO) through grant 12N9920N and is supported in part by the Australian Research Council through a Discovery Early Career Researcher Award (DE230100183).
K.T.W. acknowledges support from the European Research Council (ERC) under the European Union's Horizon 2020 research and innovation programme (Grant agreement no. 883867, project EXWINGS).
D. G. was funded by the grant "The Origin and Fate of Dust in our Universe" from the Knut and Alice Wallenberg Foundation. 
T.J.M. thanks the Leverhulme Trust for the award of an Emeritus fellowship. 
Calculations in Section~7 used the DiRAC Data Intensive service at Leicester, operated 
by the University of Leicester IT Services, which forms part of the STFC 
DiRAC HPC Facility (www.dirac.ac.uk). The equipment was funded by BEIS capital 
funding via STFC capital grants ST/K000373/1 and ST/R002363/1 and 
STFC DiRAC Operations grant ST/R001014/1. DiRAC is part of the National e-Infrastructure.
S.H.J.W. acknowledges support from the Research Foundation Flanders (FWO) through grant 1285221N. M.M. acknowledges funding from the Programme Paris Region fellowship supported by the Region Ile de France. This project has received funding under the Framework Program for Research and Innovation "Horizon 2020" under the Marie Sk?odowska-Curie grant agreement no. 945298.

\end{acknowledgements}

\bibliographystyle{aa}
\bibliography{45193corr_1a}

%%%%%%APPENDICES BEGIN    

\begin{appendix} 
\section{OH $\Lambda$$-$doubling transitions in the $\varv=0$ and 1 states}
 \label{sec:Lambda_doubling_freqs}
 
We have mentioned  in Sect. \ref{sec:identif_OH} the short-coming of the JPL catalog to calculate 
$\Lambda$-doubling transitions for high rotational quantum numbers. Explanations are given in this Appendix. 
The $\Lambda$-doubling transitions connect transitions with the same 
$N$ and $J$ in Hund's case $(b)$ or the same $\Omega$ and $J$ in Hund's case $(a)$ and 
their spacing is dominated by the $\Lambda$-doubling parameters $p$ and $q$. 
The $^2\Pi_{3/2}$ $\varv = 0$ fundamental transitions near 1.7~GHz were determined 
with a few Hertz accuracy by \citet{hudson2006} and \citet{lev2006}. 
Their intensities  decrease rapidly at room temperature and at 
modest quantum numbers. The highest rotational transitions in $\varv = 0$ from 
laboratory measurements are $J = 19/2$ ($^2\Pi_{3/2}$) and 15/2 ($^2\Pi_{1/2}$) near 66 and 71~GHz, respectively   \citep{kolbe1981,drouin2013}. In  $\varv = 1$ the quantum numbers are limited to $J = 13/2$ for both $^2\Pi_{3/2}$ and $^2\Pi_{1/2}$ near 18 and 67~GHz, respectively \citep{andresen2000} and there are even more limitations for $\varv = 2$. 
Our knowledge on the $\Lambda$-doubling transitions with higher rotational quantum numbers comes, for example, from pure rotational transitions in the far-infrared region with $\sim$1.2$-$3 MHz accuracy for the best lines \citep{martin2012}. In addition, these transitions depend predominantly on the rotational spacing between the upper or the lower $\Lambda$-component of two adjacent $J$ levels and sample only the differences between the $\Lambda$-components which are only a very small fraction of the respective transition frequencies. We conclude that it is not surprising that observations suffer from deviations up to a few MHz between the calculated  and observed $\Lambda$-doubling transitions with high quantum numbers and increasing vibrational states.

We have also mentioned  in Sect. \ref{sec:identif_OH} that the ALMA OH line observations  can be used to  improve the calculation of the OH $\Lambda$-doubling transitions. We give details below.

\citet{drouin2013} had carried out a Dunham fit pertaining to a plethora of laboratory data. A fair fraction of the data is associated with five minor isotopic species containing $^{17}$O, $^{18}$O, or D, but this is only of minor importance for the present study. 
The $\Lambda$$-$doubling is expressed with parameters $p_{ij}$ and $q_{ij}$, where  
$p_{00}$ and $q_{00}$ are the equilibrium parameters, and $i$ and $j$ indicate the 
degree of vibrational and rotational corrections to these parameters, respectively. 
\citet{drouin2013} employed 21 $\Lambda$$-$doubling parameters for OH itself plus five 
parameters to accomodate the breakdown of the Born-Oppenheimer approximation through 
the isotopic substitutions. (We mention specifically that the parameters used for OH 
without vibrational corrections were $p_{00}$ to $p_{03}$ and $q_{00}$ to $q_{05}$.) 
The data set from \citet{drouin2013} is available in the JPL catalog 
archive\footnote{https://spec.jpl.nasa.gov/ftp/pub/catalog/archive/}. Two poorly 
determined parameters (out of 97) were omitted. We added the transition frequencies from 
our present study and from \citet{khouri2019}, summarized in Table~\ref{Astro-Lines-Fit}, 
to the line list. We then carried out the line fit using Pickett's SPFIT program 
\citep{pickett1991}, as done earlier \citep{drouin2013}. Almost all changes in the values of the  spectroscopic 
parameters were well within the $1\sigma$ uncertainties. By far the largest 
relative changes occurred for $p_{02}$ and $q_{02}$, which decreased in magnitudes 
by about five times the respective uncertainty; $p_{03}$ and $q_{03}$ decreased 
in magnitudes by about three times the respective uncertainty. In addition, 
the improvement in the parameter uncertainties was small in almost all instances; 
by far the largest improvement occurred for $q_{02}$, where the new uncertainty 
was almost one third smaller. Nevertheless, as can be seen in Table~\ref{Astro-Lines-Fit}, the changes were large enough to reproduce the $\Lambda$$-$doubling transitions from radio astronomy observations 
within uncertainties on average.

The $\Lambda$-doubling transitions derived for rotational $J$ level energies up to 
$\sim$10000~cm$^{-1}$ (or $\sim$14500~K ) for $\varv = 0$ and 1 are presented in four Tables of this Appendix (see also end of Sect. \ref{sec:identif_OH}).  
The reliability of our calculations is very difficult to assess; the transition frequencies with calculated uncertainties larger than 0.5~MHz should be viewed with caution. Further calculations of the OH rotational spectra in $\varv = 0$ to 2 are available in the CDMS catalog\footnote{https://cdms.astro.uni-koeln.de/classic/entries/archive/OH/}.

\begin{table*}
\begin{center}
\caption{Quantum numbers$^a$, transition frequency (Frequency), uncertainty (Unc.), lower state energy ($E_{\rm low}$) and $A$ value 
      of OH $\Lambda$-doubling transitions in the $\varv = 0$, $^2\Pi _{3/2}$ state from our present fit.}
\label{OH_Lambda_v0_Pi3/2}
\begin{tabular}[t]{ccccccccccr@{}lr@{}lr@{}lr@{}l}
\hline \hline
$N$ & $p$ & $J$ & $F$ & $-$ & $N$ & $p$ & $\varv$ & $J$ & $F$ & 
\multicolumn{2}{c}{Frequency} & \multicolumn{2}{c}{Unc.} & \multicolumn{2}{c}{$E_{\rm low}$} &  $A$ \\
  &    &  &   &   &  &  &  &  & &   \multicolumn{2}{c}{(MHz)}  &  \multicolumn{2}{c}{(MHz)}  &  \multicolumn{2}{c}{(cm$^{-1}$)} & \multicolumn{2}{c}{($\times$10$^{-6}$s$^{-1}$)} \\
\hline
  8 &   + &  8.5 &  8 & $-$ &  8 & $-$ & 0 &  8.5 &  8 &  91188&.2709 & 0&.0118 &  1321&.2779 &  0&.455 \\
  8 &   + &  8.5 &  9 & $-$ &  8 & $-$ & 0 &  8.5 &  9 &  91203&.2537 & 0&.0119 &  1321&.2763 &  0&.455 \\
  9 & $-$ &  9.5 &  9 & $-$ &  9 &   + & 0 &  9.5 &  9 & 113605&.1955 & 0&.0225 &  1650&.8123 &  0&.691 \\
  9 & $-$ &  9.5 & 10 & $-$ &  9 &   + & 0 &  9.5 & 10 & 113621&.0752 & 0&.0226 &  1650&.8106 &  0&.692 \\
 10 &   + & 10.5 & 10 & $-$ & 10 & $-$ & 0 & 10.5 & 10 & 137959&.2898 & 0&.0400 &  2015&.0501 &  1&.00  \\
 10 &   + & 10.5 & 11 & $-$ & 10 & $-$ & 0 & 10.5 & 11 & 137975&.8958 & 0&.0400 &  2015&.0483 &  1&.00  \\
 11 & $-$ & 11.5 & 11 & $-$ & 11 &   + & 0 & 11.5 & 11 & 164119&.0489 & 0&.0671 &  2413&.6161 &  1&.38  \\
 11 & $-$ & 11.5 & 12 & $-$ & 11 &   + & 0 & 11.5 & 12 & 164136&.2367 & 0&.0670 &  2413&.6143 &  1&.38  \\
 12 &   + & 12.5 & 12 & $-$ & 12 & $-$ & 0 & 12.5 & 12 & 191954&.3152 & 0&.1069 &  2846&.0782 &  1&.84  \\
 12 &   + & 12.5 & 13 & $-$ & 12 & $-$ & 0 & 12.5 & 13 & 191971&.9613 & 0&.1068 &  2846&.0764 &  1&.84  \\
 13 & $-$ & 13.5 & 13 & $-$ & 13 &   + & 0 & 13.5 & 13 & 221334&.0293 & 0&.1633 &  3311&.9534 &  2&.39  \\
 13 & $-$ & 13.5 & 14 & $-$ & 13 &   + & 0 & 13.5 & 14 & 221352&.0274 & 0&.1633 &  3311&.9515 &  2&.39  \\
 14 &   + & 14.5 & 14 & $-$ & 14 & $-$ & 0 & 14.5 & 14 & 252124&.9162 & 0&.2410 &  3810&.7117 &  3&.03  \\
 14 &   + & 14.5 & 15 & $-$ & 14 & $-$ & 0 & 14.5 & 15 & 252143&.1736 & 0&.2409 &  3810&.7097 &  3&.03  \\
 15 & $-$ & 15.5 & 15 & $-$ & 15 &   + & 0 & 15.5 & 15 & 284190&.8005 & 0&.3451 &  4341&.7799 &  3&.76  \\
 15 & $-$ & 15.5 & 16 & $-$ & 15 &   + & 0 & 15.5 & 16 & 284209&.2356 & 0&.3450 &  4341&.7779 &  3&.76  \\
 16 &   + & 16.5 & 16 & $-$ & 16 & $-$ & 0 & 16.5 & 16 & 317392&.3410 & 0&.4823 &  4904&.5444 &  4&.58  \\
 16 &   + & 16.5 & 17 & $-$ & 16 & $-$ & 0 & 16.5 & 17 & 317410&.8811 & 0&.4821 &  4904&.5424 &  4&.58  \\
 17 & $-$ & 17.5 & 17 & $-$ & 17 &   + & 0 & 17.5 & 17 & 351587&.0429 & 0&.6607 &  5498&.3532 &  5&.49  \\
 17 & $-$ & 17.5 & 18 & $-$ & 17 &   + & 0 & 17.5 & 18 & 351605&.6228 & 0&.6604 &  5498&.3512 &  5&.49  \\
 18 &   + & 18.5 & 18 & $-$ & 18 & $-$ & 0 & 18.5 & 18 & 386629&.4537 & 0&.8906 &  6122&.5182 &  6&.49  \\
 18 &   + & 18.5 & 19 & $-$ & 18 & $-$ & 0 & 18.5 & 19 & 386648&.0141 & 0&.8904 &  6122&.5161 &  6&.49  \\
 19 & $-$ & 19.5 & 19 & $-$ & 19 &   + & 0 & 19.5 & 19 & 422371&.4796 & 1&.1855 &  6776&.3171 &  7&.56  \\
 19 & $-$ & 19.5 & 20 & $-$ & 19 &   + & 0 & 19.5 & 20 & 422389&.9664 & 1&.1852 &  6776&.3150 &  7&.56  \\
 20 &   + & 20.5 & 20 & $-$ & 20 & $-$ & 0 & 20.5 & 20 & 458662&.7788 & 1&.5620 &  7458&.9951 &  8&.71  \\
 20 &   + & 20.5 & 21 & $-$ & 20 & $-$ & 0 & 20.5 & 21 & 458681&.1420 & 1&.5616 &  7458&.9930 &  8&.71  \\
 21 & $-$ & 21.5 & 21 & $-$ & 21 &   + & 0 & 21.5 & 21 & 495351&.2024 & 2&.0407 &  8169&.7667 &  9&.91  \\
 21 & $-$ & 21.5 & 22 & $-$ & 21 &   + & 0 & 21.5 & 22 & 495369&.3956 & 2&.0403 &  8169&.7645 &  9&.91  \\
 22 &   + & 22.5 & 22 & $-$ & 22 & $-$ & 0 & 22.5 & 22 & 532283&.2601 & 2&.6460 &  8907&.8168 & 11&.2   \\
 22 &   + & 22.5 & 23 & $-$ & 22 & $-$ & 0 & 22.5 & 23 & 532301&.2398 & 2&.6454 &  8907&.8147 & 11&.2   \\
 23 & $-$ & 23.5 & 23 & $-$ & 23 &   + & 0 & 23.5 & 23 & 569304&.5950 & 3&.4052 &  9672&.3031 & 12&.5   \\
 23 & $-$ & 23.5 & 24 & $-$ & 23 &   + & 0 & 23.5 & 24 & 569322&.3204 & 3&.4046 &  9672&.3010 & 12&.5   \\
 24 &   + & 24.5 & 24 & $-$ & 24 & $-$ & 0 & 24.5 & 24 & 606260&.4552 & 4&.3480 & 10462&.3568 & 13&.8   \\
 24 &   + & 24.5 & 25 & $-$ & 24 & $-$ & 0 & 24.5 & 25 & 606277&.8875 & 4&.3473 & 10462&.3546 & 13&.8   \\
\hline
\end{tabular}
\end{center}
\tablefoot{$^{(a)}$ $N$ is the rotational quantum number, $p$ is the parity, $J$ and $F$ are the resulting momenta after adding the electron and nuclear spin angular momenta to the rotational one; $\varv$ is the vibrational state.
}
\end{table*}

\begin{table*}
\begin{center}
\caption{Quantum numbers$^a$, transition frequency (Frequency), uncertainty (Unc.), lower state energy ($E_{\rm low}$)   and $A$ value 
      of OH $\Lambda$-doubling transitions in the $\varv = 0$, $^2\Pi _{1/2}$ state from our present fit.}
\label{OH_Lambda_v0_Pi1/2}
\begin{tabular}[t]{ccccccccccr@{}lr@{}lr@{}lr@{}l}
\hline \hline
$N$ & $p$ & $J$ & $F$ & $-$ & $N$ & $p$ & $\varv$ & $J$ & $F$ & 
\multicolumn{2}{c}{Frequency} & \multicolumn{2}{c}{Unc.} & \multicolumn{2}{c}{$E_{\rm low}$} & $A$ \\
  &    &  &   &   &  &  &  &  & &   \multicolumn{2}{c}{(MHz)}  &  \multicolumn{2}{c}{(MHz)}  &  \multicolumn{2}{c}{(cm$^{-1}$)} & \multicolumn{2}{c}{($\times$10$^{-6}$s$^{-1}$)} \\
\hline
 11 & $-$ & 10.5 & 11 & $-$ & 11 &   + & 0 & 10.5 & 11 &  85665&.9971 & 0&.0267 &  2450&.2579 & 0&.105 \\
 11 & $-$ & 10.5 & 10 & $-$ & 11 &   + & 0 & 10.5 & 10 &  85703&.3790 & 0&.0268 &  2450&.2553 & 0&.105 \\
 12 &   + & 11.5 & 12 & $-$ & 12 & $-$ & 0 & 11.5 & 12 & 107036&.8197 & 0&.0438 &  2880&.4897 & 0&.179 \\
 12 &   + & 11.5 & 11 & $-$ & 12 & $-$ & 0 & 11.5 & 11 & 107073&.0719 & 0&.0441 &  2880&.4872 & 0&.179 \\
 13 & $-$ & 12.5 & 13 & $-$ & 13 &   + & 0 & 12.5 & 13 & 130078&.2350 & 0&.0701 &  3344&.4717 & 0&.281 \\
 13 & $-$ & 12.5 & 12 & $-$ & 13 &   + & 0 & 12.5 & 12 & 130113&.4391 & 0&.0705 &  3344&.4691 & 0&.281 \\
 14 &   + & 13.5 & 14 & $-$ & 14 & $-$ & 0 & 13.5 & 14 & 154660&.6445 & 0&.1087 &  3841&.6102 & 0&.418 \\
 14 &   + & 13.5 & 13 & $-$ & 14 & $-$ & 0 & 13.5 & 13 & 154694&.8639 & 0&.1093 &  3841&.6077 & 0&.419 \\
 15 & $-$ & 14.5 & 15 & $-$ & 15 &   + & 0 & 14.5 & 15 & 180652&.6153 & 0&.1643 &  4371&.2832 & 0&.594 \\
 15 & $-$ & 14.5 & 14 & $-$ & 15 &   + & 0 & 14.5 & 14 & 180685&.8990 & 0&.1650 &  4371&.2806 & 0&.594 \\
 16 &   + & 15.5 & 16 & $-$ & 16 & $-$ & 0 & 15.5 & 16 & 207920&.2054 & 0&.2433 &  4932&.8383 & 0&.811 \\
 16 &   + & 15.5 & 15 & $-$ & 16 & $-$ & 0 & 15.5 & 15 & 207952&.5907 & 0&.2440 &  4932&.8358 & 0&.812 \\
 17 & $-$ & 16.5 & 17 & $-$ & 17 &   + & 0 & 16.5 & 17 & 236326&.7051 & 0&.3548 &  5525&.5932 & 1&.07  \\
 17 & $-$ & 16.5 & 16 & $-$ & 17 &   + & 0 & 16.5 & 16 & 236358&.2198 & 0&.3557 &  5525&.5907 & 1&.07  \\
 18 &   + & 17.5 & 18 & $-$ & 18 & $-$ & 0 & 17.5 & 18 & 265732&.6519 & 0&.5115 &  6148&.8354 & 1&.38  \\
 18 &   + & 17.5 & 17 & $-$ & 18 & $-$ & 0 & 17.5 & 17 & 265763&.3164 & 0&.5123 &  6148&.8329 & 1&.38  \\
 19 & $-$ & 18.5 & 19 & $-$ & 19 &   + & 0 & 18.5 & 19 & 295996&.0281 & 0&.7297 &  6801&.8227 & 1&.74  \\
 19 & $-$ & 18.5 & 18 & $-$ & 19 &   + & 0 & 18.5 & 18 & 296025&.8563 & 0&.7305 &  6801&.8203 & 1&.74  \\
 20 &   + & 19.5 & 20 & $-$ & 20 & $-$ & 0 & 19.5 & 20 & 326972&.5743 & 1&.0298 &  7483&.7842 & 2&.14  \\
 20 &   + & 19.5 & 19 & $-$ & 20 & $-$ & 0 & 19.5 & 19 & 327001&.5751 & 1&.0306 &  7483&.7818 & 2&.14  \\
 21 & $-$ & 20.5 & 21 & $-$ & 21 &   + & 0 & 20.5 & 21 & 358516&.1794 & 1&.4363 &  8193&.9209 & 2&.59  \\
 21 & $-$ & 20.5 & 20 & $-$ & 21 &   + & 0 & 20.5 & 20 & 358544&.3573 & 1&.4370 &  8193&.9184 & 2&.59  \\
 22 &   + & 21.5 & 22 & $-$ & 22 & $-$ & 0 & 21.5 & 22 & 390479&.3150 & 1&.9770 &  8931&.4066 & 3&.07  \\
 22 &   + & 21.5 & 21 & $-$ & 22 & $-$ & 0 & 21.5 & 21 & 390506&.6708 & 1&.9776 &  8931&.4042 & 3&.07  \\
 23 & $-$ & 22.5 & 23 & $-$ & 23 &   + & 0 & 22.5 & 23 & 422713&.4928 & 2&.6823 &  9695&.3897 & 3&.60  \\
 23 & $-$ & 22.5 & 22 & $-$ & 23 &   + & 0 & 22.5 & 22 & 422740&.0243 & 2&.6828 &  9695&.3873 & 3&.60  \\
 24 &   + & 23.5 & 24 & $-$ & 24 & $-$ & 0 & 23.5 & 24 & 455069&.7308 & 3&.5839 & 10484&.9933 & 4&.16  \\
 24 &   + & 23.5 & 23 & $-$ & 24 & $-$ & 0 & 23.5 & 23 & 455095&.4332 & 3&.5843 & 10484&.9910 & 4&.16  \\
\hline
\end{tabular}
\end{center}
\tablefoot{$^{(a)}$ $N$ is the rotational quantum number, $p$ is the parity, $J$ and $F$ are the resulting momenta after adding the electron and nuclear spin angular momenta to the rotational one; $\varv$ is the vibrational state.
}
\end{table*}

\begin{table*}
\begin{center}
\caption{Quantum numbers$^a$, transition frequency (Frequency), uncertainty (Unc.), lower state energy ($E_{\rm low}$)  and $A$ value 
      of OH $\Lambda$-doubling transitions in the $\varv = 1$, $^2\Pi _{3/2}$ state from our present fit.}
\label{OH_Lambda_v1_Pi3/2}
\begin{tabular}[t]{ccccccccccr@{}lr@{}lr@{}lr@{}l}
\hline \hline
$N$ & $p$ & $J$ & $F$ & $-$ & $N$ & $p$ & $\varv$ & $J$ & $F$ & 
\multicolumn{2}{c}{Frequency} & \multicolumn{2}{c}{Unc.} & \multicolumn{2}{c}{$E_{\rm low}$} & $A$ \\
&    &  &   &   &  &  &  &  & &   \multicolumn{2}{c}{(MHz)}  &  \multicolumn{2}{c}{(MHz)}  &  \multicolumn{2}{c}{(cm$^{-1}$)} & \multicolumn{2}{c}{($\times$10$^{-6}$s$^{-1}$)} \\
\hline
  8 &   + &  8.5 &  8 & $-$ &  8 & $-$ & 1 &  8.5 &  8 &  86177&.8025 & 0&.0554 &  4840&.2064 &  0&.387 \\
  8 &   + &  8.5 &  9 & $-$ &  8 & $-$ & 1 &  8.5 &  9 &  86191&.7190 & 0&.0554 &  4840&.2048 & 0&.388 \\
  9 & $-$ &  9.5 &  9 & $-$ &  9 &   + & 1 &  9.5 &  9 & 107485&.6400 & 0&.0845 &  5157&.1087 & 0&.591 \\
  9 & $-$ &  9.5 & 10 & $-$ &  9 &   + & 1 &  9.5 & 10 & 107500&.4151 & 0&.0845 &  5157&.1070 & 0&.591 \\
 10 &   + & 10.5 & 10 & $-$ & 10 & $-$ & 1 & 10.5 & 10 & 130638&.8841 & 0&.1235 &  5507&.3314 & 0&.854 \\
 10 &   + & 10.5 & 11 & $-$ & 10 & $-$ & 1 & 10.5 & 11 & 130654&.3551 & 0&.1235 &  5507&.3296 & 0&.855 \\
 11 & $-$ & 11.5 & 11 & $-$ & 11 &   + & 1 & 11.5 & 11 & 155506&.9363 & 0&.1748 &  5890&.5079 & 1&.18  \\
 11 & $-$ & 11.5 & 12 & $-$ & 11 &   + & 1 & 11.5 & 12 & 155522&.9647 & 0&.1748 &  5890&.5060 & 1&.18  \\
 12 &   + & 12.5 & 12 & $-$ & 12 & $-$ & 1 & 12.5 & 12 & 181960&.7090 & 0&.2414 &  6306&.2157 & 1&.58  \\
 12 &   + & 12.5 & 13 & $-$ & 12 & $-$ & 1 & 12.5 & 13 & 181977&.1759 & 0&.2414 &  6306&.2138 & 1&.58  \\
 13 & $-$ & 13.5 & 13 & $-$ & 13 &   + & 1 & 13.5 & 13 & 209870&.4018 & 0&.3279 &  6753&.9818 & 2&.06  \\
 13 & $-$ & 13.5 & 14 & $-$ & 13 &   + & 1 & 13.5 & 14 & 209887&.2044 & 0&.3279 &  6753&.9798 & 2&.06  \\
 14 &   + & 14.5 & 14 & $-$ & 14 & $-$ & 1 & 14.5 & 14 & 239104&.1684 & 0&.4406 &  7233&.2873 & 2&.60  \\
 14 &   + & 14.5 & 15 & $-$ & 14 & $-$ & 1 & 14.5 & 15 & 239121&.2166 & 0&.4406 &  7233&.2853 & 2&.61  \\
 15 & $-$ & 15.5 & 15 & $-$ & 15 &   + & 1 & 15.5 & 15 & 269527&.3940 & 0&.5879 &  7743&.5703 & 3&.23  \\
 15 & $-$ & 15.5 & 16 & $-$ & 15 &   + & 1 & 15.5 & 16 & 269544&.6085 & 0&.5879 &  7743&.5683 & 3&.23  \\
 16 &   + & 16.5 & 16 & $-$ & 16 & $-$ & 1 & 16.5 & 16 & 301002&.3858 & 0&.7804 &  8284&.2290 & 3&.94  \\
 16 &   + & 16.5 & 17 & $-$ & 16 & $-$ & 1 & 16.5 & 17 & 301019&.6957 & 0&.7804 &  8284&.2270 & 3&.94  \\
 17 & $-$ & 17.5 & 17 & $-$ & 17 &   + & 1 & 17.5 & 17 & 333388&.3372 & 1&.0304 &  8854&.6237 & 4&.71  \\
 17 & $-$ & 17.5 & 18 & $-$ & 17 &   + & 1 & 17.5 & 18 & 333405&.6788 & 1&.0305 &  8854&.6216 & 4&.72  \\
 18 &   + & 18.5 & 18 & $-$ & 18 & $-$ & 1 & 18.5 & 18 & 366541&.4751 & 1&.3526 &  9454&.0788 & 5&.57  \\
 18 &   + & 18.5 & 19 & $-$ & 18 & $-$ & 1 & 18.5 & 19 & 366558&.7903 & 1&.3526 &  9454&.0767 & 5&.57  \\
 19 & $-$ & 19.5 & 19 & $-$ & 19 &   + & 1 & 19.5 & 19 & 400315&.3257 & 1&.7632 & 10081&.8849 & 6&.48  \\
 19 & $-$ & 19.5 & 20 & $-$ & 19 &   + & 1 & 19.5 & 20 & 400332&.5615 & 1&.7632 & 10081&.8828 & 6&.48  \\
\hline
\end{tabular}
\end{center}
\tablefoot{$^{(a)}$ $N$ is the rotational quantum number, $p$ is the parity, $J$ and $F$ are the resulting momenta after adding the electron and nuclear spin angular momenta to the rotational one; $\varv$ is the vibrational state.
}
\end{table*}

\begin{table*}
\begin{center}
\caption{Quantum numbers$^a$, transition frequency (Frequency), uncertainty (Unc.), lower state energy ($E_{\rm low}$)  and $A$ value  
      of OH $\Lambda$-doubling transitions in the $\varv = 1$, $^2\Pi _{1/2}$ state from our present fit.}
\label{OH_Lambda_v1_Pi1/2}
\begin{tabular}[t]{ccccccccccr@{}lr@{}lr@{}lr@{}l}
\hline \hline
$N$ & $p$ & $J$ & $F$ & $-$ & $N$ & $p$ & $\varv$ & $J$ & $F$ & 
\multicolumn{2}{c}{Frequency} & \multicolumn{2}{c}{Unc.} & \multicolumn{2}{c}{$E_{\rm low}$} & $A$ \\
&    &  &   &   &  &  &  &  & &   \multicolumn{2}{c}{(MHz)}  &  \multicolumn{2}{c}{(MHz)}  &  \multicolumn{2}{c}{(cm$^{-1}$)} & \multicolumn{2}{c}{($\times$10$^{-6}$s$^{-1}$)} \\
\hline
 11 & $-$ & 10.5 & 11 & $-$ & 11 &   + & 1 & 10.5 & 11 &  80715&.3788 & 0&.1322 &  5927&.9484 & 0&.087 \\
 11 & $-$ & 10.5 & 10 & $-$ & 11 &   + & 1 & 10.5 & 10 &  80751&.0251 & 0&.1324 &  5927&.9459 & 0&.087 \\
 12 &   + & 11.5 & 12 & $-$ & 12 & $-$ & 1 & 11.5 & 12 & 101017&.4816 & 0&.2116 &  6341&.3638 & 0&.148 \\
 12 &   + & 11.5 & 11 & $-$ & 12 & $-$ & 1 & 11.5 & 11 & 101052&.0236 & 0&.2120 &  6341&.3612 & 0&.148 \\
 13 & $-$ & 12.5 & 13 & $-$ & 13 &   + & 1 & 12.5 & 13 & 122902&.5572 & 0&.3245 &  6787&.1800 & 0&.234 \\
 13 & $-$ & 12.5 & 12 & $-$ & 13 &   + & 1 & 12.5 & 12 & 122936&.0717 & 0&.3249 &  6787&.1775 & 0&.234 \\
 14 &   + & 13.5 & 14 & $-$ & 14 & $-$ & 1 & 13.5 & 14 & 146242&.5151 & 0&.4804 &  7264&.8140 & 0&.349 \\
 14 &   + & 13.5 & 13 & $-$ & 14 & $-$ & 1 & 13.5 & 13 & 146275&.0618 & 0&.4808 &  7264&.8115 & 0&.350 \\
 15 & $-$ & 14.5 & 15 & $-$ & 15 &   + & 1 & 14.5 & 15 & 170907&.6132 & 0&.6903 &  7773&.6543 & 0&.497 \\
 15 & $-$ & 14.5 & 14 & $-$ & 15 &   + & 1 & 14.5 & 14 & 170939&.2382 & 0&.6908 &  7773&.6518 & 0&.497 \\
 16 &   + & 15.5 & 16 & $-$ & 16 & $-$ & 1 & 15.5 & 16 & 196765&.7454 & 0&.9672 &  8313&.0601 & 0&.678 \\
 16 &   + & 15.5 & 15 & $-$ & 16 & $-$ & 1 & 15.5 & 15 & 196796&.4837 & 0&.9677 &  8313&.0576 & 0&.680 \\
 17 & $-$ & 16.5 & 17 & $-$ & 17 &   + & 1 & 16.5 & 17 & 223682&.1369 & 1&.3258 &  8882&.3608 & 0&.901 \\
 17 & $-$ & 16.5 & 16 & $-$ & 17 &   + & 1 & 16.5 & 16 & 223712&.0146 & 1&.3263 &  8882&.3584 & 0&.901 \\
 18 &   + & 17.5 & 18 & $-$ & 18 & $-$ & 1 & 17.5 & 18 & 251519&.3118 & 1&.7823 &  9480&.8560 & 1&.16  \\
 18 &   + & 17.5 & 17 & $-$ & 18 & $-$ & 1 & 17.5 & 17 & 251548&.3477 & 1&.7829 &  9480&.8536 & 1&.16  \\
 19 & $-$ & 18.5 & 19 & $-$ & 19 &   + & 1 & 18.5 & 19 & 280137&.2392 & 2&.3549 & 10107&.8161 & 1&.46  \\
 19 & $-$ & 18.5 & 18 & $-$ & 19 &   + & 1 & 18.5 & 18 & 280165&.4459 & 2&.3555 & 10107&.8137 & 1&.46  \\
\hline
\end{tabular}
\end{center}
\tablefoot{$^{(a)}$ $N$ is the rotational quantum number, $p$ is the parity, $J$ and $F$ are the resulting momenta after adding the electron and nuclear spin angular momenta to the rotational one; $\varv$ is the vibrational state.
}
\end{table*}

%%%%%%%%%%%%%%%%%%%
%Show all spectra
%%%%%%%%%%%%%%%%%%%

\section{Water line spectra}
 \label{plot_H2O_lines}
 
We have gathered here all spectra extracted from our extended configuration data cubes for the H$_2$O lines 1, 5, 6, 7, 8, 10, 12, 13 and 14 listed in Table~\ref{H2O-line-list}. For lines 12 and 14 we have also added the mid configuration spectra. The circular extraction aperture selected for each line can vary  from one spectral transition to another in each source. It has been chosen so that all the  emission visible in the channel maps is included in the spectra but is not too large to degrade the S/N. For some very weak detections, we used  0\farcs08 by default for the extended configuration spectra.

 The very weak line 4 at 236.805~GHz identified in R~Hya, and perhaps in S~Pav,  is presented for the combined extended and mid arrays in Fig.~\ref{rhya_236}.

  %%%%%%%%%%%%%% 
 \begin{figure*}
\centering
\includegraphics[width=15.1cm,angle=0]{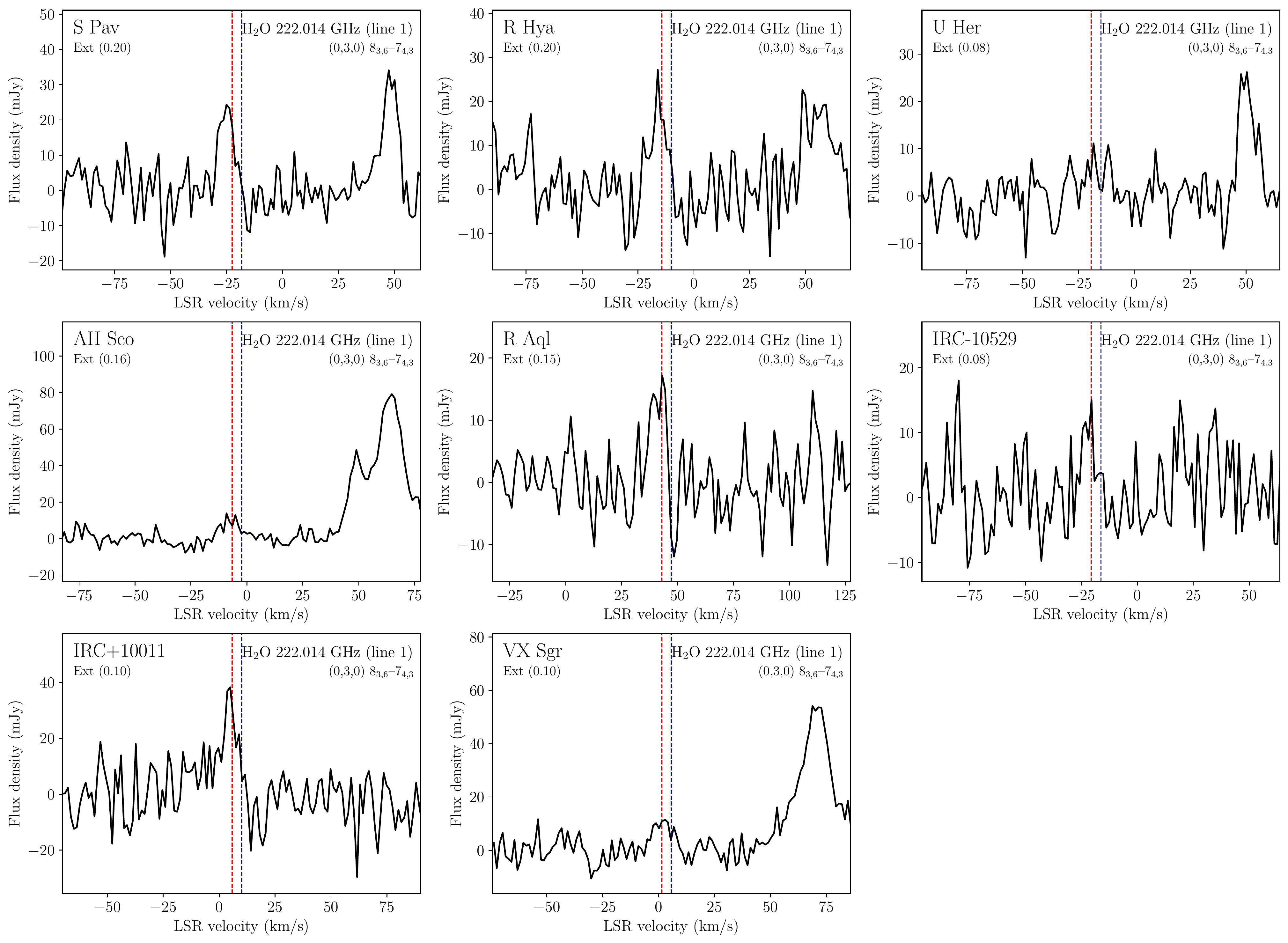}
 
\caption{Spectra of line~1 at 222.014~GHz (Table \ref{H2O-line-list}) extracted from the extended configuration array. The extraction diameter (in arc~sec) is given in parentheses below the source name in the upper left corner of each spectrum. The observed frequency is  converted  to the LSR frame using the catalog line rest frequency given in Table \ref{H2O-line-list}. The blue vertical line indicates the adopted new LSR systemic velocity (see Table\ref{primarysource_list}). The red vertical line shows the LSR velocity for the slightly different frequency determined in this work (see Table \ref{H2O-line-list}).  
} 
\label{line01_allsrc_ext}
\end{figure*}

%%%%
\begin{figure*}
\centering
\includegraphics[width=15.1cm,angle=0]{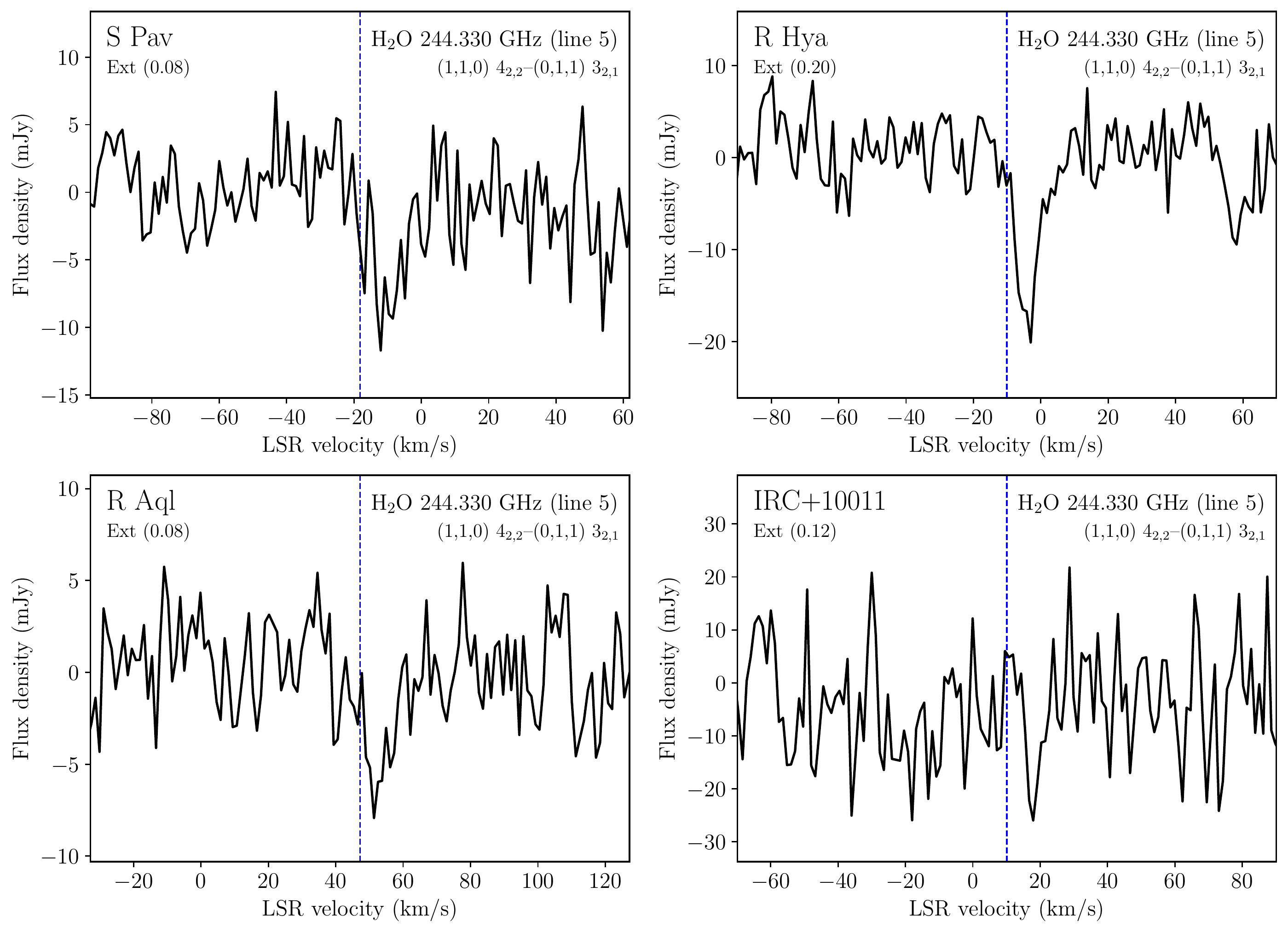}
 
\caption{Spectra of line~5 at 244.330~GHz (Table \ref{H2O-line-list}) extracted from the extended configuration array. The extraction diameter (in arc~sec) is given in parentheses below the source name in the upper left corner of each spectrum. The observed frequency is converted  to the LSR frame using the catalog line rest frequency given in Table \ref{H2O-line-list}. The blue vertical line indicates the adopted new LSR systemic velocity (see Table\ref{primarysource_list}). 
} 
\label{line05_allsrc_ext}
\end{figure*}

%%%%%%%%

\begin{figure*}
\centering
\includegraphics[width=15.1cm,angle=0]{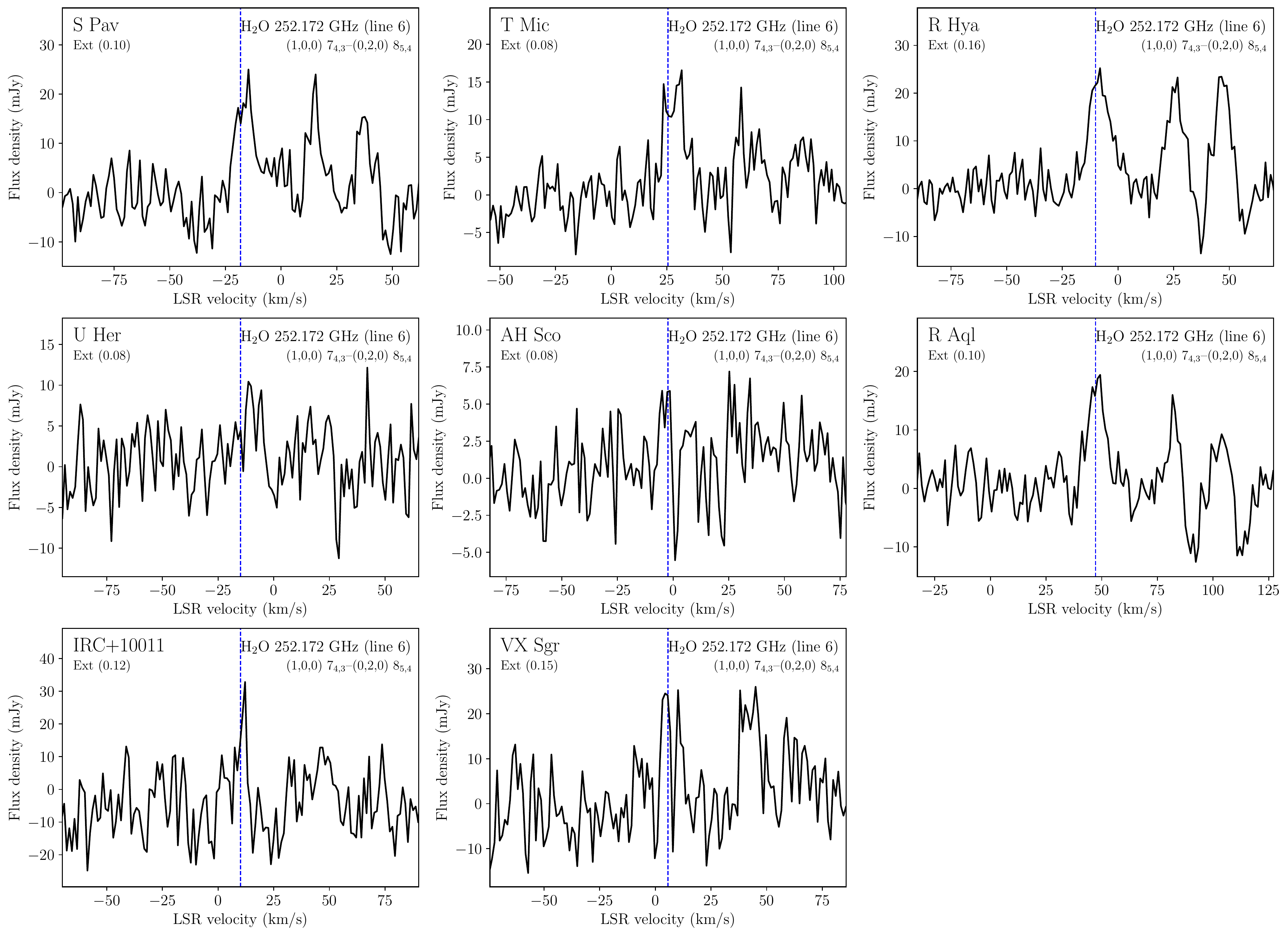}
 
\caption{Spectra of line~6 at 252.172~GHz (Table \ref{H2O-line-list}) extracted from the extended configuration array. The rest of the figure caption is as in Fig.~\ref{line05_allsrc_ext}.} 
\label{line06_allsrc_ext}
\end{figure*}

%%%%%%
\begin{figure*}
\centering
\includegraphics[width=15.1cm,angle=0]{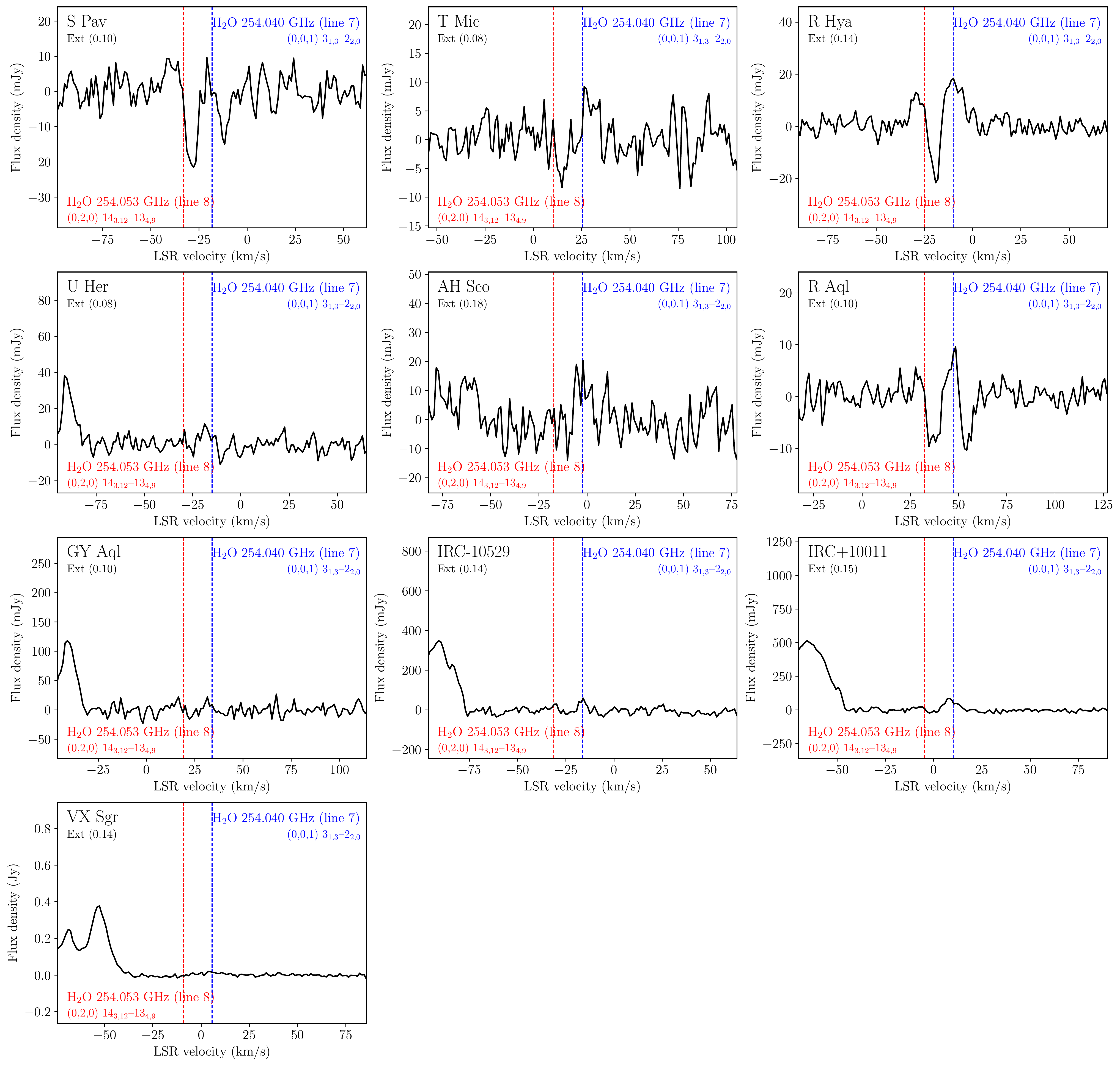}
 
\caption{Spectra of lines~7  and 8 at 254.040 and 254.053~GHz (Table \ref{H2O-line-list}) extracted from the extended configuration array. The extraction diameter (in arc~sec) is given in parentheses below the source name in the upper left corner of each spectrum. The blue and red vertical lines (lines~7 and 8) indicate the adopted new LSR systemic velocity (see Table\ref{primarysource_list}) using the catalog rest frequencies (Table~\ref{H2O-line-list}).} 
\label{line07_allsrc_ext}
\end{figure*}

%%%%%%%%
\begin{figure*}
\centering
\includegraphics[width=15.1cm,angle=0]{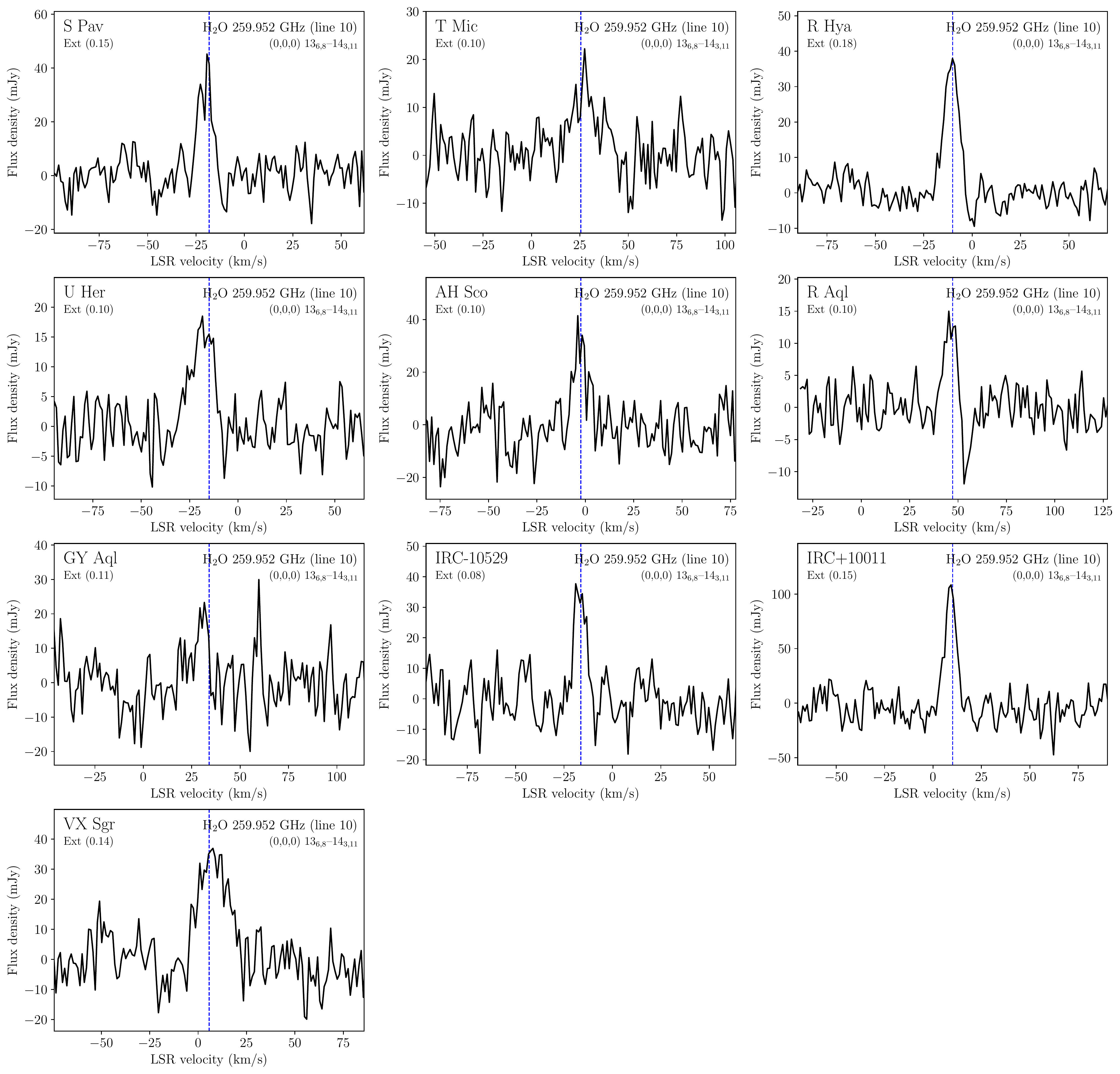}
 
\caption{Spectra of line~10 at 259.952~GHz (Table \ref{H2O-line-list}) extracted from the extended configuration array. The rest of the figure caption is as in Fig.~\ref{line05_allsrc_ext}..} 
\label{line10_allsrc_ext}
\end{figure*}

%%%%%%%%
\begin{figure*}
\centering
\includegraphics[width=15.1cm,angle=0]{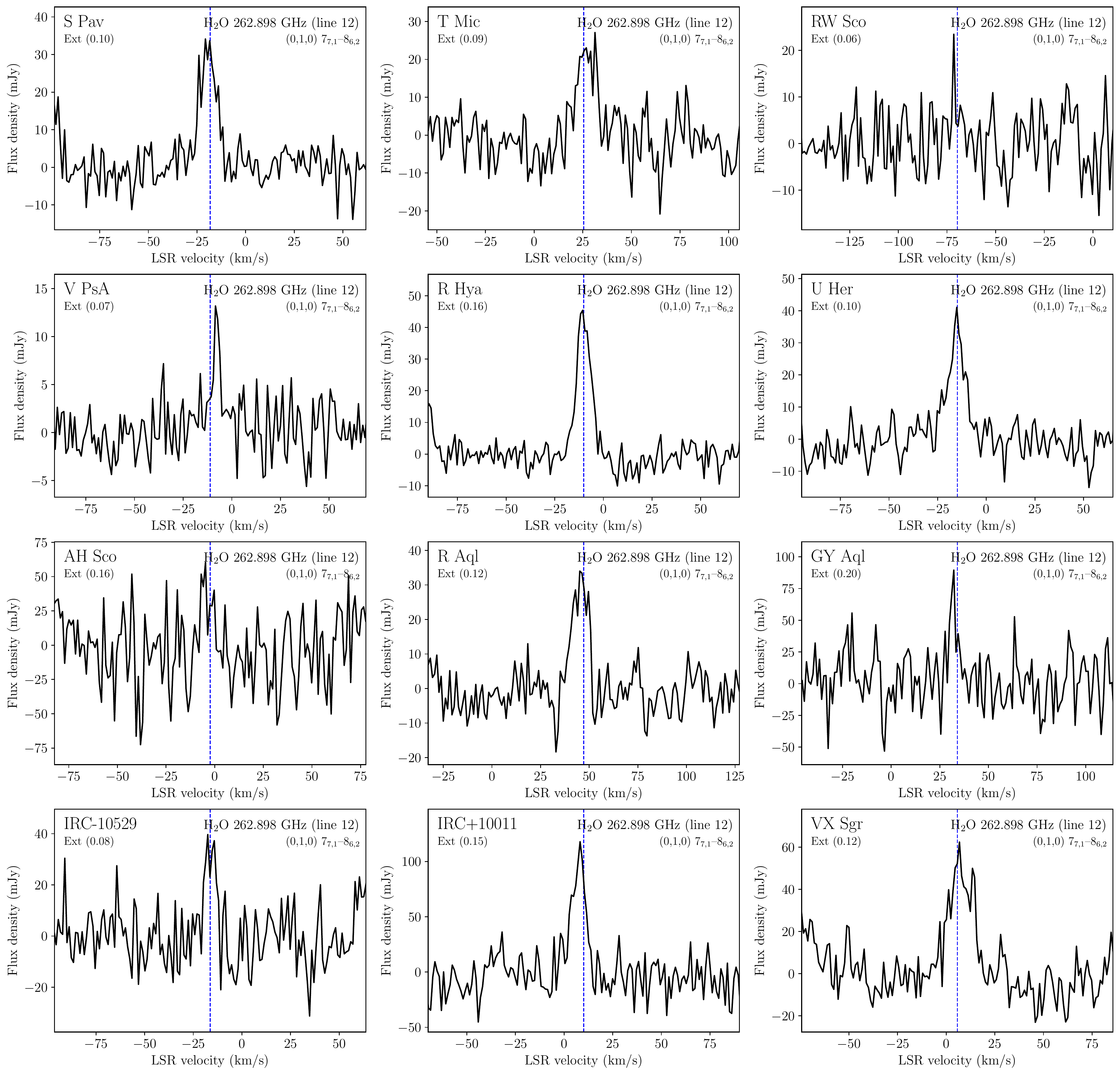}
 
\caption{Spectra of line~12 at 262.898~GHz (Table \ref{H2O-line-list}) extracted from the extended configuration array. The rest of the figure caption is as in Fig.~\ref{line05_allsrc_ext}.} 
\label{line12_allsrc_ext}
\end{figure*}

%%%%%%%%
\begin{figure*}
\centering
\includegraphics[width=15.1cm,angle=0]{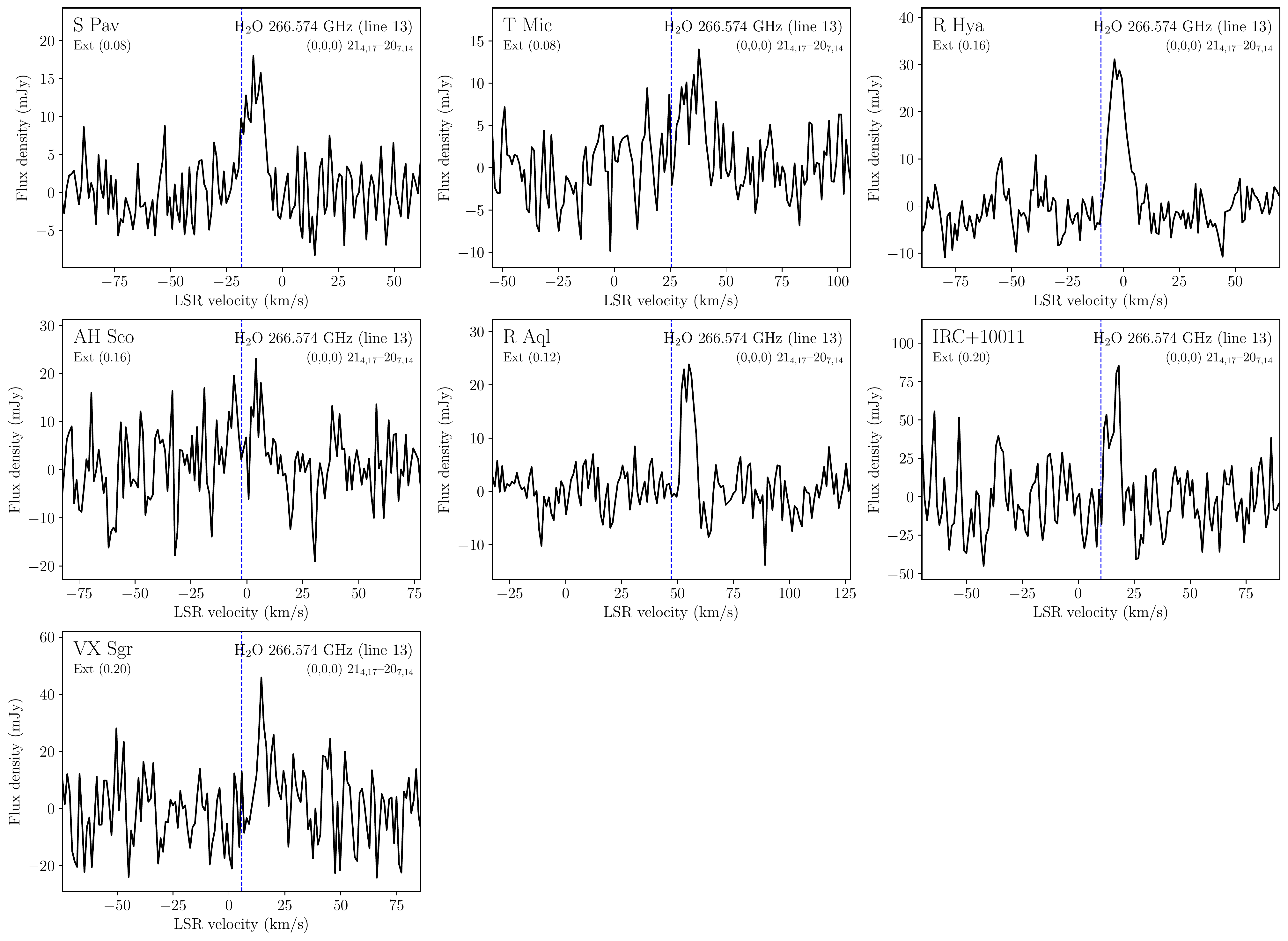}
 
\caption{Spectra of line~13 at 266.574~GHz (Table \ref{H2O-line-list}) extracted from the extended configuration array. The rest of the figure caption is as in Fig.~\ref{line05_allsrc_ext}.} 
\label{line13_allsrc_ext}
\end{figure*}

%%%%%%%%
\begin{figure*}
\centering
\includegraphics[width=15.1cm,angle=0]{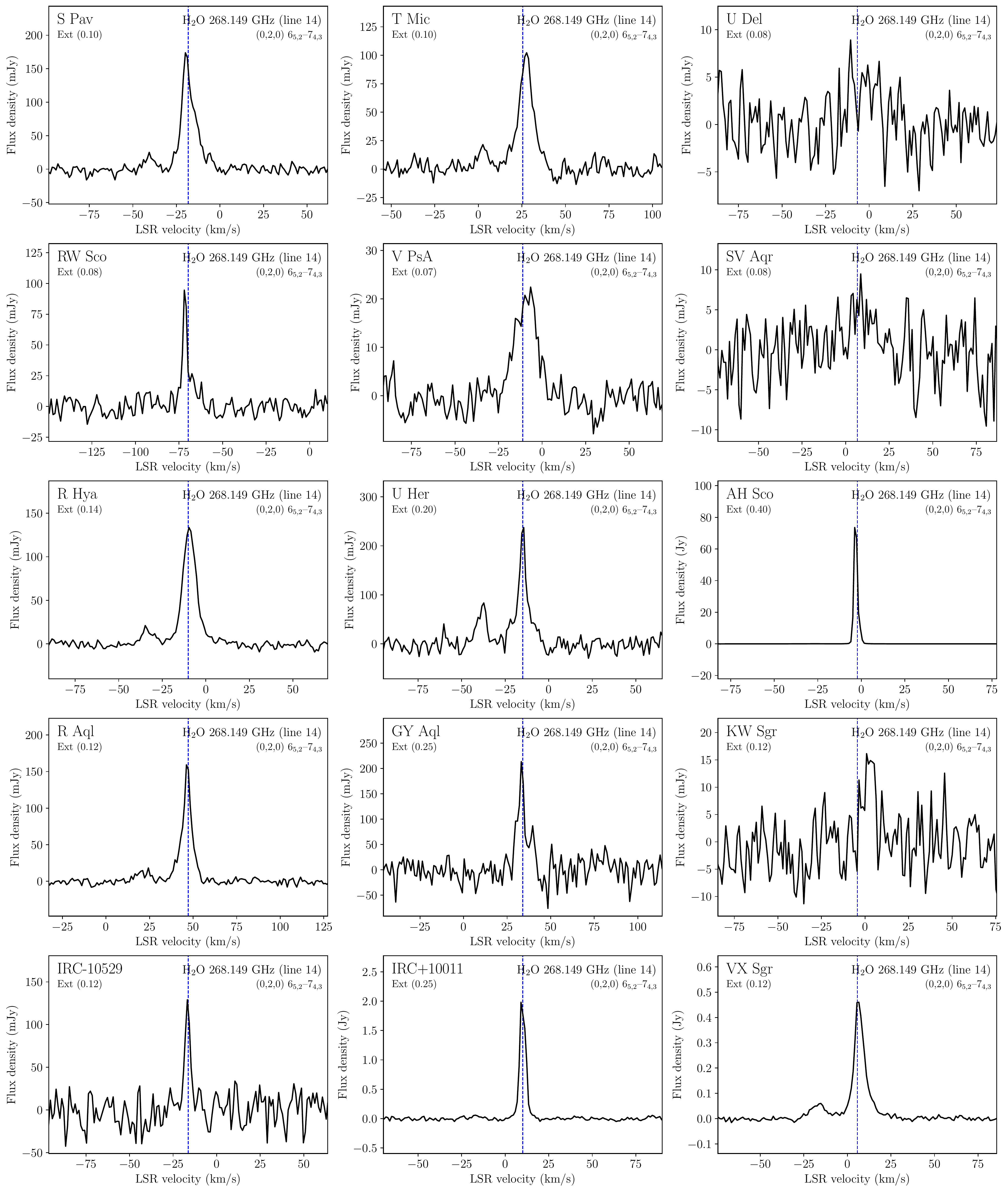}
 
\caption{Spectra of line~14 at 268.149~GHz (Table \ref{H2O-line-list}) extracted from the extended configuration array. The rest of the figure caption is as in Fig.~\ref{line05_allsrc_ext}.} 
\label{line13_allsrc_ext}
\end{figure*}

%%%%%%%%
\begin{figure*}
\centering
\includegraphics[width=15.1cm,angle=0]{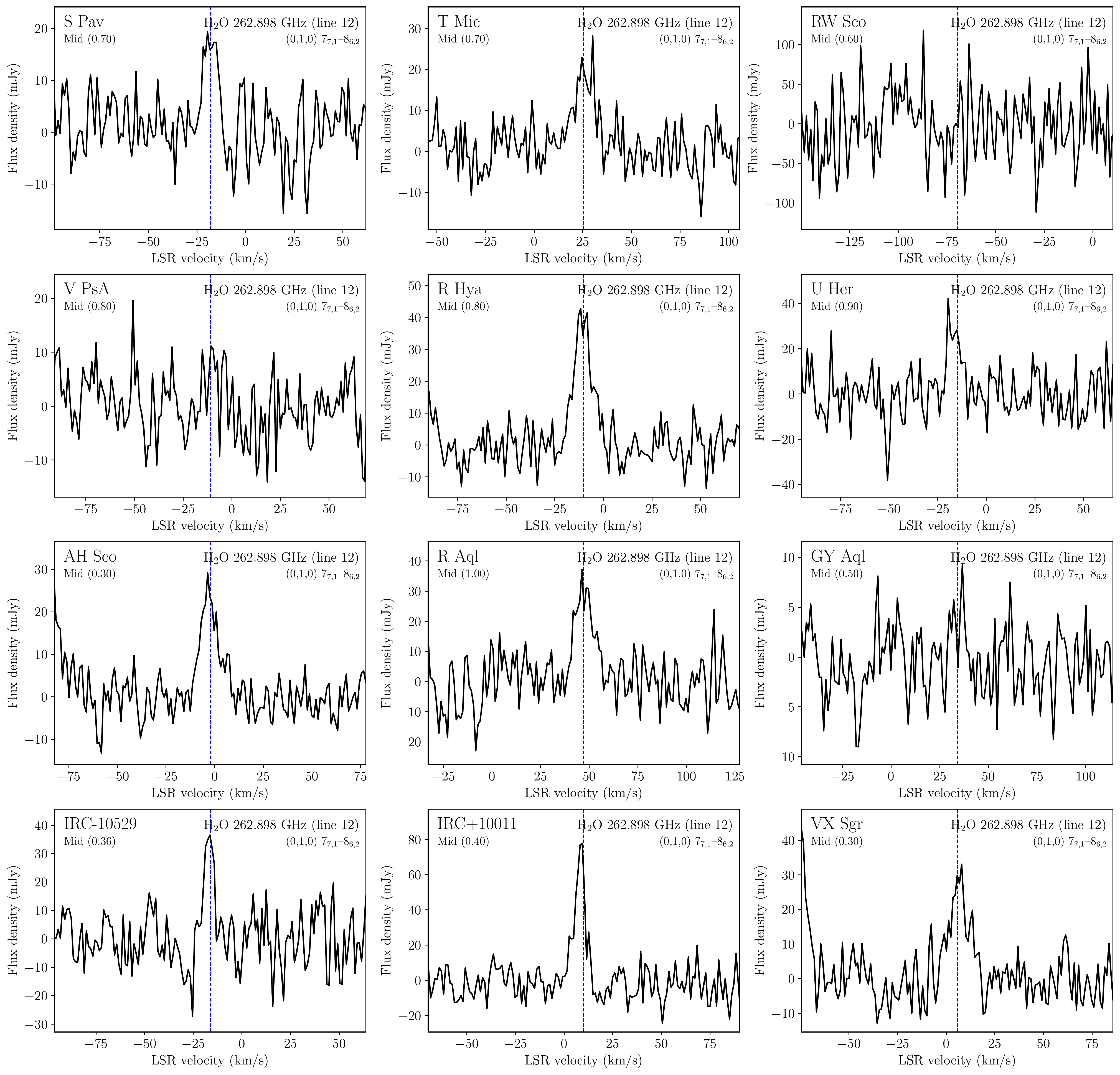}
 
\caption{Spectra of line~12 at 262.898~GHz (Table \ref{H2O-line-list}) extracted from the $\it{mid}$ configuration array. The rest of the figure caption is as in Fig.~\ref{line05_allsrc_ext}.} 
\label{line12_allsrc_mid}
\end{figure*}

%%%%%%%%
\begin{figure*}
\centering
\includegraphics[width=15.1cm,angle=0]{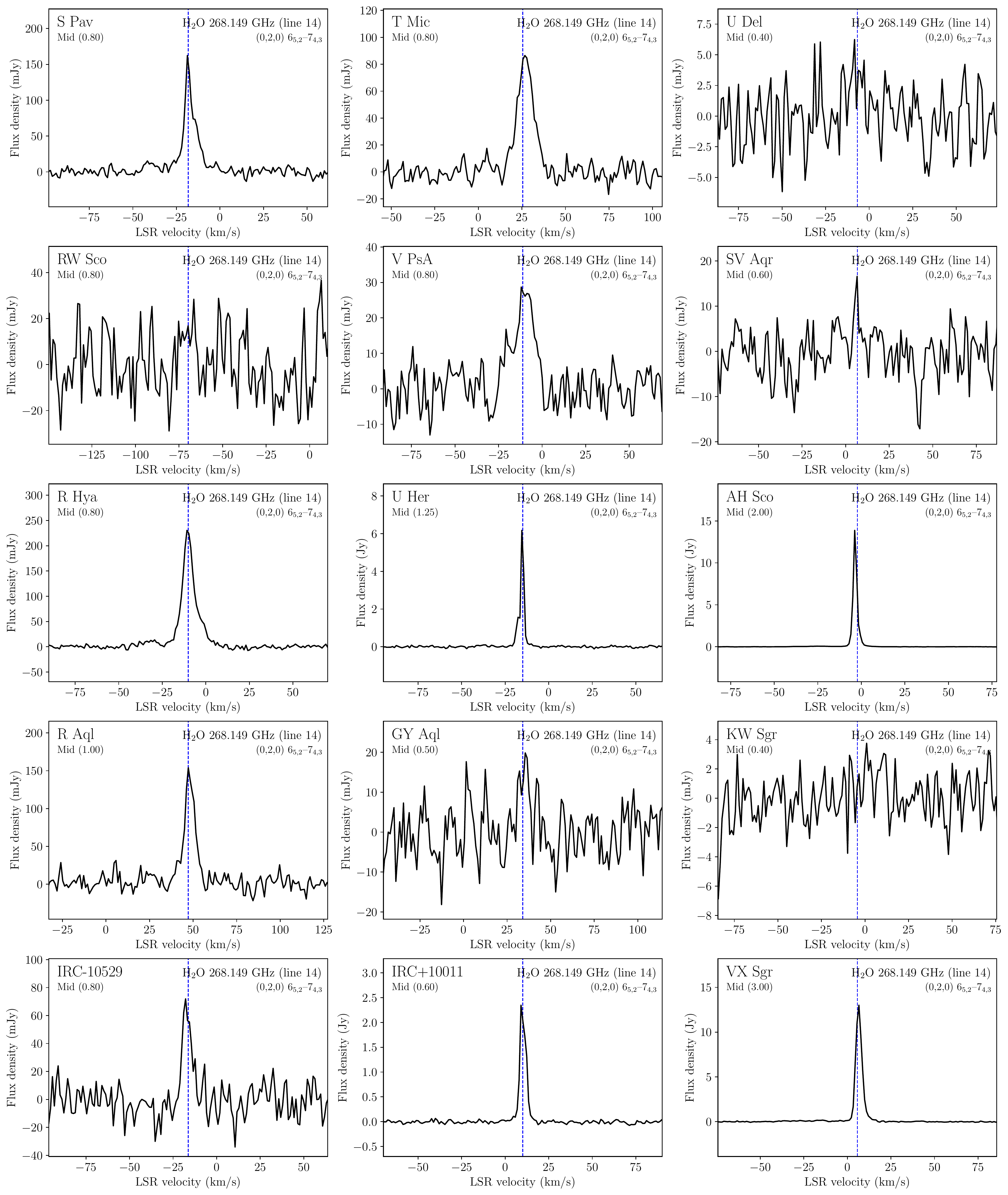}
 
\caption{Spectra of line~14 at 268.149~GHz (Table \ref{H2O-line-list}) extracted from the $\it{mid}$ configuration array. The rest of the figure caption is as in Fig.~\ref{line05_allsrc_ext}.} 
\label{line14_allsrc_mid}
\end{figure*}
%%%%%%%

\section{236.805~GHz H$_2$O line, 254.17$-$254.32~GHz line emission region} 
\label{sec:254.23_vxsgr_ahsco}
%%%%

Fig.~\ref{rhya_236}, upper panel, shows the water emission observed in R~Hya around 236.805~GHz from the highest energy levels identified in this work (line~4 in Table~\ref{H2O-line-list}) for different extraction diameters of the combined high and mid arrays. The spectra   peak at 3.7~$\pm$1.5~mJy for an aperture diameter of  0\farcs16; the uncertainty is at the 2$\sigma$ level. Peak emission around 10 mJy is also visible with the extended array alone in the same frequency range and for an  aperture diameter of 0\farcs16. 
There is also an uncertain detection of the 236.805~GHz transition toward S Pav (Fig. ~\ref{rhya_236}, lower panel). The spectrum is rather noisy but  we observe  a peak flux density of $\sim$10~mJy with the combined array for an aperture diameter of 0\farcs16. Detection is less convincing around 236.805~GHz with the extended array alone. 

\begin{figure}
\centering
\includegraphics[width=8.8cm,angle=0]{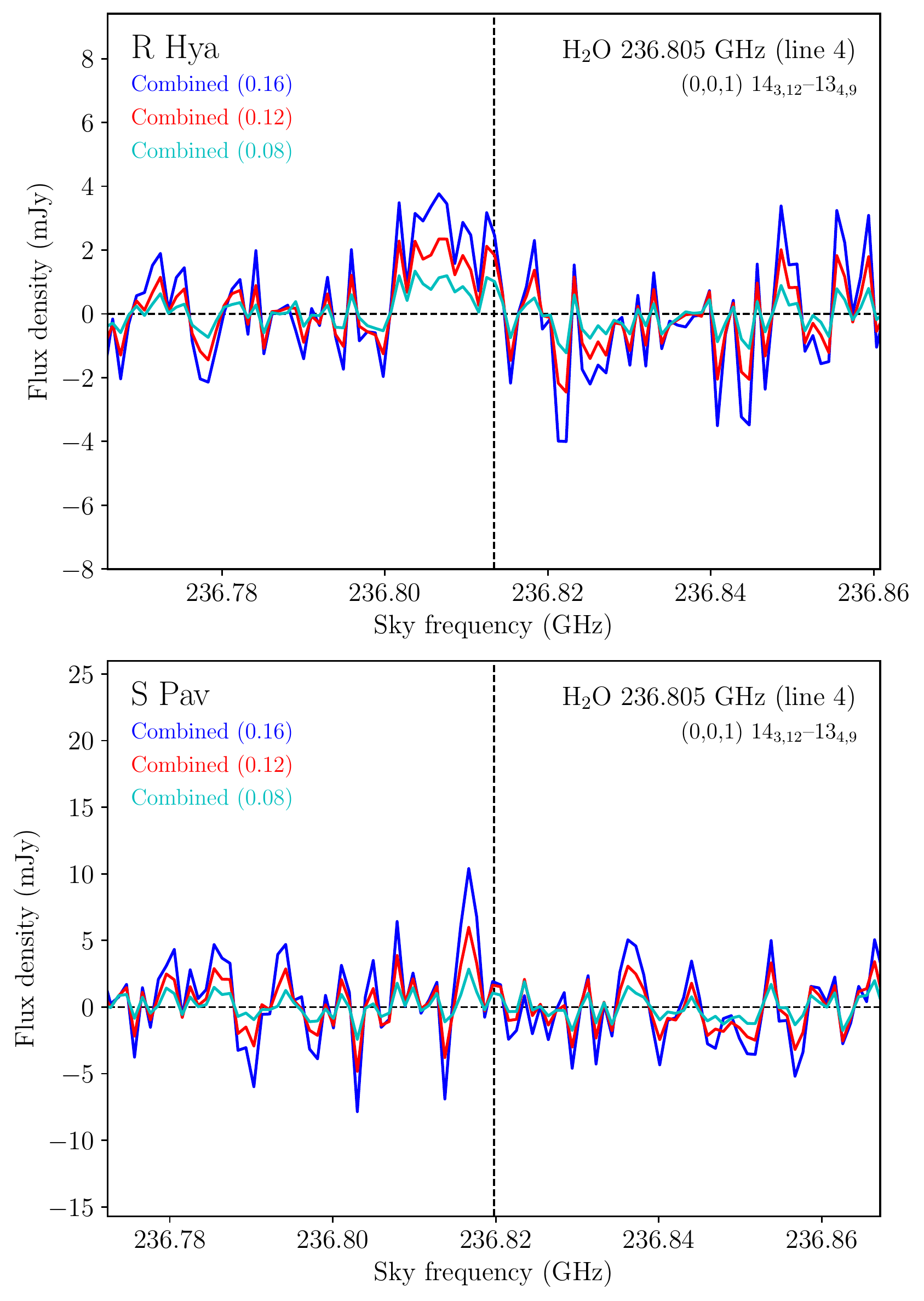}
 
\caption{Spectrum of the (0,0,1)~$14_{3,12}$--$13_{4,9}$ transition  of para H$_2$O in R~Hya (upper panel). Following the JPL catalog,  we assume a rest frequency of 236.8054~GHz (line 4 in Table \ref{H2O-line-list}). The line profiles are extracted from the combined high and mid  resolution data cubes for circular apertures with  0\farcs08,  0\farcs12 and 0\farcs16 diameters. The vertical dashed line marks the water line frequency shifted by the systemic velocity of the star ($\varv$$^{\rm new}_{\rm LSR }$ in Table~\ref{primarysource_list}).  The lower panel shows the same spectrum toward S~Pav. }
\label{rhya_236} 
\end{figure}

Fig.~\ref{vxsgr_ahsco_254} presents the emission line profile toward two supergiants, AH Sco and VX Sgr, in the 254.17 to 254.32 GHz frequency range. There are two major  features in this range, one near 254.217 GHz for the $\varv=0$, $J= 6-5$ transition of $^{30}$SiO, and another one around 254.28~GHz corresponding to the $\varv=0$, $J_{K_a,K_c} = 6_{3,3}-6_{2,4}$ transition of SO$_{2}$, or to a blend of the latter transition with SO$_{2}$, $\varv=0$, $J_{K_a,K_c} = 24_{2,22}-24_{1,23}$. The vertical line at 254.235~GHz marks the (0,1,1) $7_{3,4}$ $-$ (1,1,0) $6_{5,1}$ transition of para H$_2$O as predicted  in \citet{H2O_W2020} (line 9 in Table \ref{H2O-line-list}). 
The weak feature observed near 254.235~GHz for the high resolution data and an aperture diameter of 0\farcs2  falls within $\sim$3~MHz of  the predicted water line 9 in both sources. However, the detection of line 9 is uncertain because the apparent H$_2$O emission could just be part of the complex emission and absorption features in the blueshifted wing of the $\varv=0$,  $^{30}$SiO line profile. We note that the $^{30}$SiO blueshifted line wing is rather steep and that the dip observed near 254.233~GHz in the line profile  is visible in the extended configuration data when the diameter of the extraction aperture is $\la$0\farcs12. This dip could thus be interpreted as the absorption signature of $^{30}$SiO material at large distances along the line of sight to the central star. Such features have been seen in the ALMA data of other evolved stars \citep[e.g.,][]{takigawa2017, decin2018, hoai2021}, especially when the beam size is small compared to the angular size of the stellar disk.  
In addition, the radiative transfer models of  \citet{schoenberg1988} show that a distinct blueshifted emission feature may arise from an optically thick line with extended scattering zones and, depending on the adopted turbulence, an enhanced blue line wing may also be reproduced in the CO line models \citep[e.g.,][]{debeck2012}. Hence, our line profile in the 254.18--254.25~GHz range could just be pure $^{30}$SiO emission that is extended and optically thick. 
However, since both AH Sco and VX Sgr have similar terminal expansion velocities of $\sim$34$-$35~km\,s$^{-1}$ \citep{gottlieb2022}, it is not possible to exclude that the 254.235~GHz feature is due to water by examining these two spectra alone. We note that this feature was not observed in any AGBs of the {\sc atomium} sample. Finally, we also note that the unidentified signal observed in R~Hya and R~Aql at 254.244~MHz (Wallstr\"om et al., in preparation) lies at about 9~MHz from the 254.235~GHz H$_2$O (line~9); this could perhaps be due to the infall of water toward the central star.

 %%%%%%%%%
\begin{figure}
\centering
\includegraphics[width=9.5cm,angle=0]{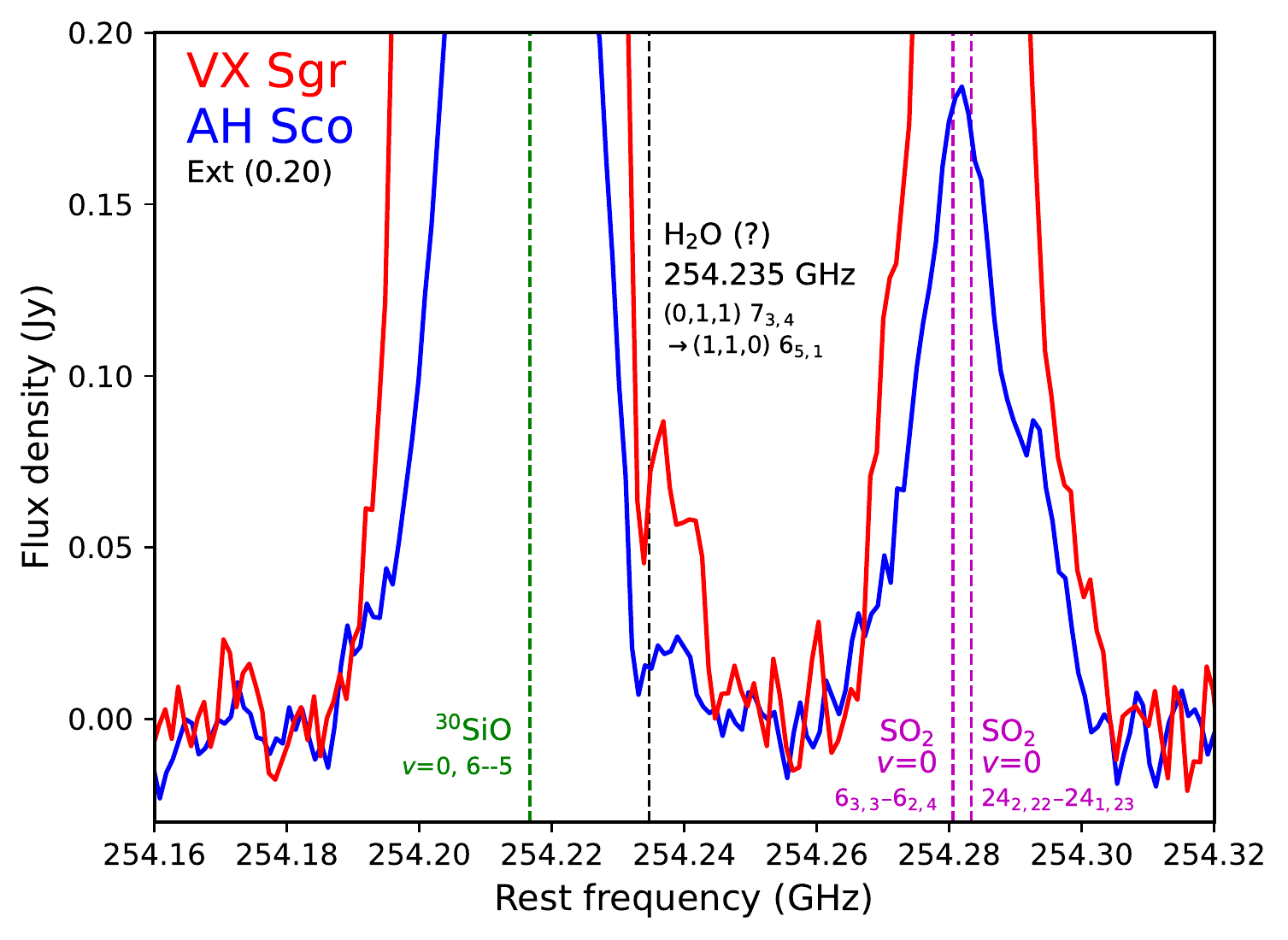}
 
\caption{Emission line profiles toward AH Sco (blue)  and VX Sgr (red) in the vicinity of the $\varv=0$, $J = 6$$-$5 transition of $^{30}$SiO (at 254.217~GHz) and of the (0,1,1)~$7_{3,4}$--$(1,1,0)~6_{5,1}$  transition of para water near  
 254.235~GHz (line 9, Table \ref{H2O-line-list}). The spectra are extracted from the high resolution data for a circular aperture  of 0\farcs2  diameter and plotted  in the rest frequency frame. The dotted black vertical line is the expected water frequency according to \citet{H2O_W2020} (see  Table \ref{H2O-line-list}). Identification  of the water transition at 254.235~GHz is uncertain (see discussion in Appendix~\ref{sec:254.23_vxsgr_ahsco}).
 }
\label{vxsgr_ahsco_254} 
\end{figure}

%%%%%%%%%%%%%%%%%%%%
\section{Water channel maps and zeroth moment maps} 
\label{sec:h2o_channel_maps}

A few channel maps obtained for the extended configuration are shown and discussed  in Sect.~\ref{sec:channmaps_H2O}. 
Additional channel maps are presented here at: 268.149~GHz (line~14)  in S~Pav (Fig.~\ref{chan_spav_h2o_268}), IRC$+$10011 (Fig.~\ref{chan_irc+10011_h2o_268}) and the two RSGs, VX~Sgr and AH~Sco (Figs.~\ref{chan_vxsgr_h2o_268} and \ref{chan_ahsco_h2o_268}); 
262.898~GHz (line~12) in R~Hya and U~Her (Fig.~\ref{chan_rhya_uher_262}), S~Pav and IRC$+$10011 (Fig.~\ref{chan_spav_i+10011_262}) and,  VX~Sgr and AH~Sco (Fig.~\ref{chan_vxsgr_ahsco_262});
259.952~GHz (line~10) in R~Hya and S~Pav (Fig.~\ref{chan_rhya_spav_259}) and, in IRC$+10011$ and VX~Sgr (Fig.~\ref{chan_i10011_vxsgr_259}); 222.014~GHz (line~1) in R~Hya and S~Pav (Fig~\ref{chan_rhya_spav_222}), 
IRC$+$10011 and VXSgr (Fig.~\ref{chan_i10011_vxsgr_222}) and, R~Aql and AH~Sco (Fig.~\ref{chan_raql_ahsco_222}). 

Additional mom~0 maps are also gathered in this Appendix at 222.014~GHz in R~Hya, IRC$+10011$, VX~Sgr and S~Pav (Fig.~\ref{line1mom0_222_h2o}),  and at 254.053~GHz in S~Pav and R~Hya (Fig.~\ref{abso_emission_254-spav_rhya}).

%%%SPav 268 GHz
\begin{figure*}
\centering
\includegraphics[width=16.5cm,angle=0]{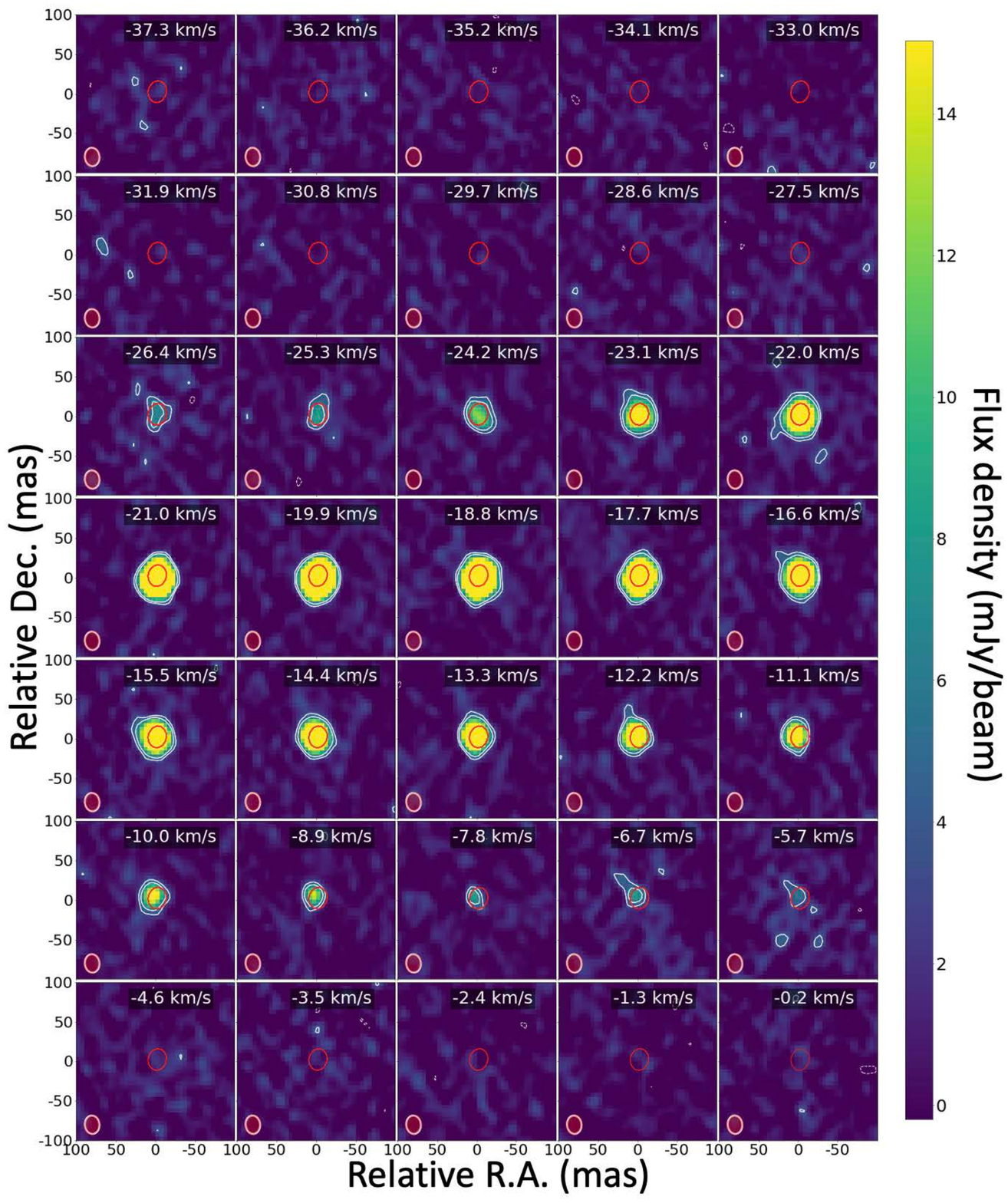}
 
\caption{High resolution channel map of S Pav in the (0,2,0)~$6_5,_2$--$7_4,_3$ rotational transition of water at 268.149 GHz. Caption as in Fig.~\ref{chan_rhya_uher_h2o_268} except for the velocity range and the line peak flux density, 75 mJy/beam; the typical r.m.s. is 1~mJy/beam. The HPBW  is (23$\times$18)~mas at PA~4$^{\circ}$ and (25$\times$20)~mas at PA~$-$13$^{\circ}$ for the line and  continuum, respectively. 
}
\label{chan_spav_h2o_268} 
\end{figure*}
%%%%%%

%%%IRC+10011 268 GHz
\begin{figure*}
\centering
\includegraphics[width=16.5cm,angle=0]{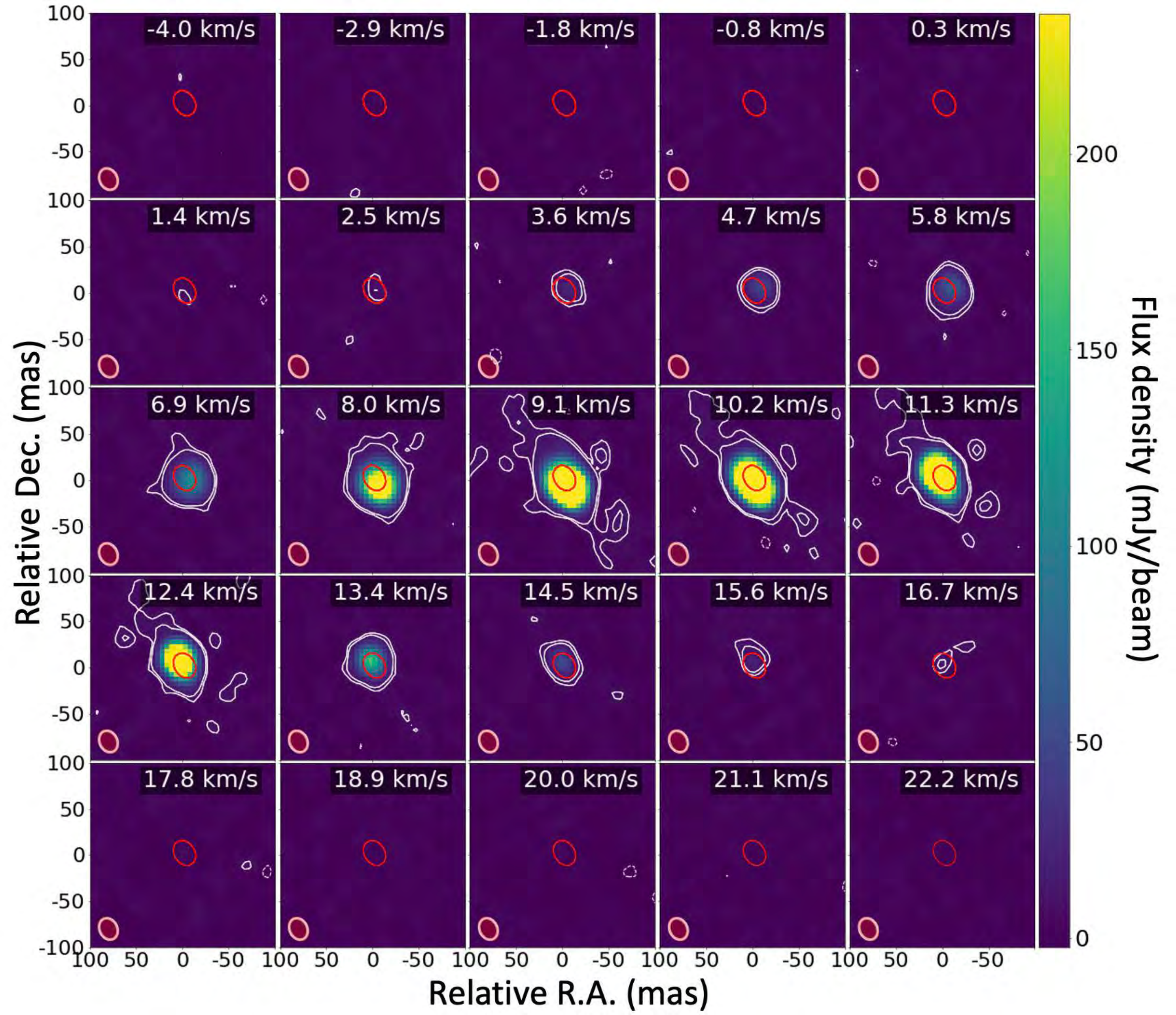}
 
\caption{High resolution channel map of IRC$+10011$ in the (0,2,0)~$6_5,_2$--$7_4,_3$ rotational transition of water at 268.149~GHz. 
Caption as in Fig.~\ref{chan_rhya_uher_h2o_268} except for the velocity range and the line peak flux density, 1178 mJy/beam; the typical r.m.s. is 2~mJy/beam. The HPBW  is (24$\times$19)~mas at PA~26$^{\circ}$ and (27$\times$19)~mas at PA~31$^{\circ}$ for the line and  continuum, respectively. 
}
\label{chan_irc+10011_h2o_268} 
\end{figure*}
%%%%

%%%VX Sgr 268 GHz
\begin{figure*}
\centering
\includegraphics[width=14.0cm,angle=270]{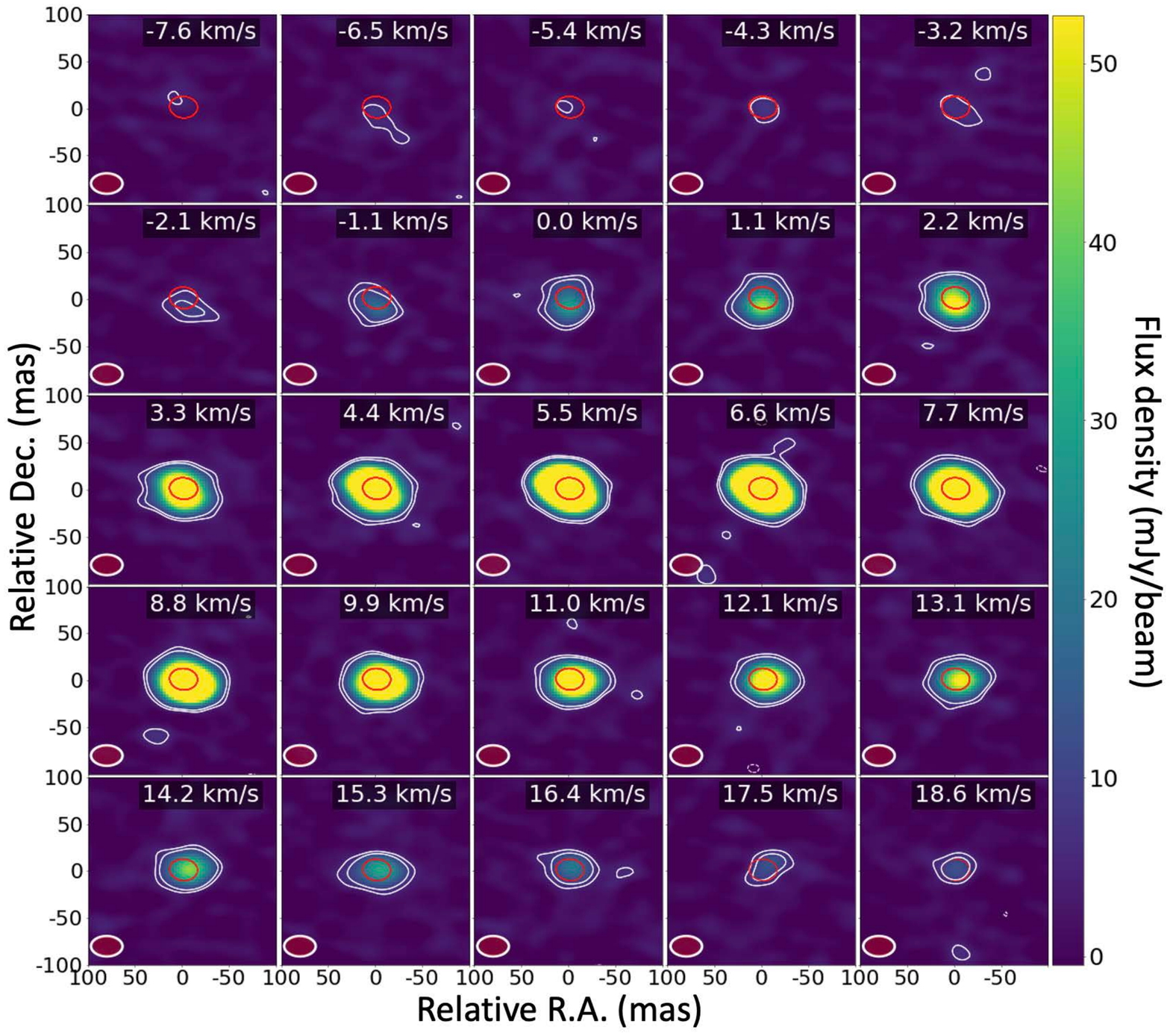}
 
\caption{High resolution channel map of VX~Sgr in the (0,2,0)~$6_5,_2$--$7_4,_3$ rotational transition of water at 268.149~GHz. 
Caption as in Fig.~\ref{chan_rhya_uher_h2o_268} except for the velocity range and the line peak flux density, 263 mJy/beam; the typical r.m.s. is 2~mJy/beam. The HPBW  is (33$\times$21)~mas at PA~89$^{\circ}$ and (28$\times$20)~mas at PA~89$^{\circ}$ for the line and  continuum, respectively. 
}
\label{chan_vxsgr_h2o_268} 
\end{figure*}

%%%

%%%AH Sco 268 GHz
\begin{figure*}
\centering
\includegraphics[width=16.5cm,angle=0]{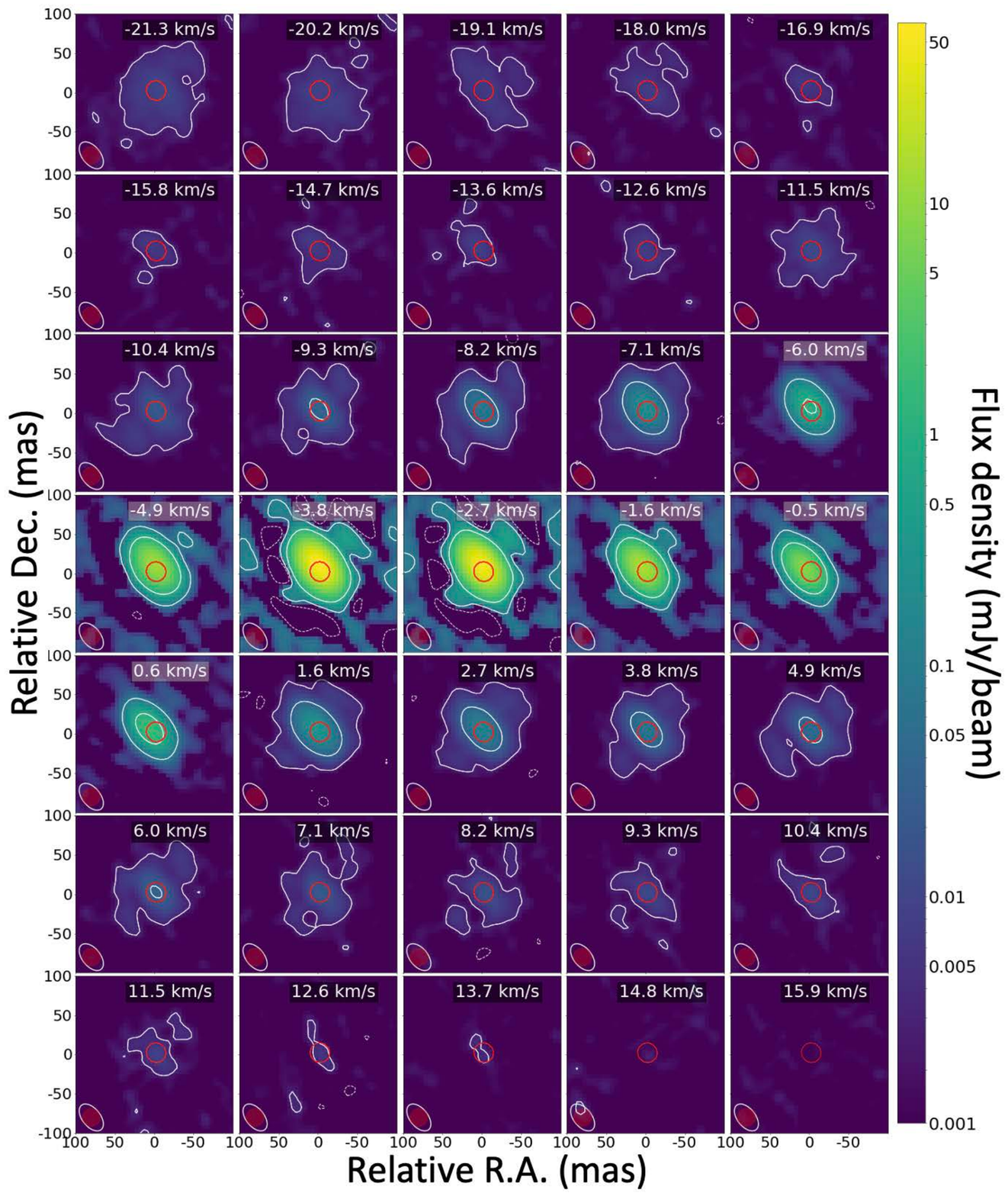}
 
\caption{High resolution channel map of AH~Sco in the (0,2,0)~$6_5,_2$--$7_4,_3$ rotational transition of water at 268.149~GHz. 
Caption as in Fig.~\ref{chan_rhya_uher_h2o_268} except for the velocity range and the line peak flux density, 60344 mJy/beam; the typical r.m.s. is 1~mJy/beam. The HPBW  is (40$\times$23)~mas at PA~38$^{\circ}$ and (23$\times$23)~mas at PA~70$^{\circ}$ for the line and  continuum, respectively. 
}
\label{chan_ahsco_h2o_268} 
\end{figure*}

%%%%%%%%%%%%%
%Figs at 262.898 GHz
\begin{figure*}
\centering
\includegraphics[width=15.5cm,angle=0]{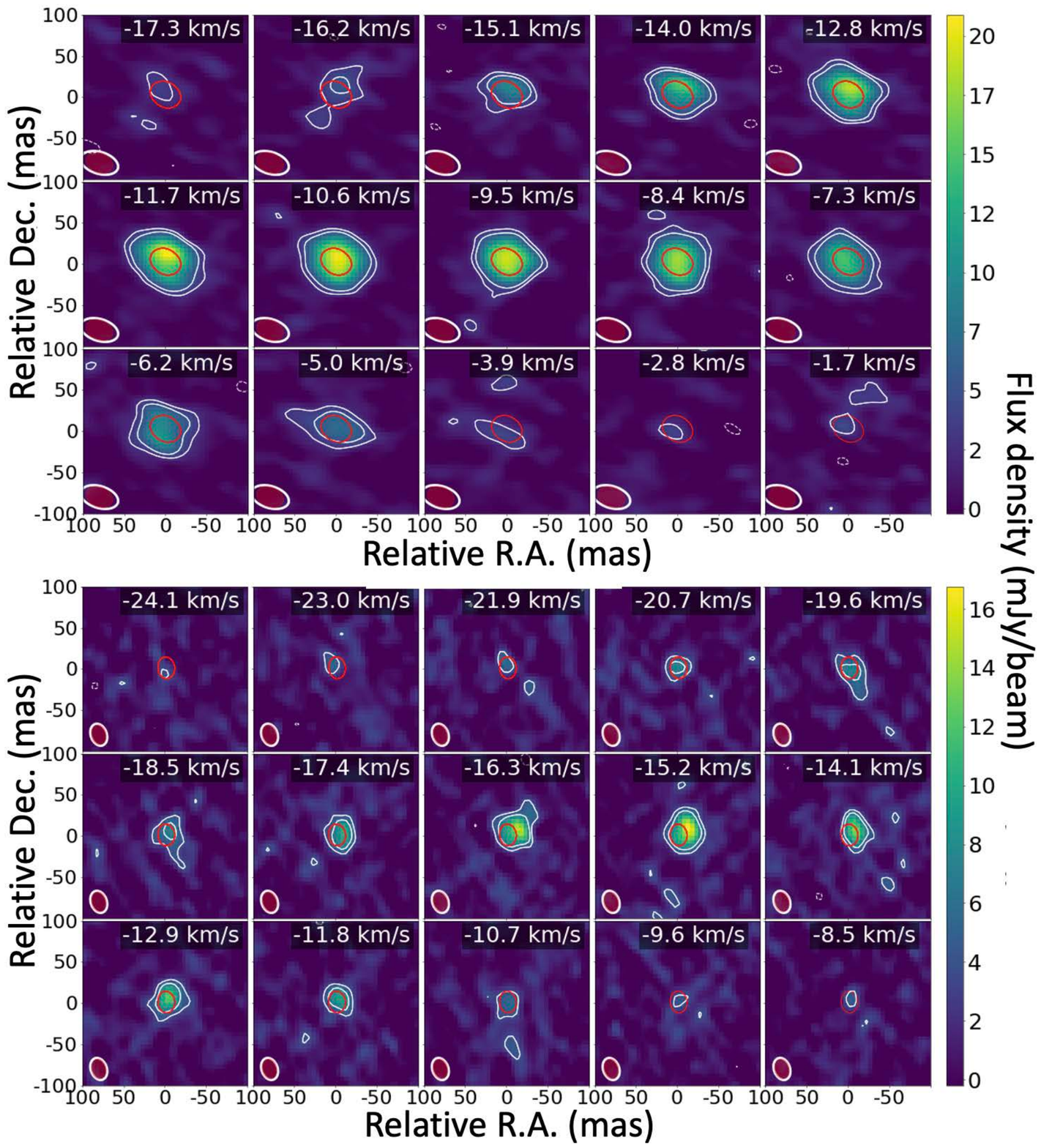}
 
\caption{High resolution channel maps of R~Hya and U~Her in the (0,1,0)~$7_7,_1$--$8_6,_2$ rotational transition of water (line~12) at 262.898~GHz  (upper and lower panels).  Caption as in Fig.~\ref{chan_rhya_uher_h2o_268} except for the line frequency, velocity range and the line peak flux density, 20.9~mJy/beam (R~Hya) and 16.7~mJy/beam (U~Her); the typical r.m.s. is 1~mJy/beam in both stars. 
The line HPBW  is (47$\times$27)~mas at PA~73$^{\circ}$ (R Hya) and (28$\times$20)~mas  at PA~20$^{\circ}$ (U~Her). 
The continuum HPBW is (34$\times$25)~mas at PA~67$^{\circ}$ (R Hya) and (24$\times$18)~mas  at PA~8$^{\circ}$ (U~Her). 
}
\label{chan_rhya_uher_262} 
\end{figure*}

%S Pav and IRC+10011 at 262 GHz
\begin{figure*}
\centering
\includegraphics[width=15.5cm,angle=0]{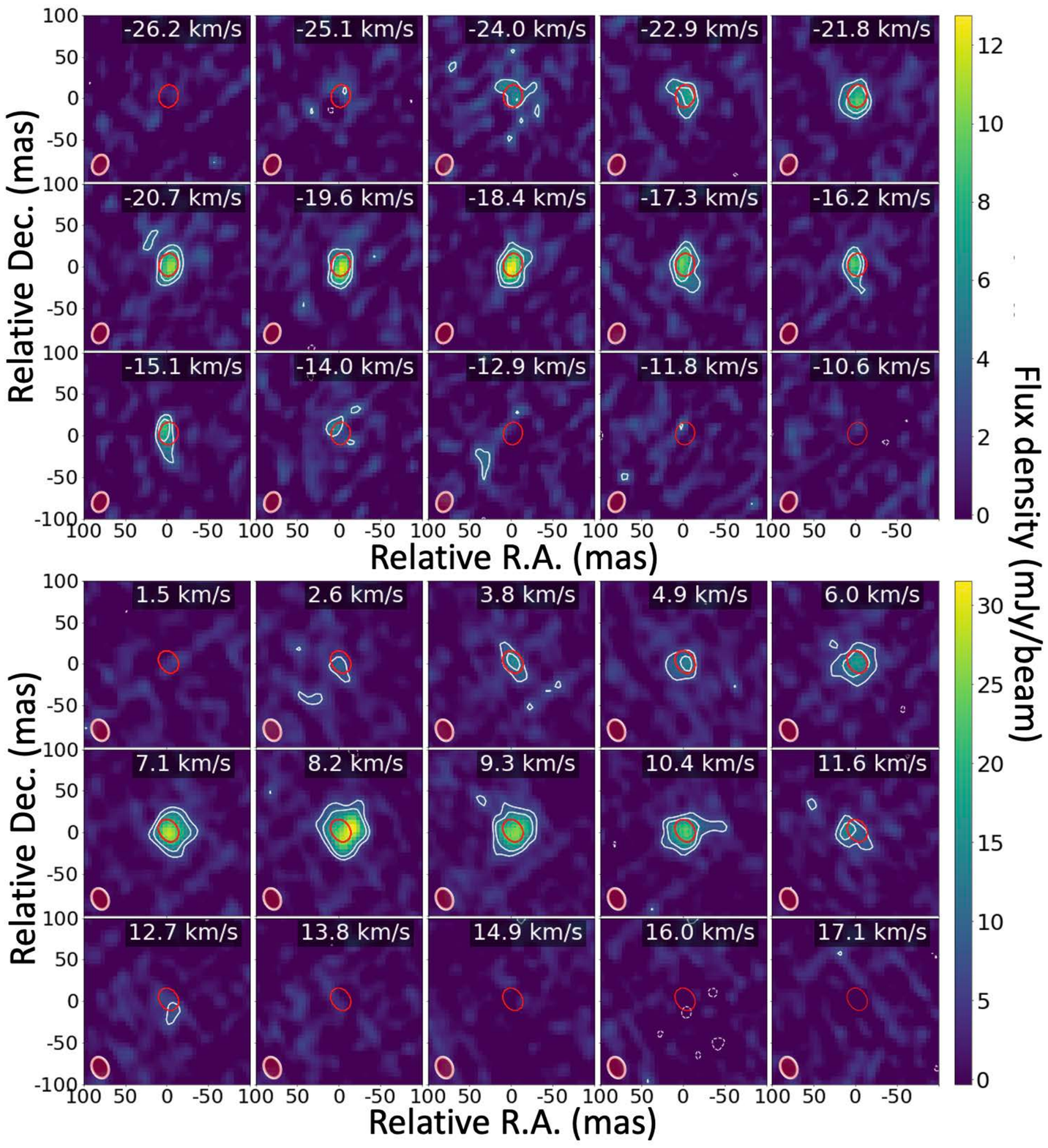}
 
\caption{High resolution channel map of S~Pav and IRC$+$10011  in the (0,1,0)~$7_7,_1$--$8_6,_2$ rotational transition of water (line~12) at 262.898~GHz (upper and lower panels). Caption as in Fig.~\ref{chan_rhya_uher_h2o_268} except for the line frequency, velocity range and the line peak flux density, 12.8~mJy/beam (S~Pav) and 31.5~mJy/beam (IRC$+10011$); the typical r.m.s. is 1 and 2~mJy/beam, respectively. 
The line HPBW  is (24$\times$18)~mas at PA~$-$20$^{\circ}$ (S~Pav) and (26$\times$20)~mas  at PA~22$^{\circ}$ (IRC$+$10011). 
The continuum HPBW is (25$\times$20)~mas at PA~$-$13$^{\circ}$ ( S~Pav) and (27$\times$19)~mas  at PA~31$^{\circ}$ (IRC$+$10011). 
}
\label{chan_spav_i+10011_262} 
\end{figure*}

%%%% VX Sgr, AH Sco at 262 GHz
\begin{figure*}
\centering
\includegraphics[width=15.5cm,angle=0]{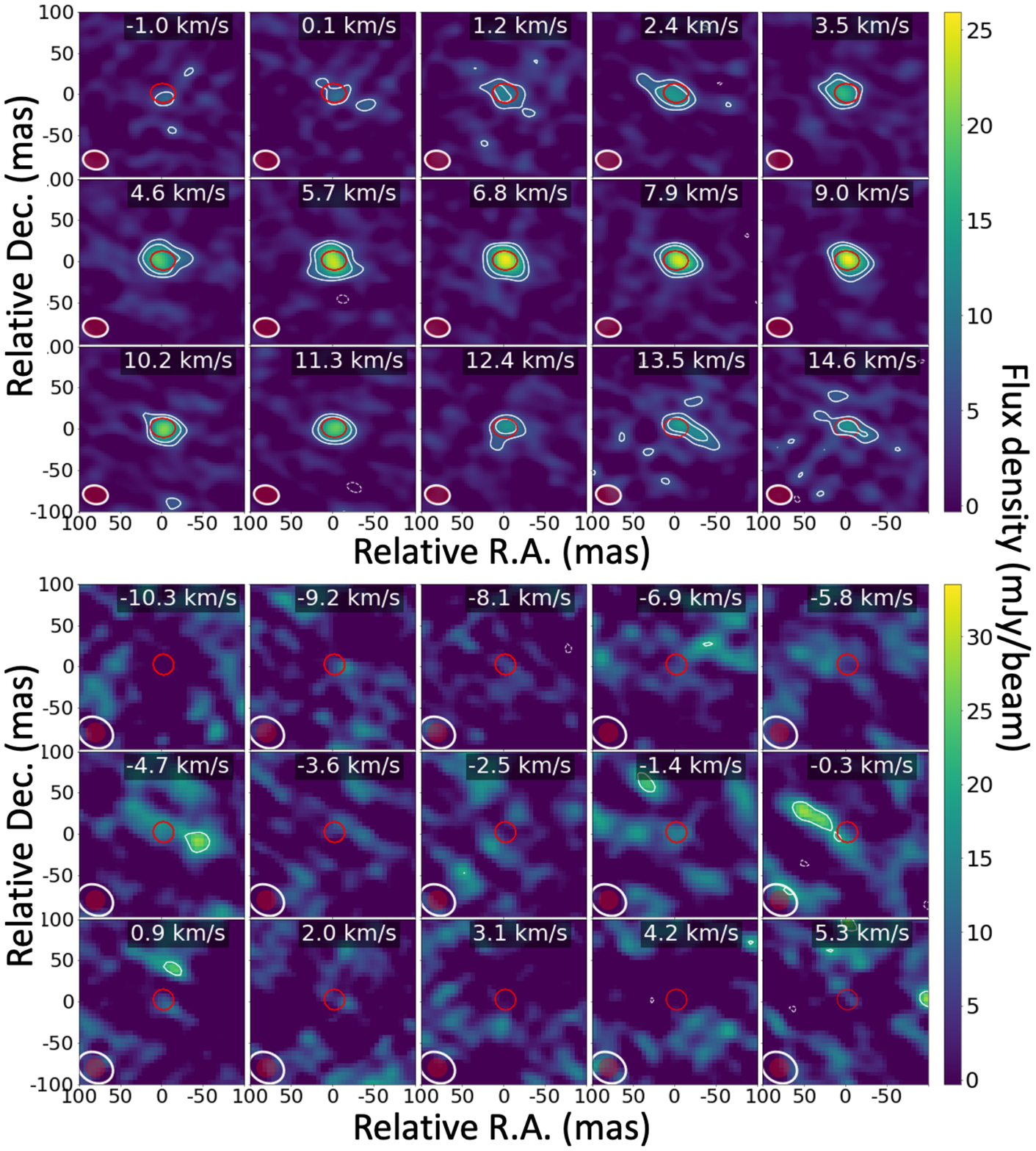}
 
\caption{High resolution channel map of VX~Sgr and AH~Sco in the (0,1,0)~$7_7,_1$--$8_6,_2$ rotational transition of water (line~12) at 262.898~GHz (upper and lower panels)  . Caption as in Fig.~\ref{chan_rhya_uher_h2o_268} except for the line frequency, velocity range and the line peak flux density, 26~mJy/beam (VX~Sgr) and 33.5~mJy/beam (AH~Sco); the typical r.m.s. is 2 and 6.5~mJy/beam, respectively. 
The line HPBW  is (30$\times$22)~mas at PA~80$^{\circ}$ (VX~Sgr) and (45$\times$36)~mas  at PA~59$^{\circ}$ (AH~Sco). 
The continuum HPBW is (28$\times$20)~mas at PA~89$^{\circ}$ ( VX~Sgr) and (23$\times$23)~mas  at PA~70$^{\circ}$ (AH~Sco). 
}
\label{chan_vxsgr_ahsco_262} 
\end{figure*}

%%%%%%%%
\begin{figure*}
\centering
\includegraphics[width=13.0cm,angle=270]{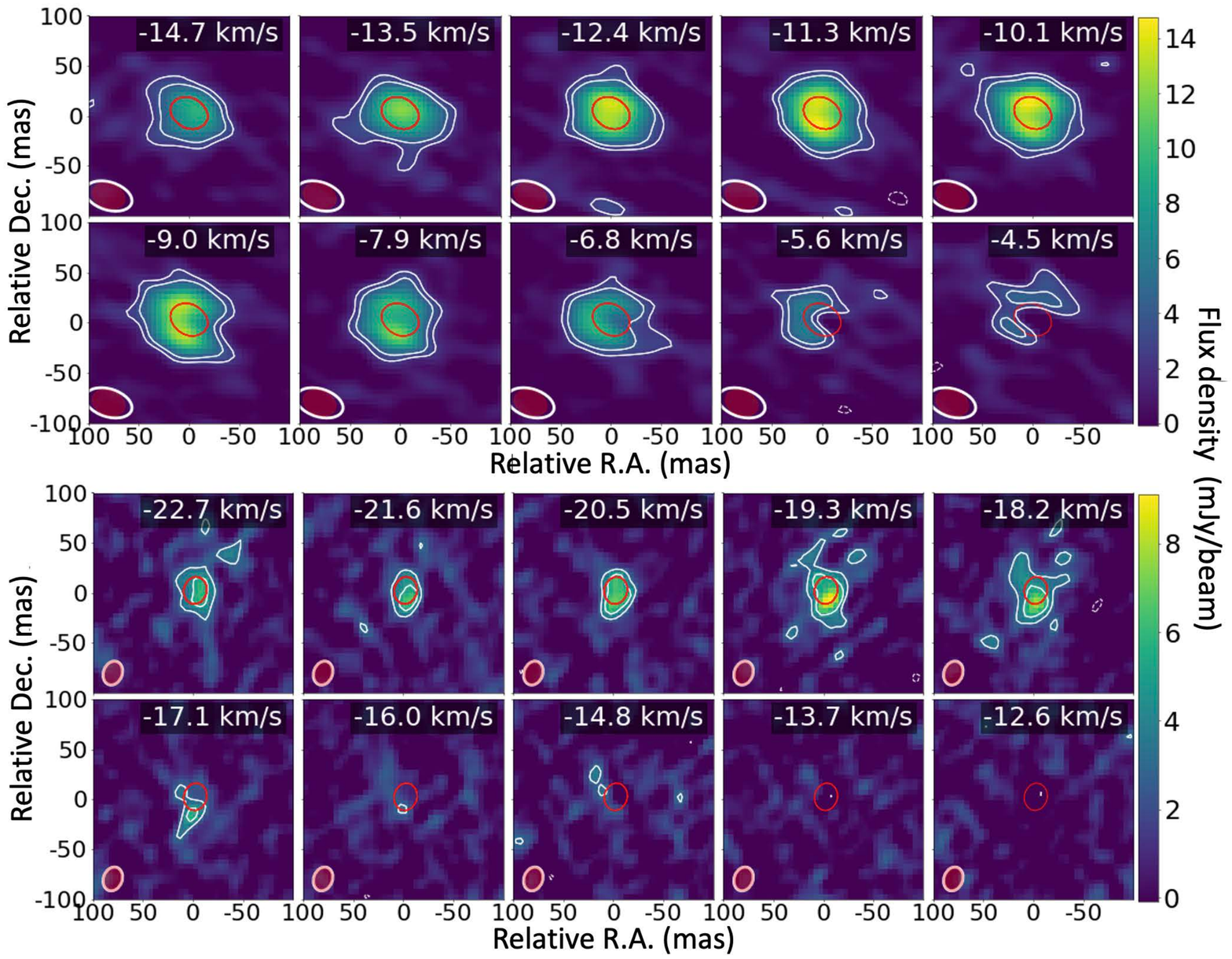}

\caption{High resolution channel maps of relatively low energy (3954 K or 2748 cm$^{-1}$) (0,0,0)~$13_{6,8}$--$14_{3,11}$ transition of water at 259.952~GHz in R~Hya and S~Pav (upper and lower panels). 
Caption as in Fig.~\ref{chan_rhya_uher_h2o_268} except for the velocity range and the line peak flux density,  15 and 9~mJy/beam in R~Hya and S~Pav, respectively; the typical r.m.s. noise is 1 mJy/beam for both stars. The HPBW  is (48$\times$28)~mas at PA~72$^{\circ}$ and (34$\times$25)~mas at PA~67$^{\circ}$ for the line and  continuum in R~Hya, respectively, 
and (24$\times$19~mas) at PA~$-$19$^{\circ}$ (line)  and (25$\times$20)~mas at PA~$-$13$^{\circ}$ (continuum) in S~ Pav. 
}
\label{chan_rhya_spav_259}

\end{figure*}

%%%%%%%%%%%%%%

\begin{figure*}
\centering
\includegraphics[width=13.0cm,angle=270]{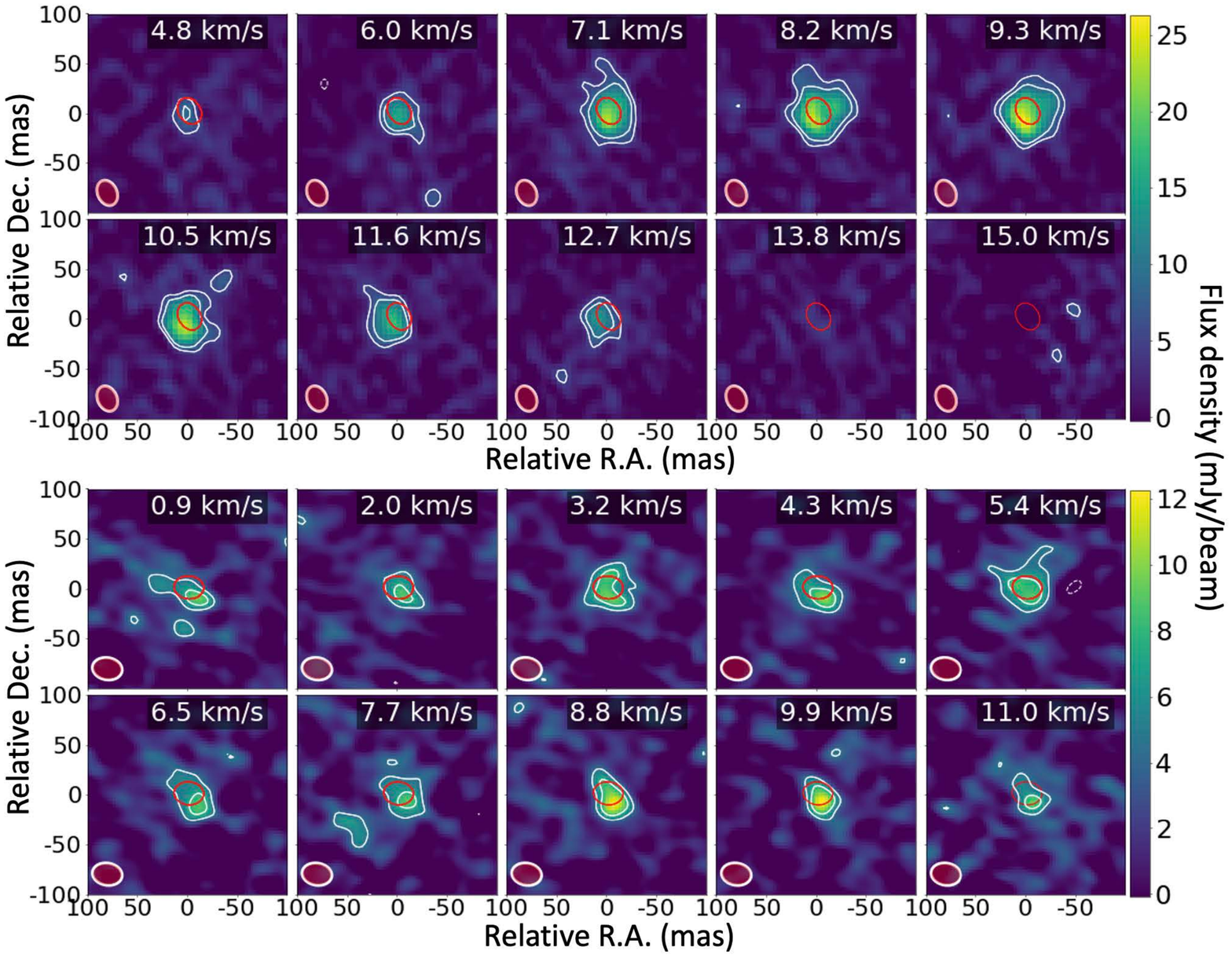}
 
\caption{High resolution channel maps of water at 259.952~GHz in IRC$+$10011 and VX Sgr (upper and lower panels). 
Caption as in Fig.~\ref{chan_rhya_uher_h2o_268} except for the velocity range and the line peak flux density,  26.5 and 12.2~mJy/beam in
 IRC$+$10011 and VX Sgr, respectively; the typical r.m.s. noise is 1.7~mJy/beam (IRC$+10011$) and 1.5~mJy/beam (VX~Sgr). The HPBW  is (26$\times$19)~mas at PA~24$^{\circ}$ and (27$\times$19)~mas at PA~31$^{\circ}$ for the line and  continuum in IRC$+$10011, respectively, 
and (30$\times$22)~mas at PA~79$^{\circ}$ (line)  and (28$\times$20)~mas at PA =~89$^{\circ}$ (continuum) in VX~Sgr. 
}
\label{chan_i10011_vxsgr_259} 
\end{figure*}

%%%%%%%%

\begin{figure*}
\centering
\includegraphics[width=15.5cm,angle=0]{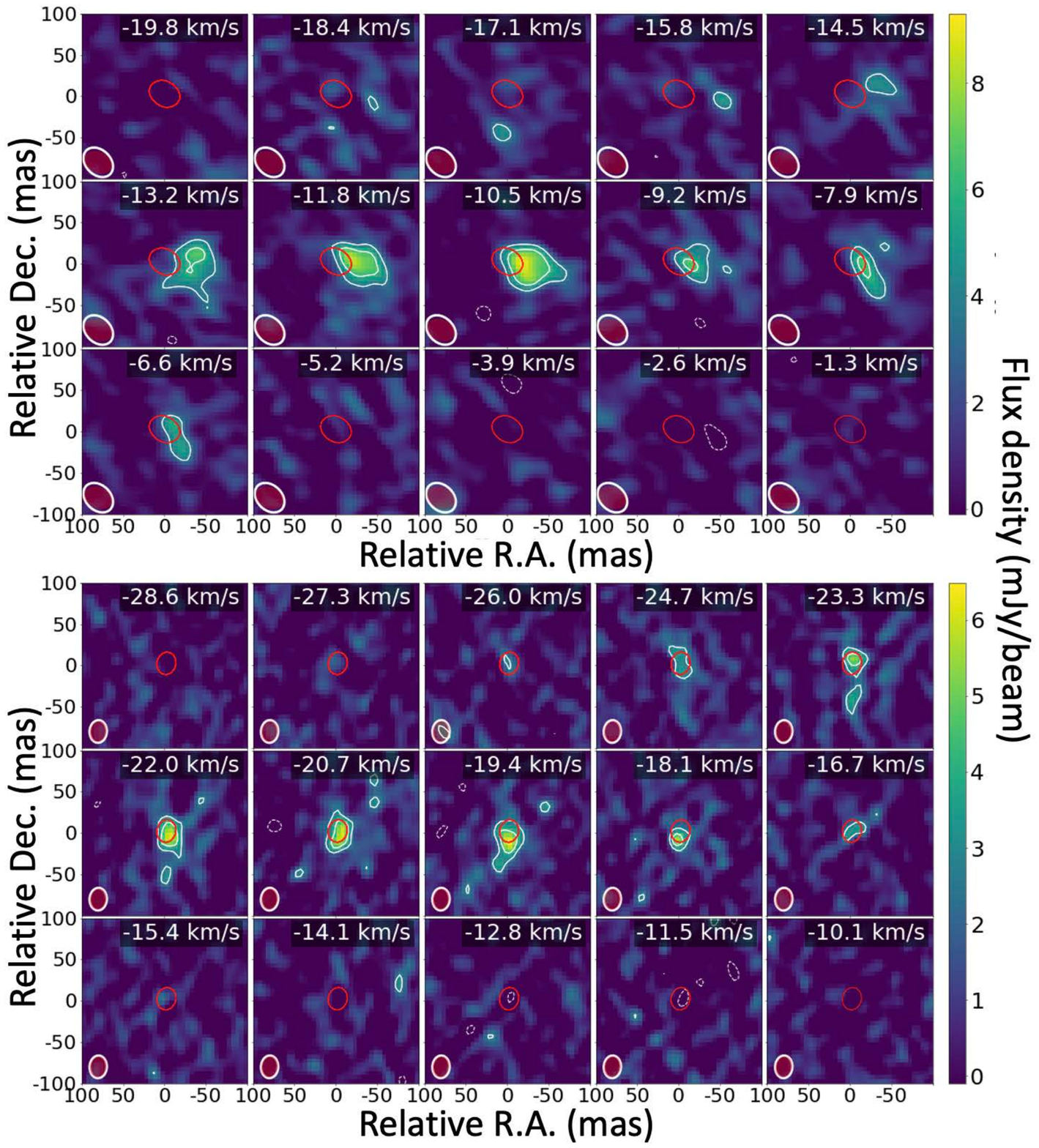}

\caption{High resolution channel maps of  high energy (8331 K)  (0,3,0)~$8_{3,6}$--$7_{4,3}$ transition of H$_2$O at 222.014~GHz in R~Hya and S~Pav (upper and lower panels). 
Caption as in Fig.~\ref{chan_rhya_uher_h2o_268} except for the velocity range and the line peak flux density,  9.3 and 6.5~mJy/beam in R~Hya and S~Pav, respectively; the typical r.m.s. noise is $\sim$1 mJy/beam in both stars. The LSR velocity in each panel is determined from our own rest frequency, 222017.31~MHz, whose uncertainty is low compared to that in W2020 (see Table~\ref{H2O-line-list}). The HPBW  is (41$\times$30)~mas at PA =~45$^{\circ}$ and (34$\times$25)~mas at PA =~67$^{\circ}$ for the line and  continuum  in R~Hya, respectively, 
and (27$\times$21)~mas at PA~$-$3$^{\circ}$ (line)  and (25$\times$20)~mas at PA~$-$13$^{\circ}$ (continuum) in S~ Pav. 
}
\label{chan_rhya_spav_222}

\end{figure*}

%%%%%%%%%%%%%%%%

\begin{figure*}
\centering
\includegraphics[width=15.5cm,angle=0]{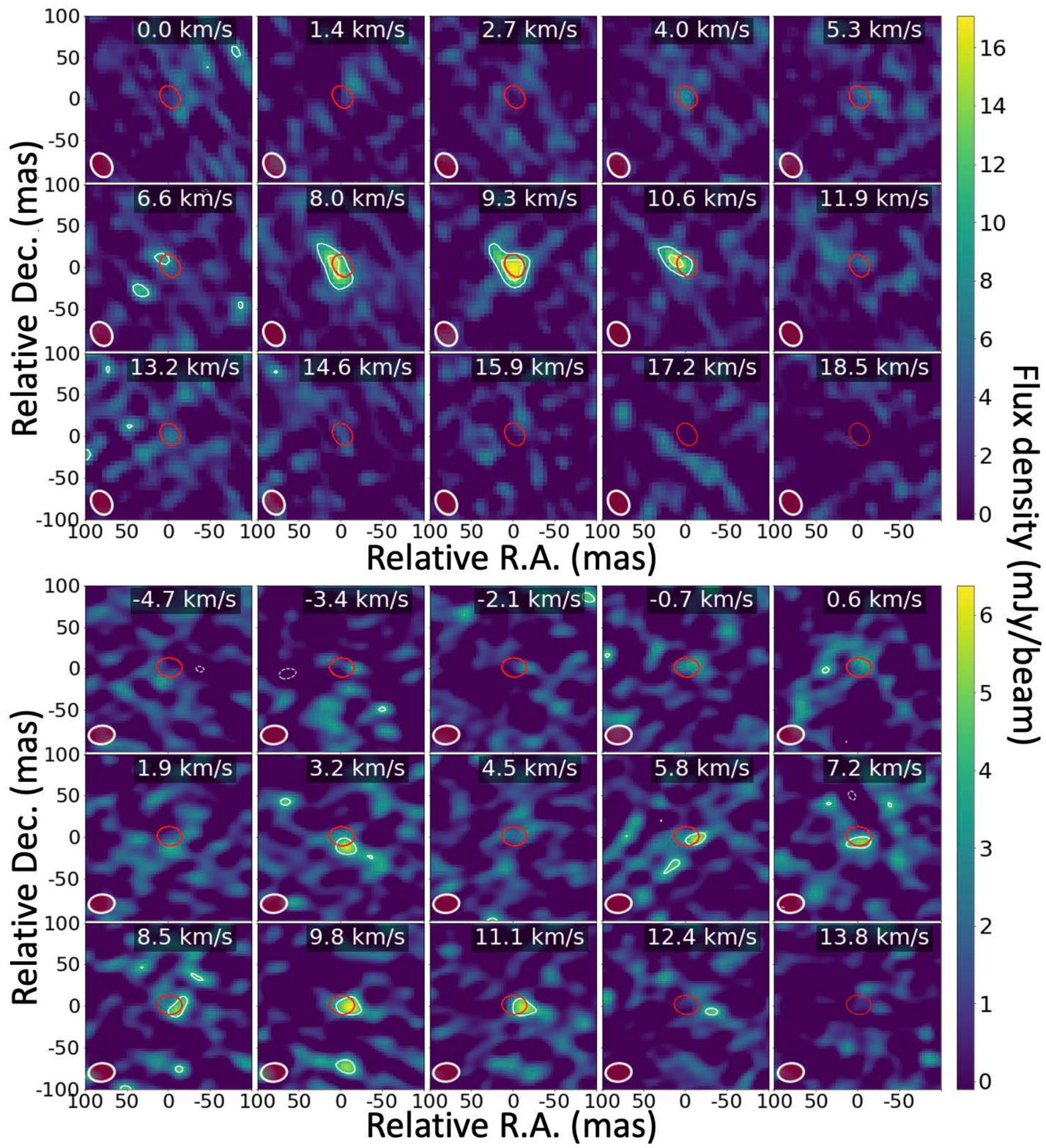}
 
\caption{High resolution channel maps of water at 222.014~GHz in IRC$+$10011 and VX~Sgr (upper and lower panels). Caption as in Fig.~\ref{chan_rhya_spav_222}  
except for the velocity range and the line peak flux density, 17.1 and 6.4 mJy/beam in IRC$+$10011 and VX~Sgr, respectively; the typical r.m.s. noise is 3~mJy/beam (IRC$+$10011) and 1.5~mJy/beam (VX~Sgr). The HPBW  is (30$\times$24)~mas at PA~30$^{\circ}$ and (27$\times$19)~mas at PA =~31$^{\circ}$ for the line and  continuum  in IRC$+$10011, respectively,  
and (31$\times$22)~mas at PA~$-$84$^{\circ}$ (line)  and (28$\times$20)~mas at PA~89$^{\circ}$ (continuum) in VX~ Sgr. 
}
\label{chan_i10011_vxsgr_222} 
\end{figure*}

%%%%%%%%%%
%%% 222 GHz spectra, R Aql and AH Sco
%%%%%%%%%%

\begin{figure*}
\centering
\includegraphics[width=15.5cm,angle=0]{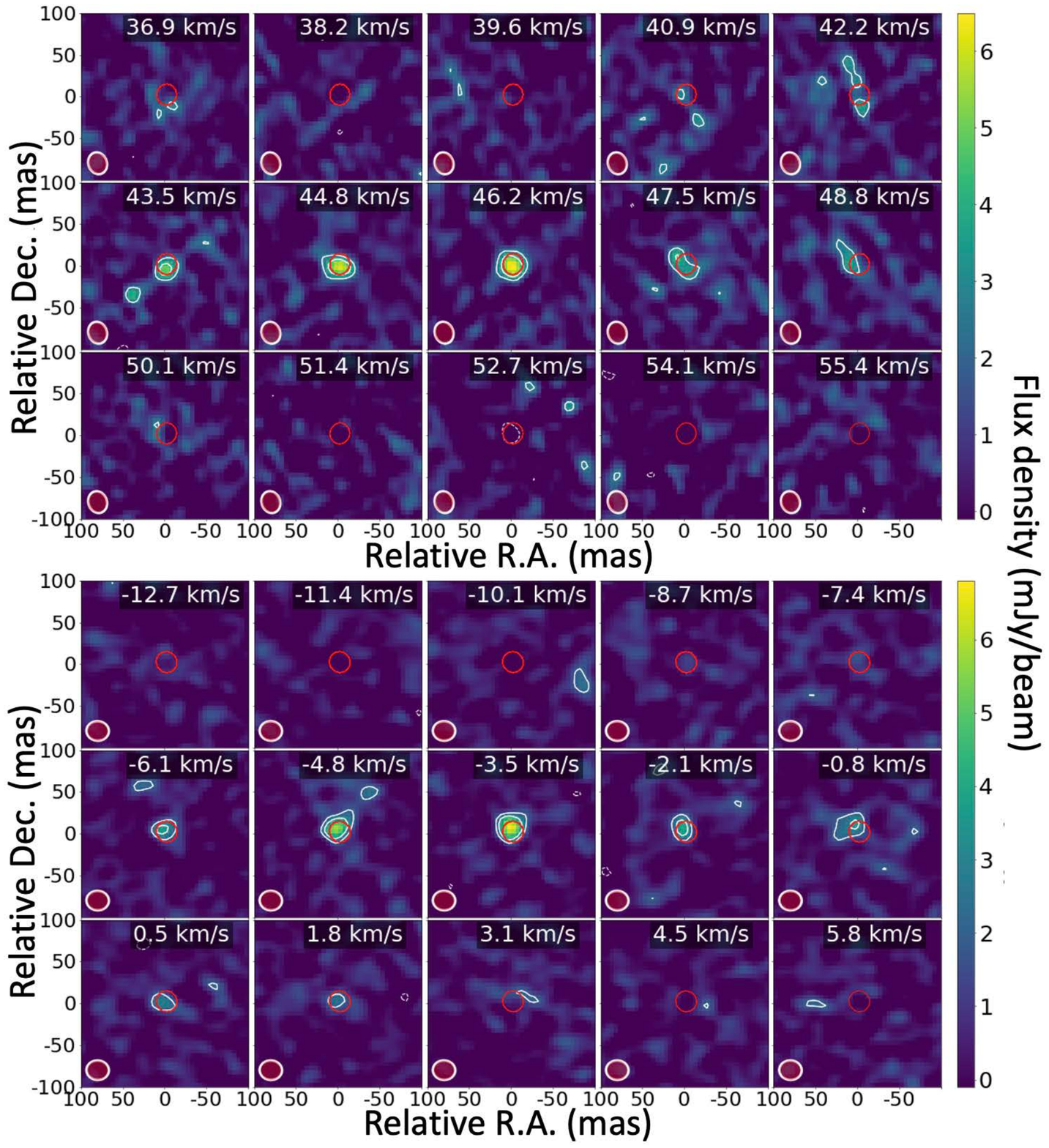}
 
\caption{High resolution channel maps of water at 222.014~GHz in R~Aql and AH~Sco (upper and lower panels). Caption as in Fig.~\ref{chan_rhya_spav_222} except for the velocity range and the line peak flux density, 6.5 and 6.8 mJy/beam in R~Aql and AH~Sco, respectively; the typical r.m.s. noise is 1~mJy/beam (R~Aql) and 0.6~mJy/beam (AH~Sco). 
The HPBW  is (27$\times$22)~mas at PA~18$^{\circ}$ and (24$\times$22)~mas at PA =~13$^{\circ}$ for the line and  continuum in R~Aql, and (27$\times$23)~mas at PA~86$^{\circ}$ (line)  and (23$\times$23)~mas at PA~70$^{\circ}$ (continuum) in AH~Sco. 
}
\label{chan_raql_ahsco_222} 
\end{figure*}

%%%%%%%%%%%

\begin{figure*}
\centering 
\includegraphics[width=13.5 cm,angle=270]{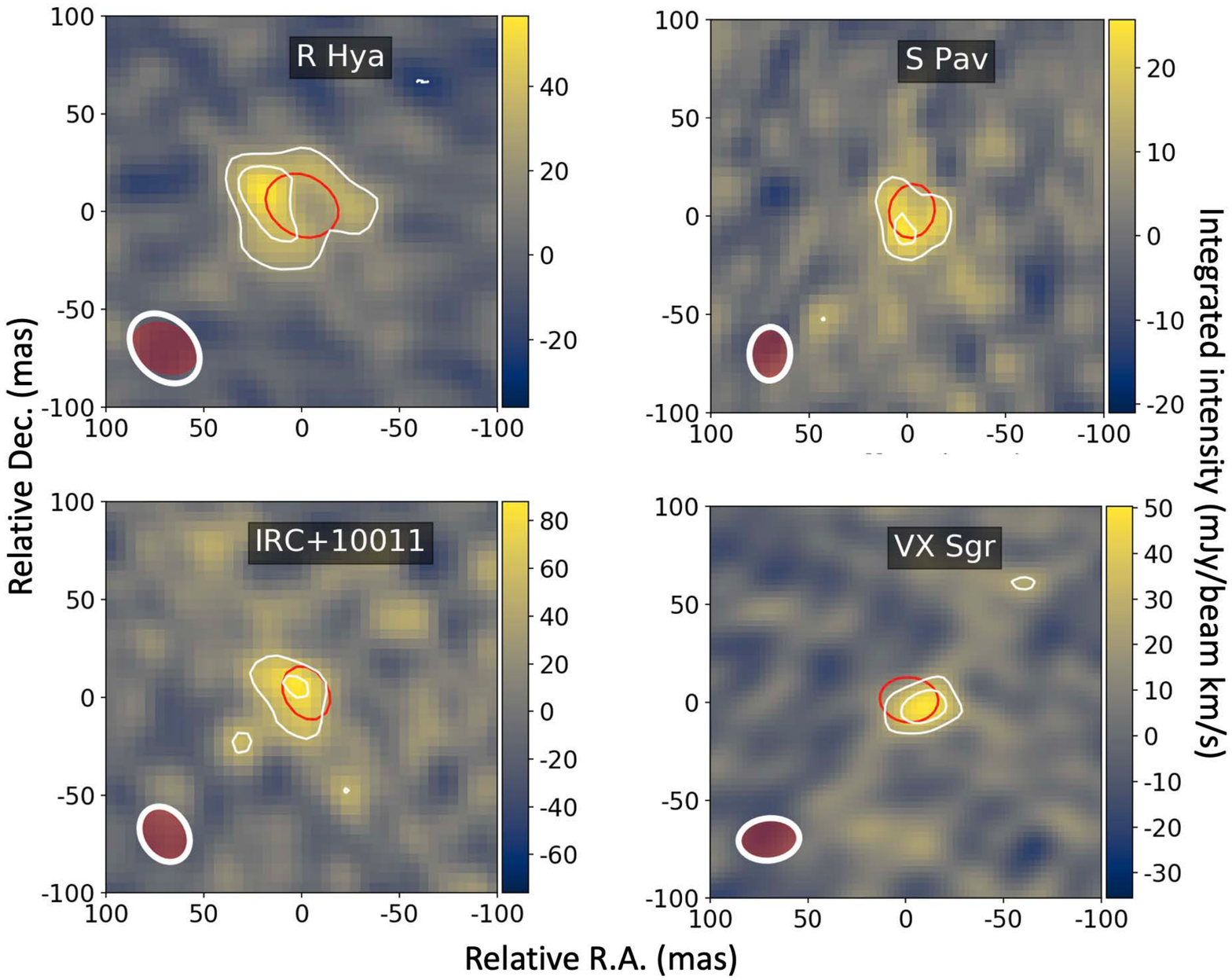}

\caption{Zeroth moment emission maps of the (0,3,0)~$8_{3,6}$--$7_{4,3}$ transition of ortho H$_2$O at 222.014~GHz toward R Hya, IRC$+$10011, VX Sgr and S Pav for the extended array configuration. The map field of view is 100$\times$100 mas for all sources. The white contours are for 3 and 5$\sigma$  emission. The red contour at the map center delineates the extent at half peak intensity of the continuum emission. The noise level is 7.4, 16.0, 7.8 and 4.5 mJy/beam.km/s for R~Hya, IRC$+10011$, VX~Sgr and S~Pav, respectively. The line HPBW (white ellipse) is, (41$\times$30)~mas at PA~45$^{\circ}$, (30$\times$24)~mas at PA~30$^{\circ}$, (31$\times$22)~mas at PA~$-$84$^{\circ}$ and (27$\times$21)~mas at PA~$-$3$^{\circ}$ in R~Hya, IRC$+$10011, VX~Sgr and S~Pav, respectively. The associated 
continuum HPBW (dark-red ellipse) is 
(34$\times$25)~mas at PA~67$^{\circ}$, 
(27$\times$19)~mas at PA~31$^{\circ}$, 
(28$\times$20)~mas at PA~89$^{\circ}$ 
and (25$\times$20)~mas at PA~$-$13$^{\circ}$. 
(Maps are integrated over $-$19.8 to $-$5.2, 1.4 to 15.9, $-$0.7 to 13.8 and $-$26.0 to $-$14.1~km\,s$^{-1}$ for R Hya, IRC$+$10011, VX~Sgr and S~Pav, respectively.) 
}
\label{line1mom0_222_h2o}
\end{figure*}

%%%%%%%%%%%%%%%%%%

\begin{figure*}
\centering \includegraphics[width=12.5cm,angle=270]{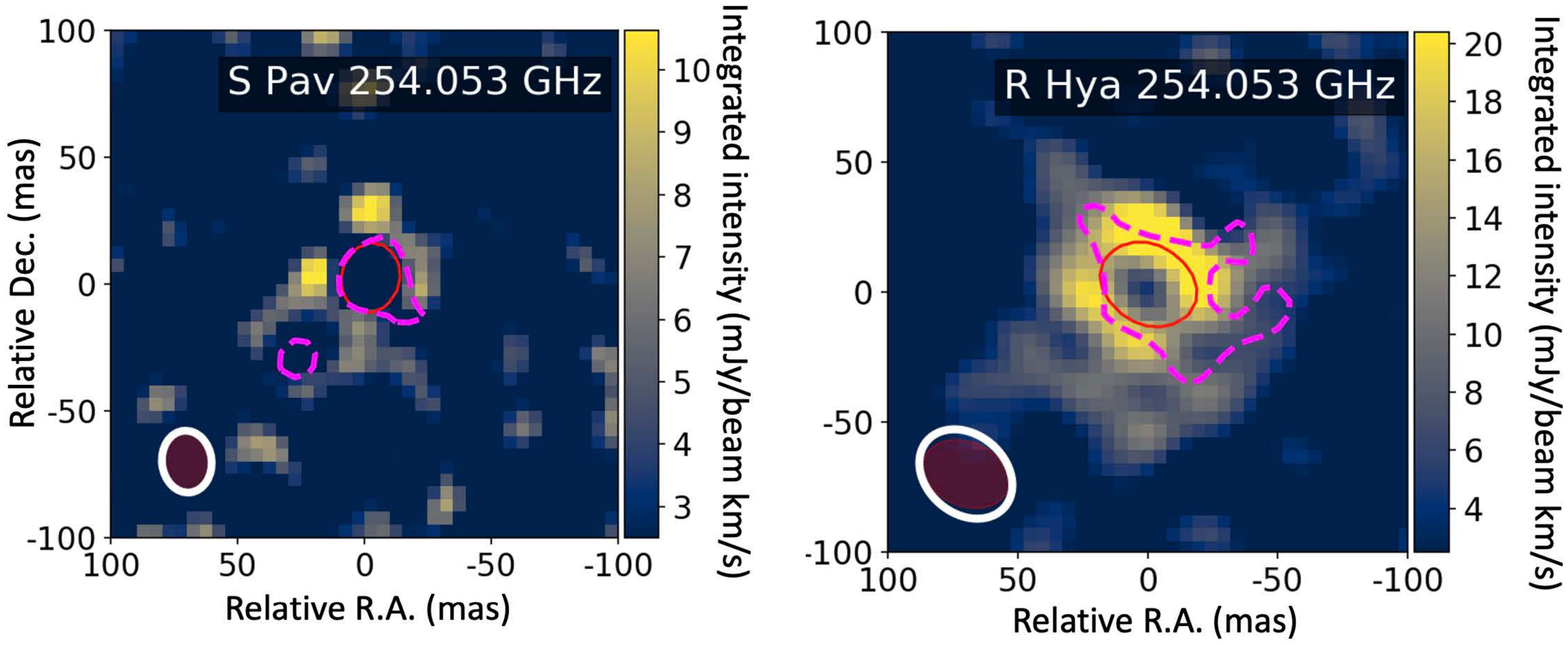}
\caption{Comparison of the mom~0 absorption and emission of water at  254.053~GHz (line~8 in Table~\ref{H2O-line-list}). 
 $\it Left$ $panel$:  Absorption in S~Pav is delimited by the dotted $-5\sigma$ contour while the  ring-like emission is shown in yellow.  The noise level is 3 mJy/beam~km\,s$^{-1}$. 
 $\it Right$ $panel$: As for left panel but  in R~Hya. The noise level is 2.5 mJy/beam~km\,s$^{-1}$  
 In both panels the red contour at the map center delineates the extent at half peak intensity of the continuum emission. 
 The HPBW  is (24$\times$19)~mas at PA~6$^{\circ}$  and (25$\times$20)~mas at PA~$-$13$^{\circ}$) for the line (white ellipse) and continuum (dark-red ellipse) in S~Pav, and 
(39$\times$30)~mas at PA~49$^{\circ}$ (line) and (34$\times$25)~mas at PA~67$^{\circ}$ (continuum) in R~Hya. 
(The velocity intervals are:  $-2.6$ to 6.5~km\,s$^{-1}$ and $-9.3$ to $-3.9$~km\,s$^{-1}$ for S~Pav absorption  and emission; $-9.2$ to $-2.6$~km\,s$^{-1}$ and $-19.1$ to $-10.5$~km\,s$^{-1}$ for R~Hya  absorption and emission.)
 }
\label{abso_emission_254-spav_rhya}
\end{figure*}

%%%%%%%%%%%

\section{OH channel maps and zeroth moment maps} 
\label{sec:oh_channel_maps}

Additional OH channel maps  obtained from the extended resolution data cubes are presented here for the $J = 27/2 $ rotational level  (Figs.~\ref{rhya_channmap_OH_27} and \ref{raql_channmap_OH_27} in R~Hya and R~Aql) and the 29/2 level (Fig.~\ref{spav_channmap_OH_29} in S~Pav).

Fig.~\ref{Fig_OH_emission_29_2_raql_vxsgr} shows the $J=29/2$  zeroth moment maps for R~Aql and VX~Sgr and, zeroth moment maps in the same $J = 29/2 $ rotational level are presented in the main text for R~Hya and S~Pav (Fig.\ref{Fig_OH_emission_29_2}). 
        OH zeroth moment maps in the $J =27/2$ level  are shown in Fig. \ref {Fig_OH_emission_27_2} for R Hya, S Pav, R Aql and VX Sgr.

%%%%%%%%%%%%%%%
\begin{figure*}
\centering
\vspace{-1.5cm}
\includegraphics[width=16.3cm,angle=0]{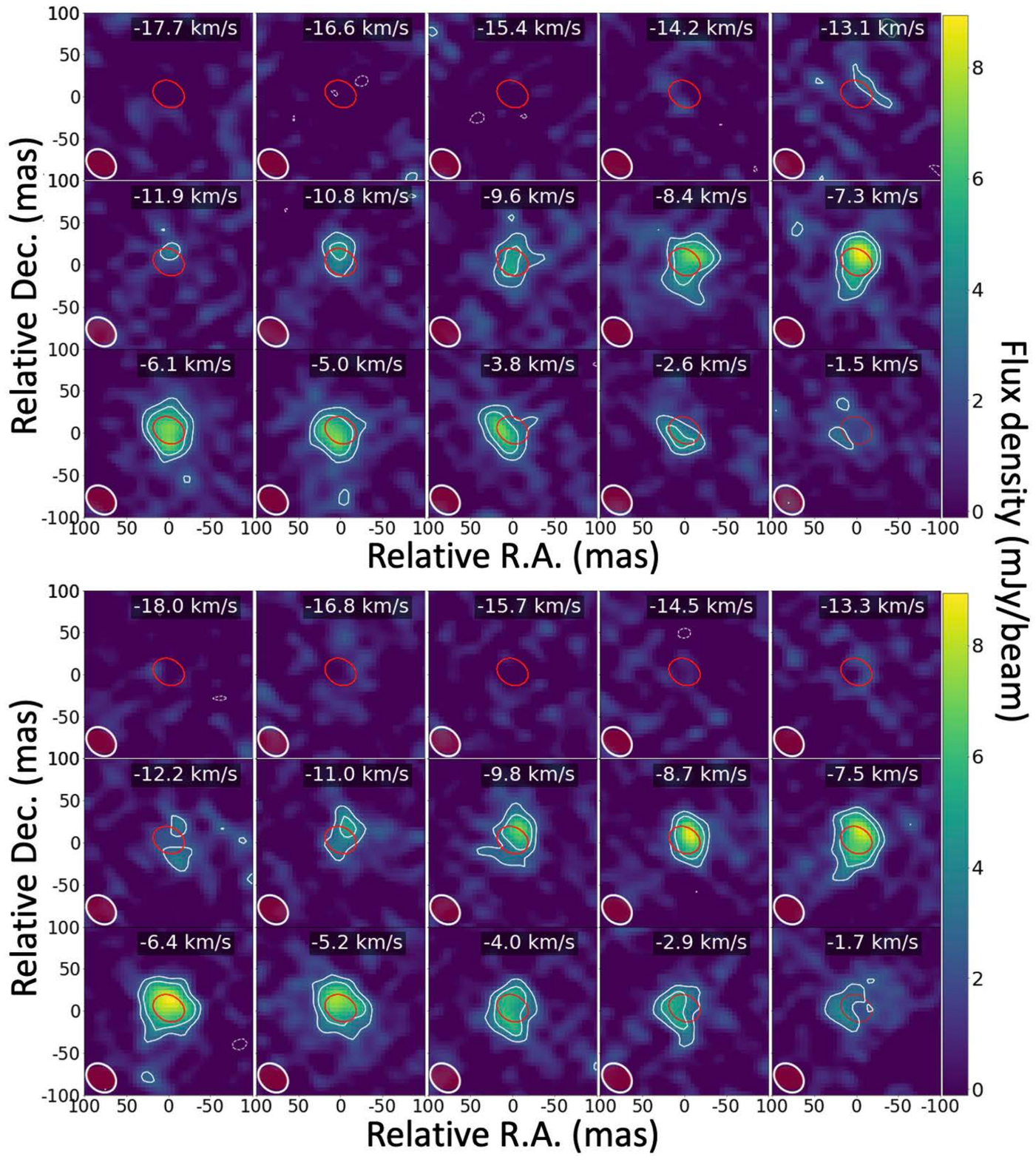}
 
\caption{High resolution channel map of $J$ = 27/2, $F'-F'' = 13-13$ and 14$-$14 transitions of OH in R Hya (upper and lower panels, respectively). Figure caption as in Fig. \ref {rhya_channmap_OH_29} except for the velocity range. 
The line peak flux density and the r.m.s. noise level are identical to those in Fig. \ref{rhya_channmap_OH_29}. 
The HPBW is (41$\times$30)~mas at PA~45$^{\circ}$ and (34$\times$25)~mas at PA~67$^{\circ}$ for line and continuum, respectively. 
}
\label{rhya_channmap_OH_27}
\end{figure*}

%%%%%% 

\begin{figure*}
\centering
\vspace{-1.5cm}
\includegraphics[width=16.3cm,angle=0]{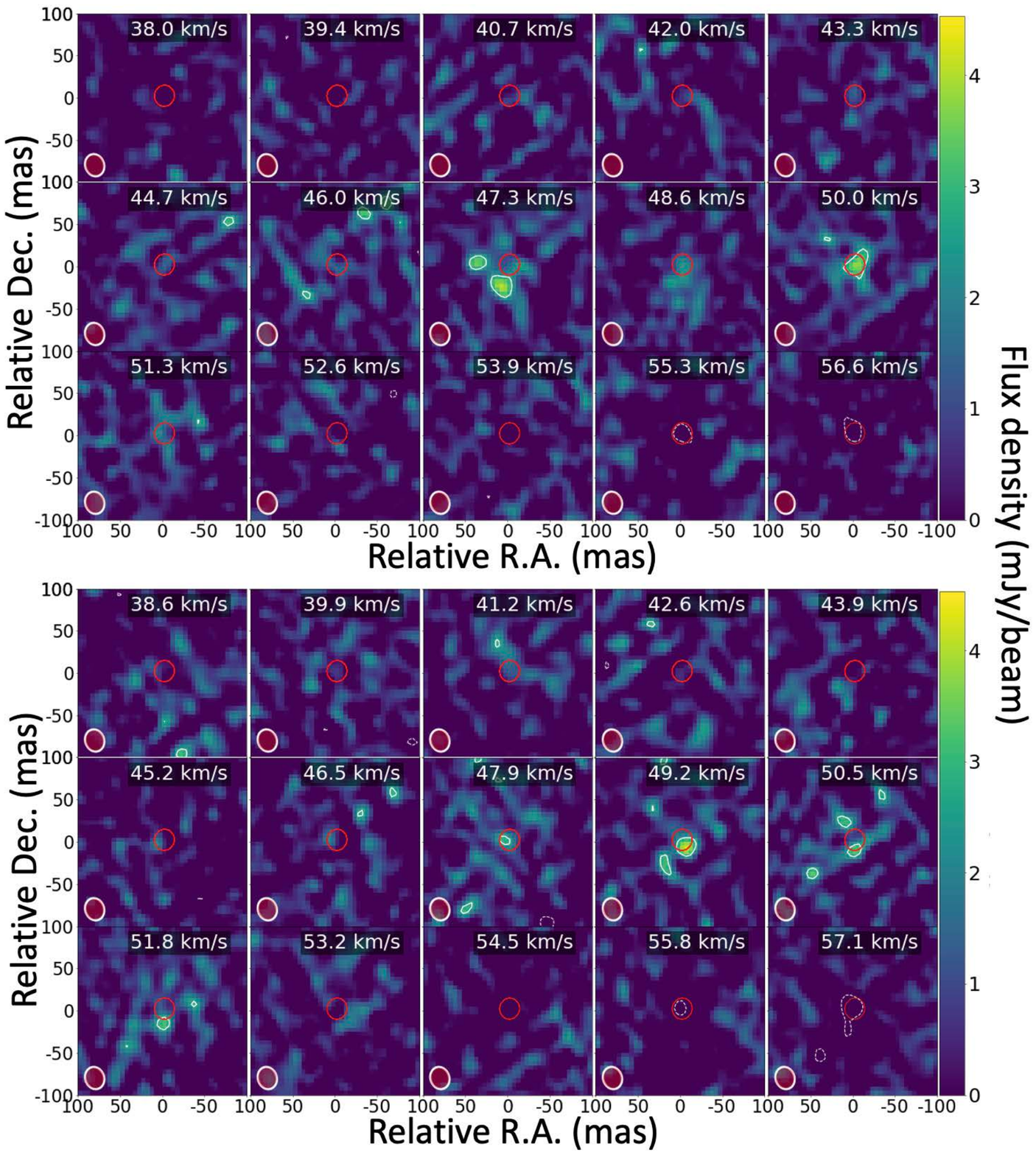}
 
\caption{High resolution channel map of $J$ = 27/2, $F'-F'' = 13-13$ and 14$-$14 transitions of OH in R Aql (upper and lower panels, respectively). Figure caption as in Fig. \ref {rhya_channmap_OH_29} except for the beam widths and the velocity range. 
The line peak flux density and r.m.s. noise levels are $\sim$5 and 1 mJy/beam, respectively. The line and continuum HPBWs are (27$\times$22)~ mas at PA~18$^{\circ}$ and (24$\times$22)~ mas at PA~$-$13$^{\circ}$, respectively.
}
\label{raql_channmap_OH_27}
\end{figure*}

%%%%%%

\begin{figure*}
\centering
\vspace{-1.5cm}
\includegraphics[width=16.3cm,angle=0]{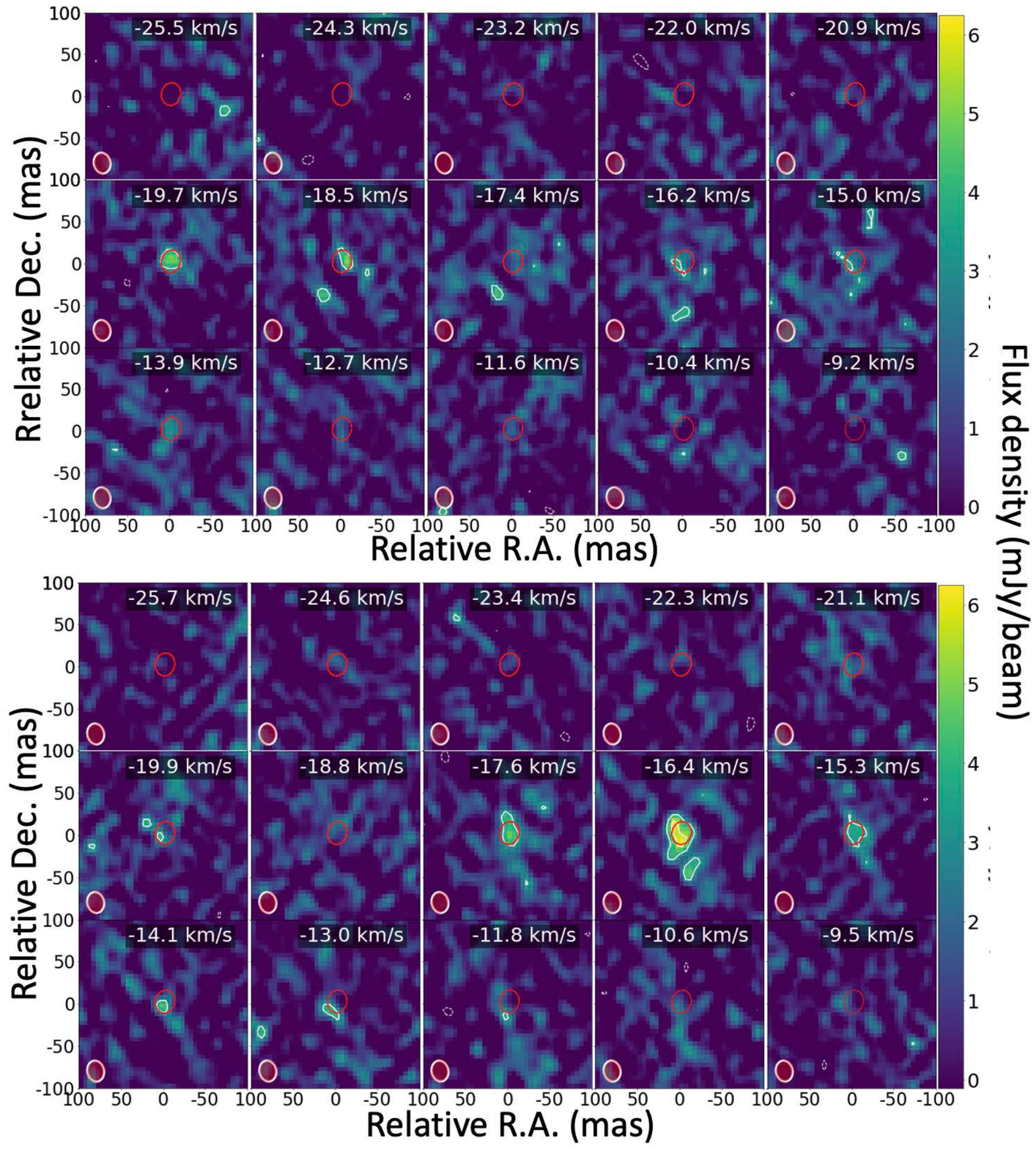}
 
\caption{High resolution channel map of $J$ = 29/2, $F'-F'' = 14-14$ and 15$-$15 transitions of OH in SPav (upper and lower panels, respectively). Fig. caption as in Fig. \ref {rhya_channmap_OH_29} except for the velocity range. 
The line peak flux density and r.m.s. noise level are $\sim$6 and 1~mJy/beam, respectively. 
The line and continuum HPBWs are (25$\times$20)~ mas at PA =~11$^{\circ}$ and (25$\times$20)~mas at PA~$-$13$^{\circ}$, respectively.
}
\label{spav_channmap_OH_29}
\end{figure*}

%%%%%%%
%%OH zeroth moment maps below

\begin{figure*}
\centering
\includegraphics[width=13.0cm,angle=270]{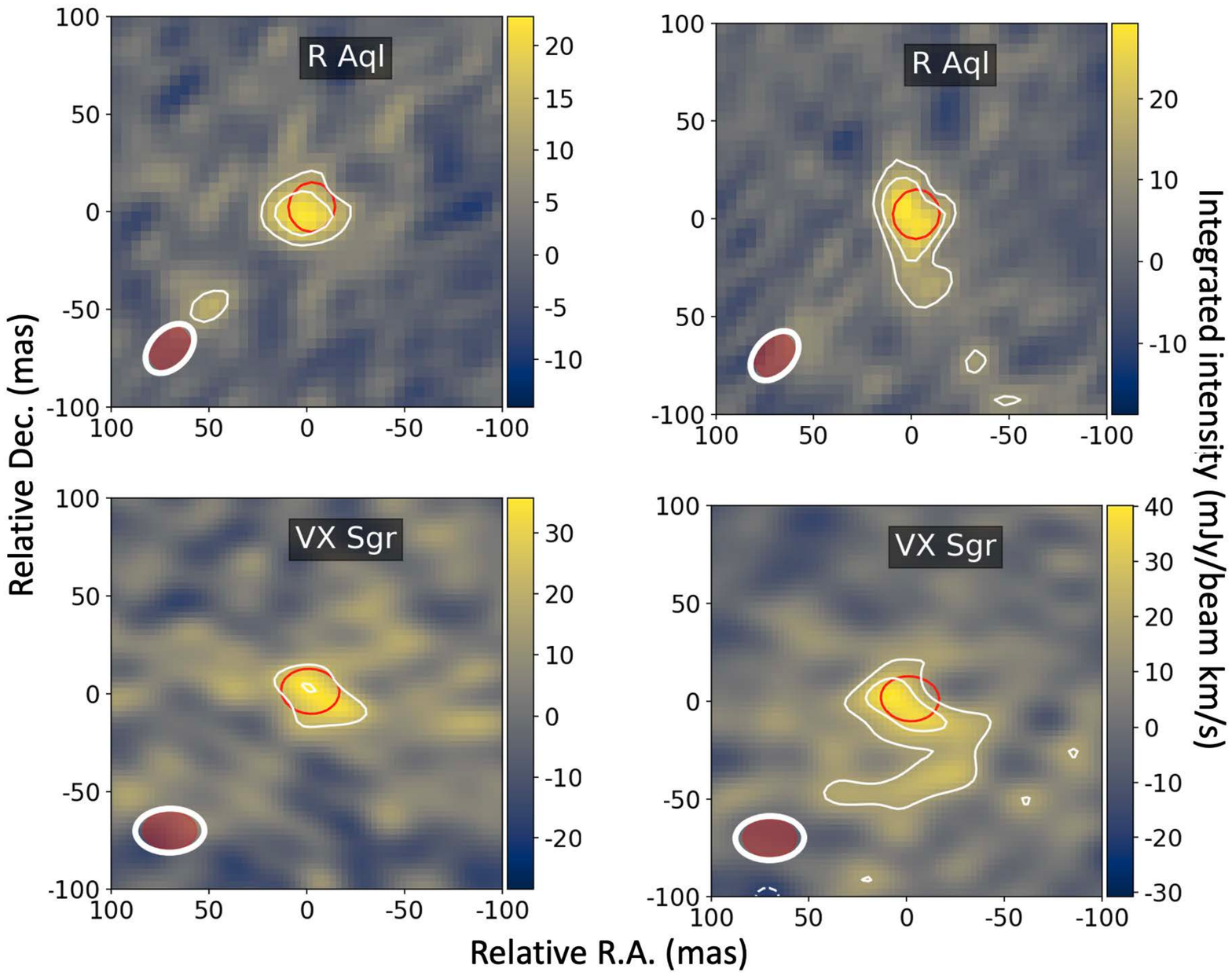}
 
\caption{Zeroth moment maps of OH emission in the $J=29/2, F'-F'' = 14-14$ (left panels) and $15-15$ (right panels) transitions in R~Aql and VX~Sgr as marked. Figure caption as in Fig. \ref{Fig_OH_emission_29_2} except for the velocity intervals in the $F'-F'' =14-14$ and 15$-$15 transitions which are: 46.8 to 53.8~km\,s$^{-1}$ and 46.4 to 53.4~km\,s$^{-1}$ in R~Aql; 3.0 to 15.8~km\,s$^{-1}$ and 6.1 to 17.8~km\,s$^{-1}$ in VX~Sgr. 
The line HPBWs are  (29$\times$20)~mas at PA~$-$43$^{\circ}$ and (35$\times$23)~mas at PA~$-$89$^{\circ}$ in R~Aql and VX~Sgr, respectively. The continumm HPBWs are (24$\times$22)~ mas at PA~$-$13$^{\circ}$ (R~Aql) and (28$\times$20)~mas at PA~89$^{\circ}$ (VX~Sgr). 
 }
\label{Fig_OH_emission_29_2_raql_vxsgr} 
\end{figure*}

%%%%%%%

\begin{figure*}
\centering
\includegraphics[width=15.3cm,angle=0]{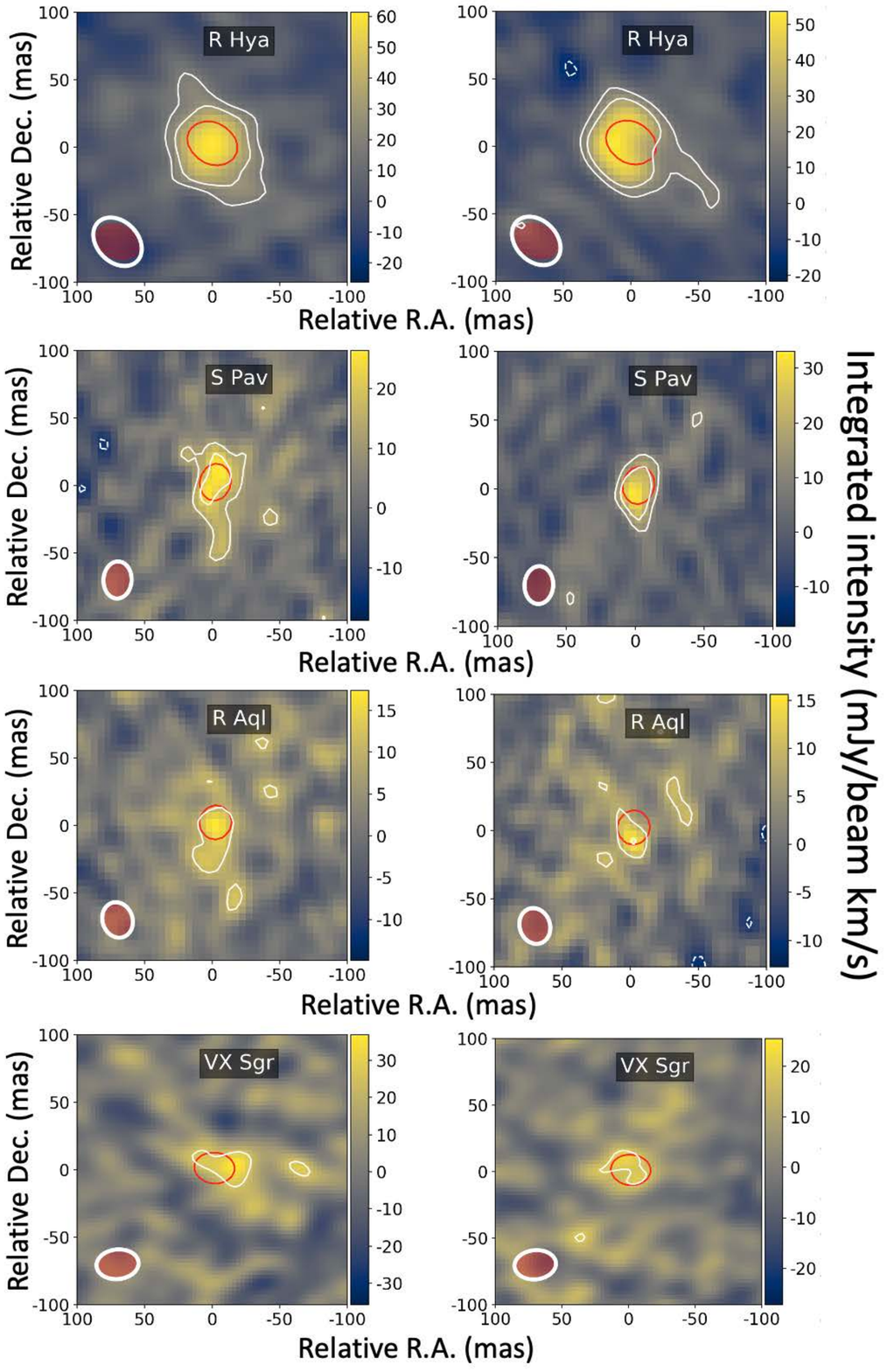}
 
\caption{Zeroth moment maps of OH emission in the $J=27/2, F'-F'' = 13-13$ (left panels) and $14-14$ (right panels) transitions in R~Hya, S~Pav, R~Aql and VX~Sgr as marked. Figure caption as in Fig. \ref{Fig_OH_emission_29_2} except for the velocity intervals. The velocity intervals in the $F'-F'' =13-13$ and 14$-$14 transitions are: $-$10.7 to $-$4.1~km\,s$^{-1}$ and $-11.5$ to $-2.2$ km\,s$^{-1}$ in R~Hya; $-21.0$ to $-10.4$~km\,s$^{-1}$ and $-21.8$ to $-12.5$ km\,s$^{-1}$ in S~Pav; 43.3 to 51.3~km\,s$^{-1}$ and 46.6 to 51.7~km\,s$^{-1}$ in R Aql; 2.9 to 21.5~km\,s$^{-1}$ and 7.5 to 16.8~km\,s$^{-1}$ in VX~Sgr. 
 The line HPBW is 
(41$\times$30)~ mas, (27$\times$21)~mas, (27$\times$22)~mas and (31$\times$22)~mas 
at PA~45$^{\circ}$, $-$3$^{\circ}$, 18$^{\circ}$ and $-$84$^{\circ}$ in R~Aql, S Pav, R Aql and VX~Sgr, respectively.
The continuum HPBWs are: (34$\times$25)~mas at PA~67$^{\circ}$ (R~Hya), (25$\times$20)~mas at PA~13$^{\circ}$ (S~Pav), 
(24$\times$22)~ mas at PA~$-$13$^{\circ}$ (R~Aql) and (28$\times$20)~mas at PA~89$^{\circ}$ (VX~Sgr).
}
\label{Fig_OH_emission_27_2} 
\end{figure*}

%%%%%%%%%%%%%
%%%%%%%%%%%%%

\section{Bands 6 and 7 OH lines toward omi~Cet (Mira)} 
\label{sec:append_mira_ohspec}

\begin{figure*}
\centering
\includegraphics[width=14.0cm,angle=0]{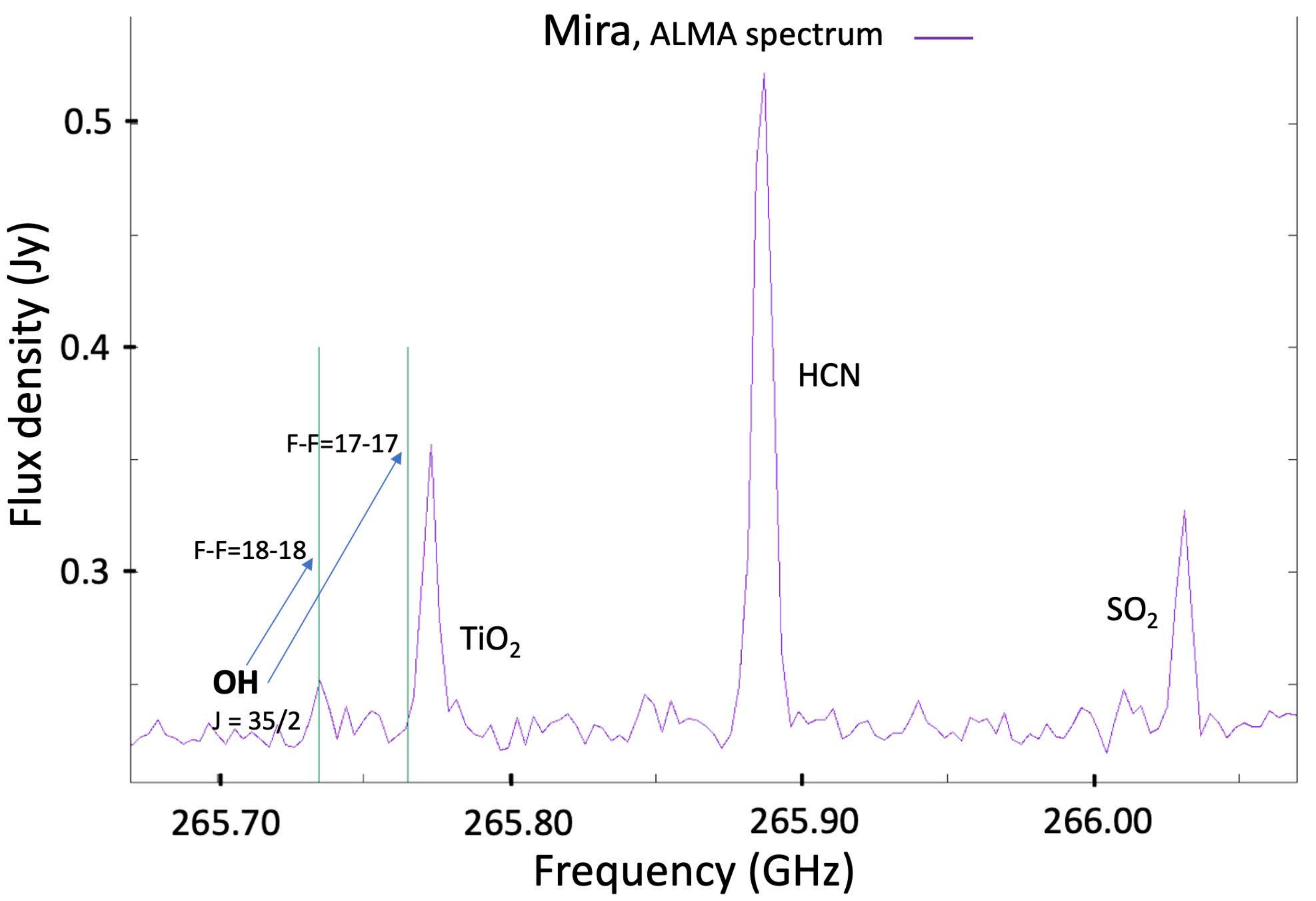}
 
\caption{OH spectrum and nearby transitions of TiO$_2$, HCN and SO$_2$ observed in the range 265.65 to 266.05 GHz toward omi~Cet (Mira). The frequencies of the two hyperfine transitions of OH in the $\varv=0$, $J=35/2$ rotational level are indicated with two light green vertical lines. The $F'-F'' = 17 - 17$ hyperfine transition of the $J=35/2$ level is  blended with the $24(2,22)-24(1,23)$ transition of TiO$_2$. This spectrum is extracted from a 0\farcs2$\times$0\farcs2 region containing all the OH emission. }
\label{OH_Mira_Band6}
\end{figure*}

%%%%
 \begin{figure*}
\centering
\includegraphics[width=14.0cm,angle=0]{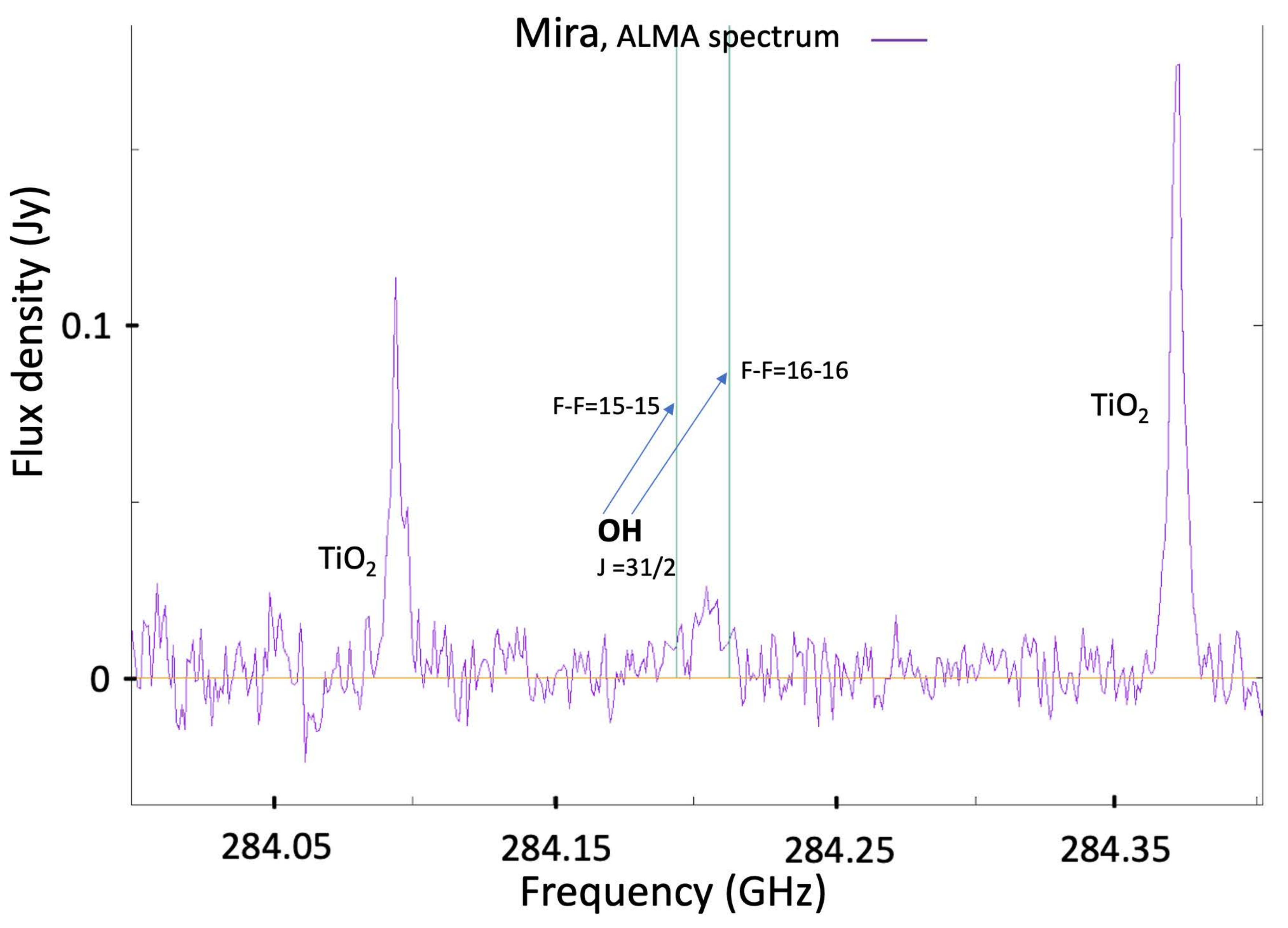}
 
\caption{Line spectrum in the vicinity of the $\varv=0$, $J=31/2$ rotational level of OH and of two nearby TiO$_2$ transitions observed in the 284.00 to 284.40 GHz range toward omi~Cet (Mira). The expected $F'-F'' = 15 - 15$ and $16-16$ hyperfine transitions of OH in the $J=31/2$ $\Lambda$-doublet are indicated with two light green vertical lines at the expected frequencies. The two TiO$_2$ lines correspond to the $27(7,21)-27(6,22)$  and $21(1,21)-26(0,20)$ transitions at 284.0943 and 284.3719 GHz, respectively. Spectrum extracted from a 0\farcs25$\times$0\farcs20  region containing all the OH emission. }
\label{OH_Mira_Band7}
\end{figure*}
%%%%%

Using the ALMA archive,  we report here on the detection of high-$J$ hyperfine transitions of OH in the atmosphere of omi~Cet (Mira). In the Band 6 data acquired with 20 mas resolution (project 2017.1.00393.S) we have identified a weak feature (S/N~$\sim$2.5$-$3)  near 265.735~GHz (Fig.~\ref{OH_Mira_Band6}) coinciding, within the frequency uncertainties, with  the $\varv=0$, $J=35/2$, $F'-F'' =18-18$ hyperfine transition of OH (see  Table~\ref{OHline_list}).   
We have not been able to identify any other molecular line carrier at this frequency. Unfortunately, the second hyperfine transition of the $J$ = 35/2 $\Lambda$-doublet (see second thin green vertical line in Fig.~\ref{OH_Mira_Band6}) is blended with the relatively strong TiO$_2$ $24(2,22)-24(1,23)$ line emission at 265.7705 GHz;  
 this suggests that the OH $J$~=~35/2, $F'-F'' = 17-17$ transition is shifted by more than 1~MHz with respect to the frequency given in Table~\ref{OHline_list}.

Using the Band 7 data (project 2018.1.00749.S), we have observed in Mira another weak signal, $\sim$20 mJy, nearly coinciding with the $\varv=0$, $J = 31/2$, $\Lambda$-doublet of OH at 284.2032~GHz (Fig.~\ref{OH_Mira_Band7}). However, the two hyperfine transitions of the $J=31/2$ level expected near 284.2032~GHz (see the two green vertical lines  in Fig.~\ref{OH_Mira_Band7}) are not separated here and we consider this identification as still uncertain although we have not been able to identify any other possible line carrier at this frequency. On the other hand, the two nearby TiO$_2$ transitions are clearly identified in Fig. ~\ref{OH_Mira_Band7} and well separated from the $\varv=0$, $J = 31/2$ transition. The same ALMA project also covers the $J=37/2, \Lambda$-doublet of OH at 295.99877 and 296.02859 GHz but, unfortunately, an SO$_2$ line coincides with this doublet. 

\end{appendix} 
\end{document}